\documentstyle[12pt,epsf]{report}
\newcommand{\nll}{\nonumber \\}

\newcommand{\bq}{\begin{equation}}
\newcommand{\eq}{\end{equation}}
\newcommand{\ba}{\begin{eqnarray}}
\newcommand{\ea}{\end{eqnarray}}

\newcommand{\nobody}{\rule{0ex}{1ex}}

\newcommand{\nobodyfrac}{\frac{\nobody}{\nobody}}
\newcommand{\req}[1]{(\ref{#1})}
\begin{document}
\newtheorem{fig}[figure]{Figure}
\newtheorem{tab}[table]{Table}
\voffset -2cm
\begin{flushleft}
LMU-03/98
\end{flushleft}
\thispagestyle{empty}
\nobody\vspace{2cm}
\begin{center}
\vspace{1.5cm}\hfill\\ 
{\Large\bf The Phenomenology of Extra Neutral Gauge Bosons}\vspace{1cm}\\
A. Leike\\
{\it Ludwigs--Maximilians-Universit\"at, Sektion Physik, Theresienstr. 37,\\
D-80333 M\"unchen, Germany}\\
E-mail: leike@graviton.hep.physik.uni-muenchen.de\\
\end{center}
\nobody\vspace{4cm}\\
\centerline{\bf Abstract}\vspace{0.5cm}

\noindent
The phenomenological constraints on extra neutral gauge bosons at
present and at future colliders are reviewed.
Special attention is paid to the influence of radiative corrections,
systematic errors, and kinematic cuts on the $Z'$ constraints.
Simple estimates of the $Z'$ constraints from different reactions are derived.
They make the physical origin of these constraints transparent.
The results existing in the literature are summarized and compared with
the estimates. 
The consequence of model assumptions on the $Z'$ constraints is
discussed.
The paper starts with an overview of $Z'$ parameters and the possible links
between them by model assumptions.
It continues with a discussion of $Z'$ limits and $Z'$ measurements in
different reactions at $e^+e^-$ and $\mu^+\mu^-$ colliders.
It follows an overview of the corresponding limits at proton colliders.
Possible $Z'$ constraints from other reactions as
$ep$ collisions, atomic parity violation, neutrino scattering and
cosmology are briefly mentioned.
%
%
\vfill\newpage
\tableofcontents
\vfill\newpage

\chapter{Introduction} \label{intro}
%
A neutral gauge boson is a spin--one particle without charge.
It transmits forces in gauge theories.
One well known neutral gauge boson is the photon.
In 1923, A. H. Compton found the direct experimental confirmation that
the photon is an elementary particle.
The photon is connected with the $U(1)_{em}$ gauge symmetry of
electrodynamics.
Noether's theorem states that this symmetry must correspond to a
conserved quantity, the electric charge.
The $U(1)_{em}$ gauge symmetry is exact.
Therefore, the mass of the photon is zero.
It can mediate interactions to infinite distances.

Matter interacts not only through electromagnetic forces.
In particular, weak decays as the $\beta$--decay violate
quantum numbers respected by electrodynamics. 
These processes indicate that there must be additional fundamental
interactions. 
In 1961, S.L. Glashow, S. Weinberg and A. Salam proposed the unified
description of electromagnetic and weak phenomena in a gauge theory
based on the gauge group $SU(2)_L\times U(1)_Y$.
At low energies, we can only observe the $U(1)_{em}$ symmetry.
Therefore, the $SU(2)_L\times U(1)_Y$ gauge symmetry must be broken at some
energy scale $E_{weak}$.
Processes with energies $E\ll E_{weak}$ feel only the $U(1)_{em}$
symmetry of electrodynamics.
At energies $E > E_{weak}$, the interaction has the full
$SU(2)_L\times U(1)_Y$ gauge symmetry.
In the language of particles, this means that the weak interaction
must be mediated by gauge bosons, which have masses of the order of
$E_{weak}$. 
We know experimentally that $E_{weak}=O(100)\,GeV$. 

The theory of weak interactions explains $\beta$ decay by the
virtual exchange of a heavy positively or negatively charged gauge
boson, the $W^\pm$.
These charged gauge bosons are associated to the $SU(2)_L$ gauge group.
The $SU(2)_L$ gauge symmetry has also one neutral (diagonal) generator.
The particle associated with this neutral generator must be a
neutral gauge boson with a mass of the order of $E_{weak}$.
Because the electromagnetic symmetry $U(1)_{em}$ is different from the
$U(1)_Y$ symmetry, the mass eigenstates $\gamma$ and $Z$ of the
neutral gauge bosons are a linear combination of the two neutral gauge bosons 
associated to the $SU(2)_L$ and $U(1)_Y$ gauge groups.
Their properties are predicted in the electroweak theory.
The predictions for the $Z$ boson are confirmed
by the experiments at LEP and SLC with an incredible precision.

In addition to electroweak interactions, there exist at
least two other fundamental interactions, the strong interaction and
gravity.

Many physicists believe that all fundamental interactions must have
one common root.
They don't like the complicated gauge group of the SM.
They suppose that strong and electroweak interactions can be described
by one simple gauge group $G$ at very high energies $E>E_{GUT}$.
Such theories are called grand unified theories (GUT's) \cite{langgut}.
For energies $E\ll E_{GUT}$ the gauge group $G$ must be broken
\cite{slansky} to retain the SM gauge symmetry $SU(3)_c\times
SU(2)_L\times U(1)_Y$.
One can imagine this symmetry breaking similar to the breaking of the
$SU(2)_L\times U(1)_Y$ symmetry to $U(1)_{em}$ in the SM.

As was shown by H. Georgi and S.L. Glashow in 1974, the smallest
simple gauge group $G$, which can contain the SM, is $G=SU(5)$.
The number $n$ of neutral gauge bosons of a GUT is given by
$n=rank[G]$.
We have $rank[SU(5)]=4$.
Therefore, there is no room for additional neutral gauge bosons in the
$SU(5)$ GUT.

GUT's make predictions which can be tested in experiments.
In particular, they predict that the proton must decay.
This decay is mediated by the exchange of gauge bosons with a mass
$O(E_{GUT})$. 
It is the analogue of the $\beta$ decay described in the electroweak
theory.
To be consistent with present experiments on proton decay, we get the
condition $E_{GUT}>10^{15}\,GeV$.
This energy is much larger than $E_{weak}$.
It is important that it is smaller than the Planck mass,
$M_P=\sqrt{\hbar c/G_N}\approx 1.2\cdot 10^{19}\,GeV$. 
At energies above the Planck mass, gravity is expected to become as
strong as the other interactions.
At energies well below $M_P$, as it happens in GUT's, the effects of
gravity can be neglected. 
$E_{GUT}$ is also predicted as the energy where the three running
gauge coupling constants of the SM gauge group become equal.
The value of $E_{GUT}$ obtained experimentally by this matching
condition predicts a proton lifetime, which contradicted the
measurement already several years ago.
Only the recent precision measurements at LEP and SLC 
could prove that the three running gauge couplings do not meet in one
point if they run as predicted in the SU(5) GUT.
Therefore, one must add something else if one wants to describe
all SM interactions by one simple gauge group.
One popular direction of research is supersymmetry.

We are interested here in another solution of the problem, the
consideration of larger unification groups.
All GUT's with gauge groups larger than $SU(5)$ predict at least one
extra neutral gauge boson ($Z'$).
It was shown by H. Fritzsch and P. Minkowski in 1975 that the next
interesting gauge group larger then $SU(5)$ is $SO(10)$.
The $SO(10)$ theory predicts one extra neutral gauge boson because
$rank[SO(10)]=5$. 
It is a non--trivial fact that all SM fermions of one generation fit
in only one multiplet of $SO(10)$.
To complete the multiplet, one new fermion with the quantum numbers of
the right--handed neutrino must be added.
The $SO(10)$ GUT is not in contradiction with present experiments.
GUT's with gauge groups larger than $SO(10)$ predict more than
one extra neutral gauge bosons and many new fermions.
These new (exotic) fermions must be heavy to make the theories
consistent with present experiments.

The mass of the $Z'$ is not constrained by theory.
A priori, it can be anywhere between $E_{weak}$ and $E_{GUT}$.
As was shown in references \cite{9707451,9707451add}, it has naturally
a mass of about one TeV in some supersymmetric GUT's.
Then, a $Z'$ can be observed at the next generation of colliders.
An observation of a $Z'$ would provide information on the GUT group and
on its symmetry breaking.
It is of special interest for experimental physicists because a $Z'$
would serve as a calibration point for future detectors.
The study of the $Z'$ phenomenology is therefore an important part of the
scientific program of every present and future collider.

As the SM $Z$ boson, the $Z'$ is expected to be a very short--lived particle.
It can only be observed through its decay products or through indirect
interference effects.
It can be detected either in very high energy processes or in
high precision experiments at lower energies.
In the first kind of processes, the energy of the colliding particles
must be high enough to produce a $Z'$. 
The decay products of the $Z'$ must be then detected above the SM
background.
Such a background is always present because the SM $Z$ boson or the
photon are produced by the same processes, which create a $Z'$.
In precision experiments, the experimental errors and
the errors due to the theoretical predictions of the observables must be
smaller than the expected deviations due to a $Z'$. 

There are several previous reviews on $Z'$ physics.
References \cite{e6first,e6} give an overview of $Z'$ physics in
the $E_6$ GUT. 
In reference \cite{e6add} the search for a $Z'$ in high precision
experiments is reviewed.
A recent short review is given in \cite{cvetrev}.

In the recent years, extensive studies on the sensitivity of future
$e^+e^-$ \cite{ee}, $pp(p\bar p)$ \cite{pp} and $e^\pm p$ \cite{ep}
colliders to a $Z'$ have been completed.  
Among the new developments are
\begin{description}
\item[$\bullet$] the inclusion of radiative corrections, realistic
experimental cuts and systematic errors in $Z'$ studies and 
the development of codes allowing direct $Z'$ fits to experimental data;
\item[$\bullet$] the discussion of a $Z'$ not only in the context
of a model but also without model assumptions;
\item[$\bullet$] the experimental successes during the last years of
precision measurements, heavy flavour tagging and highly polarized beams
allow the investigation of new observables.
\end{description}

In this paper, we review the main results of these new developments.
Special attention is paid to the {\it mechanisms} leading to $Z'$ limits in
the different reactions. 
The resulting approximate formulae make the dependence of $Z'$
constraints and $Z'$ measurements on experimental conditions and model
parameters transparent. 
The approximate formulae are compared with the existing results in the
literature.\hfill\vspace{1ex}\\ 

We divide the paper into five chapters and several appendices.
In the first chapter, we introduce the parameters describing
extra neutral gauge bosons.
The $E_6$ GUT is considered as an example of a theory containing extra
neutral gauge bosons.
Far below the $Z'$ resonance, the $Z'$ can be described as a special
case of four fermion contact interactions.
A general formalism of the inclusion of $Z'$ effects in Standard Model cross
sections are form factors.
We emphasize that it is necessary to distinguish between model
dependent and model independent $Z'$ analyses.
We conclude the first chapter with some general remarks on the data
analysis.

In the second chapter, we list $Z'$ constraints at
$e^+e^-,\ e^-e^-$ and $\mu^+\mu^-$ colliders.
The constraints from various reactions are considered in individual sections.  
Every section is organized in the same pattern.
First, the relevant observables are discussed in the Born
approximation. 
It is followed by a discussion of radiative corrections.
Finally, different $Z'$ constraints are considered.
If necessary, different cases of the center--of--mass energy are distinguished.

In the third chapter, the $Z'$ constraints obtained at hadron colliders are
considered.

$Z'$ constraints from other experiments; electron--proton collisions,
atomic parity violation, neutrino scattering and cosmology are mentioned
in chapter four. 

Chapter five contains our summary and conclusions.

Two appendices complete this paper.
The main notation is collected in appendix A.
Appendix B contains an overview of the available {\tt FORTRAN} codes
suitable for $Z'$ fits in different experiments.

\chapter{Preliminary considerations} 
%
\section{Parameters describing extra neutral gauge bosons}\label{guts1}  
Information on extra neutral gauge bosons can be
obtained through experimental constraints {\it on} or measurements
{\it of} its parameters.
These $Z'$ parameters are introduced in this section.

After fixing our assumptions and notation, we consider effects of
gauge boson mixing.
We then deal with the $Z'$ couplings to SM fermions and with $Z'$
decays to SM particles.
\subsection{Assumptions and limitations}\label{guts11}
Suppose that a $Z'$ exists in nature.
Then, its parameters depend on many unknown details of the theory.

We assume that at most one extra neutral gauge boson is light enough to
give the first signal in future experiments and that no other signals
from a GUT are found at that time.
In the case where there are additional signals of new physics, they would
give interesting extra information on the parameters of the new theory.

We assume that the underlying effective gauge group at low energies is
\bq
\label{gutgauge}
SU(3)_c\times SU(2)_L\times U(1)_Y\times U'(1).
\eq

We further assume that the couplings of the $Z'$ to fermions are
universal for all generations.  
In models where this is not the case, one has to be careful about the
suppression of flavour changing neutral currents.
See reference \cite{family3} for recent constraints on such models and
for further references.

A GUT containing a $Z'$ predicts many new particles as, for
example, additional (exotic) fermions, additional charged gauge bosons
or additional Higgs bosons. 
In the simplest case of a $SO(10)$ GUT, the only additional fermion is
one right--handed neutrino. 
The number of exotic fermions rises drastically for larger gauge 
groups.
The couplings of these fermions to gauge bosons are fixed in the GUT
but their masses are not constrained by the theory.
Under our assumption that the $Z'$ gives the only signal of the GUT,
the decays of the $Z'$ to exotic fermions and Higgs bosons are
kinematically suppressed \cite{plb387}. 

We neglect the mixing of SM fermions with exotic fermions
\cite{prd3848} and possible simultaneous mixing of gauge bosons
and fermions \cite{prd55}.

In a SUSY GUT, many interactions involving supersymmetric particles
are predicted.  
We do not consider these effects \cite{9701343}.

We do not discuss either the special case of leptophobic $Z'$ models
\cite{leptophobic,plb381}, which where constructed to explain the discrepancy
of $R_b$ at LEP with the SM expectations.

BESS models \cite{bess} also predict extra neutral gauge bosons.
The corresponding $Z'$ limits derived for hadron
\cite{bessold,besshad} and lepton \cite{bessold,besslep,lepzpold}
colliders will not be discussed here. 
\subsection{Mixing}
There are no quantum numbers which forbid a mixing of neutral gauge bosons.
However, the $ZZ'$ mixing arises naturally in many models.
\subsubsection{Kinetic mixing}
We first consider the case where all neutral gauge bosons are {\it massless}.
Assume diagonal kinetic terms of the gauge fields.
The general tree level parametrization of the neutral current
Lagrangian for the gauge group \req{gutgauge} is then \cite{holdom,pacomix}
\bq
\label{curmix}
-{\cal L}_{NC} = \bar f\gamma^\beta \left\{
gT_3^fW_{3\beta} + g'_{11}Y^fB'_\beta + g'_{12}Y^fZ'_{2\beta} 
+ g'_{21}Q'^fB'_\beta + g'_{22}Q'^fZ'_{2\beta}  \right\}f,
\eq
where the summation over the fermions $f$ is understood,
$T_3^f$ is the third component of the SM isospin, $Y^f$ is 
the hypercharge and $Q'^f$ is the charge due to the new $U'(1)$.
The particles associated with the $U(1)_Y$ and $U'(1)$ gauge groups
are denoted as $B'$ and $Z'_2$, correspondingly.
$W_{3\beta}$ is the same field as defined in the SM.

The non--zero contributions proportional to $g'_{12}$ and $g'_{21}$ can
arise during the evolution from GUT energies to the weak scale.
The two by two matrix $g'_{ij}$ can be made triangular by a rotation of 
the two abelian gauge bosons around the angle 
$\theta_K,\ c_K=\cos\theta_K,\ s_K=\sin\theta_K$,
\bq
\label{currentmix}
\left( \begin{array}{c} B'_\beta \\ Z'_{2\beta} \end{array} \right)
=
\left( \begin{array}{rl}  c_K & -s_K \\
                          s_K &\ c_K \end{array} \right)
\left( \begin{array}{c} B_\beta \\ Z'_\beta \end{array} \right).
\eq
The angle $\theta_K$ describes the kinetic mixing.
After the mixing \req{currentmix}, the Lagrangian \req{curmix} transforms to 
\bq
\label{curmix2}
-{\cal L}_{NC} = \bar f\gamma^\beta \left\{
gT_3^fW_{3\beta} + g_{11}Y^fB_\beta + g_{12}Y^fZ'_\beta 
+ g_{22}Q'^fZ'_\beta\right\}f
\eq
with $g_{11}=g'_{11}c_K+g'_{12}s_K,\ g_{12}=-g'_{11}s_K+g'_{12}c_K,
\ g_{21}=g'_{21}c_K+g'_{22}s_K\equiv 0$ and 
$g_{22}=-g'_{21}s_K+g'_{22}c_K$.

If there is no additional symmetry requiring $g_{12}=0$, the term
proportional to $g_{12}$ is needed to ensure the renormalizability of
the theory \cite{pacomix}. 

Kinetic mixing plays an important role in some leptophobic models 
\cite{prd54} and can communicate SUSY-breaking to the visible sector
\cite{npb492}.

Kinetic mixing between the gauge bosons $W_{3\beta}$ and $B_\beta$ is
forbidden in the SM by $SU(2)_L$ gauge invariance.
\subsubsection{Mass mixing}
%
Gauge symmetry must be broken at low energies to describe massive gauge bosons.

In the SM, the entries of the mass matrix 
\bq
\label{smmix}
{\cal L}_M = \frac{1}{2}(B,\ W_3) M_{SM}^2
\left( \begin{array}{c} B\\ W_3\end{array} \right),\ \ \ 
M_{SM}^2 =
\left( \begin{array}{cc}  M_B^2   & -M_WM_B \\
                          -M_WM_B  & M_W^2 \end{array} \right)
\eq
are related to the vacuum expectation value $v$ of the Higgs field, 
\bq
\label{smhiggs}
\Phi = \left( \begin{array}{c} \phi^+\\ \phi^0\end{array} \right),\ \ \ 
\langle \phi^0\rangle = \frac{gv}{\sqrt{2}},\ \ \ 
M_W=\frac{gv}{2},\ \ \ M_B=\frac{g_{11}v}{2}.
\eq

$M_{SM}^2$ is diagonalized by a rotation of the symmetry eigenstates
around the Weinberg angle $\theta_W,\ 
c_W=\cos\theta_W,\ s_W=\sin\theta_W$ leading to the well-known
mass eigenstates of the  photon and the $Z$ boson, 
\bq
\label{bwmix}
\left( \begin{array}{c} \gamma \\ Z \end{array} \right)
=
\left( \begin{array}{rl}  c_W & s_W \\
                          -s_W &\ c_W \end{array} \right)
\left( \begin{array}{c} B \\ W_3 \end{array} \right).
\eq
The masses of the mass eigenstates are 
\bq
\label{mz2def}
M_Z^2=M_W^2+M_B^2,\ \ \ M_\gamma^2=0.
\eq
The mass of the photon is zero due to the exact $U(1)_{em}$ gauge
symmetry at low energies reflecting charge conservation.
This symmetry protects the photon from further mixing.

The Weinberg angle is related to the entries of the mass matrix \req{smmix},
\bq
\tan 2\theta_W=\frac{2M_WM_B}{M_W^2-M_B^2}.
\eq
We get the following relations between the Weinberg angle and the mass values,
\bq
\label{smmixconstr}
 t_W^2=\frac{s^2_W}{c^2_W} = \frac{M^2_B-M_\gamma^2}{M_Z^2-M_B^2}
=\frac{M_B^2}{M_W^2},
\ \ \ \mbox{and}\ \ \ 
M_BM_W=s_Wc_W (M_Z^2-M_\gamma^2) =s_Wc_W M_Z^2.
\eq

In a theory with the gauge group \req{gutgauge}, the mass matrix of
the $Z$ and $Z'$ receives non-diagonal entries $\delta M^2$, which are again
related to the vacuum expectation values of the Higgs fields,
\begin{equation}
\label{zznomix}
{\cal L}_M = \frac{1}{2}(Z,\ Z') M_{ZZ'}^2
\left( \begin{array}{c} Z\\ Z'\end{array} \right),\ \ \ 
M_{ZZ'}^2 =
\left( \begin{array}{rl}  M_Z^2       & \delta M^2 \\
                          \delta M^2  & M_{Z'}^2 \end{array} \right).
\end{equation}
We assume that the vacuum expectation values of the Higgs fields are real.
The mass matrix \req{zznomix} is diagonalized by a rotation of the
fields $Z$ and $Z'$ around the mixing angle $\theta_M$,  
$c_M=\cos\theta_M,\ s_M=\sin\theta_M$ leading to 
the mass eigenstates $Z_1$ and $Z_2$,
\bq
\label{zzmix}
\left( \begin{array}{c} Z_1 \\ Z_2 \end{array} \right)
=
\left( \begin{array}{rl}  c_M & s_M \\
                        - s_M & c_M \end{array} \right)
\left( \begin{array}{c} Z \\ Z' \end{array} \right).
\eq
The masses $M_1$ and $M_2$ of the mass eigenstates $Z_1$ and $Z_2$ are
\bq
M_{1,2}^2 =\frac{1}{2}\left[M_Z^2+M_{Z'}^2\pm
\sqrt{(M_Z^2-M_{Z'}^2)^2+4(\delta M^2)^2}\right].
\eq
It follows that
\bq
\label{massrel}
M_1<M_Z<M_2, \mbox{\ \ and hence\ \ } 
\rho_{mix} =\frac{M_W^2}{M_1^2c_W^2}>\frac{M_W^2}{M_Z^2c_W^2}=\rho_0=1.
\eq
We have $M_Z=M_1$ and $M_{Z'}=M_2$ for $\theta_M=0$.
LEP~1 and SLC performed precision measurements of the mass eigenstate $Z_1$.

Similar to the SM, the mixing angle $\theta_M$ is related to the
entries of the mass matrix \req{zznomix}, 
\bq
\tan 2\theta_M=\frac{2\delta M^2}{M_Z^2-M_{Z'}^2}.
\eq
We get the following relations between $\theta_M$ and the
mass values, 
\bq
\label{massconstr}
t^2_M = \frac{s^2_M}{c^2_M}=\frac{M_Z^2-M_1^2}{M_2^2-M_Z^2}
\eq
and
\bq
\label{higgsconstr}
\delta M^2=s_Mc_M (M_1^2-M_2^2).
\eq
For fixed $M_1$ and $M_Z$, equation \req{massconstr}
relates $\theta_M$ and $M_2$ independently of the Higgs sector. 
This constraint on $\theta_M$ is called the mass constraint \cite{e6first,e6}. 
It predicts $\theta_M\sim 1/M_2$ for large $M_2$.
For a fixed Higgs sector, $\delta M^2$ is also fixed leading to the
constraint on $\theta_M$ given in equation \req{higgsconstr}.
It is called the Higgs constraint \cite{e6first,e6}.
For large $M_2$, it is stronger than the mass constraint
predicting the asymptotic behaviour $\theta_M\sim 1/M_2^2$ \cite{e6first,e6}.
\subsubsection{Alternative parametrization of the mixing}
%
In left-right models \cite{lrmod} with the gauge group
\bq
\label{lrgauge}
SU(2)_L\times SU(2)_R\times U(1)_{B-L},
\eq
the mass eigenstates $\gamma, Z_1$ and $Z_2$ are obtained by a
one--step mixing through an orthogonal 3 by 3 matrix \cite{mixlr},
\bq
\label{zzmixone}
\left( \begin{array}{c} Z_1 \\ Z_2 \\ \gamma\end{array} \right)
=
\left( \begin{array}{ccc}
c_Wc_M & s_Wc_Mc_{LR}+s_Ms_{LR} & s_Wc_Ms_{LR}-s_Mc_{LR} \\
-c_Ws_M & -s_Ws_Mc_{LR}+c_Ms_{LR} & -s_Ws_Ms_{LR}-c_Mc_{LR} \\
-s_W    & c_Wc_{LR}               & c_Ws_{LR} 
 \end{array} \right)
\left( \begin{array}{c} W_{3L} \\ W_{3R}  \\ Y_{B-L}\end{array} \right),
\eq
depending on the three mixing angles $\theta_W, \theta_M$ and
$\theta_{LR}$ with $s_{LR}=\sin\theta_{LR},c_{LR}=\cos\theta_{LR}$.
In equation \req{zzmixone}, we adapted the notation to our
definitions.
Note that the fields $W_{3L},W_{3R},Y_{B-L}$ commute in the gauge group
\req{lrgauge}. 

In previous sections, we described the mixing of the neutral gauge
bosons associated with the gauge group \req{gutgauge} by two mixing
angles $\theta_W$ and $\theta_M$.
To apply this procedure to left-right models, one has to first break
the gauge group \req{lrgauge} to $SU(2)_L\times U(1)_Y\times U'(1)$.
This breaking will generate non-diagonal terms in the associated gauge
fields through fermion loops \cite{prd54} if
$\sum_{\scriptsize\mbox{chiral\ fermions\ }f}\ (Q^fQ'^f)\neq 0$. 
These non-diagonal terms can be described by a non-zero $g_{12}$ in
equation \req{curmix2}.
In the case $g_R=g_L$, no non-diagonal terms exist ($g_{12}=0$) and the
mass matrix can be diagonalized by the two angles $\theta_W$ and
$\theta_M$ \cite{lrmod} in both mixing procedures.

In this paper, we prefer the two-step mixing procedure.
It has advantages in the model independent approach where one starts
with the gauge group \req{gutgauge} and the corresponding symmetry
eigenstates $B$ and $Z'$. 
\subsection{Couplings to SM fermions}\label{veccoup}
The couplings of the symmetry eigenstate $Z'$ to fermions are fixed in a GUT.
Experiments are sensitive to the couplings of the mass eigenstates
$Z_1$ and $Z_2$.
\subsubsection{Couplings of the Symmetry Eigenstates}
The Lagrangian \req{curmix2} can be written in the form 
\bq
\label{eq31}
-{\cal L}_{NC} = eA_\beta J_\gamma^\beta +
g_1 Z_\beta J_Z^\beta + g_2 Z'_\beta J_{Z'}^\beta.
\eq
It contains the currents
\ba
\label{eq32}
J_\gamma^\beta=\sum_{f}\;\bar{f} \gamma^\beta\;v_f(0)\; f,\ \ \ 
J_Z^\beta = \sum_{f}\;\bar{f} \gamma^\beta\;[v_f - \gamma_5 a_f]\; f,\ \ \ 
J_{Z'}^\beta = \sum_{f}\;\bar{f} \gamma^\beta\;[v'_f - \gamma_5 a'_f]\; f.
\ea

To fix the notation, we give here the coupling constants of the photon and
the SM $Z$ boson to the SM fermion $f$,
\bq
\label{coupeq}
\begin{array}{rclrclrcl}
g_0 &\equiv& e=\sqrt{4\pi\alpha}=gs_W,& v_f(0) &=& Q^f,& a_f(0) &=& 0,\nll
g_1 &=&\frac{e}{2s_Wc_W}=(\sqrt{2} G_\mu M_Z^2)^{1/2},
& v_f &=& T_3^f-2Q^fs^2_W,& a_f &=& T_3^f,
\end{array}
\eq
with $T_3^e = -\frac{1}{2}$, $Q^e=-1$.
Differently from the SM, it is useful to define the electric charge as
$Q^f\equiv T_3^f + \sqrt{\frac{5}{3}}Y^f$ in GUT's.
The comparison of the two Lagrangians \req{curmix2} and \req{eq31}
then gives $g_{11}=\sqrt{\frac{3}{5}}e/c_W$.

The couplings $g_2,\ v'_f$ and $a'_f$ depend on the particular $Z'$ model.  
In the Sequential Standard Model (SSM), all couplings of the $Z'$ to
SM fermions are equal to those of the SM $Z$ boson, 
$g_2=g_1,\ \ v'_f=v_f,\ \ a'_f=a_f$. 
Although it is hard to obtain the SSM in GUT's, it is a popular
benchmark model, which is limited in different experiments.

Starting from the Lagrangian \req{curmix2} and identifying $g_2=g_{22}$, 
one obtains $v'_f$ and $a'_f$ as a function of the charge $Q'^f$ and 
the hypercharge $Y^f$ \cite{pacomix},
\ba
\label{gencoup}
v'_f &=& Q'^{fL}+Q'^{fR}+\frac{g_{12}}{g_{22}}(Y^{f_L}+Y^{f_R})
      =Q'^{fL}+Q'^{fR}+\frac{g_{12}}{g_{22}}\sqrt{\frac{3}{5}}(-T_3^f+2Q^f),
\nll
a'_f &=& Q'^{fL}-Q'^{fR} +\frac{g_{12}}{g_{22}}(Y^{f_L}-Y^{f_R}) 
      =Q'^{fL}-Q'^{fR} +\frac{g_{12}}{g_{22}}\sqrt{\frac{3}{5}}(-T_3^f).
\ea
We define the couplings to left- and right-handed fermions as
\bq
\label{lrdef}
L'_f=\frac{1}{2}(v'_f+a'_f),\ \ \ R'_f=\frac{1}{2}(v'_f-a'_f).
\eq

A possible charge $Q'^{\nu_L^c}=-Q'^{\nu_R}$ is not considered here
because we count the right handed neutrino as an exotic fermion, which
is assumed to be heavy.
Therefore, $v_\nu-a_\nu$ does not enter our analysis.
Hence, only {\it seven} of the eight couplings $v'_f$ and $a'_f$ are
independent.  

For extra neutral gauge bosons arising in the gauge group \req{gutgauge},
we have 
\bq
\label{qpt3}
[Q'^f,T_3^f]=0.
\eq
Then, the two relations 
\bq
\label{chargedef}
Q'^{u_L}=Q'^{d_L}\equiv Q'^{q_L} \mbox{\ \ and\ \ }
Q'^{e_L}=Q'^{\nu_L}\equiv Q'^{l_L}
\eq
must be fulfilled to preserve $SU(2)_L$ gauge invariance.
Relations similar to \req{chargedef} are fulfilled for the
hypercharges $Y^f$ in the SM.
Therefore, the relations \req{chargedef} remain valid if the charges
$Q'^f$ are replaced by the corresponding $Z'$ couplings to fermions.
It follows that  $v'_f$ and $a'_f$ can be parametrized by the
following {\it five} independent couplings,
\bq
\label{coup5}
L'_u=L'_d\equiv L'_q,\ L'_e=L'_\nu\equiv L'_l,\ R'_u,\ R'_d,\ R'_e.
\eq
\subsubsection{Couplings of the Mass eigenstates}
From equations \req{gencoup} and \req{zzmix}, we deduce the
couplings of the mass eigenstates $Z_1$ and $Z_2$ to fermions,
\ba
\label{vf1}
a_f(1) &=&  c_M a_f + \frac{g_2}{g_1} s_M a'_f
\equiv a_f(1-y_f),\nll
v_f(1) &=&  c_M v_f + \frac{g_2}{g_1} s_M v'_f
\equiv a_f(1) \left[1-4|Q^f|s^2_W(1-x_f)\right],\nll
a_f(2) &=&  c_M a'_f + \frac{g_1}{g_2} s_M a_f,\nll
v_f(2) &=&  c_M v'_f + \frac{g_1}{g_2} s_M v_f.
\ea
The functions $x_f$ and $y_f$ give the $ZZ'$ mixing in terms
of form factors, 
\ba
\label{formix}
x_f &=& (1-v_f/a_f)^{-1}
\left(\frac{v_f+t_Mv'_fg_2/g_1}{a_f+t_Ma'_fg_2/g_1} 
-\frac{v_f}{a_f}\right)\approx
\theta_M\frac{g_2}{g_1}\frac{a'_f}{a_f}\frac{v'_f/a'_f-v_f/a_f}{1-v_f/a_f},\nll
y_f &=& -s_M \frac{g_2a'_f}{g_1 a_f}  + (1 - c_M) 
\approx -\theta_M\frac{g_2a'_f}{g_1 a_f}.
\ea
They are convenient for a simultaneous
description of $ZZ'$ mixing and electroweak corrections at the $Z_1$
peak \cite{zefit}.
We neglect terms of higher order in the mixing angle in the last
approximation in equation \req{formix}. 
For small mixings, $x_f$ and $y_f$ are small being
proportional to $\theta_M$. 
\subsection{Decay width}\label{zndecaywidth}
We consider here only $Z'$ decays to SM particles.
We refer, for example, to reference \cite{e6} for the decay widths to
other particles present in a GUT and to \cite{prd36} for $Z'$ decays to bosons.

\subsubsection{Born approximation}
The partial decay width of the $Z_n,\ n=1,2$ to a fermion pair $f\bar f$ is
\cite{roroso10} 
\bq
\label{zpgamma}
\Gamma(Z_n\rightarrow f\bar f)\equiv \Gamma_n^{f0}=
N_f\mu M_n\frac{g_n^2}{12\pi}
\left\{\left[ v_f(n)^2+a_f(n)^2\right]
\left( 1+2\frac{m_f^2}{M_n^2}\right) - 6a_f(n)^2 \frac{m_f^2}{M_n^2}\right\},
\eq
$N_f$ is the color coefficient, i.e. $N_f=1(3)$ for $f=l(q)$.
$\mu$ is the phase space factor due to the massive final fermions, 
$\mu=\sqrt{1-4m_f^2/M_n^2}$.

A $Z'_2$ originating from the gauge group $U'(1)$ given in
\req{gutgauge} or the $Z'$ arising after kinetic mixing
\req{currentmix} have no couplings to $W$ bosons.   
However, in the case of a non-zero mass mixing, the $Z$
component contained in the mass eigenstates $Z_1$ and $Z_2$ interacts
with $W$ bosons, $g_{WWZ_1}=g_{WWZ}c_M,\ g_{WWZ_2}=g_{WWZ}s_M$,
where $g_{WWZ}=ec_W/s_W$ is the SM coupling between $W$'s and the
SM $Z$ boson.
The structure of the interaction is essentially the same as in the SM. 
The partial decay width of the $Z_2$ to a $W^+W^-$ pair is then \cite{prd36}
\ba
\label{zpgammaww}
\Gamma(Z_2\rightarrow W^+W^-)&\equiv& \Gamma_2^{W0}\\
&=&\frac{g^2_{WWZ}M_2}{192\pi}s^2_M
\left(\frac{M_2}{M_Z}\right)^4\left(1-4\frac{M_W^2}{M_2^2}\right)^{3/2}
\left(1+20\frac{M_W^2}{M_2^2}+12\frac{M_W^4}{M_2^4}\right).\nonumber
\ea
In GUT's, 
$\Gamma(Z_2\rightarrow W^+W^-)$ is kept in a reasonable range
for $M_2\rightarrow \infty$ because the potentially dangerous factor 
$M_2^4$ is compensated by small mixing angles
$s^2_M\approx 1/M_2^4$, due to the Higgs constraint \req{higgsconstr}.

The decays $Z_2\rightarrow f\bar fV,\ V=Z,W$ are of higher order.
However, they are logarithmically enhanced by soft and collinear radiation.
In the limit $M_Z^2/M_2^2\ll 1$, the decay width becomes \cite{raredecg}
\ba
\label{gamrare}
\Gamma(Z_2\rightarrow f\bar fZ)&\equiv&\Gamma_2^{ffZ0}\\
&&\nobody\hspace{-2.6cm}
=\frac{M_2g_1^2g_2^2}{192\pi^3}\left[R_f^2(2)R_f^2(1)+L_f^2(2)L_f^2(1)\right]
\left[ \ln^2\left(\frac{M_Z^2}{M_2^2}\right) +
3\ln\left(\frac{M_Z^2}{M_2^2}\right) +5 -\frac{\pi^2}{3}
+O{\left(\frac{M_Z^2}{M_2^2}\right)}\right].\nonumber 
\ea
The decay width $\Gamma(Z_2\rightarrow f\bar fW)\equiv\Gamma_2^{ffW0}$
can be obtained from equation \req{gamrare} by an appropriate
replacement of the couplings and masses.
The decays $Z_2\rightarrow f\bar f\gamma$  and $Z_2\rightarrow
W^+W^-\gamma$ are considered as radiative corrections to the decays 
$Z_2\rightarrow f\bar f$ and $Z_2\rightarrow W^+W^-$.

All other decays of the $Z_2$ to SM particles are expected to be small.
Decays of the $Z_2$ to non-standard particles depend on additional
model parameters.
We assume that they are kinematically suppressed. 

The total decay width $\Gamma_n$ of the $Z_n$ is defined as the sum over all
partial decay widths.
It is convenient to combine $M_n$ and $\Gamma_n$ in the complex mass, 
\bq
\label{complexmass}
m_n^2 = M_n^2-iM_n\Gamma_n.
\eq
%
\subsubsection{Higher order effects}
The $Z'$ width can be neglected in experiments with typical
energy transfers much lower than $M_2$ as far as $\Gamma_2\ll M_2$.
$\Gamma_2$ can be measured in $pp(p\bar p)$
collisions for $s>2(3)M_2^2$ or in $e^+e^-$ collisions for $s\ge M_2^2$.
To reach the experimental accuracy, the inclusion of higher order
effects is necessary in these experiments.
Radiative corrections to $\Gamma_2$ are absolutely necessary for
precision measurements at the $Z_2$ 
peak in $e^+e^-$ or $\mu^+\mu^-$ collisions.

Of course, tree level decays to non-standard particles have to be
included if they are not suppressed.
In this case, the corresponding parameters must be measured in
independent experiments. 
We concentrate here on known higher order effects of the decays to SM
particles.
\paragraph*{Radiative corrections}\label{gamma2rc}
Radiative corrections to $\Gamma_2^{f0}$ can be deduced from the
results known for SM $Z$ decays \cite{plb259},
\ba
\label{gffcorr}
\Gamma_n^f&=&\Gamma_n^{f0} R_{\rm QED}R_{QCD},\ \ \ n=1,2,\nll
R_{\rm QED} &=& \left(1+\frac{3}{4}\frac{\alpha}{\pi}(Q^f)^2\right),\nll
R_{\rm QCD} &=& 0\mbox{\ \ for\ }f=l,\\
R_{\rm QCD} &=& 1 + \frac{\alpha_s(M_1^2)}{\pi} 
+ 1.405\frac{\alpha_s^2(M_1^2)}{\pi^2}
- 12.8\frac{\alpha_s^3(M_1^2)}{\pi^3}
- \frac{(Q^f)^2}{4}\frac{\alpha\alpha_s(M_1^2)}{\pi^2} \mbox{\ \ for\ }f=q.
\nonumber
\ea
For $\Gamma_2^t$, the top mass should also be taken into account in the
radiative corrections \cite{plb248}. 
See reference \cite{plb195} for $O(\alpha\alpha_s)$ corrections in
that case.

The SM weak corrections to SM $Z$ decays are calculated in reference
\cite{npb276} in terms of form factors.
See also section \ref{zpeeff24}.  
The concept can be generalized to SM weak corrections of $Z_2$ decays to
SM fermions.
The full one-loop corrections in the GUT depend on the details of the theory.
They can only be calculated if all relevant new parameters are known
from independent experiments.

The main corrections to $\Gamma_2^{W0}$ are QED corrections from
radiation off the final state and the Coulomb singularity.
See section \ref{zpeeww2} or \cite{lep2ww} for further details and references.

Radiative corrections to $\Gamma_2^{ffV0}$ are expected to be a small
correction to a small quantity.
The known SM corrections to four fermion final states are summarized
in reference \cite{lep2}.
\paragraph*{Mass mixing}
Mass mixing changes the partial decay widths of the $Z_2$ because it
induces changes in the couplings.
The effect is of the order $\theta_M$ and is therefore small due to the
present experimental limits on the mixing.
It is interesting for precision measurements in $e^+e^-$ or
$\mu^+\mu^-$ collisions at $s\approx M_2^2$.
It must be taken into account together with weak corrections of the
full GUT.
\paragraph*{Energy dependent width effects}
At lowest order, a particle has no width.
It is obvious that a width is needed to describe a resonance.
The simplest approximation is to use the constant width in the
propagator, which is calculated in the previous sections.
The next step of precision is to take into account that the width
is a function of the energy. 

In general, the inclusion of a finite width violates gauge invariance
because it partially takes into account effects, which are of higher
order in perturbation theory.
It is shown in \cite{4fgammasold} that the gauge violating
terms can be enhanced by large kinematic factors $\sim s/m_e^2$ in some
processes with four fermion final states, i.e. in $e^+e^-\rightarrow
e^+W^-\nu_e$. 
The problem can be cured by the inclusion of additional higher order
contributions, which restore the gauge invariance, see
\cite{4fgammasold,4fgammas} for further details. 

The effect is under control in fermion pair production where one
should take the $s$--dependent width \cite{gammazs}. 
It is $\Gamma_2(s) \approx \frac{s}{M_2^2}\;\Gamma_2(M_2^2)$ if
the vector boson can decay only into light particles.

The $s$--dependence of $\Gamma_2$ leads to a shift  $\Delta
E_{peak}=\frac{1}{2}(\Gamma_2/M_2)^2M_2$ of the $Z_2$ peak
position \cite{gammazs}.
For $s>M_2^2$ the width should be taken $s$--dependent too because it
influences the radiative tail as explained in section \ref{zpeeff21}.
\subsection{Summary of $Z'$ parameters}
Under the assumptions of section \ref{guts11}, we are left with the
following $Z'$ parameters, 
\ba
M_2,\Gamma_2 &\ &\mbox{the\ mass\ and\ width of the mass eigenstate\ } Z_2,\nll
\theta_M     &\ &\mbox{the\ $ZZ'$\ mixing\ angle,}\nll
g_2  &\ &\mbox{the\ coupling\ strength},\nll
v_f(2),a_f(2)
&\ &\mbox{the\ vector\ and\ axial\ vector\ couplings\ to\ fermions.} 
\ea

One can choose \cite{prd50} the following quantities to parametrize
the five independent $Z_2$ couplings \req{coup5} describing $v_f(2)$
and $a_f(2)$, 
\ba
\label{coup5ee}
\epsilon_A&\equiv&
[L_l(2)-R_e(2)]^2\frac{g_2^2}{4\pi\alpha}\frac{s}{M_{Z'}^2-s}
=a_e^2(2)\frac{g_2^2}{4\pi\alpha}\frac{s}{M_{Z'}^2-s},\nll
P_V^e&\equiv&\frac{L_l(2)+R_e(2)}{L_l(2)-R_e(2)}=\frac{v_e(2)}{a_e(2)},
\nll 
P_L^q&\equiv&\frac{L_q(2)}{L_l(2)-R_e(2)}=\frac{v_q(2)+a_q(2)}{2a_e(2)},\nll
P_R^i&\equiv&\frac{R_i(2)}{L_q(2)}
=\frac{v_i(2)-a_i(2)}{v_i(2)+a_i(2)},\ \ \ i=u,d.
\ea

In collisions of unpolarized protons, one is insensitive to the
relative sign of the $Z'$ couplings.
Then, an alternative set of parameters is convenient \cite{1063},
\ba
\label{ppparm}
\gamma_L^l&\equiv&\frac{L_l^2(2)}{L_l^2(2)+R_e^2(2)}=
\frac{(v'_l+a'_l)^2}{2(v'^2_e+a'^2_e)},\nll
\gamma_L^q&\equiv&\frac{L_q^2(2)}{L_l^2(2)+R_e^2(2)}=
\frac{(v'_q+a'_q)^2}{2(v'^2_e+a'^2_e)},\nll
\tilde{U}&\equiv&\frac{R_u^2(2)}{L_q^2(2)}=
\frac{(v'_u-a'_u)^2}{(v'_q+a'_q)^2},\nll
\tilde{D}&\equiv&\frac{R_d^2(2)}{L_q^2(2)}=
\frac{(v'_d-a'_d)^2}{(v'_q+a'_q)^2}.
\ea
$\epsilon_A$ can be added to equation \req{ppparm} as the fifth parameter.

Both parameter sets are related, see reference \cite{prd52}.
\section{Extra neutral gauge bosons in $E_6$ models}\label{guts3} 
In an $E_6$ GUT \cite{e6,roroe6}, the five independent charges $Q'^f$ 
are constrained \cite{prd52} in addition to \req{chargedef}, 

\bq
\label{e6charge}
Q'^{e_L^c}=Q'^{u_L^c}=Q'^{q_L},\ \ \ Q'^{l_L}=Q'^{d_L^c}.
\eq
Therefore, the charges $Q'^f$ of a general $Z'$ in an $E_6$ GUT can be
parametrized by two independent parameters.
However, the three conditions \req{e6charge} lead to only two
relations between the $Z'$ couplings \cite{prd52} as far as the ratio
$g_{12}/g_{22}$ in equation \req{gencoup} is unknown,
\bq
\label{e6check}
2L'_q + R'_u +R'_e=0,\ \ \ L'_q - R'_d - L'_l + R'_e=0.
\eq
The experimental check of the conditions \req{e6check} allows to determine
whether a $Z'$ can belong to the $E_6$ breaking scheme or not.

There are many  breakings possible in the $E_6$ group.
See \cite{e6breaking} for an extensive discussion.
Let us not consider the general $E_6$ case but
restrict ourselves to the gauge breaking scheme 
\bq
\label{e6so10}
E_6\rightarrow SO(10)\times U(1)_\psi
\rightarrow SU(5)\times U(1)_\chi\times U(1)_\psi,
\eq
where the linear combination
\bq
\label{betadef}
Z'(\beta)=\chi\cos\beta + \psi\sin\beta
\eq
is assumed to be light. 
The special case $\eta=Z'(-\arctan\sqrt{5/3})$ is often considered.
$\beta$ is the free parameter in the breaking scheme \req{e6so10}.
For the $\chi$ arising in the breaking of the SO(10)
\cite{lrmod,roroso10,fritzsch,prd30} 
to SU(5) \cite{su5} and the $\psi$ arising in the breaking of the $E_6$ to
$SO(10)$, we get
\bq
-Q'^{q_L}_\chi=\frac{1}{3}Q'^{l_L}_\chi=\frac{1}{\sqrt{40}},\ \ \ 
Q'^{q_L}_\psi=Q'^{l_L}_\psi=\frac{1}{\sqrt{24}},\ \ \ 
g_2=g_1\sqrt{\frac{5}{3}}s_W,\ \ \ g_{12}=0.
\eq
Equation \req{gencoup} now defines the couplings of the $Z'$ to SM
fermions as given in table \ref{zecoup}.
%
\begin{table}[tbh]
\begin{center}
\begin{tabular}{|l|lcc|lcc|}\hline\rule[-2ex]{0ex}{5ex} 
$f$   &$E_6$: &\ \ \ $a'_f$ &\ $v'_f$ & LR: &\ $a'_f$ &\ \ $v'_f$\\ 
\hline 
$\nu$&& $3\frac{\cos\beta}{\sqrt{40}} + \frac{\sin\beta}{\sqrt{24}}$
    & $3\frac{\cos\beta}{\sqrt{40}} + \frac{\sin\beta}{\sqrt{24}}$
   && $\frac{1}{2\alpha}$ & $\frac{1}{2\alpha}$
\rule[-2ex]{0ex}{5ex}\\
$e$&& $\frac{\cos\beta}{\sqrt{10}} + \frac{\sin\beta}{\sqrt{6}}$
    & $2\frac{\cos\beta}{\sqrt{10}}$
   && $\frac{\alpha}{2}$ & $\frac{1}{\alpha}-\frac{\alpha}{2}$
\rule[-2ex]{0ex}{5ex}\\
$u$&& $-\frac{\cos\beta}{\sqrt{10}} + \frac{\sin\beta}{\sqrt{6}}$
    & 0 
   &&$-\frac{\alpha}{2}$&$-\frac{1}{3\alpha}+\frac{\alpha}{2}$
\rule[-2ex]{0ex}{5ex}\\
$d$&& $\frac{\cos\beta}{\sqrt{10}} + \frac{\sin\beta}{\sqrt{6}}$ 
    & $-2\frac{\cos\beta}{\sqrt{10}}$ 
   && $\frac{\alpha}{2}$ & $-\frac{1}{3\alpha}-\frac{\alpha}{2}$
\rule[-2ex]{0ex}{5ex}\\
\hline
\end{tabular}\medskip
\end{center}
{\small\it \begin{tab}\label{zecoup}
The vector and axial vector couplings of the $Z'$ to SM fermions in the $E_6$
and LR models.
\end{tab}} \end{table}

One has to be careful with the different notations for $\beta$ in the
literature. 
For example, the model parameter $\theta$ in reference \cite{e6} is
related to our $\beta$ as $\beta=\theta+\pi/2$.

Consider the gauge breaking scheme \cite{langmix,langmix2}
\bq
\label{e6su2su2}
SO(10)\rightarrow SU(3)_c\times SU(2)_L\times SU(2)_R\times U(1)_{B-L}
\eq
as a second example.
Now the $Z'$ couples to the current 
\bq
\label{lrcurrent}
J_{LR}^\beta = \sqrt{\frac{5}{3}}\left(\alpha J_{3R}^\beta -
\frac{1}{2\alpha}J_{B-L}^\beta\right),\ \ \ 
\alpha\equiv \sqrt{\frac{c_W^2}{s_W^2}\frac{g_R^2}{g_L^2} -1},
\eq
where $g_L$ and $g_R$ are the couplings to the left- and right handed
gauge groups. 
$J_{3R}$ is the current associated with the $SU(2)_R$
group and $B$ and $L$ are baryon and lepton numbers, $B-L=2(Q-T_{3L}-T_{3R})$.
$\alpha$ is constrained to lie in the range 
$\sqrt{2/3}\le\alpha\le\sqrt{c_W^2/s_W^2-1}$.
Within our conventions, we have 
\bq
Q'^f_{LR}=Q'^f_\chi\sqrt{\frac{2}{5}}\frac{\alpha^2+1}{\alpha},\ \ \ 
g_2=g_1s_W,\ \ \ 
\frac{g_{12}}{g_2}=\frac{1}{\sqrt{15}}\frac{3\alpha^2-2}{\alpha}.
\eq
The case $\alpha=\sqrt{2/3}$ is identical to $Z=\chi$.
Again, equation \req{gencoup} gives all $Z'$ couplings to SM fermions
shown in table~\ref{zecoup}.

The decay width of a $Z'$ to fermion pairs is small in $E_6$ theories.
Typical values for $\sum_f\Gamma_2^{f}/M_{Z'}$ are
between 0.5\% and 2\% if only decays to SM fermions are kinematically allowed.
If the decays to all exotic fermions and Higgs bosons are possible,
$\Gamma_2$ becomes roughly three times larger \cite{prd36,langmix}.

The entries of the mass matrix \req{zznomix} are completely defined
in a fixed model.
For example, in a model with a gauge symmetry breaking by two Higgs doublets
and one Higgs singlet \cite{prd36},
\bq
\label{guthiggs}
\Phi_1 = \left( \begin{array}{c} \phi_1^0\\ \phi_1^-\end{array} \right),\ \ \ 
\Phi_2 = \left( \begin{array}{c} \phi_2^+\\ \phi_2^0\end{array} \right),\ \ \ 
\Phi_3=\phi_3^0,\ \ \ 
\langle \phi_i^0\rangle = \frac{gv_i}{\sqrt{2}},\ i=1,2,3,\ \ \ 
v^2=v_1^2+v_2^2,
\eq
one gets \cite{prd36}
\bq
\label{deltam2}
\delta M^2=2M_Z^2s_W\frac{Q'(\phi_1)v_1^2-Q'(\phi_2)v_2^2}{v^2},\ \ \ 
M_{Z'}^2=4M_Z^2 s_W\frac{1}{v^2}\sum_{i=1}^3[Q'(\phi_i^0)v_i]^2.
\eq
$Q'_1$ and $Q'_2$ are the $U'(1)$ charges of the two Higgs doublets.
%
The entry $M_Z^2$ of the mass matrix \req{zznomix} is known from the SM.
Equation \req{deltam2} gives an example for the Higgs constraint
\req{higgsconstr}.
The values of $Q'(\phi_1)$ and $Q'(\phi_2)$ depend on the particular
symmetry breaking \cite{prd42}.

The Higgs constraint can be combined with the formula for
$\Gamma_2^{W}$. 
It gives $\Gamma_2^{W}\approx\Gamma_2^{e}$ in left-right models \cite{plb208}.
\section{Extra neutral gauge bosons and contact interactions}\label{guts5} 

Far below the resonance, $Z'$ effects can be described by four fermion
contact interactions.
The  interaction of a $Z'$ with the massless
fermions $f$ and $F$ is then given by the amplitude
\begin{eqnarray}
\label{contactzp}
{\cal M}(Z') = \frac{g_2^2}{M_{Z'}^2} \sum_{f,F}\left( 
  L'_fL'_F \bar{u}_{f,L}\gamma_\beta u_{f,L} \, 
           \bar{u}_{F,L}\gamma^\beta u_{F,L}
+ R'_fR'_F \bar{u}_{f,R}\gamma_\beta u_{f,R} \, 
           \bar{u}_{F,R}\gamma^\beta u_{F,R}
\right. \\ \nonumber \left.
+ R'_fL'_F \bar{u}_{f,R}\gamma_\beta u_{f,R} \, 
           \bar{u}_{F,L}\gamma^\beta u_{F,L}
+ L'_fR'_F \bar{u}_{f,L}\gamma_\beta u_{f,L} \, 
           \bar{u}_{F,R}\gamma^\beta u_{F,R}
\right).
\end{eqnarray}
This can be compared with four fermion contact interactions \cite{eichten},
\begin{eqnarray}
\label{contact}
{\cal M}(contact) = 4\pi \sum_{f,F}\left[ 
\frac{\eta_{LL}^{fF}}{(\Lambda^{fF}_{LL})^2}
\bar{u}_{f,L}\gamma_\beta u_{f,L}\,\bar{u}_{F,L}\gamma^\beta u_{F,L}
+  \frac{\eta_{RR}^{fF}}{(\Lambda^{fF}_{RR})^2}
\bar{u}_{f,R}\gamma_\beta u_{f,R} \, \bar{u}_{F,R}\gamma^\beta u_{F,R}
\right. \\ \nonumber \left.
+  \frac{\eta_{RL}^{fF}}{(\Lambda^{fF}_{RL})^2}
\bar{u}_{f,R}\gamma_\beta u_{f,R} \, \bar{u}_{F,L}\gamma^\beta u_{F,L}
+  \frac{\eta_{LR}^{fF}}{(\Lambda^{fF}_{LR})^2}
\bar{u}_{f,L}\gamma_\beta u_{f,L} \, \bar{u}_{F,R}\gamma^\beta u_{F,R}
\right].
\end{eqnarray}
$\Lambda^{fF}_{ij},\ i,j=L,R$ are the scales of the new physics 
 and $\eta_{ij}^{fF}=\pm 1$.

Constraints on four fermion contact interactions can always be 
interpreted as constraints on extra neutral gauge bosons,
\bq
\label{contactparm}
\frac{g_2^2L'_mL'_n}{4\pi M_{Z'}^2}\equiv\frac{L^N_mL^N_n}{s}
=\frac{\eta_{LL}^{mn}}{(\Lambda_{LL}^{mn})^2},\ \ \ m,n=f,F.
\eq
We use the definition 
\bq
\label{normcoupdef}
L^N_f\equiv L'_f\frac{g_2}{\sqrt{4\pi}}\frac{\sqrt{s}}{M_{Z'}}.
\eq
Relations similar to \req{contactparm} can be written for the other
helicity combinations.
Note that $Z'$ interactions far below the $Z'$ peak can only constrain the
ratios $L_f^N,R_f^N$ and not the $Z'$ couplings and the $Z'$ mass
separately.

Interactions of extra neutral gauge bosons are not as
general as 4 fermion contact interactions.
The Lagrangian
\req{contactzp} describes the interaction between fermions of flavours
$f$ and $F$ by the {\it four}
couplings $L^N_f,L^N_F,R^N_f$ and $R^N_F$. 
The same process is described in contact interactions by the
{\it twelve} parameters $\Lambda^{mn}_{ij},\ i,j=L,R,\ mn=ff,fF,FF$.

If a future helicity conserving experiment shows small deviations
from the SM predictions, one can always parametrize the deviation
in terms of  contact interactions. 
If the new interaction is due to a $Z'$, the parameters of the
contact interactions fulfill the relations
\ba
\label{contactzp2}
\eta^{nn}_{LL}=\eta^{nn}_{RR}=1,\ \ \ \eta^{mn}_{LR}=\eta^{mn}_{RL},\ \ \ 
(\Lambda^{mn}_{LR})^2=(\Lambda^{mn}_{RL})^2
=\sqrt{(\Lambda^{mn}_{LL})^2(\Lambda^{mn}_{RR})^2},\ mn=ff,fF,FF,\nll
(\Lambda^{fF}_{ii})^2=\sqrt{(\Lambda^{ff}_{ii})^2(\Lambda^{FF}_{ii})^2},
\ i=L,R.\hspace{3cm}\ 
\ea
The normalized $Z'$ couplings can then be calculated according to
equation \req{contactparm}.
Under the assumption of the gauge group \req{gutgauge}, the $Z'$
couplings must satisfy the additional relations \req{coup5}. 

Of course, $Z_2$ effects near the $Z_2$ resonance cannot be described
by contact interactions. 
\section{Four fermion interactions and form factors}\label{guts3plus} 
%
In many experiments, a $Z'$ can be detected through four fermion
interactions. 
In addition to the $Z'$ exchange, we always have the SM contributions
where the $Z'$ is replaced by the photon or the $Z$ boson.
The contributions due to the exchange of extra neutral gauge bosons
can be absorbed into the couplings of the SM $Z$ boson \cite{riemann91}.
The following discussion is general for four fermion interactions.
It can be directly applied to Bhabha and M{\o}ller scattering,
$pp,p\bar p$ or $ep$ scattering. 

Consider the four fermion interaction $e^+e^-\rightarrow f\bar f\
(f\neq e)$ as an example.
The amplitude for this process is
\ba
\label{fourfampl}
{\cal M}  = \sum_n\frac{g_n^2}{s-m_n^2}                            
\bar{v}(e)\gamma_\beta \left[v_e(n)-\gamma_5 a_e(n)\nobodyfrac\right] u(e)\cdot
\bar{u}(f)\gamma^\beta \left[v_f(n) -\gamma_5 a_f(n)\nobodyfrac\right] v(f).
\ea
In the SM, the summation runs over the photon and the $Z$ boson ($n=0,1$).
In a theory including a $Z'$, the extra amplitude with the $Z'$ exchange, 
${\cal M}_E={\cal M}(n=2)$, has to be added.
It is important that ${\cal M}_E$ has the same structure as the SM
amplitude. 
Then one can formally include the contribution of ${\cal M}_E$ in 
the couplings of the SM $Z$ boson leaving the couplings of the photon
unchanged. 

Consider the amplitude with $Z$ and $Z'$ exchange only,
\ba
\label{fourfampl2}
{\cal M}  &=& \sum_{n=1,2}\frac{g_n^2}{s-m_n^2}                            
\bar{v}(e)\gamma_\beta \left[v_e(n)-\gamma_5 a_e(n)\nobodyfrac\right] u(e)\cdot
\bar{u}(f)\gamma^\beta \left[v_f(n) -\gamma_5 a_f(n)\nobodyfrac\right] v(f)\nll
&\equiv& \frac{g_1^2}{s-m_1^2}                            
\bar{v}(e)\gamma_\beta u(e)\cdot\bar{u}(f)\gamma^\beta v(f)\cdot v_e(1)v_f(1)
(1+\epsilon_{vv}) \nll
&&\hspace{1.0cm}
-\bar{v}(e)\gamma_\beta u(e)\cdot\bar{u}(f)\gamma^\beta\gamma_5v(f)
\cdot v_e(1)a_f(1)(1+\epsilon_{va}) \nll
&&\hspace{1.0cm}
-\bar{v}(e)\gamma_\beta\gamma_5u(e)\cdot\bar{u}(f)\gamma^\beta v(f)
\cdot a_e(1)v_f(1)(1+\epsilon_{av}) \nll
&&\hspace{1.0cm}
+\bar{v}(e)\gamma_\beta\gamma_5u(e)\cdot\bar{u}(f)\gamma^\beta\gamma_5v(f)
\cdot a_e(1)a_f(1)(1+\epsilon_{aa}),\nll
\mbox{with\ }&& \epsilon_{xy}=\chi_{Z/Z'}
\frac{x_e(2)y_f(2)}{x_e(1)y_f(1)},\ \ \  x,y=a,v,\ \ \ 
\chi_{Z'/Z}=\frac{g_2^2(s-m_1^2)}{g_1^2(s-m_2^2)}.
\ea
The coefficients $\epsilon_{xy}$ contain all information of the
amplitude ${\cal M}_E$.
Various additional amplitudes ${\cal M}_E$  arising, for example, from weak
corrections or $ZZ'$ mixing, can be written in the form
\req{fourfampl2} if the quantities  $\epsilon_{xy}$ are specified
\cite{leikeustron}.  

Following the tradition of electroweak corrections,
we want to parametrize the contributions $\epsilon_{xy}$ by (complex) form
factors $\rho_{ef},\kappa_e,\kappa_f,\kappa_{ef}$, which are
introduced by replacements of the couplings, 
\ba
\label{formrepl}
v_e(1)v_f(1)&\rightarrow& a_e(1) a_f(1)
\left[\nobodyfrac 1-4|Q^e|s^2_W\kappa_e
-4|Q^f|s^2_W\kappa_f + 16|Q^eQ^f|s^4_W\kappa_{ef}\right],\nll
v_e(1)&\rightarrow& a_e(1)\left[\nobodyfrac 1-4|Q^e|s^2_W\kappa_e\right],\nll
v_f(1)&\rightarrow& a_f(1)\left[\nobodyfrac 1-4|Q^f|s^2_W\kappa_f\right],\nll
a_e(1),a_f(1)&\rightarrow&\mbox{unchanged},\nll
g_1^2=&\rightarrow& g_1^2\rho_{ef}.
\ea
Comparing with equation \req{fourfampl2}, the form factors can be expressed
through $\epsilon_{xy}$, 
\ba
\label{fourfampl3}
\rho_{ef}&=&1+\epsilon_{aa}\nll
\kappa_f&=&\frac{1}{\rho_{ef}}
\left[1+\frac{\epsilon_{av}v_f(1)-\epsilon_{aa}a_f(1)}
{v_f(1)-a_f(1)}\right]\\
\kappa_{ef}&=&\frac{1}{\rho_{ef}}
\left[1+\frac{\epsilon_{vv}v_e(1)v_f(1)+\epsilon_{aa}a_e(1)a_f(1)
             -\epsilon_{av}a_e(1)v_f(1)-\epsilon_{va}v_e(1)a_f(1)}
{[v_e(1)-a_e(1)][v_f(1)-a_f(1)]}\right].\nonumber
\ea
In particular, the additional $Z'$ amplitude can be included in the
$Z$ couplings specifying $\epsilon_{xy}$ as given in equation \req{fourfampl2},
\ba
\label{formzprepl}
\rho_{ef}&=& 
1+\chi_{Z'/Z}\frac{a_e(2)a_f(2)}{a_e(1)a_f(1)},\nll
\kappa_f&=&\frac{1}{\rho_{ef}}
\left[1+\chi_{Z'/Z}\frac{a_e(2)[v_f(2)-a_f(2)]}
                        {a_e(1)[v_f(1)-a_f(1)]}\right],\nll
\kappa_{ef}&=&\frac{1}{\rho_{ef}}
\left[1+\chi_{Z'/Z}\frac{a_e(2)[v_f(2)-a_f(2)][v_e(2)-a_e(2)]}
                        {a_e(1)[v_f(1)-a_f(1)][v_e(1)-a_e(1)]}\right].
\ea
The result agrees with the formulae given in reference \cite{zfitter}.
As mentioned there, this method of the inclusion of the $Z'$
contributions has the advantage that it can be easily implemented in computer
codes designed for SM calculations.
The form factors \req{fourfampl3} and \req{formzprepl} work equally
well for any four fermion process. 
They include the $Z'$ contribution without any approximation.
Of course, they are $s$-dependent, in general not small and even
resonating near the $Z'$ peak.

Consider the case where two additional amplitudes are added to ${\cal M}_{SM}$,
\bq
\label{fourfampl4}
{\cal M} = {\cal M}_{SM} + {\cal M}_1 +{\cal M}_2.
\eq 
Suppose that both additional amplitudes are parametrized in the way
described above, i.e. assume that the form factors
$\rho_{ef}^i,\kappa_e^i,\kappa_f^i,\kappa_{ef}^i,\ i=1,2$ are known.
Then, the combined form factors can be calculated by taking into account
$\epsilon_{xy}^\Sigma=\epsilon_{xy}^1+\epsilon_{xy}^2$,
\ba
\label{fourfampl5}
\rho_{ef}^\Sigma=\rho_{ef}^1+\rho_{ef}^2-1,\ \ \ 
\kappa_f^\Sigma=\frac{\kappa_f^1\rho_{ef}^1+\kappa_f^2\rho_{ef}^2-1}
{\rho_{ef}^1+\rho_{ef}^1-1},\ \ \ 
\kappa_{ef}^\Sigma=\frac{\kappa_{ef}^1\rho_{ef}^1+\kappa_{ef}^2\rho_{ef}^2-1}
{\rho_{ef}^1+\rho_{ef}^1-1}.
\ea
The summation rules \req{fourfampl5} are exact.
In many applications, the form factors are not very different from one.
In this case, the approximate summation rules
\bq
\label{fourfampl6}
\rho_{ef}^\Sigma =\rho_{ef}^1\rho_{ef}^2,\ \ \ 
\kappa_f^\Sigma =\kappa_f^1\kappa_f^2,\ \ \ 
\kappa_{ef}^\Sigma =\kappa_{ef}^1\kappa_{ef}^2
\eq
are often used. In their derivation, contributions proportional to 
$\epsilon_{xy}\epsilon_{x'y'}$ are neglected.

The case \req{fourfampl4} arises, for example, in the simultaneous
description of $ZZ'$ mixing and electroweak corrections.
The functions $\epsilon^m_{xy}$ describing the $ZZ'$ mixing are
\ba
\label{epsweak}
\epsilon_{xy}^m&=&\chi_{Z_1/Z}
\left(c_M+\frac{g_{Z'}}{g_Z}s_M\frac{x'_e}{x_e}\right)
\left(c_M+\frac{g_{Z'}}{g_Z}s_M\frac{y'_f}{y_f}\right)-1,\ \ \ x,y=a,v,\nll
\chi_{Z_1/Z}&=&\frac{s-m_Z^2}{s-m_1^2}.
\ea
Again, the resulting form factors can be calculated \cite{leikeustron}
using equation \req{fourfampl3},
\ba
\label{formmix}
\rho_{ef}^m&=&
\chi_{Z_1/Z}\left(c_M+\frac{g_{Z'}}{g_Z}s_M\frac{a'_e}{a_e}\right)
\left(c_M+\frac{g_{Z'}}{g_Z}s_M\frac{a'_f}{a_f}\right),\nll
\kappa_f^m&=&\frac{1+\frac{s_Mg_{Z'}}{c_Mg_Z}\frac{v'_f-a'_f}{v_f-a_f}}
                  {1+\frac{s_Mg_{Z'}}{c_Mg_Z}\frac{a'_f}{a_f}},\nll
\kappa_{ef}^m&=&\kappa_e^m\kappa_f^m.
\ea

The form factors $\rho_{ef}^m, \kappa_f^m$ and $\kappa_{ef}^m$ are
related to the functions $x_f$ and $y_f$ introduced in equation \req{vf1},
\bq
\rho_{ef}^m=(1-y_e)(1-y_f),\ \ \ \kappa_f^m=1-x_f.
\eq
\section{Model dependence of $Z'$ constraints}\label{guts2} 
Future experiments will either be consistent with the SM
or show deviations from the SM predictions.
In the first case, the data can be used to constrain
extensions of the SM, for example, contact interactions or
theories predicting extra neutral gauge bosons. 
In the case of a deviation, one can try to interpret it in terms
of $Z'$ parameters.
This procedure could either fail or favor some $Z'$ models compared
to others.
The analysis can be done with or without assumptions on the $Z'$ model.
We call these procedures {\it model dependent} and {\it model
independent} analyses. 

Of course, there are several steps from the model independent $Z'$
analysis to the model dependent $Z'$ analysis.
As far as the model assumptions are consistent with the experimental
data, a model dependent analysis is justified
and welcome to learn more details about the underlying theory.

For example, helicity conserving processes can be parametrized by four
fermion contact interactions. 
Only little can be learned about the origin of the new interaction in
this case.
If there are deviations from the SM and the conditions
\req{contactzp2} and \req{coup5} are fulfilled, the new
interaction is consistent with a $Z'$ coming from the gauge group
\req{gutgauge}.  
This theoretical assumption
increases our knowledge about the origin of the new interaction.
If the couplings fulfill the relations \req{e6check}, the new
interaction is consistent with a $Z'$ coming from the $E_6$ group. 
The subsequent assumption that the new interaction {\it is} due
to a $Z'$ from an $E_6$ breaking, increases the model dependence but
allows to probe further details of the assumed model.
The experimental verification that $-3L'_q=L'_l$ is also fulfilled
supports the hypothesis that the interaction is due to a $Z'$
from a $SO(10)$ breaking.
Finally, the new interaction could be tested for compatibility
with a definite $Z'$, e.g. $Z'=\chi$.
This hypothesis contains most model assumptions but allows more
detailed tests of the theory and the best fits to the remaining free
parameters. 

Examples of model dependent $Z'$ constraints are the lower bounds
on $M_{Z'}$ or the regions of $\theta_M$ allowed for certain $E_6$ models
quoted in the Particle Data Book. 

For a {\it model independent} analysis, one has to pay a price.
Usually, one cannot constrain single $Z'$ parameters but only certain
combinations of them. 
Often, only a limited set of observables is useful for the model
independent analysis.
Examples of model independent $Z'$ constraints are the allowed regions of
$v_f^M=v'_f\theta_M$ and $a_f^M=a'_f\theta_M$ from LEP\,1 data
\cite{zpc53,layssac,leikez1}, the constraints on $v_l^N\approx
v'_l/M_{Z'}$ and  $a_l^N\approx a'_l/M_{Z'}$ from LEP2 data
\cite{L31961} or the constraints on $\sigma_T^l=\sigma(p\bar
p\rightarrow Z')\cdot Br_2^l$ from Tevatron data
\cite{prl79}. 

Model independent $Z'$ constraints can always be converted into model
dependent $Z'$ constraints specifying the $Z'$ model.
The constraints on model parameters obtained in such a two--step
procedure are in general weaker than the constraints that would be
obtained by a direct model dependent fit to the same data. 

Model dependent and model independent $Z'$ analyses are complementary.
Both analyses have advantages and disadvantages, which are summarized in
table \ref{mdmitab}. 
Throughout this paper, we explicitly mention whether a constraint is
obtained in a model dependent or in a model independent analysis.

%
\begin{table}[tbh]
\begin{center}
\begin{tabular}{|ll|ll|}\hline\rule[-2ex]{0ex}{5ex} 
& Model dependent $Z'$ analysis && Model independent $Z'$ analysis\\ 
\hline 
{\bf --}& The constraints are a mixture & 
{\bf +}& The constraints result from data only.\rule[0ex]{0ex}{3ex}\\
{\bf }& of experimental results &
{\bf }& They are not biased \\
{\bf }& and theoretical assumptions. &
{\bf }& by theoretical assumptions.\\
{\bf --}& A separate data analysis is needed &
{\bf +}& $Z'$ limits for a new $Z'$ model can be\\
{\bf }& for every new $Z'$ model. &
{\bf }& deduced without a new data analysis.\\
{\bf +}&  Single $Z'$ parameters&
{\bf --}& Only combinations of $Z'$ parameters\\
{\bf }&  can be constrained. &
{\bf }&  can be constrained. \\
{\bf +}& $Z'$ limits from different experiments &
{\bf --}& $Z'$ limits from different experiments\\
{\bf }& can always be compared. &
{\bf }& cannot always be compared.
\rule[-2ex]{0ex}{4ex}\\
\hline
\end{tabular}\medskip
\end{center}
{\small\it \begin{tab}\label{mdmitab}
Advantages ({\bf +}) and disadvantages ({\bf --}) of a model dependent
and model independent $Z'$ analysis.
\end{tab}} 
\end{table}
\section{Extracting $Z'$ limits from data}\label{guts6} 
In any analysis, one selects observables $O_i$ from the data and compares 
them with the predictions $O_i($SM$+Z')$ in a theory including a $Z'$. 
This allows to exclude or to confirm $Z'$ models at a given confidence
level.
The procedure is different in experiments with indirect and with
direct $Z'$ limits.
\subsection{Indirect $Z'$ limits}\label{guts61} 
%
In experiments with indirect $Z'$ signals, for example in 
$e^+e^-\rightarrow \mu^+\mu^-$ below the $Z'$ resonance, the SM
already predicts a large number of events of the given signature. 
A $Z'$ gives a signal in the observable $O_i$ if it produces a
deviation $\Delta^{Z'}O_i$ from the SM prediction $O_i($SM$)$, which is
larger than the experimental error $\Delta O_i$.
Here and in the following, $\Delta O_i$ stands for the error of
an asymmetry, the error of a ratio of cross sections or for the relative
error of a cross section.
Because of the large number of events, these errors can be assumed to be
Gaussian distributed. 
Neglecting correlations, we define
\bq
\label{chisq}
\chi^2 = \sum_{i}
\left[\frac{O_i-O_i(\mbox{SM}+Z')}{\Delta O_i}\right]^2.
\eq
For $\chi^2>\chi^2_{min}+\chi_{cl}^2$, the considered model is excluded
at a certain confidence level depending on $\chi_{cl}$.

In a real experiment, all possible observables would be measured and
contribute to the final result. 
In a {\it theoretical} analysis, the additional information due to
measurements of two observables $O_1$ and $O_2$ 
related by theory (i.e.  $O_1(SM+Z')=O_2(SM+Z')$) could be taken into
account by the inclusion of {\it one} of these observables but with a
smaller effective error $\Delta O_\Sigma$. 
This error can be estimated as
\ba
\label{comberror}
\chi^2&=&\sum_{i=1,2}
\left[\frac{O_i-O_i(\mbox{SM}+Z')}{\Delta O_i}\right]^2
=\left[\frac{O_1-O_1(\mbox{SM}+Z')}{\Delta O_\Sigma}\right]^2,\nll
\frac{1}{(\Delta O_\Sigma)^2}&=&\sum_{i=1,2}\frac{1}{(\Delta O_i)^2}.
\ea 

The generalization to more observables is straight forward.

Experimental errors consist of statistical and systematic contributions.
We assume in the estimates of the following sections that the combined
error is the quadratic sum, 
\bq
\label{rdef}
\Delta O = \sqrt{(\Delta^{stat} O)^2+(\Delta^{syst} O)^2}
=\Delta^{stat}O\sqrt{1+r^2}.
\eq
We define the ratio $r=\Delta^{syst}O/\Delta^{stat}O$. 

Optimal observables \cite{prd457,zpc62} can be constructed to get the
maximum sensitivity to $Z'$ parameters.
To measure (small) $Z'$ parameters $\lambda(Z')$ in indirect $Z'$
analyses, one can look for deviations in differential cross sections,
\bq
d\sigma(\phi) = d\sigma_{SM}(\phi)+ d(\Delta^{Z'}\sigma(\phi))
=d\sigma_{SM}(\phi)+ \lambda(Z')d\tilde\sigma(\phi).
\eq
Examples for $\lambda(Z')$ are $\theta_M$ or $\sqrt{s}/M_2$.
The $Z'$ parameter is measured by an integration over the phase space
$d\phi$ with a weight function $f(\phi)$ \cite{prd457},
\bq
f^{(1)}\left(\lambda(Z')\right)=\int f(\phi)d\sigma(\phi).
\eq
The weight function can be chosen in such a way that the sensitivity 
to $\lambda(Z')$ becomes maximum.
For one parameter, one gets  \cite{prd457}, 
$f(\phi)=\Delta^{Z'}\sigma(\phi)/\sigma_{SM}(\phi)$.
The generalization to several parameters is given in reference \cite{zpc62}.
\subsection{Direct $Z'$ limits}\label{guts62} 
%
Direct $Z'$ production is possible in
$e^+e^-$ or $\mu^+\mu^-$ collisions for $s\approx M_2^2$ or in hadron
collisions for $s>M_2^2$.

In direct production in $e^+e^-$ or $\mu^+\mu^-$ collisions, many $Z'$
events are expected at future colliders.
It can be assumed that the events are Gaussian distributed allowing a
$\chi^2$ analysis \req{chisq}.
In  contrast to indirect $Z'$ limits, the number of $Z'$ events at
the $Z_2$ peak is much larger than the SM background allowing
precision measurements.

In direct $Z'$ production at hadron colliders, the number of SM
background events is expected to be very small or zero.
A few $Z'$ events serve as a signal.
Then, one cannot assume that these events are Gaussian distributed.
Constraints on $Z'$ models at a given confidence level are obtained for all
models, which predict the same number of $Z'$ events.
For example, in the case where the signal is Poisson distributed, the
SM background is zero and zero events are observed, all theories predicting
$N_{Z'}=3$ events are excluded at 95\% confidence \cite{PDB}.
For a non-zero background, $N_{Z'}$ depends on it \cite{PDB}.
\chapter{$Z'$ search at $e^+e^-,\ e^-e^-$ and $\mu^+\mu^-$ colliders}
\label{zpee}
%
Lepton colliders have the advantage that different observables can be
detected above a small background.
As has been shown by the LEP and SLC experiments, different
fermions as $e,\mu,\tau,c,b$ can be tagged in the final state. 
The polarization of $\tau'$s and likely of top quarks can be measured.
Highly polarized electron beams are available.
One can also hope for a reasonable positron polarization.
At muon colliders, one expects some polarization of both beams \cite{mumuproc}.

$e^+e^-$ and $\mu^+\mu^-$ collisions yield several interesting
reactions which can probe different properties of extra $Z$ bosons.

{\it Fermion pair production} allows for a measurement of a large number of
different observables.
It is assumed that the final fermions are not electrons or electron neutrinos.
All couplings of the $Z'$ to charged SM fermions can be constrained separately.
This is a unique property of this reaction.

{\it Bhabha} and {\it M{\o}ller scattering} have much larger
event rates than fermion pair production. 
In addition, M{\o}ller scattering could profit from two highly
polarized electron beams. 
Of course, these reactions are sensitive to gauge boson couplings to
electrons only. 
The sensitivity to these couplings competes with fermion pair production.

{\it W pair production} is very sensitive to $ZZ'$ mixing.
This sensitivity is enhanced for large energies because a non--zero
$ZZ'$ mixing destroys the gauge cancellation between the different
amplitudes present in the SM.
The sensitivity to other $Z'$ parameters cannot compete with
fermion pair production.

All {\it other reactions} in  $e^+e^-$ or $e^-e^-$ collisions can not
add useful information on extra neutral gauge bosons.

Although we will explicitly mention mostly $e^+e^-$ and $e^-e^-$
collisions, all results presented in this chapter are also applicable to 
$\mu^+\mu^-$ or $\mu^-\mu^-$ collisions.
One important difference arises for measurements on and beyond the $Z_2$ peak.
As known from $Z_1$ physics, a precise measurement of the mass and
the width of the resonance relies crucially on the accurate
monitoring of the beam energy.
Here, a muon collider would have clear advantages compared to an electron
collider \cite{mumuproc} because the latter suffers from a large beam
energy spread.   
\section{$Z'$ search in $e^+e^-\rightarrow f\bar f$ }\label{zpeeff}
The sensitivity of fermion pair production to different $Z'$ parameters
depends crucially on the ratio of the center--of--mass energy $\sqrt{s}$ to
the masses $M_1$ and $M_2$ of the gauge bosons $Z_1$ and $Z_2$.
We distinguish four different cases,
\ba
\label{scases}
\mbox{case\ 1:\ }&&s\approx M_1^2,\nll
\mbox{case\ 2:\ }&&M_1^2\neq s<M_2^2,\nll
\mbox{case\ 3:\ }&&s\approx M_2^2,\nll
\mbox{case\ 4:\ }&&s>M_2^2,
\ea
where $s\approx M_n^2$ means $(M_n-\Gamma_n)^2 < s < (M_n+\Gamma_n)^2$.

{\it Case 1} is very important in setting constraints on the $ZZ'$ mixing
angle.
The $Z_2$ propagator, i.e. the sensitivity to $M_2$ is usually
suppressed by a factor $\Gamma_1/M_1$.

{\it Case 2} would give the first signals of a $Z'$ in
$e^+e^-$ collisions. 
The SM parameters are already precisely known
from measurements at the $Z_1$ peak leading to accurate predictions for 
observables at higher energies. 
Information about a $Z'$ is obtained from the differences to these predictions.
If there is agreement with the SM predictions, lower bounds on the
$Z'$ mass can be set for a fixed model. 
The sensitivity to the $ZZ'$ mixing angle is suppressed by a factor
$\Gamma_1/M_1$ compared to measurements at the $Z_1$ peak.

{\it Case 3} is certainly the best possibility to get precise and detailed
information about a $Z'$.
The corresponding measurements have much in common with the LEP and SLC
experiments at the $Z_1$ peak. 

{\it Case 4} is interesting because it can constrain a $Z'$ with couplings
to SM fermions much weaker than predicted in the usual GUT's.  
Such a $Z'$ could escape detection in experiments below its resonance.

This section is organized as follows.
After giving the relevant observables in the Born approximation, we discuss 
the different radiative corrections.
A discussion of constraints on $Z'$ parameters in the four cases \req{scases}
follows in different subsections.
Model dependent and model independent constraints are distinguished.

Every subsection on $Z'$ constraints is organized by the same pattern.
For every  constrained $Z'$ parameter, the physical origin of the
constraint is explained first and a simple estimate is given.
The estimate is then confronted with $Z'$ constraints obtained from present
experiments and with constraints obtained in theoretical analyses for future
colliders. 
\subsection{Born Approximation}\label{zpeeff1}

\subsubsection{Amplitude}\label{zpeeff11}
The amplitude for $e^+e^-\rightarrow (\gamma,Z,Z',\dots)\rightarrow
f\bar f,\ \ f\neq e$ is
\ba
\label{link}
{\cal M}  = \sum_n\frac{g_n^2}{s-m_n^2}                            
\bar{v}(e)\gamma_\beta \left[v_e(n)-\gamma_5 a_e(n)\nobodyfrac\right] u(e)\, 
\bar{u}(f)\gamma^\beta \left[v_f(n) -\gamma_5 a_f(n)\nobodyfrac\right] v(f).
\ea
The summation runs over the exchanged gauge bosons. 
In contrast to equation \req{contactzp}, we use here the parametrization in
vector and axial vector couplings.
This is useful because the photon has a pure vector coupling while the $Z$
boson has almost a pure axial vector coupling to the $e^+e^-$ pair in the
initial state. 

Only products of an {\it even} number of couplings of the $Z'$ to fermions
appear in the amplitude \req{link}. Hence, the process $e^+e^-\rightarrow
f\bar f$ cannot distinguish between $Z'$ models, which differ only by the
signs of all $Z'$ couplings to fermions. 

Measurements at one energy point sufficiently far away from the $Z_1$
and $Z_2$ resonances can only restrict the normalized couplings $a_f^N$
and $v_f^N$ \cite{zpmi}, 
\bq
\label{normcoup}
a_f^N\equiv a_f(2)\sqrt{\frac{g_2^2 }{4\pi}\frac{s}{m_2^2-s}}
\mbox{\ \ and\ \ } 
v_f^N\equiv v_f(2)\sqrt{\frac{g_2^2 }{4\pi}\frac{s}{m_2^2-s}}
\eq
and not the $Z_2$ couplings $a_f(2),v_f(2)$ and the $Z_2$ mass $m_2$
separately. 
Therefore, a heavy $Z_2$ with large couplings cannot be distinguished from
a light $Z_2$ with small couplings as far as $M_2^2\gg s$.
Below the resonance, fermion pair production by a $Z'$ can
be described by effective four fermion contact interactions, see
section \ref{guts5}.
Note that the definition \req{normcoup} differs slightly from the definition
\req{normcoupdef}. 
We will use definition \req{normcoup} in this section
because it is natural in fermion pair production.
For comparisons with other reactions the definition \req{normcoupdef} must be
used.
Note that the difference between the two definitions is $O(s/M_2^2)$.
This is a small quantity for $Z'$ masses at the detection limit 
$M_{Z'}\approx 3\sqrt{s}\mbox{\ to\ }7\sqrt{s}$. 

Several widths $\Gamma_2$ below the $M_2$ resonance, the width
$\Gamma_2$ does not influence the $Z'$ limits and can be neglected.
Therefore, the indirect $Z'$ limits from fermion pair production
remain valid for extra $Z$ bosons which have for some reason a width
much larger than predicted in usual GUT's.
This is an important difference with $Z'$ limits from hadron colliders,
which depend on $\Gamma_2$ through branching ratios as $Br_2^\mu$.

The amplitude can be decomposed as 
${\cal M} = {\cal M}_{SM} + g_2\theta_M{\cal M}_M$
on the $Z_1$ resonance in the limit of small $ZZ'$ mixing.
According to equation \req{vf1}, ${\cal M}_M$ is proportional to either 
$a'_f$ or $v'_f$. 
Therefore, measurements at the $Z_1$ peak constrain the combinations
\bq
\label{coupmdef}
a_f^M=\theta_Mg_2a'_f \mbox{\ \ and\ \ } v_f^M=\theta_Mg_2v'_f,
\eq
and not  $a'_f,v'_f$ and $\theta_M$ separately \cite{plb381}. 
Similar to the off--resonance case, a $Z'$ with large couplings
and small $ZZ'$ mixing cannot be distinguished from a $Z'$
with small couplings and large $ZZ'$ mixing as far as
$\theta_M\ll 1$.
\subsubsection{Cross section}\label{zpeeff12}
The total and the forward--backward cross sections are defined as
\begin{equation}
\label{eqafb}
\sigma_{T} = \int_{-1}^1 dc \frac{d \sigma}{dc},\ \ \  
\sigma_{FB} = \int_0^1 dc \frac{d \sigma}{dc}
- \int_{-1}^0 dc \frac{d \sigma}{dc},\ \ \ 
c=\cos\theta.
\end{equation}
$\theta$ is the angle between the outcoming anti-fermion $\bar f$ and
the incoming positron. 
At the Born level, the cross sections $\sigma_A^f,\ A=T,FB$ are
\bq
\label{born}
\sigma_A^f = d_A\sum_{m,n=0}^N\;\sigma_A^0(s;m,n) 
= d_AN_f\;\frac{\pi\alpha^2}{s}\;\sum_{m,n=0}^N\;
C_{A}(m,n)\;\chi_m(s)\;\chi_n^*(s)
\eq
with $d_T=\frac{4}{3}$ and $d_{FB}=1$.
$N_f$ is due to color, $N_f=1(3)$ for $f=l(q)$.

The summation runs over all interferences.
$\chi_n(s)$ is the propagator of the vector boson $Z_n$ with the invariant
energy squared $s$, 
\bq
\label{eq28}
\chi_n(s) = \frac{g_n^2}{4\pi\alpha}\frac{s}{s-m_n^2}.
\eq
$C_{T}(m,n)$ and $C_{FB}(m,n)$ contain the vector and axial vector couplings
of the gauge boson $Z_n$ to the fermion $f$, $v_f(n)$ and $a_f(n)$, 
and the helicities of the initial ($\lambda_+,\lambda_-$) and final ($h_+,
h_-$) fermions,
\ba
\label{tfb}
C_{T}(m,n;\lambda_1,\lambda_2,h_1,h_2) &=& 
C_T(\mbox{initial\ fermions})\times C_T(\mbox{final\ fermions}),\nll
C_T(\mbox{initial\ fermions})&=&\lambda_1[ v_e(m)v_e^*(n)+a_e(m)a_e^*(n)]+
\lambda_2[ v_e(m)a_e^*(n)  + a_e(m)v_e^*(n)],\nll
C_{T}(\mbox{final\ fermions}) &=& h_1[ v_f(m)v_f^*(n)  + a_f(m)a_f^*(n)]+
h_2[ v_f(m)a_f^*(n)  + a_f(m)v_f^*(n)],\nll
C_{FB}(m,n;\lambda_1,\lambda_2,h_1,h_2) &=& 
C_{T}(m,n;\lambda_2,\lambda_1,h_2,h_1)
\ea
with
\bq
\label{lambdadef}
\begin{array}{rclrcl}
\lambda_1&=&1-\lambda_+\lambda_-, & \lambda_2&=&\lambda_+-\lambda_-,\nll
h_1&=&\frac{1}{4}(1-h_+h_-), & h_2&=&\frac{1}{4}(h_+-h_-).
\end{array}
\eq
The unpolarized case corresponds to $\lambda_+=\lambda_-=h_+=h_-=0$,
i.e. $\lambda_1=h_1=1$ and $\lambda_2=h_2=0$.
The couplings of the vector bosons to fermions are given in section
\ref{veccoup}. 

The left--right cross sections are defined as
\ba
\label{lrfbdef}
\sigma_{LR} &=&
\sigma_T(\lambda_-=-1,\lambda_+=0)-\sigma_T(\lambda_-=+1,\lambda_+=0),\nll
\sigma_{LR,FB} &=&
\sigma_{FB}(\lambda_-=-1,\lambda_+=0)-\sigma_{FB}(\lambda_-=+1,\lambda_+=0),
\ea
where a summation over the polarizations of the final states is assumed.
The cross section
\bq
\label{poldef}
\sigma_{pol} =
\sigma_T(h_-=-1,h_+=0)-\sigma_T(h_-=+1,h_+=0)
\eq
is useful if the polarization of the final state can be measured.

The differential cross section can be calculated from $\sigma_T$ and
$\sigma_{FB}$ as
\bq
\frac{d\sigma}{dc} =\frac{3}{8}(1+c^2)\cdot \sigma_T +
\frac{1}{2}c\cdot\sigma_{FB}. 
\eq

We would like to mention here that cross sections depending on
transverse polarizations can be considered if the polarization of
the final state is measurable or if transverse beam polarization is
available, see \cite{tsai} for a general discussion and
\cite{npb379} for applications to LEP~1.
Simple transverse asymmetries are suppressed as $m_f/\sqrt{s}$, where
$m_f$ is the mass of the polarized particle, while double transverse
asymmetries don't have such a suppression.
We do not consider these potentially interesting observables here
because they suffer from experimental difficulties.

For $b$-- and top quark production, the finite fermion mass $m_f$ must
be taken into account in the cross section \req{tfb},
\ba
\label{bornm}
C_T(\mbox{final\ fermions})\! &=&  
[v_f(m)v_f^*(n) + a_f(m)a_f^*(n)]\left[
1+\frac{1}{2}h_1+\frac{\beta^2}{2}(-2+h_1)\right]\beta\\
&&-a_f(m)a_f^*(n)\frac{1}{2} (2+h_1)(1-\beta^2)\beta
  + [v_f(m)a_f^*(n) + a_f(m)v_f^*(n)]h_2\beta^2,\nll
C_{FB}({\rm final\ fermions})\! &=&  
[v_f(m)a_f^*(n) + a_f(m)v_f^*(n)]h_1\beta^2 +
v_f(m)v_f^*(n)h_2\beta -  a_f(m)a_f^*(n)h_2\beta^3.\nonumber
\ea
We have $\mu^2=4m_f^2/s$ and $\beta = \sqrt{1 - \mu^2}$.
%
\subsubsection{$ZZ'$ mixing}\label{zpeeff13}
The effect of $ZZ'$ mixing is already included in the couplings
$v_f(n),a_f(n)$, $n=1, 2$ by definition \req{vf1}.
One can rewrite equation \req{tfb} in terms of the unmixed couplings
and form factors $x_f, y_f$, which take into account the mixing,
\ba
\label{mixrepl}
v_f(1)&\rightarrow& a_f(1) \left[1-4|Q^f|s^2_W(1-x_f)\right],\nll
a_f(1)&\rightarrow& a_f,\nll
\frac{\pi\alpha}{2s^2_Wc^2_W}&\rightarrow&
\frac{\pi\alpha}{2s^2_Wc^2_W}(1-y_e)(1-y_f).
\ea
This formalism is very useful at the $Z_1$ peak where $ZZ'$ mixing
and electroweak corrections must be described simultaneously \cite{zefit}.
The replacements \req{mixrepl} must also be applied to the couplings
in the total width $\Gamma_1$ appearing in the propagator.

\subsubsection{Observables}\label{zpeeff14}
Helicity conservation allows non-zero cross sections for only the 
four different spin configurations
$\lambda_+=-\lambda_-=\pm 1$ and $h_+=-h_-=\pm 1$.
Furthermore, we see from equation \req{tfb} that the forward--backward 
cross section $\sigma_{FB}$ is uniquely related to the total cross section
$\sigma_T$. 
Therefore, only four of the cross sections and asymmetries, which one can
construct from $\sigma_T^f$ and $\sigma_{FB}^f$, are independent.
An example of independent observables are the total cross section 
and simple asymmetries,
\ba
\label{observables1}
\sigma_T^f && \mbox{total\ cross\ section},\nll
A_{FB}^f=\frac{\sigma_{FB}^f}{\sigma_T^f}
&&\mbox{forward--backward\ asymmetry},\nll
A_{LR}^f=\frac{\sigma_{LR}^f}{\sigma_T^f}
&&\mbox{left--right\ asymmetry},\nll
A_{pol}^f=\frac{\sigma_{pol}^f}{\sigma_T^f}
&&\mbox{polarization\ asymmetry\ of\ the\ final\ state}.
\ea
They can be constructed for every final fermion flavour $f$.

Not all observables listed above are measurable. 
All four observables require the detection of the flavour $f$.
Forward--backward asymmetries require discrimination between particles
and antiparticles.
Left--right asymmetries require beam polarization.
$A_{pol}^f$ requires a measurement of the polarization of one final particle.

One can measure combined asymmetries for some final states.
\ba
\label{observables2}
A_{LR,FB}^f&&\mbox{combined\ left--right\ forward--backward\ asymmetry},\nll
A_{pol,FB}^f&&\mbox{combined\ polarization\ forward--backward\ asymmetry},\nll
A_{LR,pol}^f&&\mbox{combined\ left--right\ polarization\ asymmetry}.
\ea
As mentioned above, the combined asymmetries must be related to simple
asymmetries, 
\bq
\label{observables3}
A_{LR,pol}^f=\frac{4}{3}A_{FB}^f,\ \ \ 
A_{pol,FB}^f=\frac{3}{4}A_{LR}^f,\ \ \ 
A_{LR,FB}^f=\frac{3}{4}A_{pol}^f.
\eq
Assuming lepton universality, we find a further relation between
observables with leptons in the final state,
\bq
\label{observables4}
A_{LR}^l=A_{pol}^l.
\eq

Of course, the relations \req{observables3} and \req{observables4}
do not hold for $m_f\neq 0$. 
They are modified by radiative corrections.
However, as we will see in the next section, they are a good 
approximation for light final state fermions
and radiative corrections with appropriate kinematic cuts.

In a theoretical analysis, the additional information due to
measurements of several related observables as
 $A_{LR}^l$ and $A_{pol}^\tau$ 
could be taken into account by the inclusion of only {\it one}
of these observables but with a smaller effective error, see equation
\req{comberror}. 

Optimal observables \cite{prd457,zpc62} can be constructed to enhance the
sensitivity to $Z'$ parameters.

Two parameters, the axial vector and the vector coupling of the $Z'$ to
charged {\it leptons}, $a'_l, v'_l$, are measured by three independent
observables.  
Contradictory signals in all three observables could disprove a $Z'$
as the origin of these signals. 
A consistent measurement of non-zero $a'_l$ or $v'_l$ allows the measurement of
couplings to the other fermions $f$ by the four additional independent
observables \req{observables1}.
Therefore, two additional relations between the four observables must
be fulfilled if the interaction is due to a $Z'$.
At the $Z'$ peak, they are $A_{FB}^f = \frac{3}{4} A_{pol}^f A_{LR}^e$
and $A_{LR}^f = A_{LR}^e$ for $m_f=0$ in the Born approximation.
Again, the hypothesis of a new vector boson could be disproved.
If the deviation turns out to be inconsistent with $Z'$ interactions, one
can try to describe the new physics in the more general framework
of four fermion contact interactions \req{contact}.

To eliminate the systematic errors from the luminosity measurement
and flavour detection and to reduce the sensitivity to radiative
corrections, ratios of  total cross sections are considered as well as
observables depending on the sum of all light 5 quark flavours,
\bq
\label{ohad}
R^{had}=\frac{\sigma_T^{u+d+s+c+b}}{\sigma_T^\mu},\ \ \ 
A_{LR}^{had}=A_{LR}^{u+d+s+c+b},\ \ \ 
R_b=\frac{\sigma_T^b}{\sigma_T^{u+d+s+c+b}},\ \ \ 
R_c=\frac{\sigma_T^c}{\sigma_T^{u+d+s+c+b}}.
\eq
Similar observables are defined by the sum over all lepton flavours 
(considering only the $s$--channel contribution for final electrons). 
We denote them with the flavour index $l$.

At the peak of a gauge boson $Z_n$, the partial decay widths 
$\Gamma_n^f=\Gamma(Z_n\rightarrow f\bar f)$ 
are important additional observables.

As mentioned before, we do not consider asymmetries involving
transverse polarizations.
There is one potentially interesting observable, the $P$ and $T$ odd
transverse-normal spin correlation, which is proportional to the
imaginary part of the product of the propagators of the exchanged
gauge bosons. 
This observable gives bounds on $M_2$ from measurements at the $Z_1$
peak, which are not suppressed by the factor $\Gamma_1/M_1$.
Unfortunately, this potential sensitivity is completely killed by the
loss of statistics in the measurements of the two transverse
correlations of the final $\tau's$ \cite{plb280}.
Contributions proportional to the imaginary part of the $Z_1$
propagator are suppressed as $\Gamma_1/M_1$ off the $Z_1$ peak.
\subsection{Radiative corrections}\label{zpeeff2}

All observables entering the $Z'$ search must be predicted with 
theoretical errors smaller than the expected experimental error. 
This demands the inclusion of radiative corrections.
Fortunately, not all radiative corrections are of equal importance.
This allows simplifying approximations.
The $O(\alpha)$ corrections to the SM process are presented in
references \cite{npb160,convolt,ffoalf,whfort,666}, for a review see
\cite{yellowbibel}. 

\subsubsection{QED corrections}\label{zpeeff21}
Among the complete $O(\alpha)$ corrections to
$e^+e^- \longrightarrow (\gamma,Z,Z',\dots) \longrightarrow f\bar f,$
the numerically largest QED corrections are a gauge invariant subset.
Furthermore, initial state corrections, final state corrections, and the
interference between them are separately gauge invariant.
The QED corrections can be calculated in a model independent way. 
They depend on the kinematics as the scattering angles and the energies
of all final particles.
We focus here on QED corrections to light fermion production. 

The final state corrections and the interference between initial and
final state corrections to the new $Z'$ interferences can be obtained from
the SM result. 
The initial state corrections to the $Z'$ interferences are
calculated in \cite{zet} and \cite{zefb} for massless
and in \cite{inim} for massive final fermions.

We discuss here the initial state radiation because it is of
major importance for $Z'$ tagging.
Initial state corrections can be calculated in the structure function
approach \cite{kuraev},
\bq
\label{sfapproach}
\sigma^{ISR}(s) 
=\hspace{-0.1cm}\int_{x_1^-}^1{\rm d}x_1\int_{x_2^-}^1{\rm d}x_2
                  D(x_1,s)D(x_2,s)\sigma^0(x_1x_2s).
\eq
The structure function approach assumes that the colliding electron
and positron have energies degraded by radiated photons,
which is described by the structure function $D(x,s)$.
The structure function is independent of the particular observable. 
To calculate the corrections to distributions, one has to boost the
final particle pair to the laboratory system for every choice of $x_1$
and $x_2$. 
This is easy in Monte Carlo algorithms but impossible in analytical
calculations.  

Alternatively, initial state corrections can be calculated in the flux
function approach, 
\bq
\label{convol}
\sigma_{A}^{ISR}(s) = \left[\nobodyfrac 1 + S(\epsilon)\right]\sigma_A^0(s)
+\int_{\epsilon}^{\Delta}dv\;\sigma_{A}^0\left(\nobodyfrac s(1-v)\right)\;
H_{A}^e(v).
\eq
The flux functions $H_{A}^e(v)$ depend on the particular observable. 
They describe the probability of the emission of a photon with a
certain energy fraction. 
$v$ is the energy of the emitted photon in units of the beam energy.
For $\sigma_T, \sigma_{FB}$ and $d\sigma/d c$ they can
be found in references \cite{convolt}, \cite{softexp} and \cite{666}.
To order $O(\alpha)$, we have
\ba
\label{hard}
H_{T}^e(v) &=& \bar{H}_{T}^e(v) + \frac{\beta_e}{v} = 
\beta_e\;\frac{1+(1-v)^2}{2v},\nll
H_{FB}^e(v) &=& \bar{H}_{FB}^e(v) + \frac{\beta_e}{v} = 
\frac{\alpha}{\pi}\;(Q^e)^2\;\frac{1+(1-v)^2}{v}\;\frac{1-v}
{(1-\frac{v}{2})^2}\;\left[ L_e - 1 -\ln\frac{1-v}{(1-\frac{v}{2})^2}
 \right],\nll
\beta_e &=& \frac{2\alpha}{\pi}\;(Q^e)^2\;(L_e-1),\ \ \
L_e = \ln\frac{s}{m_e^2},\ \ \ Q^e = -1.
\ea

The quantity $[1 + S(\epsilon)]$ in equation \req{convol}
describes the Born term plus corrections due to soft  and virtual photons. 
To order $O(\alpha)$, we have
\ba
\label{soft}
S(\epsilon) = \bar{S} + \beta_e \ln\epsilon = 
\beta_e\;\left(\ln\epsilon + \frac{3}{4}\right) + \frac{\alpha}
{\pi}\;(Q^e)^2\;\left(\frac{\pi^2}{3}-\frac{1}{2}\right).
\ea

The structure function and the flux function approach give an
equivalent description of QED corrections. 
See references \cite{lep2ww,vandermark} for an extensive discussion and
further references.

Starting from the convolution \req{convol}, the origin of the
radiative tail and its magnitude can be estimated.
We do this ignoring details of the radiator functions $S(\epsilon)$ and
$H_A^e(v)$. 
The $s'$ dependence of the Born cross section is
\bq
\label{decomp}
\sigma_A\approx\frac{1}{s'}\chi_m(s')\;\chi_n^*(s')
=\frac{s}{m_n^{*2}-m_m^2}\frac{1}{s}
\left[\frac{s'}{s'-m_n^{*2}}-\frac{s'}{s'-m_m^2}\right].
\eq
For $m=n$, the first factor  of the last expression becomes
\bq
\frac{s}{m_n^{*2}-m_n^2}=-\frac{i}{2}\frac{M_n}{\Gamma_n}\frac{s}{M_n^2}.
\eq
This imaginary quantity must be met by another imaginary multiplier
to give contributions to the cross section.
Because we average over transverse polarizations, it can arise only
from the $v$ integration \req{convol} over the remaining factors of equation 
\req{decomp}.
Keeping only the relevant term after partial fraction
decomposition, one gets
\bq
\label{radtail}
\sigma_T^{ISR}(s)-\sigma_T^0(s)\approx
\frac{i}{2}\frac{M_n}{\Gamma_n}\frac{s}{M_n^2}\frac{M_n^2}{s}
\int_0^\Delta d v\;\frac{1}{1-v-m_n^{*2}/s}
\approx\frac{i}{2}\frac{M_n}{\Gamma_n}
\ln\frac{m_n^{*2}/s-1+\Delta}{m_n^{*2}/s-1}.
\eq
The real part of the argument of the logarithm  is negative for
 $s>M_n^2$ and $\Delta>1-M_n^2/s$.
These are the necessary conditions for the development of the
radiative tail; the center--of--mass energy must be larger then
the mass of the resonance, and the radiation of
photons must be allowed, which are sufficiently hard to ensure a
``radiative return'' to the resonance.

\begin{figure}[tbh]
\ \vspace{1cm}\\
\begin{minipage}[t]{7.8cm} {
\begin{center}
\hspace{-1.7cm}
\mbox{
\epsfysize=7.0cm
\epsffile[0 0 500 500]{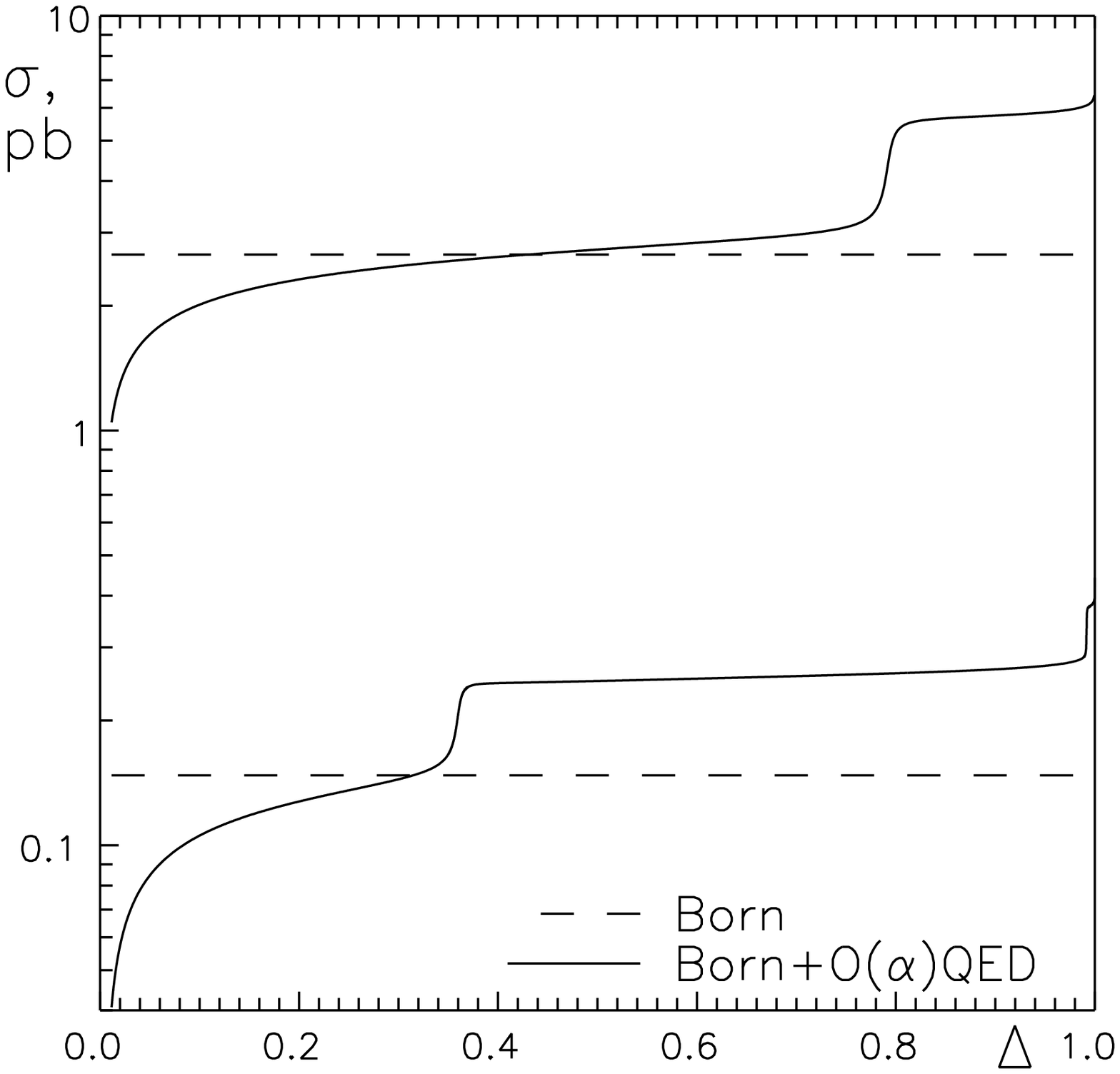}%
}
\end{center}
\vspace*{-0.5cm}
\noindent
{\small\it
\begin{fig} \label{sigmadelta} 
The total cross section $\sigma_T^\mu$ as function of the cut
on the photon energy $\Delta$ in units of the beam energy 
for $M_{Z'}=M_\eta=800\,GeV$. The upper 
(lower) set of curves corresponds to $\sqrt{s}=200(1000)\,GeV$.
This is figure~1 from reference~\cite{leikeustron}.
\end{fig}}
}\end{minipage}
\hspace*{0.5cm}
\begin{minipage}[t]{7.8cm} {
\begin{center}
\hspace{-1.7cm}
\mbox{
\epsfysize=7.0cm
\epsffile[0 0 500 500]{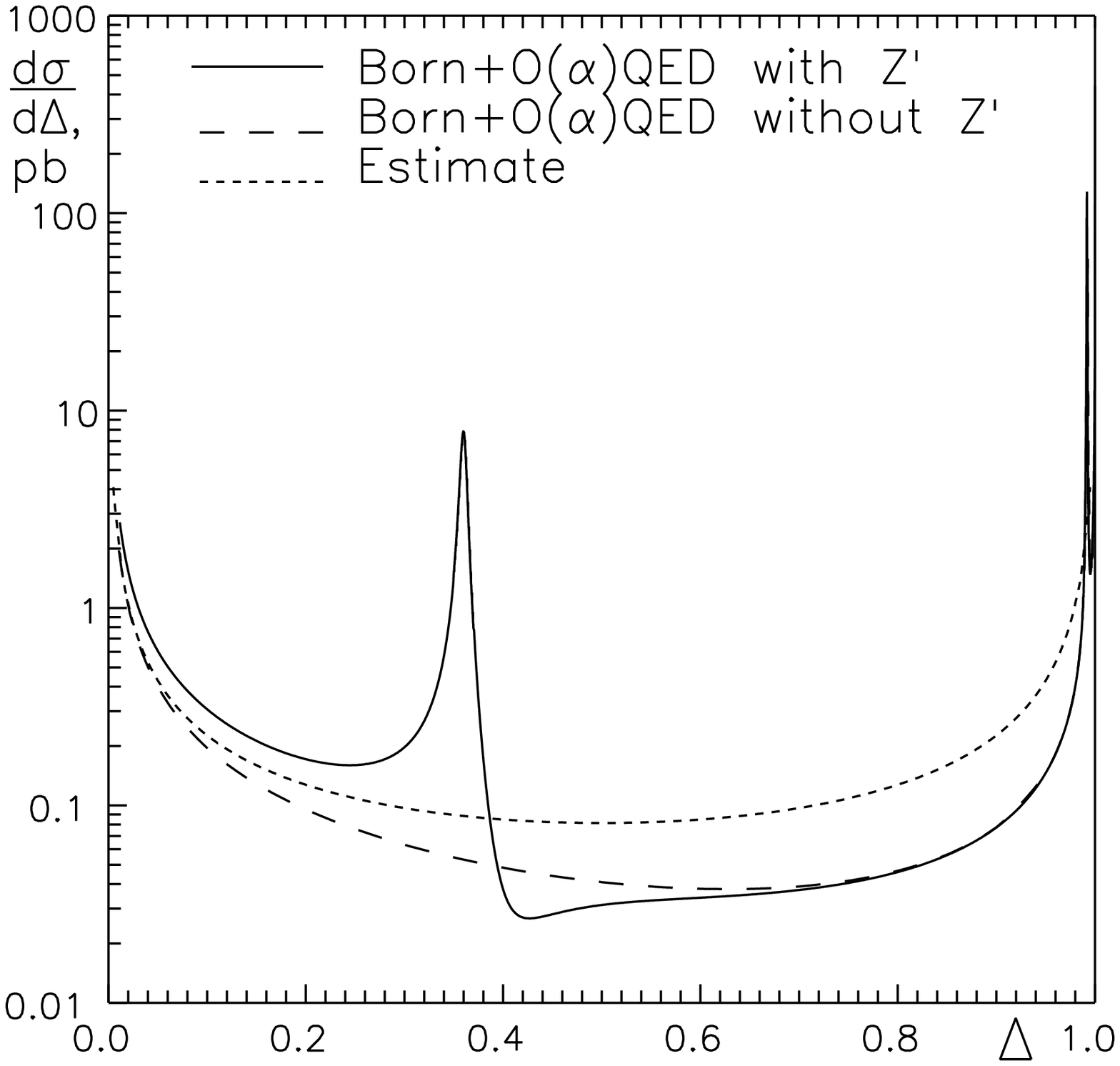}
}
\end{center}
\vspace*{-0.5cm}
\noindent
{\small\it
\begin{fig} \label{dsigmaddelta} 
$d\sigma_T^\mu/d\Delta$ as function of the photon energy $\Delta$ in
units of the beam energy. 
For the solid line, we choose $M_{Z'}=M_\eta=800\,GeV$ and
$\sqrt{s}=1000\,GeV$. 
The dashed line is the SM.
The dotted line is the function 
$\sigma_T^{\mu 0}\beta_e/[\Delta(1-\Delta)]$, compare estimate \req{ngasm}.
\end{fig}}
}\end{minipage}
\end{figure}

We now estimate the magnitude of the radiative tail by restoring the
missing factors, 
\bq
\label{radtail2}
\mbox{rad.\ tail\ }\approx \sigma_T^{ISR}(s)-\sigma_T^0(s)\approx
\sigma_T^0(s;n,n)\cdot\beta_e\frac{\pi}{2}\frac{M_n}{\Gamma_n}.
\eq
Only the contribution of the exchange of the vector boson $n$ appears in
equation \req{radtail2}.
Therefore, the other interferences are not enhanced by the radiative tail.
Putting $s_W^2=\frac{1}{4}$ and $M_n/\Gamma_n=M_1/\Gamma_1$, one gets 
$\sigma_T^l(s;0,1)= 0$ and $\sigma_T^l(s;1,1)/\sigma_T^l(s;0,0)= 1/9$
in the limit $s\gg M_Z^2$. 
This gives
\bq
\mbox{rad.\ tail\ }\approx 7\cdot\sigma_T^l(s;1,1)\approx 7/10\cdot\sigma_T^l,
\eq
which is in reasonable agreement with the exact calculation and with
figure~\ref{sigmadelta}.  
For $b$ quark production, where the $Z$ boson exchange
$\sigma_T^b(s;1,1)$ dominates over the photon exchange $\sigma_T^b(s;0,0)$,
the effect of the radiative tail is much more pronounced.

The radiative tail enhances SM cross sections, while  for
$M_{Z'}>\sqrt{s}$ the $Z'$ signal is not enhanced.
Therefore, the radiative tail must be removed for a $Z'$ search below
the $Z'$ resonance.
This can be done by removing events with hard photons by demanding
$\Delta<1-M_1^2/s$. 

The dependence of the cross section on $\Delta$
is shown in figure~\ref{sigmadelta} for two different energies. 
The upper curve corresponds to an energy above the $Z$ peak but below
the $Z'$ peak, the lower curves to an energy above the $Z$ and $Z'$ peaks.
One recognizes the step--like behaviour for photon energies where the
radiative tail(s) are ``switched on''. 
We see that the radiatively corrected cross section is numerically 
similar to the Born prediction only for a certain cut, 
which rejects all hard photons from the radiative returns to resonances.
This is the reason why $Z'$ analyses at the Born level give limits,
which are numerically similar to those obtained with radiative
corrections and appropriate kinematic cuts. 
Of course, radiative corrections must be included in fits to real data.

The radiative tail is due to the emission of photons with energies
$E_\gamma$ in the (narrow) interval
\bq
\label{deltalim}
\Delta^-= 1-\frac{M_n^2}{s}-\frac{M_n\Gamma_n}{s} < \frac{E_\gamma}{E_{beam}}
< 1-\frac{M_n^2}{s}+\frac{M_n\Gamma_n}{s} =\Delta^+.
\eq
Therefore, experiments with energies above resonances have sharp
peaks in the photon energy spectrum.
This is illustrated in figure~\ref{dsigmaddelta}.
There, the $Z_1$ and $Z_2$ resonances are represented by peaks 
at $E_\gamma/E_{beam}=\Delta_n=1-M_n^2/s$ with $\Delta_1=0.992$ and
$\Delta_2=0.36$. 
All final states $f\bar f$ contribute to the peaks. 
Their heights and widths  depend on the width of the related gauge
boson as indicated in equations \req{radtail} and \req{deltalim}.

\subsubsection{Weak corrections}\label{zpeeff22}
The precision of present and future $e^+e^-$ colliders is high enough
to be sensitive to weak corrections.
They can be implemented by form factors
\cite{zfitter,formfactors,formbard} applying the following formal
replacements of the coupling constants in the Born cross section,
\ba
\label{weakrepl}
v_f(0)&\rightarrow& v_f(0)F_A(q^2),\nll
v_e(1)v_f(1)&\rightarrow& a_e(1) a_f(1)
\left[\nobodyfrac 1-4|Q^e|s^2_W\kappa_e
-4|Q^f|s^2_W\kappa_f + 16|Q^eQ^f|s^4_W\kappa_{ef}\right],\nll
v_e(1)&\rightarrow& a_e(1)\left[\nobodyfrac 1-4|Q^e|s^2_W\kappa_e\right],\nll
v_f(1)&\rightarrow& a_f(1)\left[\nobodyfrac 1-4|Q^f|s^2_W\kappa_f\right],\nll
a_e(1),a_f(1)&\rightarrow&\mbox{unchanged},\nll
g_1^2=\frac{4\pi\alpha}{2s^2_Wc^2_W}&\rightarrow&
\sqrt{2}G_\mu M_Z^2\rho_{ef}.
\ea
The complex functions $\rho_{ef},\kappa_e,\kappa_f$ and $\kappa_{ef}$
contain all information about the SM weak corrections, while the complex
function $F_A(q^2)$ takes into account the effects of the vacuum
polarization of the photon. Although the largest contribution to
$F_A(q^2)$ comes from QED, we feel that it should be mentioned in this
section. 
If the new functions are set to one, we recover the Born formulae.
$\kappa_{ef}$ is different from $\kappa_{e}\kappa_{f}$ due to box
contributions, which enforce the additional replacement rule for the
combination $v_e(1)v_f(1)$. 
The conjugated couplings in the Born cross section have
to be replaced with the corresponding complex conjugated relations
\req{weakrepl}. 
The functions $\kappa_e,\kappa_f$ and
$\kappa_{ef}$ can be  absorbed in effective Weinberg angles.

$ZZ'$ mixing effects and weak corrections have to be treated
simultaneously for predictions at the $Z_1$ peak. 
According to equations \req{mixrepl} and \req{fourfampl6}, it can be
done \cite{zefit} by the following  replacements in equation \req{weakrepl},
\ba
\label{mixrepl2}
\kappa_f&\rightarrow& \kappa_f^M=\kappa_f(1-x_f),\nll
\kappa_{ef} &\rightarrow& \kappa_{ef}^M=\kappa_{ef}(1-x_e)(1-x_f),\nll
\rho_{ef} &\rightarrow& \rho_{ef}^M=\rho_{ef}\rho_{mix}(1-y_e)(1-y_f),\nll
M_Z^2&\rightarrow& M_1^2.
\ea
In \req{mixrepl2}, we have neglected terms, which are proportional to
the SM weak corrections times the mixing angle $\theta_M$.
The multiplier $\rho_{mix}=M_Z^2/M_1^2$ takes into account that the
replacement \req{weakrepl} of $g_1^2$ is valid only for the mass of
the symmetry eigenstate $M_Z$.
Alternatively, one could calculate 
the form factors \req{mixrepl2} by the sum rule
\req{fourfampl5} where $\kappa_f^1,\kappa_{ef}^1$ and $\rho_{ef}^1$
are the weak form factors $\kappa_f,\kappa_{ef}$ and $\rho_{ef}$ and
$\kappa_f^2,\kappa_{ef}^2$ and $\rho_{ef}^2$ are the mixing form
factors $\kappa_f^m,\kappa_{ef}^m$ and $\rho_{ef}^m$ given in equation
\req{formmix}. 
The replacements \req{mixrepl2} must also
be made to the $Z_1$ width entering the propagator of the mass eigenstate.

Non-standard one loop corrections cannot be calculated without knowledge
about the underlying theory.
If the $Z'$ is the first signal of physics beyond the SM, these
corrections will probably not play an important role.
They are expected to be a small correction to a small deviation from
the SM prediction.

See references \cite{pacomix,prd40} for the renormalization of $SU(2)_L\times
U(1)_Y\times U'(1)$ gauge theories.
\subsubsection{QCD corrections}\label{zpeeff23}
QCD corrections have to be taken into account in the case of hadronic 
final states.
They don't feel the gauge boson exchanged before. 
Therefore, the known SM results for massless \cite{plb259}  and massive
\cite{plb248,jlzabl,koerner} final state fermions can be used.
The $O(\alpha_s)$ QCD corrections can be obtained from the $O(\alpha)$ final
state QED corrections by the 
replacement $\alpha\rightarrow \frac{4}{3}\alpha_s$. 
They depend on the mass of the final state fermion and on the maximal
allowed energy $E_g=\Delta_g\sqrt{s}/2$ of the radiated gluon.
For $m_f=0$ and $\Delta_g=1$, the lowest order QCD corrections vanish for 
forward--backward asymmetries and reduce to the well known factor 
$1+\alpha_s/\pi$ for total cross sections.

As known from SM calculations, radiative corrections due to spin
asymmetries of the final state need special care. 
Due to spin-flip contributions induced by gluon radiation, the result
of a calculation with zero fermion masses $m_f=0$ does not coincide with the
result calculated for $m_f\neq 0$ in the limit $m_f\rightarrow 0$
\cite{koernerm0}. 
The problem also arises in final state QED corrections, however, it is
numerically less important because $\alpha$ is numerically smaller
than $\alpha_s$. 
\subsubsection{Corrections to $\Gamma_1$}\label{zpeeff24}
As is known from the experiments at LEP and SLC, the measurements on
the $Z_1$ resonance are sensitive to radiative corrections to the
$Z_1$ width.
In contrast to the corrections to $\Gamma_2$ described in section
\ref{gamma2rc}, we give here the formulae including the weak corrections to
$\Gamma_1$,
\bq
\label{z1width}
\Gamma_1^f = \frac{N_fG_\mu\sqrt{2}M_1^3}{12\pi}\rho_f^Z\mu
R_{\rm QED}R_{\rm QCD}(M_1^2)\left\{\left[ v_f(1)^2+a_f(1)^2\right]
\left( 1+2\frac{m_f^2}{M_1^2}\right) - 6a_f(1)^2 \frac{m_f^2}{M_1^2}\right\}.
\eq
The functions $R_{\rm QED}$ and $R_{\rm QCD}$ describing the QED and QCD
corrections \cite{plb259} are given in equation \req{gffcorr}.

The function $\rho_f^Z$ absorbs $\Delta r$ arising during the
replacement of the weak coupling constant by the muon decay constant,
\bq
\label{g2repl}
g_1^2=\frac{4\pi\alpha}{s^2_Wc^2_W}=
\frac{G_\mu\sqrt{2}M_Z^2}{1-\Delta r}
=\frac{G_\mu\sqrt{2}M_W^2}{(1-\Delta r)c^2_W}.
\eq
As in $e^+e^-\rightarrow f\bar f$, the remaining weak corrections can
be taken into account \cite{npb276} by a replacement of the vector couplings,
\bq
v_f(1)\rightarrow a_f(1)\left[1-4|Q^f|s^2_W\kappa_f\right].
\eq
The functions $\kappa_f$ are often absorbed in $s^2_W$ defining
effective Weinberg angles.
In comparison to the weak corrections to $e^+e^-\rightarrow f\bar f$,
we have no box contributions in the $Z_1$ decay. Therefore, a special
replacement rule \req{weakrepl} for the product $v_e(1)v_f(1)$ is absent.

In the case of a non-zero $ZZ'$ mixing, we have to apply additional
replacements similar to the rule \req{mixrepl2}, 
\ba
\label{mixrepl3}
\rho_Z^f &\rightarrow& \rho_Z^f\rho_{mix}(1-y_f)^2,\nll
\kappa_f&\rightarrow& \kappa_f=\kappa_f(1-x_f).
\ea
\subsection{$Z'$ constraints at $s \approx M_1^2$}\label{zpeeff3}
The $Z'$ signal in measurements at the $Z_1$ peak is a deviation of the 
couplings of the mass eigenstate $Z_1$ to fermions from the SM prediction.
This prediction depends on the SM parameters, which are also defined by
measurements at the $Z_1$ peak.
If one wants to constrain $Z'$ parameters from the same data,
the cleanest analysis would be a simultaneous fit of SM and
$Z'$ parameters. We call this procedure {\it direct} $Z'$ analysis.

Alternatively, the data distributed around the $Z_1$ peak can be fitted
first in a model independent procedure.  
The results of such a fit are $M_1$, partial and 
total decay widths $\Gamma_1^f$ and cross sections and asymmetries at the peak.
This output is the input of a second fit, which determines the SM and 
$Z'$ parameters. 
We call this procedure {\it indirect} $Z'$ analysis.
The first direct analysis was done in \cite{zefit}. 
The indirect analyses are shown to agree with this direct analysis.

As mentioned in the introduction, we will not consider the mixing of
the new fermions with SM fermions.
See reference \cite{nardi} for such an analysis based on LEP\,1 data.

In the following subsections, we discuss different $Z'$ constraints. 
We always start with a derivation of a simple estimate.
This estimate shows the scaling of the constraint with different
parameters of the experiment as integrated luminosity,
center--of--mass energy and systematic errors. 
We then discuss present constraints and possible constraints from
future experiments and prove the quality of our estimates.
\subsubsection{Model independent constraints on $v_f^M$ and $a_f^M$}
\paragraph*{Estimate}
The quantities $v_f^M$ and $a_f^M$ defined in equation \req{coupmdef} 
can be constrained at the $Z_1$ peak independently of the $Z'$ model.

Consider the partial decay width and different asymmetries at the $Z_1$ peak,
\ba
\label{peakobs}
\Gamma_1^f &=& M_1\frac{g^2}{12\pi}\left[\nobodyfrac v_f^2(1)+a_f^2(1)\right]
N_f\approx 
\Gamma_1^{f0}\left\{1+2\frac{v_fv_f^M+a_fa_f^M}{g_1(v_f^2+a_f^2)}\right\}
=\Gamma_1^{f0}+\Delta^{Z'}\Gamma_1^f,\nll
A_{FB}^f &=& \frac{3}{4}A_eA_f
\approx A_{FB}^{f0} + \frac{3}{4}A_f^0\Delta A_e +\frac{3}{4}A_e^0\Delta A_f
=A_{FB}^{f0}+\Delta^{Z'}A_{FB}^{f},\nll
A_{LR}^f &=& A_e \approx A_e^0 +\Delta A_e=A_{LR}^{0}+\Delta^{Z'}A_{LR},\nll
A_{pol}^f &=& \frac{4}{3}A_{LR,FB}=A_f \approx A_f^0 +\Delta A_f
=A_{pol}^{f0}+\Delta^{Z'}A_{pol}^f
\ea
\bq
\label{mconstr0}
\mbox{with\ }A_f\equiv\frac{2a_f(1)v_f(1)}{a_f(1)^2 + v_f(1)^2},\mbox{\ and\ } 
\Delta A_f \approx 2\frac{v_fa_f^M+a_fv_f^M}{g_1(v_f^2+a_f^2)}-
4\frac{a_f^2v_fa_f^M+v_f^2a_fv_f^M}{g_1(v_f^2+a_f^2)^2}.
\eq
The index zero denotes the observables without mixing.

If the deviation $\Delta^{Z'}O_i$ in the observable $O_i$ is larger
than the experimental error $\Delta O_i$, 
\bq
\label{mconstr}
\Delta^{Z'}\Gamma_1^f > \Delta\Gamma_1^f,\ \ \ 
\Delta^{Z'}A_{FB}^f > \Delta A_{FB}^f,\ \ \ 
\Delta^{Z'}A_{LR}^f > \Delta A_{LR}^f,\ \ \ 
\Delta^{Z'}A_{pol}^f > \Delta A_{pol}^f,
\eq
one can see a signal.
The relations \req{mconstr} and \req{fourfampl6} predict that the
different observables are blind to $Z'$ models predicting $v_f^M$ and
$a_f^M$ between two parallel lines.
Neglecting systematic errors, the $Z'$ constraints \req{mconstr} 
scale like $1/\sqrt{L}$. 
For $f=e, \nu$ and $s_W^2=\frac{1}{4}$, equations \req{peakobs} and
\req{mconstr} transform to
\bq
\label{estimavm}
|a_e^M| > \frac{g_1}{4}\frac{\Delta\Gamma_1^e}{\Gamma_1^{e0}},\ \ \ 
|a_\nu^M+v_\nu^M| > \frac{g_1}{2}\frac{\Delta\Gamma_1^\nu}{\Gamma_1^{\nu0}},
\ \ \ 
|v_e^M| > \frac{g_1}{6}\frac{\Delta A_{FB}^e}{A_e^0},\ \ \ 
|v_e^M| > \frac{g_1}{4}\Delta A_{LR}^e.
\eq
The constraint from $A_{FB}^e$ linear in $\theta_M$ exists only for
$s_W^2\neq\frac{1}{4}$ leading to $A_e^0\neq 0$.
The equations \req{estimavm} allow an independent constraint or
measurement of $a_\nu^M+v_\nu^M=L_\nu^M$ and $L_e^M$ and therefore an
experimental check \cite{zpc53,layssac} of the relations \req{coup5}.
\paragraph*{Present constraints}\label{amvmconstr}

Model independent $Z'$ constraints based on the combined data of LEP
and SLC \cite{warschau} are discussed in references
\cite{lepzpold,zpc53,layssac,leikez1}.  
Before confronting $\Gamma_1^f,A_{FB}^f$ and $A_{LR}^f$ with the data,
radiative corrections have to be included.
This implies a substitution of $g_1^2$ with $G_\mu$ in the expression
\req{weakrepl} for $\Gamma_1^f$.
The substitution induces a dependence on $\rho_{mix}=M_Z^2/M_1^2$.
This spoils the model independent limits on $v_f^M,a_f^M$ making them
dependent on the additional $Z'$ parameter $M_2$. 

The problem is solved in references \cite{zpc53,layssac} by considering special
{\it combinations} of observables where the leading dependence on
$\rho_{ef}$ and $\rho_{mix}$ drops out. 
As a result, these special combinations are also much less sensitive to
the top and the Higgs mass.

With the recent experimental data \cite{warschau} on $M_W,G_\mu$,
$s^2_W$, $\alpha(M_Z^2)$ and $M_t$, one can follow another
procedure and calculate $\rho_{mix}$ according to equation \req{g2repl}.
One gets $\rho_{mix}=1\pm0.003$, compare \cite{hollik96}.
The main sources of the uncertainty of $\rho_{mix}$ are the
experimental error of the $M_W$ measurement and the theoretical error
of $\Delta r$ arising due to the unknown Higgs mass and to a
less extent due to the experimental error of the top mass.  
In the two--$\sigma$ errors, the experimental error in the $M_W$
measurement dominates because
the theoretical error in $\Delta r$ is not doubled.
We assumed that the symmetry is broken by one Higgs doublet only. 
In general, one would obtain different results for  extended Higgs sectors.
In this sense, the error of $\rho_{mix}$ still contains a model dependence,
which is expected to be small.
The expression of $\Gamma_1^f$ is now independent of $M_2$.
The price one has to pay is that the uncertainty of $\rho_{mix}$ must 
be added to the experimental error of $\Gamma_1^f$. 

\begin{figure}[tbh]
\ \vspace{1cm}\\
\begin{minipage}[t]{7.8cm} {
\begin{center}
\hspace{-1.7cm}
\mbox{
\epsfysize=7.0cm
\epsffile[0 0 500 500]{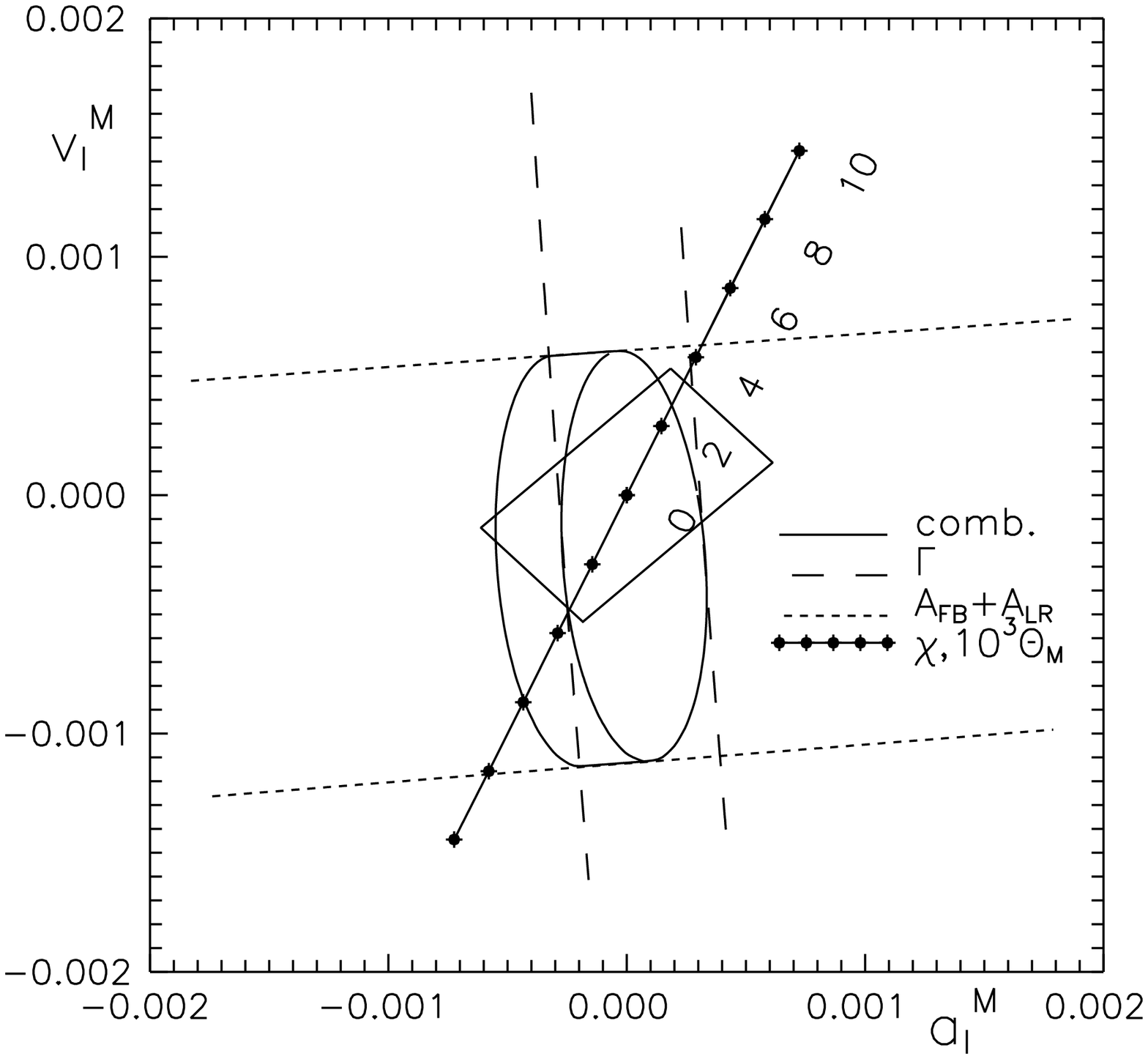}
}
\end{center}
\vspace*{-0.5cm}
\noindent
{\small\it
\begin{fig} \label{zpmix1} 
Areas of $(a_l^M,\ v_l^M)$, for which the extended gauge theory's
predictions are indistinguishable from the SM (95\% CL).
Models between the dashed (dotted) lines cannot
be detected with $\Gamma_l,$\  ($A_{FB}^l$ and $A_{LR}^l$ together). 
The regions surrounded by the solid lines cannot be resolved by
all three observables combined, see text.
The numbers at the straight line indicate the value of $\theta_M$ in
units of $10^{-3}$ for the $\chi$ model. 
The dots for $\theta_M<0$ are not labeled.
The rectangle is calculated from figure~\ref{wwfig1}.
\end{fig}}
}\end{minipage}
\hspace*{0.5cm}
\begin{minipage}[t]{7.8cm} {
\begin{center}
\hspace{-1.7cm}
\mbox{
\epsfysize=7.0cm
\epsffile[0 0 500 500]{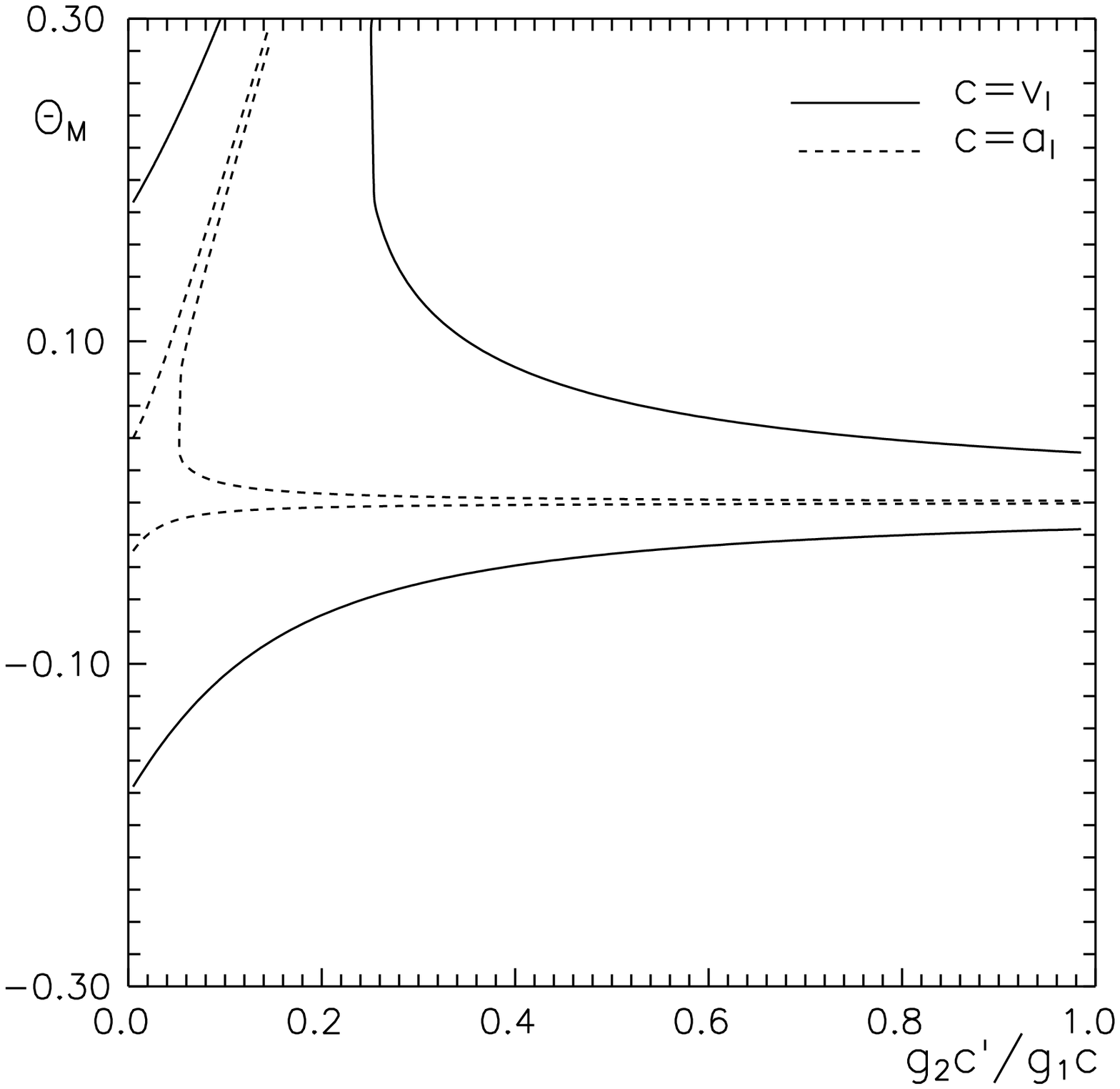}
}
\end{center}
\vspace*{-0.5cm}
\noindent
{\small\it
\begin{fig} \label{zpmix4} 
Areas of $(g_2a'_l/(g_1a_l),\theta_M)$ and $(g_2v'_l/(g_1v_l),\theta_M)$,
for which the extended gauge theory's 
predictions are indistinguishable from the SM (95\% CL).
This figure is an update of figure~4 in reference \cite{leikez1}.
\end{fig}}
}\end{minipage}
\end{figure}

Figure~\ref{zpmix1} illustrates the constraints on $a_l^M$ and $v_l^M$ obtained
from the data \cite{warschau},
$\Gamma_1^l=(83.91\pm 0.11)\,MeV,\ A_{FB}^l=0.0174\pm 0.0010,\
A_\tau =0.1401\pm 0.0067$.
In the SM, we have $A_\tau=A_{pol}^\tau=A_{LR}$.
The plotted regions correspond to a $\chi^2=\chi_{min}^2+5.99$. 
In contrast to the demonstration in reference \cite{leikez1}, we take
into account the deviations of the measurements from the theoretical
prediction in figures \ref{zpmix1} and \ref{zpmix4}.
The constraints from $\Gamma_l$ and from $A_{FB}^l$ and $A_\tau$
combined are shown  separately for $\rho_{mix}=1$.
The quantitative agreement with the estimates \req{estimavm} is very good.
The excluded regions are slightly rotated relative to the axes because
$s_W^2\neq \frac{1}{4}$.
The deviation of the experimental value from the theoretical prediction
leads to a parallel shift of the exclusion region of the corresponding
observable. 
All three observables combined cannot exclude the region inside the ellipse.
The uncertainty of $\rho_{mix}$ yields a shift of the ellipse, which
results in the larger solid region.
This shift is possible only in one direction because we always have
$M_1 < M_Z$ according to equation \req{massrel}. 
Future improved measurements of $M_W$ and $M_t$ and a determination of $M_H$
would reduce the shift.

The present data are consistent with the assumption \req{qpt3},
i.e. with $L_e^M=L_\nu^M$.
One could then interpret the constraint \req{estimavm} from the
invisible width as a constraint to $L_e^M$. 
Unfortunately, this gives no improvement to the combined region shown
in figure \ref{zpmix1}.

The {\it model independent} constraint shown in figure \ref{zpmix1}
can be interpreted as a constraint on the $ZZ'$ mixing angle $\theta_M$
for any fixed model with known couplings $a'_l$ and $v'_l$.
Varying the mixing angle, one moves on a
straight line on figure \ref{zpmix1}.
This line intersects the model
independent exclusion limit for certain values of $\theta_M$.
The values of $\theta_M$ at the intersection points define the
excluded regions of $\theta_M$.
Graphically, one obtains $-0.006<\theta_M< 0.0025$ for the $\chi$ model. 
The model independent constraints obtained from a one--parameter fit
are expected to be stronger. 

Model independent limits on $v_q^M,a_q^M$ can be obtained in a similar
procedure, see reference \cite{leikez1}. 

A global fit to LEP data would allow to constrain the 5 independent
couplings $L_l^M$, $R_e^M$, $L_q^M$, $R_u^M$ and $R_d^M$ simultaneously.
These couplings are defined in analogy to $v_f^M,a_f^M$,
\bq
\label{coupmlrdef}
L_f^M=\theta_Mg'L'_f,\ \ \ R_f^M=\theta_Mg'R'_f.
\eq
\paragraph*{Future constraints}
The present constraints on $a_e^M$ and $v_e^M$ obtained from LEP data
can be improved by future measurements of the reaction
$e^+e^-\rightarrow W^+W^-$. 
The rectangle in figure~\ref{zpmix1} is calculated from figure~1 of
reference \cite{pankovnew}. 
See section \ref{zpeeww} for details.

The constraints on $a_q^M$ and $v_q^M$ from the $Z_1$ peak cannot be
improved by measurements at $M_1^2\neq s<M_2^2$.
Future measurements at the $Z_2$ peak (if it exists) allow a separate
measurement of the couplings $a'_f,v'_f$ and of the mixing angle $\theta_M$.
\subsubsection{Model independent constraint on $\theta_M$}\label{tetmicon}
Figure~\ref{zpmix1} is independent of $\theta_M$ as far as $\theta_M$ is small.
This allows the derivation of a model independent constraint on $\theta_M$.
In the simplest approximation, where only the linear terms in
$\theta_M$ are kept in equation \req{vf1}, we obtain the estimate
\bq
\label{mitheta}
|\theta_M|<\frac{\Delta c}{c}\frac{g_1c}{g_2c'},
\mbox{\ \ \ where\ \ \ } c=a_l,v_l \mbox{\ \ and\ \ } c'=a'_l,v'_l.
\eq
$\Delta c$ is the bound on $a_l^M$ or $v_l^M$ taken from
figure~\ref{zpmix1}.  
In particular, the estimate \req{mitheta} gives $|\theta_M|<0.003$ for
$g_2a'_l/(g_1a_l)=0.62$ as it is the case in many GUT's.

The exact numerical result for $\theta_M$ as a function of
$g_2a'_l/(g_1a_l)$ is shown in figure~\ref{zpmix4}.
The approximate bound \req{mitheta} is recovered at large $g_2c'/(g_1c)$.
In contrast to \req{mitheta}, the exact calculation gives a constraint on
$\theta_M$ also for a $Z'$ with zero coupling, i.e. for $g_2c'/(g_1c)=0$.
It is $|\theta_M|<0.035$ for $c=a_l$.
This can easily be understood from equation \req{vf1} where the deviations of
the couplings $a_f(1)$ or $v_f(1)$ from $a_f$ or $v_f$ with increasing
$\theta_M$ eventually become larger than the 
experimental error even for the case $g_2=0$.

If one allows a large $ZZ'$ mixing, there is one particular $Z'$
with {\it all} couplings proportional to those of the SM $Z$ boson
{\it and} a fine-tuned overall coupling strength,
\bq
\label{gmix}
g_2c'_f=g_1c_f\frac{1-c_M}{s_M},\ c=a,v,\ f=l,c,b,
\eq
which will not produce a deviation of $c_f(1)$ from $c_f$.
The beginning of this region of insensitivity can be recognized in
figure~\ref{zpmix4}. 
Models with fine tuning \req{gmix} in all couplings can only be detected
by effects of the $Z'$ propagator.
\subsubsection{Model dependent constraint on $\theta_M$ and $M_2$}
\paragraph*{Estimate}
$\theta_M$ can be much better constrained in particular $Z'$ models
because they link the $Z'$ couplings to leptons and to quarks by model
parameters. 
Assuming $v'_f\approx v_f$ and $a'_f\approx a_f$, we obtain from 
equation \req{estimavm},
\bq
\label{thetamlim}
|\theta_M|<\theta_M^{lim}
\approx\frac{g_1}{2g_2}\frac{\Delta\Gamma_1^f}{\Gamma_1^f}.
\eq

The constraint \req{thetamlim} scales
with the integrated luminosity $L$ as
\bq
\label{thetamscale}
\theta_M^{lim} \sim \left[\frac{1+r^2}{L}\right]^{1/2}. 
\eq
$r$ is the ratio of the systematic and statistical error as defined in
equation \req{rdef}. 

Consider now the sensitivity to $M_2$.
For simplicity, we assume $\theta_M=0$ identifying $Z_2=Z'$ and
$Z_1=Z$. 
Again, we assume that the $Z$ and $Z'$ couple with 
the strengths $g_1$ and $g_2$ to SM fermions setting $v'_f\approx v_f,
a'_f\approx a_f$. 
Only the $ZZ'$ interference is important near the $Z$ resonance,
Then, we can approximate the relative shift of cross sections,
$\Delta^{Z'}O/O_{SM}=\Delta\sigma_T/\sigma_T$ by a ratio of propagators, 
\bq
\label{onres1}
\frac{\Delta^{Z'}O}{O_{SM}} \approx
\frac{g_2^2}{g_1^2}\frac{|\Re e{\chi_Z\chi^*_{Z'}}|}{|\chi_Z|^2} = 
\frac{g_2^2}{g_1^2}
\frac{s-M_Z^2}{M_{Z'}^2-s}.
\eq
The deviation $\Delta^{Z'}O$ has to be compared with the experimental
error $\Delta O$.
Choosing $\sqrt{s}=M_Z+\Gamma_Z/2$, we conclude that a $Z'$ with a mass
\bq 
\label{onres}
M_{Z'} < M_{Z'}^{lim} =
M_Z\left( 
1+\frac{g_2^2}{g_1^2}\frac{O}{\Delta O}\frac{\Gamma_Z}{M_Z}
\right)^{1/2}
\equiv 
M_Z\left( 1+\frac{1}{\Delta o}\frac{\Gamma_Z}{M_Z} \right)^{1/2},\ \ \ 
\Delta o=\frac{g_1^2}{g_2^2}\frac{\Delta O}{O}
\eq
would be detected by the observable $O$.
The factor $\Gamma_Z/M_Z$ reflects the suppression of the $ZZ'$
interference relative to the resonating contribution.

One exception is the transverse-normal spin asymmetry mentioned in section
\ref{zpeeff14}. 
It is proportional to the imaginary part of the product of the
propagators.
Hence, it has no suppression factor $\Gamma_1/M_1$.
Unfortunately, this potential sensitivity is compensated by the loss
of statistics in the measurement of this asymmetry \cite{plb280}.
The result is a net sensitivity to $M_2$ weaker than \req{onres}.
\paragraph*{Present constraints}\label{presconstr}
Measurements at the $Z_1$ resonance give the best present limits
on $ZZ'$ mixing \cite{PDB}. 

The analysis of LEP data requires the inclusion of weak corrections.
This induces a dependence of the limits on the Higgs and the top mass.
Sometimes, $\theta_M$ and $M_2$ are fitted as independent parameters, which
corresponds to a Higgs sector consisting of an arbitrary number of Higgs
doublets and singlets. 
One should note that in this case the weak corrections calculated within the
minimal SM are only approximate.
For a specified Higgs sector, $\theta_M$ and $M_2$
are related by the Higgs constraint \req{higgsconstr}.
This relation transforms the tight limits on $\theta_M$ to limits on
$M_2$, which are much better than those obtained for unconstrained
Higgs sectors, compare also figure~7 in reference \cite{hollikmix}.

\begin{figure}[tbh]
\ \vspace{1cm}\\
\begin{minipage}[t]{7.8cm} {
\begin{center}
\hspace{-1.7cm}
\mbox{
\epsfysize=7.0cm
\epsffile[-30 0 330 360]{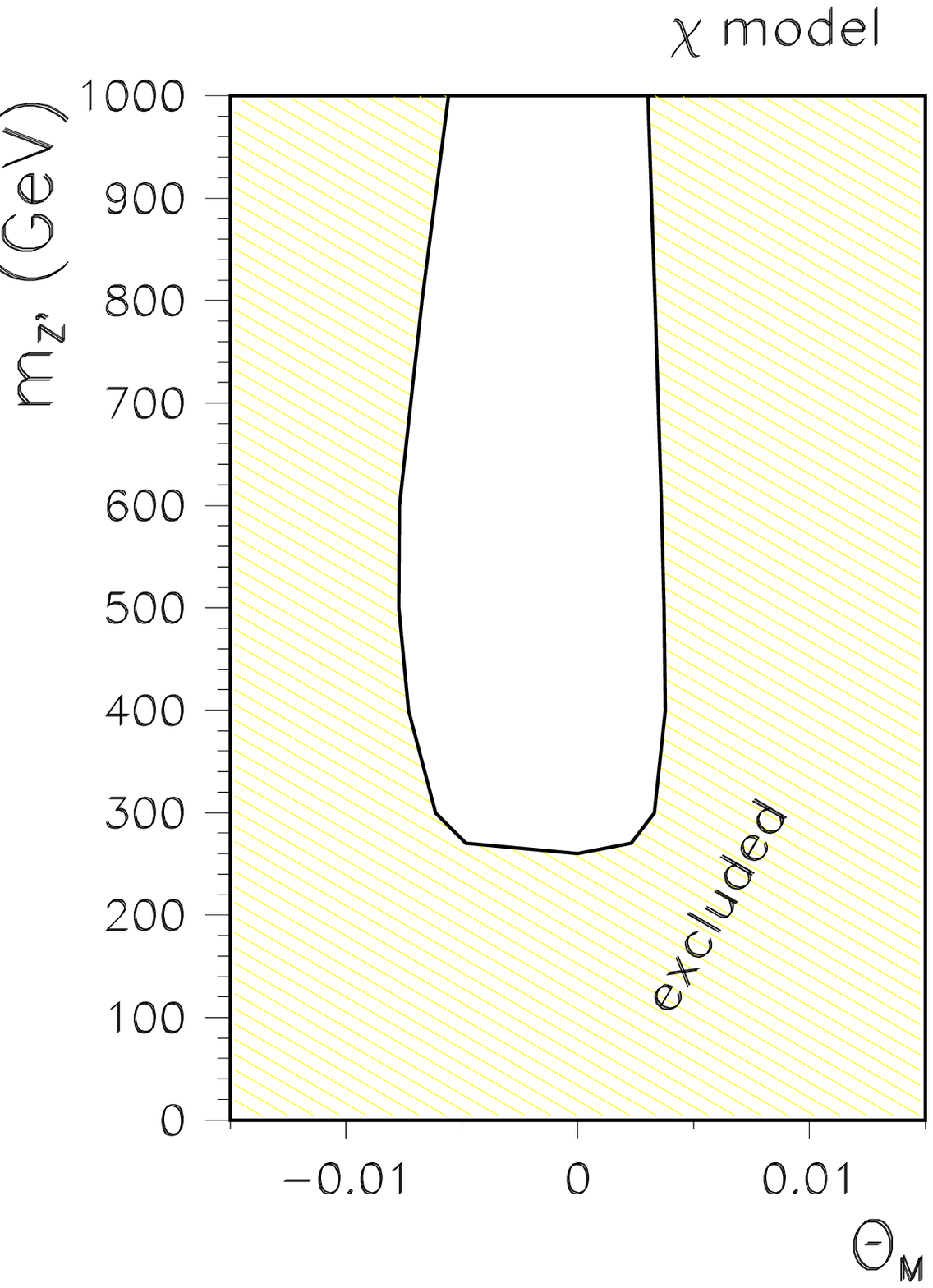}
}
\end{center}
\vspace*{-0.5cm}
\noindent
{\small\it
\begin{fig} \label{stiefel} 
The 95\% CL allowed regions in the $\theta_M-M_{Z'}$ plane for the
$\chi$ model. 
The input of this figure is $M_Z=(91.1863\pm
0.0019)\,GeV,\ M_t=(175\pm 6)\,GeV,\ M_H=150\,GeV$ and
$\alpha_s=0.118\pm 0.003$.
This figure is a preliminary result taken from reference \cite{sr1997}.
\end{fig}}
}\end{minipage}
\hspace*{0.5cm}
\begin{minipage}[t]{7.8cm} {
\begin{center}
\hspace{-1.7cm}
\mbox{
\epsfysize=7.0cm
\epsffile[0 0 360 360]{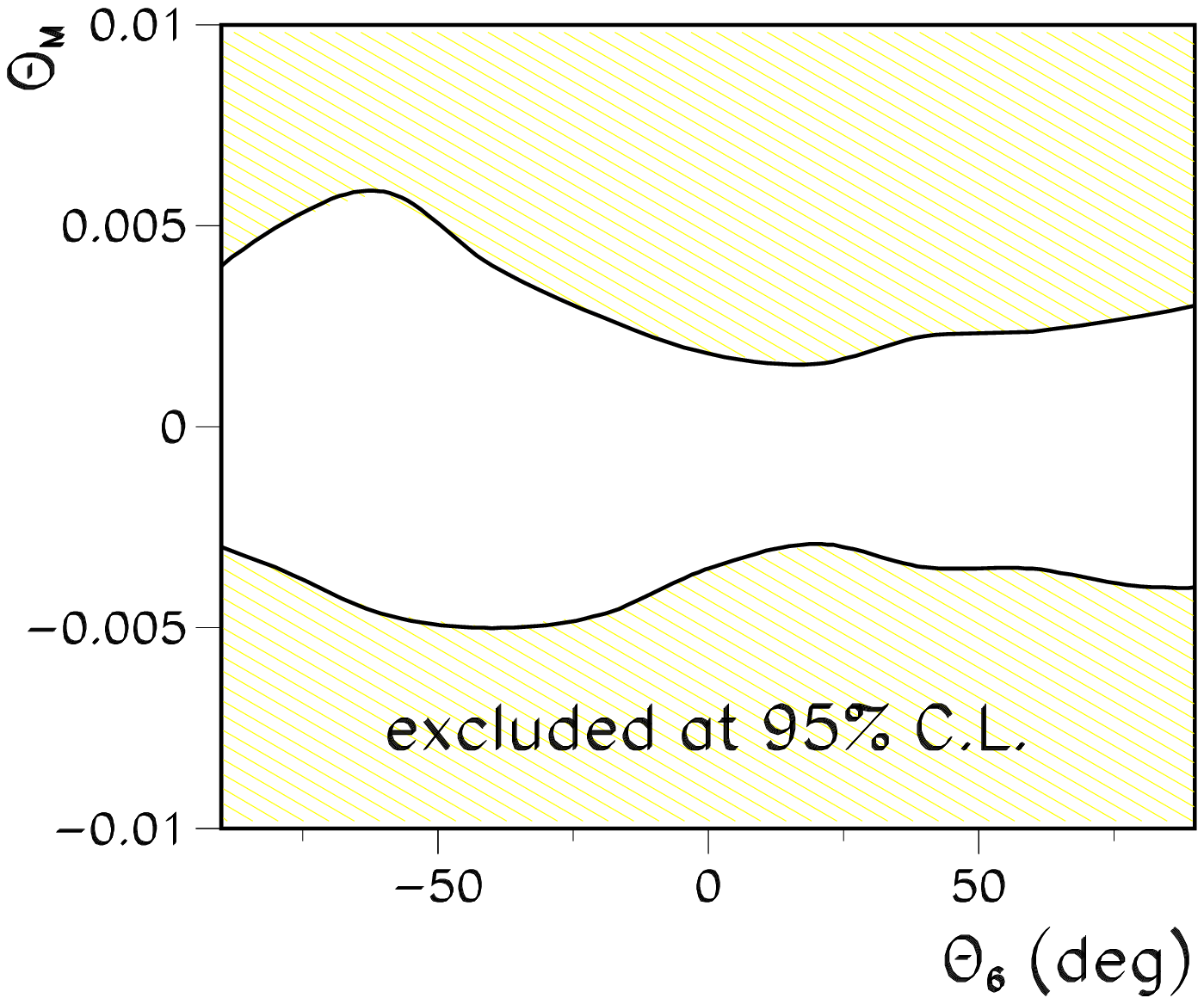}
}
\end{center}
\vspace*{-0.5cm}
\noindent
{\small\it
\begin{fig} \label{slauch} 
The 95\% CL bounds on the $ZZ'$ mixing angle, $\theta_M$, as a
function of the $E_6$ parameter $\beta=\theta_6$.
The input of this figure is $M_Z=(91.1863\pm
0.0019)\,GeV,\ M_t=(175\pm 6)\,GeV,\ M_H=150\,GeV,\ \alpha_s=0.118\pm
0.003$ and $M_{Z'}>550\,GeV$.
This is figure~1 from reference \cite{srcont1997}.
\end{fig}}
}\end{minipage}
\end{figure}

The present two-dimensional constraint of the parameter space
$\theta_M, M_2$ from L3 data is shown in figure~\ref{stiefel}.
The figure is based on LEP data published in 1997 \cite{paus}.
A $\chi^2$ fit to the observables $\sigma_T^l,\ A_{FB}^l,\ 
A_{pol}^\tau,\ R_b$ and $A_{FB}^b$ is performed.
The bound on $\theta_M$ is dominated by the data from the $Z_1$ peak,
while the bound on $M_2$ is dominated by the data above the $Z_1$ peak.

Constraints on $\theta_M$ for a $Z'$ in $E_6$ models obtained in the
same analysis \cite{srcont1997} are shown in figure~\ref{slauch}.
We see that the data give tight constraints on $\theta_M$
for all models considered.

Several indirect
\cite{lepzpold,zpc53,layssac,nardi,hollikmix,altmix,valle,renton,9412361} 
and direct \cite{zefit,L31961,zmixl3,zmixdelphi,srcont1997} $Z'$ 
analyses have been performed recently.
We can only comment on some of them and compare the results with the naive
estimates derived in the previous section.
\begin{description}
\item[Reference \cite{hollikmix}], DELAGUILA92 \ \ \ 
This analysis is based on LEP data published in \cite{moriond} and on $\nu q,
\nu_\mu e, eH$ data \cite{pdb90}, atomic parity violation data \cite{atom90}
and on data on $M_W$ \cite{pdb90,cdf}, see table~1 of reference
\cite{hollikmix}. 
The limit on $\theta_M$, see figure~7 of reference \cite{hollikmix},
is dominantly set by the LEP data. 
Special attention is paid to the constraints on the breaking
parameters of the $E_6$ theory by the data. 
Furthermore, the correlation between $M_{Z'}$ and $M_t$ is considered in
detail. 
We show in table ~\ref{thetmix} the bounds taken from figure~7 for
$M_{Z'}=700\,GeV, M_t=130\,GeV$ and $M_H=100\,GeV$.
We select $\Gamma_1=(2485\pm 9)\,MeV$ from the data \cite{moriond} for
the estimate \req{thetamlim} of $\theta_M$.
We multiply the $90\%\ CL$ numbers given in \cite{hollikmix} by
1.195 to estimate the $95\%\ CL$ bounds.
\item[Reference \cite{zefit}], LEIKE92 \ \ \ 
This analysis is based on the data from reference \cite{lep92},
which includes the LEP data from 1989 and 1990.
It is the first {\it direct} $Z'$ analysis.
It is shown there that the results of earlier {\it indirect} analyses
\cite{layssac,luo,altmix2} agree with the results of the direct analysis.
The values of $\theta_M$ given in table~\ref{thetmix} are taken from figure~3
of \cite{zefit} for $M_{Z'}=700\,GeV, M_t=150\,GeV$ and $M_H=300\,GeV$.
We use $\Gamma_Z=(2487\pm 10)\,MeV$ based on the data \cite{lep92} for our
estimate. 
\item[Reference \cite{altmix}], ALTARELLI93\ \ \ 
This analysis is based on LEP data \cite{LEP93} and on CDF and UA2 data.
The limits on $\theta_M$ are dominated by the LEP data.
In addition, the data are interpreted in terms of the $\epsilon$ parameters
defined in \cite{npb369} and in models with specified Higgs sectors.
Note that the angle $\theta_2$ parametrizing the $E_6$ models introduced
there is connected with $\beta$ introduced in equation \req{betadef} by the
relation $\beta = \theta_2-\arctan\sqrt{5/3}$. 
We take $\theta_M$ from table~5 for $M_t=150\,GeV$ and $M_H=100\,GeV$
choosing $\theta_2=0,50,-30$ degrees for $Z'=\eta, \chi, \psi$.
These values $\theta_2$ are only approximate for $Z'=\chi(\psi)$, where
the exact values should  be 52.2(-37.8) degrees. Furthermore, we
exploited the insensitivity of the reaction $e^+e^-\rightarrow f\bar
f$ relative to the simultaneous change of the signs of all $Z'$ couplings to
fermions, i.e. to a shift $\theta_2\rightarrow \theta_2+\pi$. 
We multiplied the numbers from table~5 in reference \cite{altmix} by
1.96 to estimate $95\%\ CL$ numbers from the values given at one--$\sigma$. 
We select $\Gamma_Z=(2489\pm 7)MeV$ from the data \cite{LEP93} for our
estimate. 
\item[Reference \cite{zmixdelphi}], ABREU95 \ \ \ 
This {\it direct} $Z'$ analysis of the DELPHI collaboration is based
on LEP data from 1990 to 1992.
The limits on $\theta_M$ are obtained for $M_t=150\,GeV$ and $M_H=300\,GeV$.
The effect of $\alpha_s$ is small as far as it is chosen between
$0.118$ and $0.128$.
\item[Reference \cite{9707451}], CVETI\v{C}97 \ \ \ 
This analysis is based on LEP and SLD data \cite{lastlep1}, \cite{SLD96}.
The CDF constraint $M_t=(175\pm 6)\,GeV$ is included.
We take the $95\% CL$ bounds on $\theta_M$ from table~2.
Our estimate is based on $\Gamma_Z$ from the data \cite{lastlep1},
\cite{SLD96}. 
\item[Reference \cite{sr1997}], S.RIEMANN97 \ \ \ 
This analysis is based on L3 data \cite{L31997} and LEP data
\cite{lepewwg9701} published in 1996 and 1997.
It includes the recent measurements beyond the $Z_1$ peak.
The measurements on the $Z_1$ peak define the limits on $\theta_M$,
while the measurements beyond the $Z_1$ peak define $M_{Z'}^{lim}$. 
The values for $\theta_M$ shown in table~\ref{thetmix} are 
obtained for $M_t=(175\pm 6)\,GeV,\ \alpha_s(M_Z)=0.118\pm0.003$ and
$M_{Z_2}>550\,GeV$. 
\end{description}

We now compare the results of the analyses listed above with
the results on $\theta_M$ obtained from a model independent analyses based on
leptonic observables only and specified to specific models in a second step.
\begin{description}
\item[Reference \cite{layssac}], LAYSSAC92\ \ \ 
This analysis is based on LEP data published in \cite{moriond92}.
We show in table \ref{thetmix} the bounds on $\theta_M$ taken from figure~3.
We multiply the bounds given there at one--$\sigma$ with
1.96 to estimate the $95\%\ CL$ bounds.

\item[This paper], section \ref{amvmconstr}\ \ \ 
This analysis is based on combined LEP and SLC data from 1990 to 1995
published in \cite{warschau}. 
We show in table ~\ref{thetmix} the bounds on $\theta_M$ obtained from
figure~\ref{zpmix1} by the same procedure as explained for the $\chi$ model.
\end{description}

%
\begin{table}[tbh]
\begin{center}
\begin{tabular}{|llllll|}\hline\rule[-2ex]{0ex}{5ex} 
Analysis &$\chi$&$\psi$ &$\eta$ &$LR$ & $\theta_M^{lim}$\\ 
\hline
{\bf\cite{hollikmix}}&$-0.007\ , 0.005\ $&$-0.007\ , 0.006\ $&
$-0.006\ , 0.008\ $ &$-0.008\ , 0.003\ $ &$\pm 0.006$\rule[0ex]{0ex}{3ex}\\
{\bf\cite{zefit}}&$-0.006\ , 0.008\ $ &$-0.009\ , 0.006\ $ &
$-0.011\ , 0.009\ $&$-0.004\ , 0.008\ $ &$\pm 0.007$\rule[-1ex]{0ex}{3ex}\\
{\bf\cite{altmix}}&$-0.0035, 0.0035$ &$-0.0051, 0.0043$ &$-0.013\ , 0.0090$
&$-0.0029, 0.0033$ &$\pm 0.005$\rule[-1ex]{0ex}{3ex}\\
{\bf\cite{zmixdelphi}}&$-0.0070, 0.0078$ &$-0.0075,0.0095$ &$-0.029\ ,0.029\ $
&$-0.0057, 0.0077$ &$\pm 0.005$\rule[-1ex]{0ex}{3ex}\\
{\bf\cite{9707451}}&$-0.0029, 0.0011$ &$-0.0022, 0.0026$ &$-0.0055, 0.0021$
&$-0.0013, 0.0021$ &$\pm 0.002$\rule[-1ex]{0ex}{3ex}\\
{\bf\cite{sr1997}}&$-0.0036, 0.0017$ &$-0.0039, 0.0029$ &$-0.0049, 0.0055$
&$-0.0053, 0.0033$ &$\pm 0.002$\rule[-1ex]{0ex}{3ex}\\ \hline
{\bf\cite{layssac}}&$-0.022\ ,0.012\ $ &$-0.008\ , 0.014\ $ &
$-0.024\ , 0.040\ $&$-0.006\ ,0.011\ $ &\rule[-1ex]{0ex}{3ex}\\
{\bf\ref{amvmconstr}}&$-0.006\ ,0.003\ $ &$-0.011\ , 0.006\ $ &
$-0.018\ , 0.014\ $&$-0.008\ , 0.005\ $ &\rule[-1ex]{0ex}{3ex}\\
\hline
\end{tabular}\medskip
\end{center}
{\small\it  \begin{tab} \label{thetmix} The 95\% CL ranges of the $ZZ'$
mixing angle $\theta_M$ for different $Z'$ models obtained in the
analyses listed in the text.  
The estimate $\theta_M^{lim}$ is calculated using equation \req{thetamlim} 
with $g_2/g_1 = \sqrt{\frac{5}{3}}s_W\approx 0.62$.
\end{tab} } 
\end{table}

The limits from the model independent analyses are weaker than those
obtained in the model dependent analyses.
The main reason of this difference is that the model independent
analyses are based on data from leptons in the final state only.
In the model dependent analyses, all couplings of the $Z'$
to fermions are linked by model parameters. 
This allows the inclusion of leptonic and hadronic observables in the analysis.


We list only the results of one analysis on $M_2$ because the limits
from precision measurements at the $Z_1$ peak alone
are rather poor in the case of an unconstrained Higgs sector.
\begin{description}
\item[Reference \cite{zefit}], LEIKE92\ \ \ 
See description of limits on $\theta_M$ above for details of this analysis.
We show in table~\ref{zplimlep1} the limits on $M_{Z'}$ quoted in
section 4.2 of reference \cite{zefit}.
The estimate \req{onres} is based on $\Gamma_Z=(2487\pm 10)\,MeV$
selected from the data \cite{lep92} used in \cite{zefit}.
\end{description}
%
\begin{table}[tbh]
\begin{center}
\begin{tabular}{|lrrrrr|}\hline
Analysis   &$\chi$ &$\psi$ &$\eta$ &$LR$&$M_{Z'}^{lim}$\rule[-2ex]{0ex}{5ex}\\ 
\hline
{\bf \cite{zefit}}    & 148 & 122 & 118 &  -  & 138\rule[-1ex]{0ex}{3ex}\\
\hline
\end{tabular}\medskip
\end{center}
{\small\it \begin{tab}\label{zplimlep1}
The lower bounds on $Z'$ masses
$M_{Z'}^{lim}$ in GeV excluded with 95\% CL by the 
analysis explained in the text.
The estimate $M_{Z'}^{lim}$ from \req{onres} is added with
$g_2/g_1 = \sqrt{5/3}s_W\approx 0.62$.
\end{tab}} \end{table}

We see that the predictions of the formulae \req{thetamlim} and
\req{onres} are in reasonable agreement with the numbers obtained in
the exact analyses of real data. 
Of course, they cannot reproduce details of the different models. 
\paragraph*{Future constraints}
The limits on $\theta_M$ from the data \cite{warschau} could be improved only
by future measurements of the reaction $e^+e^-\rightarrow W^+W^-$.
For details, we refer to section \ref{zpeeww}. 

The present limits on $M_2$ for models with unconstrained Higgs
sectors are much weaker than those from LEP\,2 or from the Tevatron.
Future $e^+e^-$ and $pp(p\bar p)$ experiments will further improve the
limits on $M_2$.
With constrained Higgs sectors, the indirect limits on
$M_2$ from the measurements at the $Z_1$ peak compete with the 
limits from the other experiments \cite{PDB}.
\subsection{$Z'$ limits at $M_1^2\neq s<M_2^2$} \label{zpeeff4}
Below the $Z_2$ resonance and off the $Z_1$ peak, 
different cross sections and asymmetries are predicted by the SM.
The SM parameters are known very precisely from the measurements at
the $Z_1$ peak. 
If future measurements differ from these predictions, one can try to 
interpret the differences as effects due to an extra $Z$ boson. 
The $Z'$ signal arises through interferences of the
$Z_2$ with the photon or $Z_1$ boson.
The deviations due to these interferences can be detected if they
are larger than the experimental error.

Compared to measurements at the $Z_1$ peak, the sensitivity to
$\theta_M$ is suppressed by a factor $\Gamma_1/M_1$ due to statistics.
Therefore, the dependence on $\theta_M$ can be neglected putting 
$\theta_M=0$ and identifying $Z_2=Z'$ and $Z_1=Z$.

Early $Z'$ analyses can be found in references
\cite{roroso10,roroe6,eeffhol,prd22}. 
We consider here results for $Z'$ constraints obtained by different
recent analyses. 
\subsubsection{Model independent constraints on $v_l^N$ and $a_l^N$}\label{min}
\paragraph*{Estimate}
As shown in section \ref{zpeeff1}, the amplitude of off--resonance
fermion pair production depends only on the normalized couplings
$v_f^N$ and $a_f^N$ and not on $a'_f,v'_f$ and $M_{Z'}$ separately. 

Consider the constraints arising from the independent observables
$\sigma_T^l,\ A_{FB}^l$ and $A_{LR}^l$ introduced in section \ref{zpeeff1}.
The measurement of each of these observables
excludes a certain domain of the $Z'$ couplings $v^N_l$ and $a^N_l$. 
This domain can be calculated analytically in the Born approximation 
taking into account the contributions of the $\gamma Z'$ and $ZZ'$
interferences, neglecting the $Z'Z'$ contribution.
For simplicity, we set $s^2_W=\frac{1}{4}$.
The three considered observables detect a signal if the following
conditions are fulfilled \cite{zpmi},
\ba
\label{bornexcl}
\begin{array}{rrclrl}
\sigma_T^l:&
\displaystyle{\left|\left(\frac{v_l^N}{H_T}\right)^2
+\left(\frac{a_l^N}{H_T}\right)^2\frac{\chi_Z(s)}{4}\right|} &\ge& 1,&
 H_T=&\sqrt{\frac{\alpha\chi_{cl}}{2}\frac{\sigma_T^l}{\sigma_T^l(QED)}
\frac{\Delta\sigma_T^l}{\sigma_T^l}},\\ 
A_{FB}^l:&
\displaystyle{\left|\left(\frac{v_l^N}{H_{FB}}\right)^2
-\left(\frac{a_l^N}{H_{FB}}\right)^2
\frac{(3 - A_{FB}^l \chi_Z(s))\frac{1}{4}}{A_{FB}^l-\frac{3}{16}\chi_Z(s)}
\right|} &\ge& 1,&
H_{FB}=&\sqrt{\frac{
\frac{\alpha\chi_{cl}}{2}\frac{\sigma_T^l}{\sigma_T^l(QED)}\Delta 
A_{FB}^l} 
{ A_{FB}^l-\frac{3}{16}\chi_Z(s)}},\\
A_{LR}^l:&
\displaystyle{\left|\left(\frac{v_l^N}{H''_v}\right)
\left(\frac{a_l^N}{H_{LR}}\right)\right|}&\ge& 1,&   
H_{LR}=&\sqrt{\frac{
\frac{\alpha\chi_{cl}}{2}\frac{\sigma_T^l}{\sigma_T^l(QED)}
\Delta A_{LR}}{1+\frac{1}{4}\chi_Z(s)}}.
\end{array}
\ea
$\chi_Z(s)$ is defined in \req{eq28} and
$\sigma_T^l(QED)=\frac{4\pi\alpha^2}{3s}$. 
$\sigma_T^l$ detects a $Z'$ with couplings outside an ellipse above
the $Z_1$ peak and outside a set of hyperbolas below the $Z_1$ peak. 
The forward--backward asymmetry is sensitive to a $Z'$ with couplings outside
a set of hyperbolas above the $Z_1$ peak and to a $Z'$ with
couplings outside an ellipse below the $Z_1$ peak.
The left--right asymmetry detects a $Z'$ with couplings outside a
different set of hyperbolas for all energies.
As explained in section \ref{zpeeff21}, the quantitative prediction
\req{bornexcl} is only changed a  little by radiative corrections if
appropriate kinematic cuts are applied. 
The axes of the ellipse and the hyperbolas
$H_T,H_{FB}$ and $H_{LR}$ do not depend on the $Z'$ model. 
They are proportional to the root of the experimental errors
$\Delta\sigma_T^l,\ \Delta A_{FB}^l$ and $\Delta A_{LR}$.
For further details, we refer to \cite{zpmi}.

Alternatively, the {\it differential} cross sections of left- and
right-handed beams can be considered.
The Cramer-Rao bound \cite{eadie} for the reaction $e^+e^-\rightarrow
\mu^+\mu^-$ is derived in the second reference of \cite{moeller} in the limit 
$M_{Z'}^2\gg s\gg M_Z^2$ and $s_W^2=\frac{1}{4}$.
In our notation, we have
\bq
\label{cramerrao1}
\chi^2_\infty\approx 
\frac{L}{s}\pi [10(v_e^N)^4+22(v_e^N)^2(a_e^N)^2+7(a_e^N)^4].
\eq
The bound \req{cramerrao1} gives the the theoretical limit of a
constraint of the $Z'$ parameters by the observables considered. 

For comparison, we give the bound \req{bornexcl} deduced from
$\sigma_T^l$ in the same limit as \req{cramerrao1},
\bq
\chi^2\approx 
\frac{L}{s}\pi\frac{24}{5} [(v_e^N)^2+\frac{1}{3}(a_e^N)^2]^2.
\eq
As it should be, it is worse than \req{cramerrao1}.

\begin{figure}[tbh]
\ \vspace{1cm}\\
\begin{minipage}[t]{7.8cm} {
\begin{center}
\hspace{-1.7cm}
\mbox{
\epsfysize=7.0cm
\epsffile[0 0 275 275]{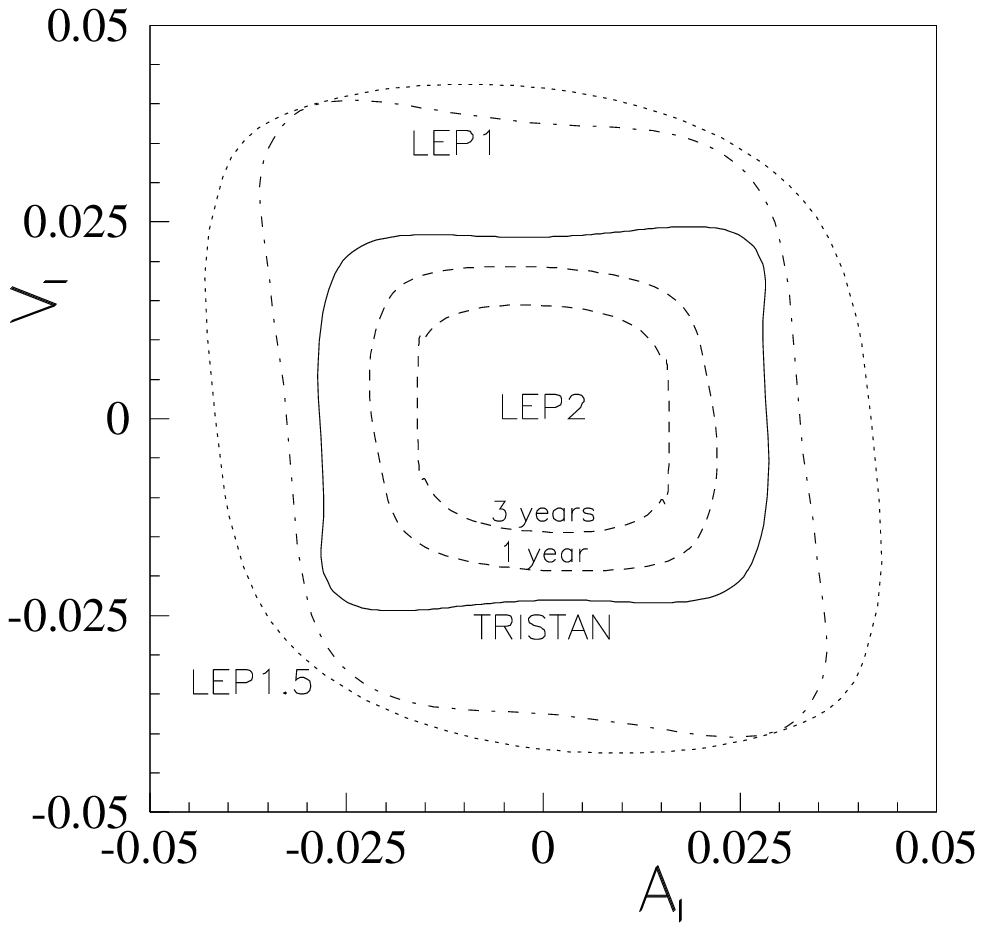}
}
\end{center}
\vspace*{-0.5cm}
\noindent
{\small\it
\begin{fig} \label{panktris} 
Combined regions allowed by $\sigma_T^\mu$ and $A_{FB}^\mu$ for
$\chi^2=\chi^2_{min}+4$ in the ($A_l,V_l$) plane for TRISTAN, LEP1,
LEP1.5 and LEP2 colliders.
Two bounds are shown for LEP2 corresponding to $L=150\,pb^{-1}$
(one year of running) and $L=500\,pb^{-1}$ (three years of running). 
Radiative corrections are included.
This is figure~4 of reference \cite{pankovpol1}.
\end{fig}}
}\end{minipage}
\hspace*{0.5cm}
\begin{minipage}[t]{7.8cm} {
\begin{center}
\hspace{-1.7cm}
\mbox{
\epsfysize=7.0cm
\epsffile[0 0 360 360]{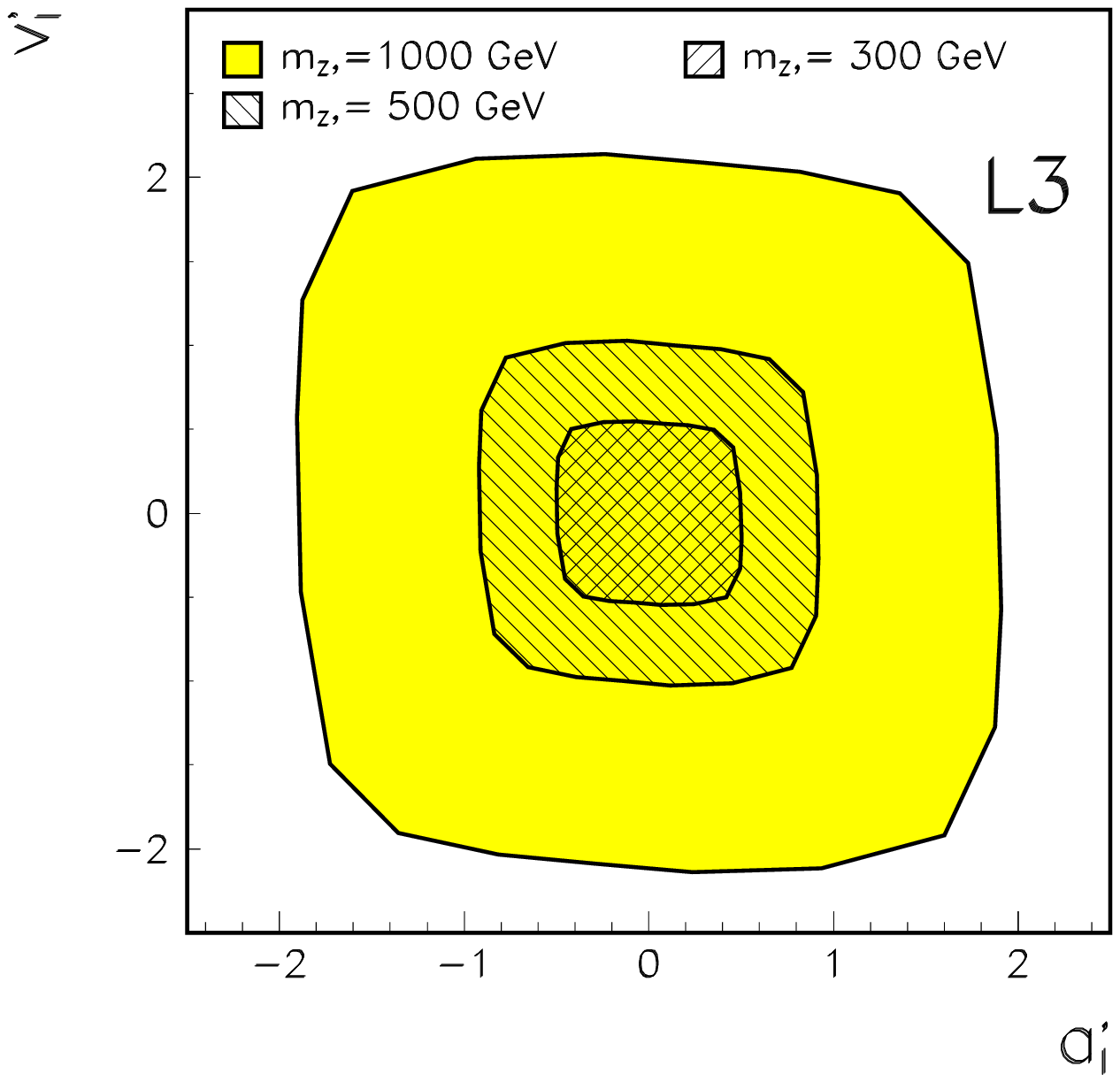}
}
\end{center}
\vspace*{-0.5cm}
\noindent
{\small\it
\begin{fig} \label{leptlep} 
The 95\% CL allowed regions in the $a'_l,v'_l$ plane from L3 data now
(large region) and at the end of LEP2 (small region).
The input of this figure is $M_Z=(91.1863\pm 0.0019)\,GeV,\
M_t=(175\pm 6)\,GeV,\ M_H=150\,GeV,\ \alpha_s=0.118\pm 0.003$ and
$M_{Z'}=500\,GeV$. 
This is figure~2 of reference \cite{srcont1997}.
\end{fig}}
}\end{minipage}
\end{figure}

\begin{figure}[tbh]
\ \vspace{1cm}\\
\begin{minipage}[t]{7.8cm} {
\begin{center}
\hspace{-1.7cm}
\mbox{
\epsfysize=7.0cm
\epsffile[0 0 500 500]{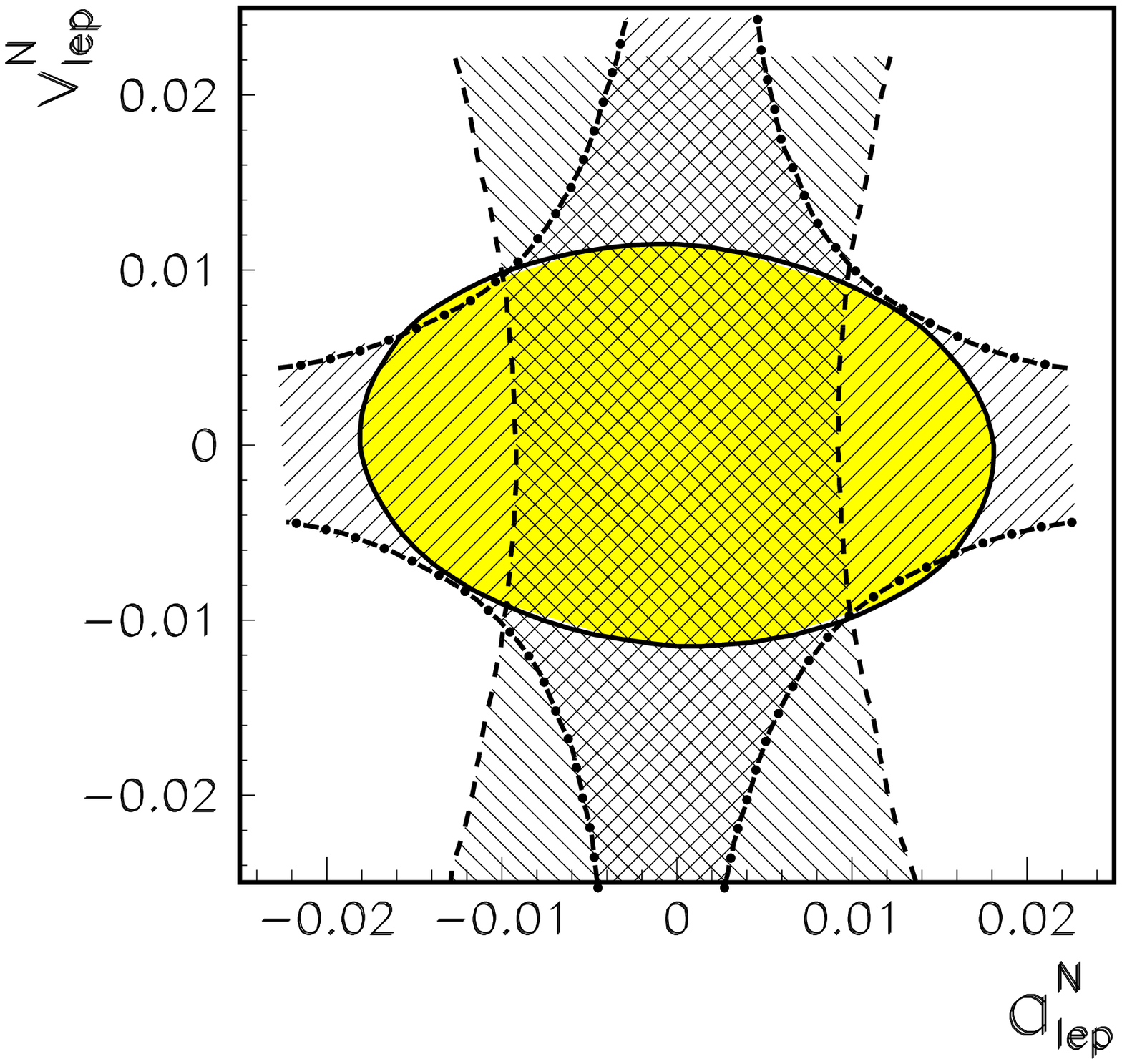}
}
\end{center}
\vspace*{-0.5cm}
\noindent
{\small\it
\begin{fig} \label{leptexcl} 
Areas of $(a_l^N,\ v_l^N)$ values, for which the extended gauge theory's
predictions are indistinguishable from the SM (95\% CL)
for $\sqrt{s}=500\,GeV$ and $L=20\,fb^{-1}$.
Models inside the  ellipse cannot be detected with $\sigma_T^l$
measurements.
Models inside the hatched areas with 
falling (rising) lines cannot be resolved with $A_{FB}^l$ ($A_{LR}^l$).
I thank S. Riemann for providing this figure.
\end{fig}}
}\end{minipage}
\hspace*{0.5cm}
\begin{minipage}[t]{7.8cm} {
\begin{center}
\hspace{-1.7cm}
\mbox{
\epsfysize=7.0cm
\epsffile[0 0 500 500]{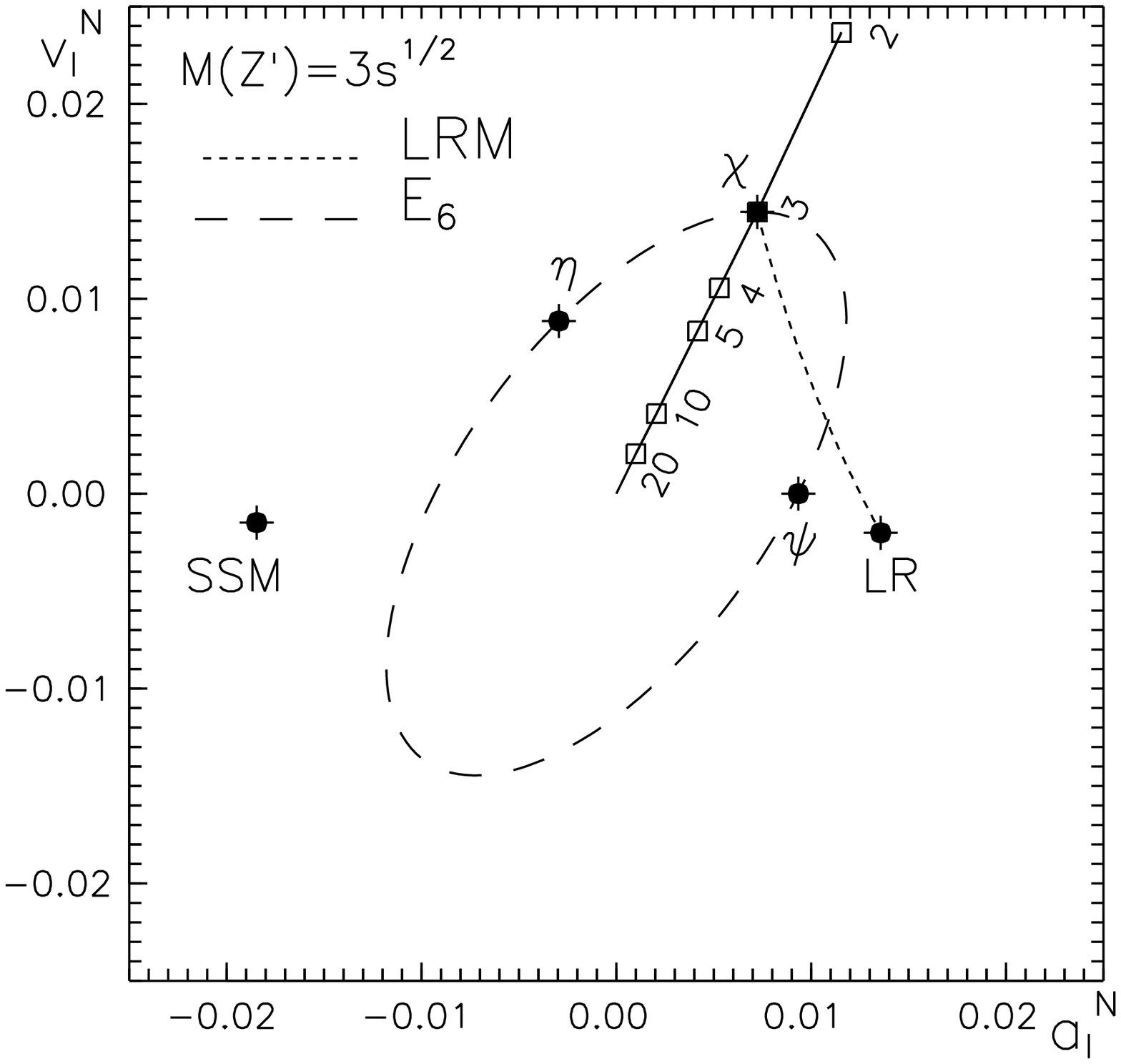}
}
\end{center}
\vspace*{-0.5cm}
\noindent
{\small\it
\begin{fig} \label{couplept} 
The normalized leptonic vector and axial vector
couplings $v_l^N$ and $a_l^N$ for  $M_{Z'} = 3 \sqrt{s}$ in typical GUT's.
For the $\chi$ model, $M_{Z'}$ is varied in units of $\sqrt{s}$.
\end{fig}}
}\end{minipage}
\end{figure}

To conclude, we remark that expressions similar
to \req{bornexcl} were obtained some time ago in reference
\cite{npb78} to constrain the interactions of the SM $Z$ boson.
\paragraph*{Present Constraints}\label{presconstrsgtmz}
%
Present constraints on $v_l^N$ and $a_l^N$ are available from
measurements at TRISTAN and at LEP\,2.
\begin{description}
\item[Reference \cite{pankovtrist}], OSLAND97 \ \ \ 
This analysis is based on measurements of $\sigma_T^\mu$ and $A_{FB}^\mu$ at 
TRISTAN \cite{tristan} ($L=300\,pb^{-1},\ \sqrt{s}=58\,GeV$) and
LEP1.5 ($L=5\,pb^{-1},\ \sqrt{s}=130-140\,GeV$).
The constraints on $a_l^N$ and $v_l^N$ are shown in figure~\ref{panktris}.
The constraints from future measurements at 
LEP2 ( ($L=150(500)\,pb^{-1},\ \sqrt{s}=190\,GeV$) are predicted too.
The systematic error of both observables is assumed to be 1\% at all
colliders (see table~1 in reference \cite{pankovtrist}.
The conventions are $A_l=a_l^N, V_l=v_l^N$.
\item[Reference \cite{srcont1997}], S.RIEMANN97\ \ \ 
This analysis is based on LEP data published in 1997 \cite{paus}.
A $\chi^2$ fit to the observables $\sigma_T^l,\ A_{FB}^l$ and
$A_{pol}^\tau$ is performed to give the allowed
regions in the $a'_l-v'_l$ plane.
The resulting constraints are shown in figure~\ref{leptlep}.
All radiative corrections and systematic errors are included in this analysis.
\end{description}
\paragraph*{Future Constraints}
Predictions for future constraints on $a_l^N$ and $v_l^N$ of a
$500\,GeV$ collider including radiative
corrections are shown in figure~\ref{leptexcl}.
The analysis includes statistical and systematic errors.
The resulting combined errors of the observables entering
figure~\ref{leptexcl} are  
$\Delta\sigma_T^l/\sigma_T^l=1\%,\ \Delta A_{FB}^l=1\%$ and $\Delta
A_{LR}^l=1.2\%$. 
We refer to \cite{lmu0296} for further details.
As predicted by the Born analysis \req{bornexcl}, the regions
indistinguishable from the SM are approximately ellipses for
$\sigma_T^l$ and areas between hyperbolas for $A_{FB}^l$ and $A_{LR}^l$.
The remaining two parts of the hyperbolas from $A_{FB}^l$ are outside
the figure. 
$A_{LR}$ gives only a marginal improvement to the $Z'$ {\it exclusion} limits.

A quantitative comparison with \req{bornexcl} shows that the
error of the Born prediction for the exclusion regions is below 10\%.
Of course, this number depends on the kinematic cuts as explained in
section \ref{zpeeff21}.
The equations \req{bornexcl} can be used to predict the changes
in figure~\ref{leptexcl} for different errors of the observables.

The model independent exclusion region predicted for final LEP2 data
($\sqrt{s}=190\,GeV, L=0.5\,fb^{-1})$ is shown in 
figure~3a of reference \cite{lmu0296}.
It looks very similar to figure~\ref{leptexcl} and agrees with figure
\ref{panktris}. 

For illustration purposes, the domains of the normalized couplings
$a_l^N,v_l^N$ are shown in figure~\ref{couplept} for $E_6$ models.
For a fixed $Z'$ model, i.e. known couplings $v'_l$ and $a'_l$, the limits on
$a_l^N$ and $v_l^N$ transform to $Z'$ mass limits.
This is illustrated in figure~\ref{couplept} for $Z=\chi$.
Superimposing figures~\ref{leptexcl} and \ref{couplept},
we predict $M_{\chi}^{lim}\approx 4\sqrt{s}$.
For a fixed $M_{Z'}$, figure \ref{leptexcl} can be drawn for $a'_l$
and $v'_l$ as done in figure \ref{leptlep}.

The couplings of the $Z'$ to quarks $a_q^N$ and $v_q^N$ can only
be constrained if the $Z'$ couplings to leptons are non-zero.
This is different from the constraints on $a_q^M$ and $v_q^M$ at the
$Z_1$ peak, which were possible also in the case $v_l^M=a_l^M=0$.

In models, where the couplings of the $Z'$ to leptons are considerably
smaller than those to quarks, one can have a signal in the
quarkonic observables without a signal in leptonic observables.
See reference \cite{cvsmallcoup} for a discussion of such a possibility. 

Assuming the relations \req{coup5}, the measurements at future colliders
would constrain the five couplings 
\bq
\label{coup5con}
L_l(2),\ R_e(2),\ L_q(2),\ R_u(2),\ R_d(2).
\eq
In the case of an agreement of these measurements with the SM
predictions, the allowed regions of the couplings \req{coup5con}
contain zero.

The {\it measurement} of non-zero couplings in the case of a $Z'$ signal is
investigated in the literature and will be discussed in section
\ref{zpeeff414}.

Finally, we mention that $a_f^N,v_f^N$ and $a_f^M,v_f^M$ are uniquely
related by the Higgs constraint \req{higgsconstr} in models where the
Higgs sector is specified.
An appropriate scaling and a superposition of figures \ref{zpmix1} and
\ref{leptexcl} would then allow a direct comparison of the model
independent limits from $e^+e^-\rightarrow f\bar f$ for  $s\approx
M_1^2$ and for $M_1^2\neq s<M_2^2$.
\subsubsection{Model dependent constraints on $M_{Z'}$}
%
\paragraph*{Estimate}\label{mzlimest}
To obtain an estimate for $M_{Z'}$, we fix the $Z'$ couplings 
$v'_f\approx v_f,\ a'_f\approx a_f$. 
Consider the $\gamma Z'$ interference in $\sigma_T^f$. 
We derive 
\bq
\label{offres}
\frac{\Delta^{Z'}O}{O_{SM}} \approx
\frac{g_2^2}{g_1^2}\frac{|\Re e{\chi_\gamma \chi^*_{Z'}}|}{|\chi_\gamma|^2} 
\approx\frac{g_2^2}{g_1^2}\frac{s}{M_{Z'}^2-s},\mbox{\ \ \ which\ gives\ \ \ }
M_{Z'}^{lim}=\sqrt{s}\left( 1+\frac{1}{\Delta o}\right)^{1/2}.
\eq
Models with $M_{Z'}<M_{Z'}^{lim}$ would give a signal in the observable $O$.
Comparing the two expressions for measurements on and off the
$Z$ peak given by the equations \req{onres} and \req{offres}, 
we see that \req{onres} has an additional suppression factor $\Gamma_Z/M_Z$.

We usually have $1/\Delta o\equiv g_2^2/g_1^2\cdot O/\Delta O \gg 1$.
This allows the further approximation in (\ref{offres}),
\bq
\label{epemlim}
M_{Z'}^{lim} \approx \sqrt{\frac{s}{\Delta o}}.
\eq
The approximation \req{epemlim} is in agreement with the
estimates\req{bornexcl}, which contains more details of the model.
Taking into account the scaling of the error $\Delta o\sim
\sqrt{s/L}$, one obtains the well known
scaling law of the $Z'$ mass limit \cite{zpmi,zepp,rizzo},
$M_{Z'}^{lim}\approx\left( s L\right)^{1/4}$.
It is valid in the absence of systematic errors.
A scaling including the systematic errors is obtained substituting $L$
by $L/(1+r^2)$, 
\bq
\label{epemlim2}
M_{Z'}^{lim} \sim \left[\frac{s L}{1+r^2}\right]^{1/4}.
\eq
The influence of (not too large) systematic errors on $Z'$ {\it
exclusion limits} is therefore rather moderate.

Radiative corrections give cross sections depending on kinematic cuts. 
We already noticed the moderate dependence of cross sections on the
photon energy cut $\Delta$ in the case where the radiative return is
forbidden, see figure \ref{sigmadelta}.
The influence of $\Delta$ on $M_{Z'}^{lim}$ is further reduced by the
fourth root in equation \req{epemlim2}.
\paragraph*{Present Constraints}
The best present mass limits on extra neutral gauge bosons predicted
in $E_6$ models come from proton collisions.
As an example of limits on $M_{Z'}$ from $e^+e^-$ collisions, we quote
the numbers in table~\ref{zplimlep2} read off from
reference \cite{sr1997} S. RIEMANN97.
More details of the analysis are given in section
\ref{presconstrsgtmz}. 
The estimate of $M_{Z'}^{lim}$ is calculated using the statistical
error of $\sigma_T^l$.  
Formula \req{offres} must be used because of the small numbers of events.
%
\begin{table}[tbh]
\begin{center}
\begin{tabular}{|lrrrrrr|}\hline
Analysis&$\chi$&$\psi$&$\eta$&$LR$&SSM &$M_{Z'}^{lim}$\rule[-2ex]{0ex}{5ex}\\ 
\hline
{\bf \cite{sr1997}}  & 300 & 220 & 230 & 310 & 520 &240\rule[-1ex]{0ex}{3ex}\\
\hline
\end{tabular}\medskip
\end{center}
{\small\it \begin{tab}\label{zplimlep2}
The lower bound on $Z'$ masses $M_{Z'}^{lim}$ excluded with 95\% CL by
the analysis explained in the text. 
\end{tab}} \end{table}
\paragraph*{Future Constraints}
Many analyses investigate the $Z'$ mass limits reachable at future colliders.
The minimal input of these analyses are assumptions about the
center--of--mass energy, the 
integrated luminosity and a list of observables used in the fit. 
Optionally, systematic errors are included. 
Radiative corrections have to be included into fits to real data.
However, they introduce only small changes in theoretical investigations, where
the ``data'' can be generated in the Born approximation and then be
fitted by Born formulae \cite{zpmi,zpsari}.
See section \ref{zpeeff2} for the reasons why this works well. 

We now comment on the recent theoretical analyses, the results of
which are collected in Table~\ref{zplimlc}.
\begin{description}
\item[Reference \cite{lmu0296}], LEIKE97 \ \ \ 
Theoretical analysis for LEP\,2 and a future linear collider.
The observables 
\bq
\label{obs0296}
\sigma_T^l,\ A_{FB}^l,\ A_{LR}^l,\ A_{pol}^\tau,\ 
R^{had}, A_{LR}^{had},\ R_b=\sigma_T^b/\sigma_T^{had},\ 
A_{FB}^b,\ A_{LR}^b
\eq
are included in the fit.
80\% polarization of the electron beam is assumed.
The efficiency of flavour tagging is included in the systematic errors.
The full SM radiative corrections are included.
The numbers giving limits at 95\% CL
are taken from table~3 of reference \cite{lmu0296}.
The mass limits without and with inclusion of systematic errors are shown.
We take the error of the most accurately measured observable for our
estimate of $M_{Z'}^{lim}$, i.e. $\Delta A_{LR}^{had}$ 
in all scenarios.
The one--$\sigma$ errors of $\Delta A_{LR}^{had}$ with and without
systematic errors are  
$0.8\%$ and $0.6\%$ ($0.7\%$ and $0.5\%$) for LEP2 (LC500).
\item[Reference \cite{9609248}], RIZZO96 \ \ \ 
Theoretical analysis for new gauge boson searches at different future
colliders. 
The observables 
\bq
\sigma_T^f,\ A_{FB}^f,\ A_{LR}^f, A_{LR,FB}^f, A_{pol}^\tau
\mbox{\ and\ } A_{pol,FB}^\tau
\eq
for different final state fermions $f=l,c,b,t$ are included in the fit. 
90\% polarization of the electron beam is assumed.
Initial state radiation, finite identification efficiencies and
systematic errors associated with luminosity and beam polarization
uncertainties are taken into account.
We selected for table~\ref{zplimlc} the numbers given in table~3 of
reference \cite{9609248}, which are given at 95\% confidence.
\item[Reference \cite{godfreysnow}], GODFREY96\ \ \ 
Analysis of new gauge boson searches at different future colliders.
Compared to the older analysis \cite{godfrey}, more observables and
the option of a $\mu^+\mu^-$ collider are included.
The 18 observables 
\bq
\sigma_f^\mu,\ A_{FB}^f,\ A_{LR}^f,\ f=\mu,\tau,c,b;\ 
\ A_{LR,FB}^f,\ f=\mu,c,b;\ A_{pol}^\tau,\ R^{had},\ A_{LR}^{had}
\eq
are included in the fit for $e^+e^-$ colliders. 
90\% electron polarization is assumed.
Only the 10 observables, which do not demand beam polarization, are
included for $\mu^+\mu^-$ colliders.
Detection efficiencies are taken into account.
Radiative corrections and systematic errors are not included.
The numbers quoted in table~\ref{zplimlc} are taken from figure~1 of
reference \cite{godfrey} multiplied with 1.15 to transform the 99\%
CL limits given there to 95\% CL limits.
The estimate \req{epemlim} in table~\ref{zplimlc} is obtained with the
statistical error of $A_{LR}^{had}$ as input.
\end{description}

%
\begin{table}[tbh]
\begin{center}
\begin{tabular}{|lllllllll|}\hline
Analysis &$\sqrt{s}/TeV$&$L\,fb$&$\chi$ &$\psi$ &$\eta$ &$LR$ &$SSM$ 
&$M_{Z'}^{lim}(E_6)$\rule[-2ex]{0ex}{5ex}\\ 
\hline
{\bf\cite{lmu0296}}  +syst.&0.19&\ \,0.5 & 0.99& 0.56& 0.62& 1.10& 1.50& 0.95\\
{\bf\cite{lmu0296}} \ stat.&0.19&\ \,0.5 & 1.10& 0.64& 0.69& 1.30& 1.70& 1.10\\
{\bf \cite{lmu0296}}  +syst.&0.5 & 20 & 2.8 & 1.6 & 1.7 & 3.2 & 4.0 & 2.6\\
{\bf \cite{lmu0296}} \ stat.&0.5 & 20 & 3.1 & 1.8 & 1.9 & 3.8 & 4.7 & 3.1\\
{\bf \cite{9609248}}        &0.5 & 50 & 3.2 & 1.8 & 2.3 & 3.7 & 4.0 & 2.6\\
{\bf \cite{godfreysnow}} $e^+e^-$  
                           &0.5 & 50 & 5.2 & 2.5 & 2.9 & 4.2 & 6.9 & 3.9\\
{\bf \cite{godfreysnow}} $\mu^+\mu^-$  
                           &0.5 & 50 & 4.0 & 2.3 & 2.5 & 3.7 & 6.9 & 3.9\\
\hline
\end{tabular}\medskip
\end{center}
{\small\it  \begin{tab}\label{zplimlc} The lower bound on $Z'$ masses
$M_{Z'}^{lim}$ in TeV excluded by the different analyses
described in the text.
The estimate \req{offres} for $M_{Z'}^{lim}$ is calculated with $g_2/g_1=0.62$.
\end{tab}} \end{table}

The estimate for $M_{Z'}^{lim}$ in table~\ref{zplimlc} gives a good
prediction of the exact exclusion limits. 
Of course, it cannot describe the differences between the $E_6$ models.
The scaling \req{epemlim2} predicts that 
systematic errors of the same magnitude as the statistical errors, 
i.e. $r=1$, should change
 $M_{Z'}^{lim}\rightarrow M_{Z'}^{lim}/\sqrt[4]{2}$, which is a reduction
of 16\% only. 
This is in agreement with table~\ref{zplimlc}.
Observables, which require flavour tagging, have systematic errors,
which usually dominate the statistical errors.
However, these observables give only a moderate contribution to $M_{Z'}^{lim}$
keeping the dependence of $M_{Z'}^{lim}$ on their systematic errors small.
For details, we refer to \cite{lmu0296,9609248}. 

The price one has to pay for model independent limits on $v_l^N$ and $a_l^N$
can be estimated comparing the exclusion limit for
$M_\chi^{lim}=2.8\,TeV$ quoted in table \ref{zplimlc} with the value
$M_\chi^{lim}=4\sqrt{s}=2\,TeV$ obtained in the model independent
analysis explained in section \ref{min}.
Both analyses are based on the same assumptions on the data.
The difference occurs because the model independent analysis is based
on a two parameter fit to $\sigma_T^l, A_{FB}^l$ and $A_{LR}^l$, while
the numbers quoted in table \ref{zplimlc} are based on a
one--parameter fit to many more observables.
The difference between the values for $M_{Z'}^{lim}$ from both
analyses is not too large.
This reflects the importance of  $\sigma_T^l, A_{FB}^l$ and $A_{LR}^l$
in the $M_{Z'}^{lim}$ constraint. 
\subsubsection{Constraints on $g_2$}
GUT's are the main motivation for the search for extra neutral gauge
bosons.
In GUT's, all gauge interactions are unified at high energies. 
In general, it is not expected that the renormalization group equations
change the gauge couplings drastically during the evolution from GUT
energies down to low energies
Therefore, one expects $g_2\approx g_1$ at low energies.
On the other hand, it is useful from the experimental stand point of
view to find all possible constraints on new particles, which can be
derived from the data.
This is the reason why we consider limits on a $Z'$ with small
couplings to all SM fermions.
The best present constraints on a $Z'$ with small couplings in the
case of a non-zero $ZZ'$ mixing come from measurements at the $Z_1$
peak, see figure~\ref{zpmix4}.
Here, we assume that there is no $ZZ'$ mixing. 
\paragraph*{Estimate}
Equation \req{offres} can be inverted to give an estimate of the
sensitivity to the coupling strength $g_2$,
\bq
\label{glim2}
g_2^{lim}=g_1\sqrt{\frac{M_{Z'}^2-s}{s}\frac{\Delta O}{O}}.
\eq
Models with couplings $g_2>g_2^{lim}$ would give a signal in the
observable $O$.
Formula \req{glim2} is not true near the $Z'$ resonance.
If one assumes that measurements with an accuracy of 1\% are performed at
energies $\sqrt{s},\sqrt{2s},\sqrt{4s},\dots$,
a $Z'$ with $g_2>g_1/7$ is excluded below its resonance with 95\% confidence . 

Alternatively, the bounds on $a'_lg_2$ and $v'_lg_2$ follow from the model
independent limits on $a_l^N,v_l^N$, compare definition \req{normcoup},
\bq
\label{g2lim}
a'_lg_2=a_l^N\sqrt{4\pi}\sqrt{\frac{|m_2^2-s|}{s}}.
\eq
The formula \req{g2lim} is also valid for energies above the $Z'$ resonance.  
As expected, the sensitivity to a weakly coupled $Z'$ increases for
the center--of--mass energies approaching the $Z'$ mass.
As it should be, the estimate \req{g2lim} agrees with \req{glim2}
after substituting $a_f^N$ according to \req{bornexcl}.
\paragraph*{Present and future Constraints}
The present bounds on $a_l^N$ and $v_l^N$ from TRISTAN can be read off
from figure \ref{panktris}, 
those from 1996 L3 data are given in references \cite{L31961,sr1997},
\ba
\label{g2liminput}
\mbox{TRISTAN:} & a_l^N=0.025 & v_l^N=0.03,\nll
\mbox{LEP\,1.5,\ }\sqrt{s}\approx 130\,GeV:  & a_l^N=0.095 & v_l^N=0.16,\nll
\mbox{LEP\,2,\ }\sqrt{s}=170\,GeV:  & a_l^N=0.10 & v_l^N=0.11.
\ea
The bounds on $a_l^N$ and $v_l^N$ expected from a next linear collider
can be taken from figure~\ref{leptexcl},
\bq
\label{nlcinput}
NLC,\ \sqrt{s}=500\,GeV,\ L=20\,fb^{-1}:\ \ \ a_l^N=0.0093,\ \ \ v_l^N=0.011.
\eq 

\begin{figure}
\begin{center}
\begin{minipage}[t]{7.8cm}{
\begin{center}
\hspace{-1.7cm}
\mbox{
\epsfysize=7.0cm
\epsffile[0 0 500 500]{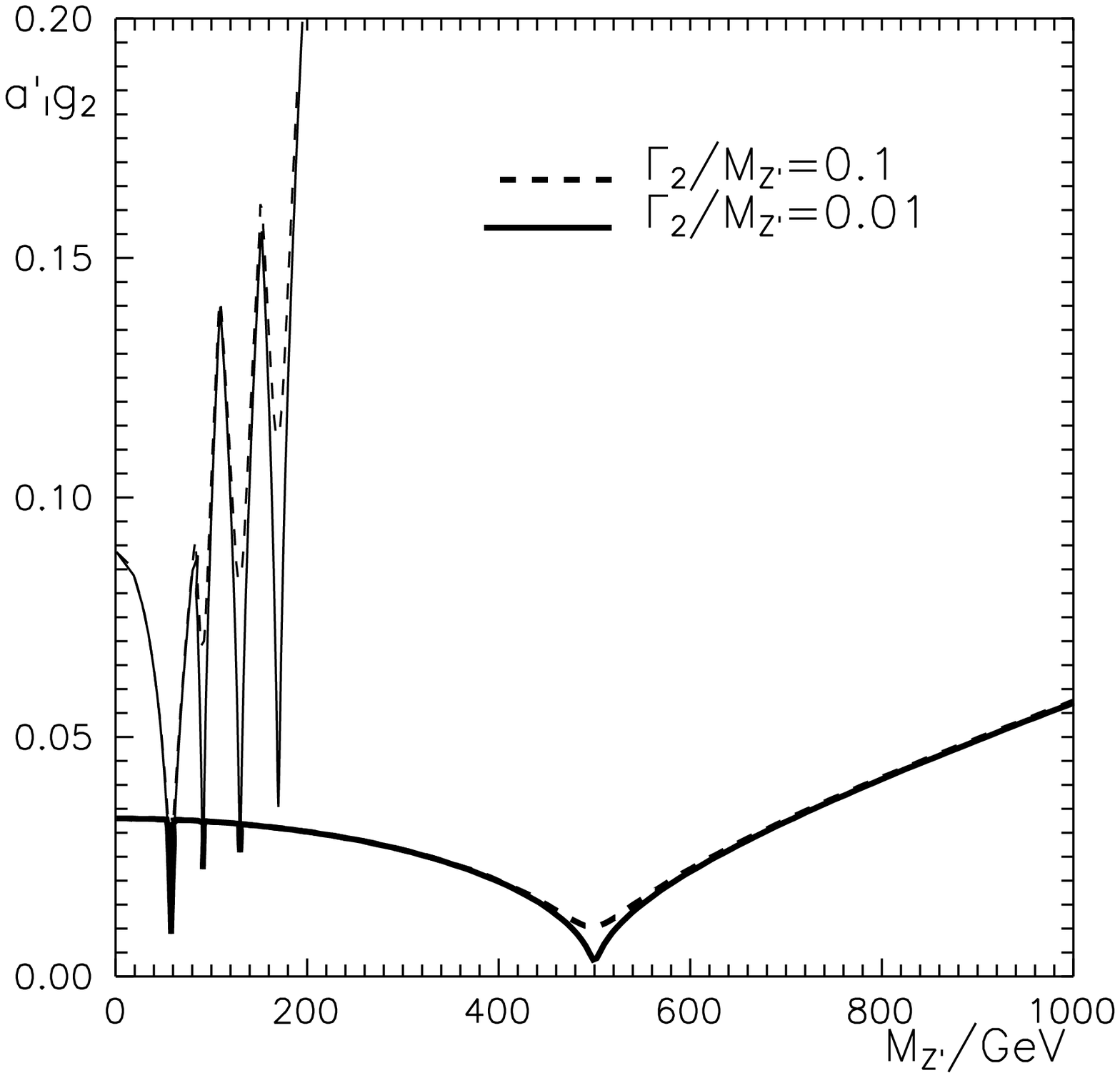}
}
\end{center}
}\end{minipage}
\end{center}
\noindent
{\small\it
\begin{fig} \label{g2lim2} 
Present (thin lines) and future (thick lines) upper bounds (95\% CL) on
$a'_lg_2$ as function of the $Z'$ mass.
Every bound is shown for $\Gamma_2/M_2=0.01$ (solid lines) and
$\Gamma_2/M_2=0.1$ (dashed lines).
See text for the input.
\end{fig}}
\end{figure}

The constraints on $a'_lg_2$ resulting from the input \req{g2liminput}
and \req{nlcinput} are shown in figure~\ref{g2lim2}. 
The limits on $g_2$ for models with other couplings are expected to
be similar because the constraints on $a_l^N$ and $v_l^N$ are
comparable.
The inclusion of PEP and PETRA data would improve the bound on $g_2$
for small $M_{Z'}$.

Although, $\Gamma_{Z'}$ is related to $g_2$ in GUT's, we consider it
as a free parameter in figure~\ref{g2lim2}.
This figure illustrates that the constraint is not very sensitive to
$\Gamma_2/M_2$.  
The same is true for the mass exclusion limits obtained in
the previous section.
This insensitivity to $\Gamma_2/M_2$ is an
important difference to $Z'$ limits from hadron collisions.

For models, where $a'_lg_2$ is known, mass limits $M_{Z'}^{lim}$ can
be derived from figure~\ref{g2lim2}.
For example, $Z'=\psi$ has pure axial couplings to electrons,
$a'_l=1/\sqrt{6},\ \ (a'_lg_2\approx 0.094)$. 
It follows that $M_{Z'}^{lim}=180\,GeV$ from
figure~\ref{g2lim2}. The value $M_{Z'}^{lim}=220\,GeV$ quoted in
table~\ref{zplimlep2} is based on the same data set.
It is better because hadronic observables also enter this number,
while only leptonic observables enter the limit from figure~\ref{g2lim2}.

As we will see later, the limit on $g_2$ for $M_{Z'}^2>s$ can be
further improved for $M_{Z'}<s$ by searching for photons from the
radiative return to the $Z'$ resonance.   
\subsubsection{Errors of model measurements}\label{zpeeff414}
In the case of deviations from the SM predictions, one has to prove
experimentally that these deviations are due to an extra neutral
gauge boson.
If this is the case, the three observables $\sigma_T^l, A_{FB}^l$
and $A_{LR}^l$ depend on the two couplings $a'_l$ and $v'_l$ only.
Therefore, there must be a relation between these observables. 
$Z'$ theories occupy only a two-dimensional subspace of the
three-dimensional space spanned by the values of these three observables.
See reference \cite{cvlc2000} for a detailed discussion of this point.
If the new interaction is due to a $Z'$, its couplings to all fermions
should be measured.
These measurements can de done with
or without model assumptions as far as these are consistent with the data.
In contrast to the $Z'$ exclusion limits, the systematic errors of
future experiments have a significant influence on measurements of
$Z'$ model parameters.
\paragraph*{Estimate}
Suppose that there exists a $Z'$ with 
$M_{Z'}=f_m M_{Z'}^{lim}<M_{Z'}^{lim}$.
We first estimate the experimental bounds on the $Z'$ mass  
$M_{Z'}^- < M_{Z'} < M_{Z'}^+$, which can be set by the observable $O$. 
Considerations similar to those of the exclusion limits in section
\ref{mzlimest} give 
\bq
M_{Z'}^\pm = M_{Z'}\left(\frac{1\mp \Delta o(1-\Delta o/f_m^2)}
{1\mp f_m^2\pm\Delta o}\right)^{1/2}\approx \frac{M_{Z'}}{\sqrt{1\mp f_m^2}}.
\eq
The last approximation is valid under the conditions
$\Delta o\ll 1-f_m^2<1$ and $\Delta o\ll f_m^2$, which are fulfilled in a
reasonable model measurement. 
For small $f_m^2$, we can further approximate,
\bq
\label{epemmea}
\frac{M^+_{Z'}-M^-_{Z'}}{M^+_{Z'}+M^-_{Z'}} \approx 
\frac{\Delta M_{Z'}}{M_{Z'}} \approx 
\frac{1}{2}\left(\frac{M_{Z'}}{M_{Z'}^{lim}}\right)^2\equiv\frac{1}{2}f_m^2.
\eq

Similar considerations can be used for a measurement of the coupling 
strength $g_2$ assuming that the  $Z'$ mass is known,
\bq
\label{epemmea2}
\frac{\Delta g_2}{g_2} \approx\frac{1}{2}\left[ f_m^2-\Delta o(1-f_m^2)\right] 
\approx\frac{1}{2} f_m^2.
\eq
Again the last sequence of the approximations relies on $\Delta o\ll f_m^2$.
In practice, the estimates \req{epemmea} and \req{epemmea2} work
satisfactorally for $f_m<\frac{1}{2}$.

The estimates \req{epemmea} and \req{epemmea2} give a general relation
between $Z'$ exclusion limits and relative errors of $Z'$ model measurements: 
They relate the amount of $(Ls)_{det}$ one has to pay for the
detection of a $Z'$ of a certain model to the amount of
$(Ls)_\varepsilon$ which is necessary for a model measurement with the
accuracy $\varepsilon$, of the 
same model by the same observables at the same confidence level
\cite{9708436},  
\bq
\label{measexcl}
(Ls)_\varepsilon\approx\frac{1}{4\varepsilon^2}\cdot(Ls)_{det}.
\eq

The influence of systematic errors on model measurements is predicted
by the estimates \req{epemmea}, \req{epemmea2} and \req{epemlim2},
\bq
\label{epemmeas}
\frac{\Delta M_{Z'}}{M_{Z'}},\ \frac{\Delta g_2}{g_2} \sim
\left[\frac{1+r^2}{sL}\right]^{1/2}.
\eq

Relations \req{epemmeas} are scaling laws similar to \req{epemlim2}.
Compared to exclusion limits, the influence of the systematic 
error is now more pronounced.
\paragraph*{Present Measurements}
There are no experimental indications for extra neutral gauge bosons.
However, in the PEP \cite{pep}, PETRA \cite{petra} and TRISTAN
\cite{tristanexp} experiments the couplings (and the mass) of the SM $Z$
boson were constrained by measurements below its resonance.
These experimental results allow the test of our estimates \req{epemmea}
and \req{epemmea2}. 
In fact, these estimates give a correct prediction of the experimental
error of the $Z$ coupling measurement $\Delta a_\mu/a_\mu$ of {\it
all} these experiments within a factor of two. 

\begin{figure}[tbh]
\ \vspace{1cm}\\
\begin{minipage}[t]{7.8cm} {
\begin{center}
\hspace{-1.7cm}
\mbox{
\epsfysize=7.0cm
\epsffile[0 0 500 500]{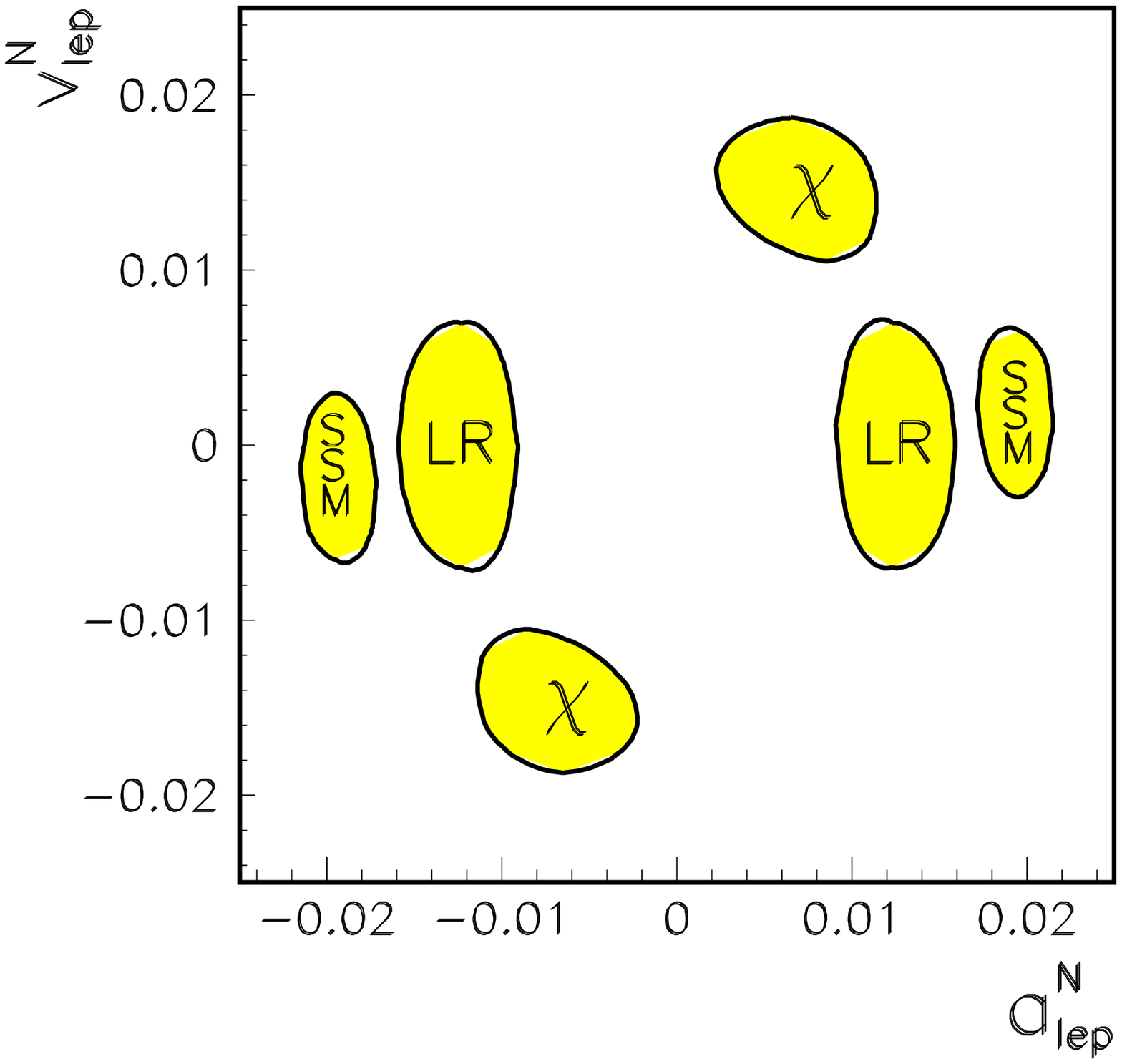}
}
\end{center}
\vspace*{-0.5cm}
\noindent
{\small\it
\begin{fig} \label{leptmeas} 
Resolution~power~of LC500
(95\% CL) for different models and $M_{Z'}=1.5\,TeV$
based on a combination of all leptonic observables.
This is figure~4b from reference \cite{lmu0296}.
\end{fig}}
}\end{minipage}
\hspace*{0.5cm}
\begin{minipage}[t]{7.8cm} {
\begin{center}
\hspace{-1.7cm}
\mbox{
\epsfysize=7.0cm
\epsffile[0 0 500 500]{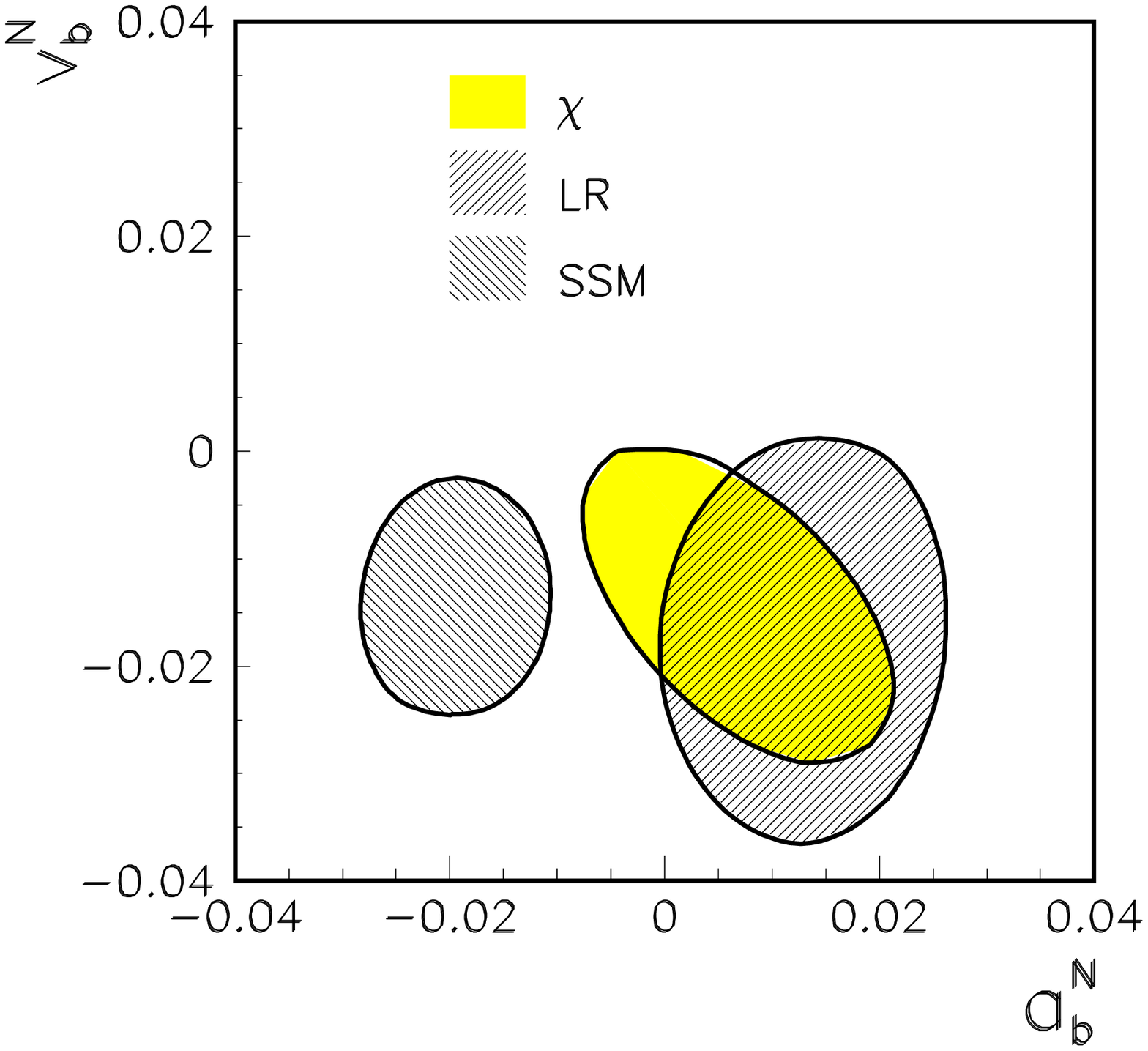}
}
\end{center}
\vspace*{-0.5cm}
\noindent
{\small\it\begin{fig} 
\label{bzpconstraint} 
Resolution power of LC500 in the $(a_b^N,v_b^N)$ plane (95\% CL)
based on a combination of all $b$--quark observables.
Different $Z'$ models are considered, $M_{Z'}$=1.5\,TeV. 
This is figure~7 from reference \cite{lmu0296}.
\end{fig}}
}\end{minipage}
\end{figure}

Let us demonstrate the estimates with the results of the AMY collaboration
\cite{tristanexp}. 
In the first step, we calculate $M_Z^{lim}\approx 350\,GeV$ using the
estimate \req{offres}. 
We took the most accurate observable $A_{FB}^\mu=-0.303\pm 0.028$ for
$\Delta o$ adding the statistical and systematic errors in quadrature.
The estimates \req{epemmea} and \req{epemmea2} predict 
$\Delta M_Z/M_Z,\Delta g/g\approx 3.4\%$.
This can be compared with the result of the AMY analysis, 
$\Delta g/g=\Delta\sqrt{a_l^2+v_l^2}/\sqrt{a_l^2+v_l^2}\approx\Delta
a_l/a_l=0.024/0.476\approx 5\%$.
\paragraph*{Future Measurements}
Figure~\ref{leptmeas} shows typical results of a fit to $a_l^N$ and
$v_l^N$ in the case of a $Z'$ signal.
It includes the full SM corrections and a cut on the photon energy.
As in the case of exclusion limits shown in figure~\ref{leptexcl},
different observables shrink different regions in the parameter space.
A two-fold sign ambiguity remains as long as fermion pair production is
the only process which detects the $Z'$. 
We would be left with a four-fold sign ambiguity without $A_{LR}^l$
because only $A_{LR}^l$ or related observables are sensitive to the sign of 
$v_l^N\cdot a_l^N$.
This underlines the essential role of beam polarization in $Z'$ {\it model
measurements}. 

A superposition of figure~\ref{leptmeas} and figure~\ref{couplept} allows
us to estimate errors of model measurements.
Assume a measurement of the overall coupling strength
$c_l^N\approx\sqrt{(a^N_l)^2+(v^N_l)^2}$ and $M_{Z'}=1.5\,TeV$. 
One finds from figure \ref{leptmeas} the errors 
$\Delta c_l^N/c_l^N= 0.27,\ 0.23$ and $0.11$ for $Z'=LR,\ \chi$ and $SSM$.
To compare this result with the estimate \req{epemmea}, one should
take $M_{Z'}^{lim}$ obtained from leptonic observables only, which are $2.0,\
2.6$ and $2.7\,TeV$ according to table~\ref{zplimlc} in \cite{lmu0296}. 
The errors predicted by the estimate \req{epemmea} are then $0.28,\ 0.17$
and $0.15$. 
The agreement of the numbers is reasonable.

%

%
In reference \cite{lmu0296}, the couplings of the $Z'$ to $b$ quarks
are constrained fixing the leptonic couplings to the
values predicted in certain $E_6$ models.
Assuming that the couplings to quarks are predicted by the same model,
one can constrain the allowed region for $a'_b,v'_b$ by observables
with $b$ quarks in the final state. 
The resulting regions are clearly off the point $(a'_b,v'_b)=(0,0)$ as
shown in figure~\ref{bzpconstraint}. 
Compared to figure~\ref{leptmeas}, the constraints on $v_q^N$ and
$a_q^N$ have a larger error, which is due to the larger systematic
errors of $b$--quark observables.

Alternatively, one can assume $Z'$ couplings to leptons, which are inside the
combined region of figure~\ref{leptexcl}  for a $500\,GeV$ collider.
Such models don't give a signal in the leptonic observables.
Taking different sets of leptonic couplings satisfying these
conditions, one can estimate the limits on $Z'$ couplings to quarks.
Such an analysis is performed in reference \cite{sr1997} for LEP2 data
and in references \cite{cvsmallcoup} for future colliders.

Assuming relation \req{qpt3}, all couplings of the $Z'$ to SM fermions
are described by the five parameters \req{coup5ee}.
One can then try to fit these parameters by all available observables
simultaneously.
This is done in reference \cite{prd50} for an $e^+e^-$ collider with
$\sqrt{s}=0.5\,TeV, L=20\,fb^{-1}$ and $M_{Z'}=1\,TeV$ for
different models using the 18 different observables
\bq
\sigma_T^l, R^{had},A_{LR}^{had},R^f=\frac{\sigma_T^f}{\sigma_T^l},
A_{FB}^f,A_{LR}^f,A_{LR,FB}^f,\ f=l,c,b,t.
\eq
In references \cite{prd50} and \cite{cvetrev}, one can find three
dimensional figures of possible constraints.
We select the result of such a fit from table~5 of the newer analysis
\cite{prd52} and present it in table \ref{mcvetee}.
No radiative corrections and no systematic errors are included in the
analysis \cite{prd52}.
The complete table~4 in reference \cite{prd52} underlines again the
crucial role of beam polarization in model measurements.
It is shown there that the errors of model measurements without beam
polarization are several times larger. 

\begin{table}[tbh]
\begin{center}
\begin{tabular}{|lrrrr|}\hline 
Analysis&$\chi$&$\psi$ &$\eta$ &$LR$\rule[-2ex]{0ex}{5ex}\\ 
\hline
$P_V^l$     &$2.00\pm 0.08$ &$0.00\pm0.04$ &$-3.0\pm 0.5$ &$-0.148\pm 0.018$\\
$P_L^q=P_l^b$&$-0.50\pm 0.04$&$0.50\pm0.10$ &$2.0\pm 0.3$ &$-0.143\pm 0.037$\\
$P_R^u$&$-1.00\pm 0.15$&$-1.00\pm0.11$ &$-1.00\pm 0.15$ &$-6.0\pm 1.4$\\
$P_R^d=P_R^b$&$3.00\pm 0.24$&$-1.00\pm0.21$ &$0.50\pm 0.09$ &$8.0\pm 1.9$\\
$\epsilon_A$&$0.071\pm 0.005$&$0.121\pm0.017$ &$0.012\pm 0.003$ 
&$0.255\pm 0.0016$\\
\hline
\end{tabular}\medskip
\end{center}
{\small\it  \begin{tab}\label{mcvetee}
Values of the couplings \req{coup5ee} for typical models and
statistical errors as determined from probes at the NLC
($\sqrt{s}=500\,GeV,\ L=20\,fb^{-1},\ M_{Z'}=1\,TeV$). 
100\% heavy flavour tagging efficiency and 100\%
longitudinal polarization of the electron beam is assumed.
The numbers are taken from table~4 of reference \cite{prd52}.
\end{tab}}
\end{table}

\begin{figure}
\begin{center}
\begin{minipage}[t]{7.8cm}{
\begin{center}
\hspace{-1.7cm}
\mbox{
\epsfysize=7.0cm
\epsffile[0 0 500 500]{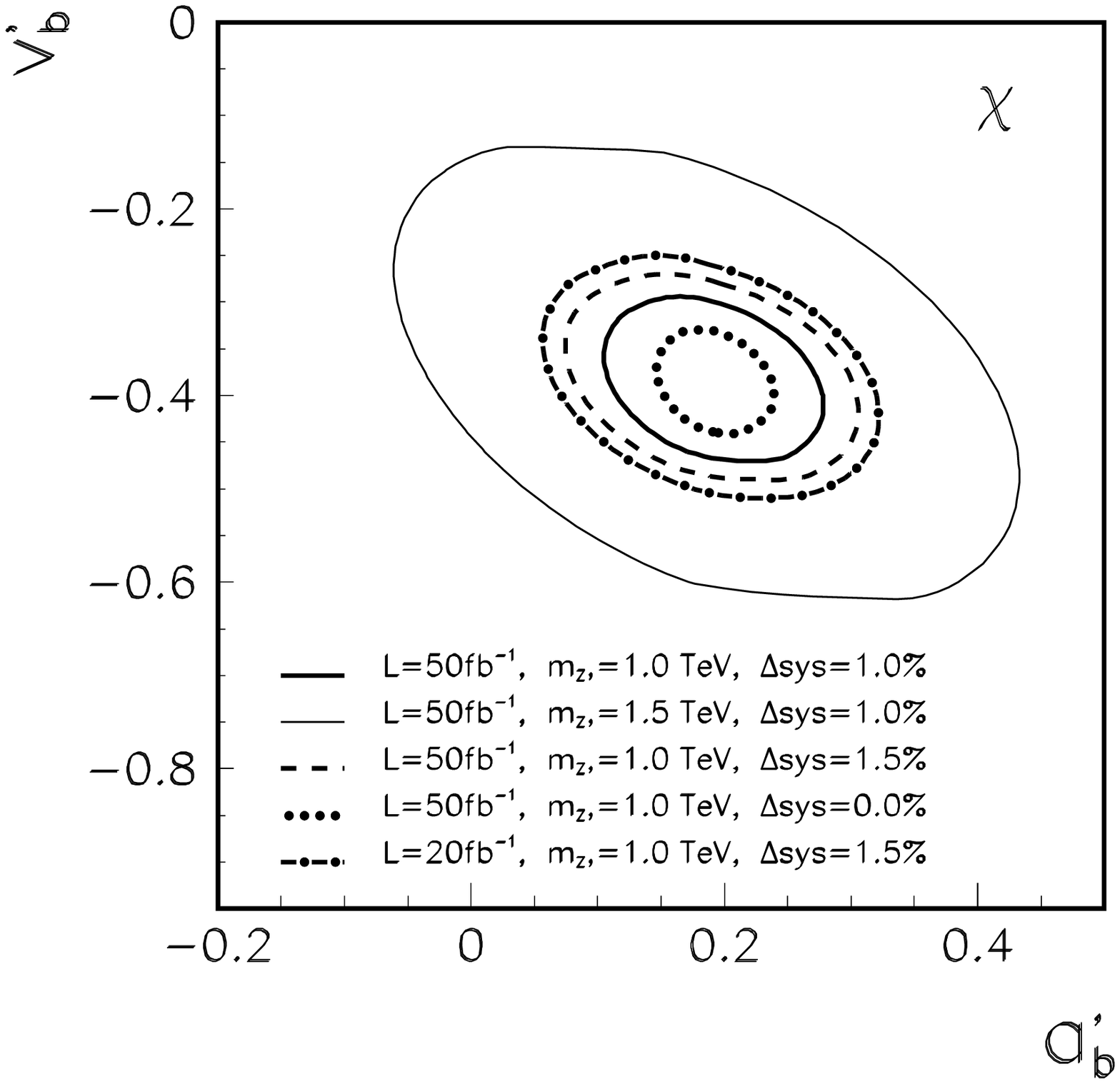}
}
\end{center}
}\end{minipage}
\end{center}
\noindent
{\small\it
\begin{fig} \label{snowfig} 
Influence~of~luminosity, $Z'$ mass, and systematic error on contours
of $Z'b\bar b$ couplings. 
A $Z'$ in the $\chi$ model is assumed; $\sqrt{s}=500\,GeV$. This is
figure~3 of reference \cite{srsnowmass}.
\end{fig}}
\end{figure}

Having made measurements of the $Z'$ couplings, one can go
one step further and check whether the signal is compatible with a
$Z'$ originating from an $E_6$  GUT, i.e. check whether the relations
\req{e6check} are fulfilled. 
If this is the case, one can try to define the underlying parameters
$g_{12}/g_2$ and $\beta$ of the breaking of the $E_6$ group from the
data \cite{prd52}.
Alternatively, one can try to constrain \cite{prd52,lmu0296,zpsari}
the parameters $\alpha$ or $\beta$ parametrizing the $E_6$ models as
given in equations \req{e6so10} and \req{lrcurrent}.
\paragraph*{Influence of systematic errors}
Let us compare the scaling \req{epemmeas} with the results of an exact
calculation shown in figure~\ref{snowfig}.
We first note that the different scenarios change mainly the size of
the constrained region but not its shape. 
The change of the size can be characterized by one number. 
Let us normalize the the size of the region without systematic errors to unity.
Then the sizes of the regions of the five scenarios in figure
\ref{snowfig} are (legend from top to down) 1.9, 5.6, 2.5,
1 and 2.9. 
Combining the curves with $\Delta^{syst}$ =0 and $\Delta^{syst}=1.5\%$
for $L=50\,fb^{-1}$ and using estimate \req{epemmeas}, we calculate
$\Delta^{syst}/\Delta^{stat}=r=2.3$, and therefore $\Delta^{stat}=0.65\%$. 
This is exactly the value, which would follow from the expected number
of $b$ quark events.
We now estimate \req{epemmeas} the sizes of all remaining 
curves in figure \ref{snowfig} as 1.8, 4.2, 2.5, 1, 2.8.
The agreement except for the second curve is surprisingly good.
The disagreement of the second curve arises because 
the accuracy $\epsilon$ of the model measurement is well above
$\frac{1}{8}$, which means that \req{epemmeas} is a bad approximation.
In a similar way, the different scenarios of the other figures in
reference \cite{srsnowmass} are reproduced by the relation \req{epemmeas}.

%

\begin{table}[htb]
\begin{tabular}{|lrrrr|}\hline
   & $\chi$ & $\psi$ & $\eta$ & LR\rule[-2ex]{0ex}{5ex}\\ 
\hline
$P_V^l$, no syst. err.   & 2.00$\pm$ 0.11         &
                          0.00$\pm$ 0.064        &
                         -3.00$^{+0.53}_{-0.85}$ &
                         -0.148$^{+0.020}_{-0.024}$ \\
$P_V^l$, syst. err. included & 2.00$\pm$ 0.15 &
                          0.00$\pm$ 0.13 &
                         -3.00$^{+0.73}_{-1.55}$  & 
                         -0.148$^{+0.023}_{-0.026}$  \\
$P_L^b$, no syst. err.   & -0.500$\pm$ 0.018 &
                           0.500$\pm$ 0.035 &
                           2.00$^{+0.33}_{-0.31}$ &
                           -0.143$\pm$0.033  \\
$P_L^b$, syst. err. included & -0.500$\pm$ 0.070 &
                           0.500$\pm$ 0.130 &
                           2.00$^{+0.64}_{-0.62}$ &
                           -0.143$\pm$0.066  \\
$P_R^b$, no syst. err.   &  3.00$^{+0.15}_{-0.14}$ &
                          -1.00$\pm$ 0.29   &
                           0.50$\pm$ 0.11  &
                           8.0$^{+2.5}_{-1.5}$      \\
$P_R^b$, syst. err. included &  3.00$^{+0.65}_{-0.50}$ &
                          -1.00$^{+0.26}_{-0.34}$  &
                           0.50$^{+0.23}_{-0.22} $ &
                          8.0$^{+6.7}_{-2.4}$  \rule[-2ex]{0ex}{5ex}\\
\hline
\end{tabular}\medskip
{\small\it  \begin{tab}\label{couplimlc}
$Z'$ coupling
combinations  $P_V^l,\ P_L^b$ and $P_R^b$
and their one--$\sigma$ errors derived from all observables with and without
systematic errors for $\sqrt{s}$=500\,GeV and  $M_{Z'}$=1\,TeV.
This is table~5 of reference \cite{lmu0296}.
\end{tab}}
\end{table}

The estimate \req{epemmeas} can be confronted with the measurements of
model parameters quoted in table \ref{couplimlc}.
The influence of systematic errors is predicted by \req{epemmeas}.
For $r=1$, as expected for leptonic observables in reference
\cite{lmu0296},  $\Delta P_V^l$ are predicted to relax by $\sqrt{2}$, 
which reproduces the tendency in table~\ref{couplimlc}. 
$P_L^b$ given in table \ref{couplimlc} is only measured by observables
with $b$ quarks in the final state. 
It is therefore dominated by the systematic errors of $b$ quark observables. 
In reference \cite{lmu0296}, these systematic errors are
roughly four times as large as the statistical errors. 
According to the estimate \req{epemmeas}, $r=4$ should enlarge $\Delta
P_L^b$ and $\Delta P_R^b$ by a factor $\sqrt{17}$. 
Indeed this happens for some models in table~\ref{couplimlc}.

To conclude, the systematic errors have a {\it large} influence on
errors of {\it model measurements}.
\paragraph*{Combining Measurements at several Energies}
The estimates obtained above for measurements at {\it one} energy
point can be generalized to measurements, which are distributed over  
{\it several} energy points $s_i$ with integrated luminosities $L_i$.
Observing that only the combination $sL$ enters all estimates,
one can rescale all measurements to one energy $s_0$ with
the luminosity $L_0= \sum_is_iL_i/s_0$ and assume
{\it one} measurement at $s_0$ with $L_0$.

A difference occurs, if one wants to use several energy points for a 
{\it simultaneous} measurement of the $Z'$ mass and the $Z'$ couplings
by a fit to the line shape as proposed in references \cite{rizzoi,1583}. 
Such a measurement was demonstrated historically for the SM $Z$ boson at
the PEP and PETRA colliders.   
It demands measurements of high accuracy at several well
separated energy points.
The luminosities $L_i$ should be chosen in such a way that the
statistical significance 
\bq
\Delta^{Z'}O/\Delta O\approx \sqrt{s_iL_i}
\eq 
is about equal for the different energy points $s_i$.

Although the error of the simultaneous fit of $M_{Z'},v'_l,a'_l$ 
scales with the product $sL$ like the error of a fit to $v_l^N,a_l^N$,
the prefactor is very different. 
One needs much more luminosity to measure the couplings and the $Z'$ mass
separately compared to a measurement of $v_f^N$ and $a_f^N$ only.
Here is one example of the amount of $sL$ needed for different
experiments $(\sqrt{s}<M_\chi=1.6\,TeV, 95\%$ CL$)$:

\begin{tabular}{ll}
Detection: & $sL\approx 0.7\,TeV^2/fb$\\
measurement of $g_2^2/M_{Z'}^2$ or $v_l^N,a_l^N$ with 15\% error: 
& $sL\approx 8\,TeV^2/fb$\\
measurement of $g_2^2$ and $M_{Z'}^2$ separately with 15\% error: 
& $sL\approx 260\,TeV^2/fb$
\end{tabular}

The first two numbers are obtained from table~\ref{zplimlc} using the
scalings \req{epemlim2} and \req{epemmea} with $r=0$, the third number is taken
from reference \cite{rizzoi}. 
\subsection{$Z'$ measurements at $s\approx M_2^2$}
\label{zpeeff5}
A precision measurement on top of the $Z_2$ resonance in $e^+e^-$ or
$\mu^+\mu^-$ collisions would be the best experiment to study the properties of
an extra neutral gauge boson. See for example \cite{e6,zppeak,pankov,9710229}
for related studies. 
Such an experiment would have much in common with the measurements at the
$Z_1$--peak at LEP\,1 and SLC, however, there are also important differences.

In the case of a non-zero $ZZ'$ mixing, in addition to the decay
$Z_2\rightarrow f\bar f$ other decay modes such as $Z_2\rightarrow W^+W^-$
or $Z_2\rightarrow Z_1H$ are allowed.
Because there are no experimental hints for a non-zero $ZZ'$ mixing,
we discuss the decay to $W^+W^-$ in a different section.
If $Z_2$--decays to exotic fermions are
kinematically allowed, the number of observables at the $Z_2$--peak is
even larger than that at the $Z_1$--peak yielding important additional
information on the breaking scheme of the underlying GUT.

A second difference to the $Z_1$--peak is the effect of beamstrahlung.
The resulting energy spread of the beams is expected to be between 0.6
and 2.5\% for the discussed $500\,GeV$ $e^+e^-$ colliders and even
larger for higher energies \cite{barklow}. 
Some precision measurements would demand an energy spread as low as
0.2\% \cite{schulte}.
A precision scan of the $Z_2$ peak is among these measurements.
Here, a $\mu^+\mu^-$ collider would have clear advantages over an
$e^+e^-$ collider.
In $\mu^+\mu^-$ collisions, the energy spread of the beams is
naturally between 0.04 and 0.08\% \cite{mumuproc}. 
It is expected that it can be further reduced to 0.01\% \cite{mumuproc}.
The absolute calibration of the center--of--mass energy is expected
to be of the same accuracy.

As in the previous cases, $Z'$ measurements at the $Z_2$ peak
can also be made with or without model assumptions.
\subsubsection{Model independent $Z'$ measurements}
\label{zpeeff51}
A model independent measurement on the $Z_2$ resonance can be
done generalizing the model independent approach to the $Z_1$
resonance \cite{smatrix,zpole}.
It has the advantage that no unknown radiative corrections are needed.
We follow here a notation close to reference \cite{zpoletr}, where
also a more extensive discussion and further references may be found.
The ansatz for the four helicity amplitudes is
\bq
\label{sampl}
{\cal M}^{fi}(s)=e^2s\sqrt{\frac{4}{3}}\left[
\frac{R_\gamma^f}{s}+\frac{R_Z^{fi}}{s-\bar m_1}
+\frac{R_{Z'}^{fi}}{s-\bar m_2}+F^{fi}(s)\right],
\ \ \ i=0,1,\dots,3.
\eq
$F^{fi}(s)$ is an analytic function without poles.
In general, the expression \req{sampl} is not unique.
However, it makes sense, if the resonances are well separated.
This is the case because we have $m_1\ll m_2$ due to the present
experimental constraints.
The QED parameter $R_\gamma^f$ can be measured at such a low energy that
all other contributions are unimportant.
Therefore, a model independent fit to the $Z_1$ resonance can be performed
{\it fixing} the QED term. 
See reference \cite{zpole} for a first fit and \cite{zpoleexp} for
recent experimental results. 
There are no hints for a $Z_2$ at the $Z_1$ resonance.
Therefore the fits to the $Z_1$ resonance are independent of a
possible term of the $Z_2$ pole.
By the same scheme, a model independent fit of the $Z_2$ resonance can 
be performed with the amplitude \req{sampl} {\it fixing} the QED {\it and} the
$Z_1$ terms.
Such a procedure is as unique as are the present model independent fits to
the $Z_1$ resonance.

The complex constants $R_\gamma^f$ and $R_Z^{fi}$ can be derived within a
given theory to a certain order of perturbation series.
At the Born level, we have
\ba
R_\gamma^f&=&Q^eQ^f,\nll
R_Z^{f0}&=&4L_eL_f\frac{g_1^2}{e^2},\ \ 
R_Z^{f1}=4L_eR_f\frac{g_1^2}{e^2},\ \ 
R_Z^{f2}=4R_eR_f\frac{g_1^2}{e^2},\ \ 
R_Z^{f3}=4R_eL_f\frac{g_1^2}{e^2},\nll 
R_{Z'}^{f0}&=&4L'_eL'_f\frac{g_2^2}{e^2},\ \ 
R_{Z'}^{f1}=4L'_eR'_f\frac{g_2^2}{e^2},\ \ 
R_{Z'}^{f2}=4R'_eR'_f\frac{g_2^2}{e^2},\ \ 
R_{Z'}^{f3}=4R'_eL'_f\frac{g_2^2}{e^2}.
\ea
The complex mass $\bar m_i=\bar M_i^2-i\bar M_i\bar\Gamma_i$ is slightly
different from the on-shell mass $m_i$ because it contains a constant
width defining the complex pole in the ansatz \req{sampl}.
If the vector boson $Z_n$ can only decay into light fermion pairs, we
have \cite{gammazs}
\bq
\bar M_n=M_n-\frac{\Gamma_n^2}{2M_n},\ \ \ 
\bar \Gamma_n=\Gamma_n-\frac{\Gamma_n^3}{2M_n^2}.
\eq

From the amplitude \req{sampl}, total cross sections can be calculated by
the standard procedure,
\bq
\label{msquare}
\sigma_i^f=\frac{1}{2s}\int|{\cal M}^{fi}(s)|^2d\Gamma
=|{\cal M}^{fi}(s)|^2\frac{1}{(4\pi)^2}\frac{\pi}{s}.
\eq
The four Born cross sections $\sigma_T^{f0}, \sigma_{FB}^{f0},
\sigma_{LR}^{f0}$ and $\sigma_{pol}^{f0}$ introduced in \req{lrfbdef}
and \req{poldef} are obtained by linear combinations of $\sigma_i^f$,
\ba
\label{sigmaadef}
\sigma_{T}^{f}  &=& +\sigma_0^f+\sigma_1^f+\sigma_2^f+\sigma_3^f,\nll
\frac{4}{3}\sigma_{FB}^{f}&=&+\sigma_0^f-\sigma_1^f+\sigma_2^f-\sigma_3^f,\nll
\sigma_{LR}^{f} &=& -\sigma_0^f-\sigma_1^f+\sigma_2^f+\sigma_3^f,\nll
\sigma_{pol}^{f} &=& -\sigma_0^f+\sigma_1^f+\sigma_2^f-\sigma_3^f.
\ea

Combining equations \req{msquare} and \req{sampl}, one arrives at a
model independent formula for cross sections,
\bq
\label{micross}
\sigma_A^{f0}(s)\approx\frac{4\pi\alpha^2}{3}
\left[\frac{r_A^{\gamma f}}{s}
+\frac{sr_A^f+(s-\bar M_1^2)j_A^f}{|s-\bar m_1^2|^2}
+\frac{sr'^f_A+(s-\bar M_2^2)j'^f_A}{|s-\bar m_2^2|^2}
\right],\ \ \ A=T,\ FB,\ LR,\ pol.
\eq
In formula \req{micross}, the functions $F^{fi}(s)$ and terms without poles
are neglected for simplicity.

At the Born level, the SM coefficients are \cite{zpole}
\ba
\label{micross2}
r_A^{\gamma f}&=&\frac{1}{4}N_f\sum_{i=0}^3[\pm 1]|R_\gamma^f|^2,\nll
r_A^f&=&N_f\left\{\frac{1}{4}\sum_{i=0}^3[\pm 1]\left|
R_Z^{fi} \right|^2 +2\frac{\bar\Gamma_1}{\bar M_1}\Im mC_A^f\right\},\nll
j_A^f&=&N_f\left\{2\Re eC_A^f-2\frac{\bar\Gamma_1}{\bar
M_1}\Im mC_A^f\right\},\nll 
C_A^f&=&\frac{1}{4}(R_\gamma^f)^*\sum_{i=0}^3[\pm 1]R_Z^{fi}.
\ea
The signs $[\pm 1]$ are the same as the cross sections $\sigma_A^f$
in equation \req{sigmaadef}.
The parameters of the $Z_2$ peak are
\ba
\label{zpmiparm}
r'^f_A&=&N_f\left\{\frac{1}{4}\sum_{i=0}^3[\pm 1]\left|
R_{Z'}^{fi} \right|^2 +2\frac{\bar\Gamma_2}{\bar M_2}\Im m(C'^f_A-C''^f_A)
\right\},\nll
j'^f_A&=&N_f\left\{2\Re eC'^f_A-2\frac{\bar\Gamma_2}{\bar M_2}\Im mC'^f_A
+2\left(1-\frac{\bar M_1^2}{\bar M_2^2}\right)\Re eC''^f_A
+2\frac{\bar\Gamma_2}{\bar M_2}\Im mC''^f_A\right\},\nll 
C'^f_A&=&\frac{1}{4}(R^f_\gamma)^*\sum_{i=0}^3[\pm 1]R_{Z'}^{fi},\nll
C''^f_A&=&\frac{1}{4}\sum_{i=0}^3[\pm 1]R_Z^{fi}(R^{fi}_{Z'})^*.
\ea
$M_1/M_2$ is a small parameter because the $Z_1$ and $Z_2$ peaks are
well separated.
We therefore neglected terms of the order $M_1^3/M_2^3,\
M_1\Gamma_1/M_2^2$ or smaller.   
This keeps the formulae \req{zpmiparm} relatively simple.

Before the formula \req{micross} can be used for fits to data, QED
(and QCD) corrections must be taken into account \cite{zpole}.
Initial state corrections can be calculated by the convolution \req{convol}.
Final state radiation and the interference between initial and final
state radiation can be included by a different convolution \cite{zpole}. 
However, these corrections don't change the pole structure.
Therefore, they could be absorbed into effective coefficients
\req{micross2} and \req{zpmiparm}.
\subsubsection{Model dependent $Z'$ measurements}
\label{zpeeff52}
$Z_2$ measurements at the $Z_2$ peak are precision measurements.
They require radiative corrections.
Unfortunately, these corrections depend on all the parameters of the
whole theory. 
If these are poorly known, theoretical uncertainties arise.
This is similar to LEP\,1 and SLC where theoretical errors of
the radiative corrections at the $Z_1$ peak arise through the unknown
Higgs mass and through the experimental errors of the top-- and the W mass.
At a $Z_2$ peak, the situation is much more uncertain. 
Today it is not known whether a $Z_2$ exists at all.
It is even more speculative to predict the underlying gauge group.
If it is known, one would still need some idea about the 
breaking scheme, the particle content and particle masses to calculate
radiative corrections. 
We have shown in the previous sections that some information
can be obtained by future experiments below the $Z_2$ peak.

Because of the difficulties mentioned above, we have to constrain
ourselves to general conclusions and estimates in the Born approximation.
The cross section at the resonance peak can be expressed through
branching ratios,
\bq
\label{sigpeak}
\sigma_T(e^+e^-\rightarrow f\bar f)\approx\sigma_T^f(M_2^2;Z_2,Z_2)
=\frac{12\pi}{M_2^2}\frac{\Gamma_2^e}{\Gamma_2}\frac{\Gamma_2^f}{\Gamma_2}
=\frac{12\pi}{M_2^2}Br_2^eBr_2^f.
\eq
For a $Z_2$ originating in usual GUT's, one expects millions of muon
pairs and tens of millions of hadron pairs per year from $Z_2$ decays 
at the proposed electron or muon colliders.
This is similar to the situation at LEP\,1 and SLC. 
The systematic errors at future colliders are expected to be at the
same level as at LEP\,1 of SLC.
Therefore, it can be expected that the $Z_2$ couplings to fermions could be
measured with a similar precision as the $Z_1$ couplings.

The measurements of $Z_2$ couplings constrain the $ZZ'$ mixing angle.
The constraints from the $Z_2$ peak are expected to be stronger than
those from the $Z_1$ peak because in a GUT the couplings of the $Z_1$
to fermions are in general larger than the couplings of the $Z_2$ to
fermions. 
A sensitivity to $\theta_M\approx 10^{-4}$ is derived in reference
\cite{pankov}.

The measurements of $Z_2$ couplings constrain parameters of the GUT.
In a naive estimate, one would expect an accuracy of a measurement of
$\cos\beta$ of an $E_6$ GUT comparable to the accuracy of the
$\sin^2\theta_W$ measurement at the $Z_1$ peak.

The measurement of $M_2$ and $\Gamma_2$ is limited by the beam energy
spread and by the error of the energy calibration $\Delta\sqrt{s}$;\ \ 
$\Delta M_2,\ \Delta\Gamma_2>\Delta\sqrt{s}$.
\subsubsection{Limit on $g_2$}
\label{zpeeff53}
The best limit one can put on a weakly interacting $Z'$ is obtained
for $M_2=\sqrt{s}$.
The cross section at the $Z_2$ peak \req{sigpeak} is independent of
the ratio $g_2/g_1$.
The sensitivity to $g_2/g_1$ is limited at an $e^+e^-$ collider because it
has a finite beam energy spread $\Delta\sqrt{s}/\sqrt{s}$.
Only cross sections of $e^+e^-$ pairs with $\sqrt{s}=M_2\pm\Gamma_2$
are enhanced. 
The observed number of $Z_2$ events can be approximated for
$\Delta\sqrt{s}>\Gamma_2$, 
\bq
\label{nffest}
N_{ff}^{Z'}\approx 
L\frac{\Gamma_2}{M_2}\frac{\sqrt{s}}{\Delta\sqrt{s}}\sigma_T^0(M_2^2;Z',Z')
\approx \frac{g_2^2}{g_1^2}L\frac{\Gamma_1}{M_1}\frac{\sqrt{s}}{\Delta\sqrt{s}}
\frac{12\pi}{M_2^2}Br_2^eBr_2^f.
\eq
In the last step of the approximation, we assumed 
\bq
\label{g2approx1}
\frac{\Gamma_2}{M_2}\approx\frac{g_2^2\Gamma_1}{g_1^2M_1}.
\eq

A $Z'$ produces a signal of $n_\sigma$ standard deviations if
\bq
\label{nffdef}
N_{ff}^{Z'}=n_\sigma\sqrt{N_{ff}^{SM}}=n_\sigma\sqrt{L\sigma_T^{f}(M_2^2)}.
\eq
An estimate for $g_2^{lim}$ can now be obtained from equation \req{nffest},
\bq
\label{gz2lim}
g_2^{lim}\approx
g_1\sqrt{n_\sigma}\left[\frac{\Delta\sqrt{s}}{\sqrt{s}}\frac{M_1}{\Gamma_1}
\frac{1}{12\pi Br_2^eBr_2^f}\right]^{1/2}
\left[\frac{\sigma_T^{f0}(M_2^2)M_2^4}{L}\right]^{1/4}.
\eq
Assuming 
\bq
\Delta\sqrt{s}/\sqrt{s}=1\%,\ \ \ 
Br_2^e=Br_2^\mu=3.36\%,\ \ \ 
L=80\,fb^{-1},\ \ \ 
\mbox{and\ \ \ }M_2=1\,TeV,
\eq
we get $g_2^{lim}\approx g_1/140$ with 95\% confidence ($n_\sigma=2$)
for the reaction $e^+e^-\rightarrow\mu^+\mu^-$.
\subsection{$Z'$ limits at $s>M_2^2$}
\label{zpeeff6}
Above the $Z_2$ resonance, the sensitivity to $ZZ'$ mixing is much lower
than on the resonance. 
We therefore neglect the mixing identifying the $Z_1$ and $Z_2$ with
the $Z$ and $Z'$.

Suppose that a $Z'$ was missed below its resonance because it has very
weak couplings.
The question we want to discuss here is, for which coupling strengths
is it possible to detect such a $Z'$ above its resonance.
If the couplings to all fermions are very small, the $Z'$ eventually
escapes detection.
A vector boson, which couples to quarks only, can still have quite
large couplings and be consistent with the present data.
See reference \cite{prl74} for a discussion of this point and for
further references and \cite{plb221b} for bounds on such a $Z'$ from
different experiments.

It turns out that the error of the photon energy measurement $\Delta
E_\gamma/E_\gamma$ is an important input of the bounds because they
arise from events with a fermion pair and one hard photon in the final state.
At a $\mu^+\mu^-$ collider, this parameter limits the constraints.
However, as mentioned in the previous section, the proposed $e^+e^-$
colliders suffer from a large beam energy spread
$\Delta\sqrt{s}/\sqrt{s}$.
Then, the error of the photon energy $\Delta E_\gamma$ in the estimates
has to be replaced by $\Delta E_\gamma+\Delta\sqrt{s}$.
\subsubsection{Model independent limits on $g_2$}
Starting from relation \req{offres}, an upper bound $g_2^{lim}$ on the
coupling strength $g_2$ of the $Z'$ to SM fermions can be derived,
\bq
\label{weakexcl}
g_2^{lim}\approx g_1 \sqrt{\frac{\Delta O}{O}\cdot\frac{s-M_{Z'}^2}{s}}.
\eq
It follows that a one percent cross section measurement of the reaction
$e^+e^-\rightarrow f\bar f$ can exclude models with $g_2>g_1/7$ at 95\%
confidence for all $M_{Z'}<\sqrt{s}$.  

The sensitivity to a $Z'$ is considerably larger than
\req{weakexcl} if  
one considers the photon energy spectrum of the reaction 
$e^+e^-\rightarrow f\bar f\gamma$.
As discussed in section \ref{zpeeff21} and shown in figure \ref{dsigmaddelta}, 
the spectrum of the photons radiated from the initial state
has sharp peaks for energies which set the $f\bar f$ subsystem
back to the resonance.
The energy $E_\gamma =\Delta\cdot E_{beam}$ of the photons responsible for the
radiative tail is distributed in the narrow range $\Delta^+<\Delta<\Delta^-$
with $\Delta^\pm = 1-M_{Z'}^2/s\pm M_{Z'}\Gamma_{Z'}/s$.
The number of events with these photons can be estimated from the
magnitude of the radiative tail \req{radtail2},
\bq
\label{ngazp}
N_\gamma^{Z'} \approx
L\sigma_T(s;Z',Z')\beta_e\frac{\pi}{2}\frac{M_{Z'}}{\Gamma_{Z'}}
\approx\frac{g_2^2}{g_1^2}
L\sigma_T(s;Z,Z)\beta_e\frac{\pi}{2}\frac{M_Z}{\Gamma_Z}. 
\eq
In the last step of approximation \req{ngazp}, we used the estimate
\req{g2approx1} and
\bq
\label{g2approx}
\frac{\sigma_T(s;Z',Z')}{\sigma_T(s;Z,Z)}\approx\frac{g_2^4}{g_1^4}.
\eq
Of course, these approximations can be improved if more details of
the model are known. 

The SM background also contributes photons to the interval
$(\Delta^-,\Delta^+)$.
The number of these events can be estimated by the convolution \req{convol},
\bq
\label{ngasm}
N_\gamma^{SM}=L\left[\sigma_T^{ISR}(s)-\sigma_T^0(s)\right]
= L\int_{\Delta^-}^{\Delta^+}dv\;
\sigma_T^0\left(\nobodyfrac s(1-v)\right)H_{A}^e(v)
\approx L\sigma_T^0(s)\frac{\beta_e(\Delta^+-\Delta^-)}{\Delta^*(1-\Delta^*)}.
\eq
$\Delta^*$ is some value between $\Delta^-$ and $\Delta^+$.
We assumed that $\sigma_T^0(s)\sim 1/s$ for energies between $1/s$ and
$1/[s(1-\Delta^*)]$. 
The estimate \req{ngasm} can be compared with the SM result in
figure~\ref{dsigmaddelta}. 
It gives a satisfactory prediction away from resonances.

The ratio of the $Z'$ signal and the SM background for photon
energies between $\Delta^-$ and $\Delta^+$ can now be estimated as
\bq
\label{sigback}
\frac{N_\gamma^{Z'}}{N_\gamma^{SM}}\approx
\frac{\pi}{4}\frac{M_Z^2}{\Gamma_Z^2}\frac{\sigma_T^0(s;Z,Z)}{\sigma_T^0(s)}
\left(1-\frac{M_{Z'}^2}{s}\right).
\eq
Again, the approximation \req{g2approx1} is used.
Note that the ratio \req{sigback} is independent of $g_2/g_1$.
Numerically, we get $N_\gamma^{Z'}/N_\gamma^{SM}\approx 40$ for
\bq
\label{figconts}
\sigma_T^\mu(1\,TeV^2)\mbox{\ \ and\ \ }M_\eta=800\,GeV.
\eq
This estimate is in a good agreement with figure \ref{dsigmaddelta}.

To detect the $Z'$ signal, two additional conditions must be fulfilled.
The luminosity must be high enough
to produce a reasonable number of events and the error of the
photon energy $\Delta E_\gamma/E_\gamma$ must be small enough to
detect the signal above the background.

Let us first assume an arbitrarily good photon energy resolution.
Assume that the events are Poisson distributed.
Then, zero observed events exclude all theories with 95\% confidence,
which predict $N_\gamma^{Z'}\ge 3$.
This can be interpreted as a limit $g_2^{lim}$ on $g_2$,
\bq
\label{g2ult}
g_2^{lim}=g_1\left[\frac{N_\gamma^{Z'}}{N_{ff}^{SM}}
\frac{\sigma_T^0(s)}{\sigma_T^0(s;Z,Z)}
\frac{2}{\pi\beta_e}\frac{\Gamma_Z}{M_Z}\right]^{1/2}.
\eq
$N_{ff}^{SM}$ is the number of fermion pairs expected in the Born approximation
as defined in equation \req{nffdef}.
The estimate \req{g2ult} gives the best bound on $g_2$, which could be reached
with a given luminosity in the reaction $e^+e^-\rightarrow f\bar f\gamma$.
Numerically, we get $g_2^{lim}\approx g_1/45$ under the assumptions
\req{figconts} and
\bq
\label{figconts2}
L=80\,fb^{-1}\mbox{\ \ and\ \ } N_\gamma^{Z'}=3.
\eq

Unfortunately, the energy resolution $\Delta E_\gamma/E_\gamma\approx
0.1/\sqrt{E_\gamma/GeV}$ of real detectors is finite \cite{desy93}.
As a result, all photons with energies 
$\Delta = (1-M_{Z'}^2/s)(1\pm\Delta E_\gamma/E_\gamma)$ are observed
in the experiment. 
For the expected luminosities at future colliders, there are photons
$N_\gamma^{SM}$ from the background even in the narrowest bin of the
photon energy.
Numerically, we get $N_\gamma^{SM}\approx 30$ from the estimate
\req{ngasm} under the assumptions \req{figconts}, \req{figconts2} and
$\Delta E_\gamma/E_\gamma =1\%$, which is in good agreement with
figure \ref{dsigmaddelta}.
With such an event number, we can assume Gaussian statistics in our estimates.
One expects a $n_\sigma-\sigma$ signal for theories predicting 
\bq
\label{resolim}
N_\gamma^{Z'}=n_\sigma\sqrt{N_\gamma^{SM}}.
\eq
The resulting expression for $g_2^{lim}$ can easily be derived from
equations \req{resolim}, \req{ngasm} and \req{ngazp},
\bq
\label{g2limphot}
g_2^{lim}=
g_1c_r\left[\frac{8}{\pi^2\beta_e}\frac{\Gamma_Z^2}{M_Z^2}\right]^{1/4}
\approx g_1\cdot0.266c_r,\ \ \ 
c_r=\left[\frac{\Delta E_\gamma}{E_\gamma}\frac{1}{N_{ff}^{SM}}\right]^{1/4}
\left[n_\sigma\frac{\sigma_T^0(s)}{\sigma_T^0(s;Z,Z)}\frac{\sqrt{s}}
{M_{Z'}}\right]^{1/2}.
\eq
Numerically, we get $g_2^{lim}\approx g_1/24$ with 95\% confidence
($n_\sigma=2$) under the same conditions as before and a photon energy
resolution of 1\% and $\sqrt{s}/M_{Z'}\approx 1$. 

As we see, the consideration of fermion pair events with one additional
hard photon gives a considerable improvement of $g_2^{lim}$
obtained from off--resonance fermion pair production.

Present upper bounds on $g_2$ from fermion pair production without
additional photons are displayed in figure \ref{g2lim2}.
Fermion pair events accompanied with hard photons are investigated at
LEP \cite{lepbelow}.
This allows to reconstruct cross sections and asymmetries of
$e^+e^-\rightarrow \mu^+\mu^-$ for energies lower than $\sqrt{s}$.
\subsubsection{Measurements of $M_{Z'}$}
If a $Z'$ signal with $M_{Z'}^2<s$ is found, the $Z'$ mass
can be measured,
\bq
M_{Z'}=\sqrt{s\left(1-\frac{E_\gamma^{peak}}{E_{beam}}\right)},\hspace{1cm} 
\frac{\Delta M_{Z'}}{M_{Z'}}=
\frac{\Delta E_\gamma}{2E_\gamma}\left(\frac{s}{M_{Z'}^2}-1\right).
\eq
We exploited formula \req{deltalim} in the derivation of this estimate.
$E_\gamma^{peak}$ is the photon energy of the hard photons from the radiative
return.
Using the knowledge of $M_{Z'}$, one can tune the energy to the resonance and
perform precision measurements there. 
\section{$Z'$ search in $e^+e^-\rightarrow e^+e^-$ and 
$e^-e^-\rightarrow e^-e^-$}
\label{zpeeee}
Bhabha and M{\o}ller scattering can probe the $Z'$
couplings to electrons only.
While Bhabha events serve as additional observables in $e^+e^-$
collisions, M{\o}ller  
scattering requires the $e^-e^-$ option of a linear collider.

Although the luminosity of $e^-e^-$ collisions is expected to be smaller than
that of $e^+e^-$ collisions because of the anti-pinch effect,
M{\o}ller  scattering has the advantage of two polarized beams and of 
a cleaner environment.

Early $Z'$ analyses can be found in references
\cite{bhabhahol,prd34,prd25} for Bhabha scattering and in
reference \cite{prd25} for M{\o}ller scattering. 
\subsection{Born Approximation}\label{zpeeee1}
\subsubsection{Amplitude}\label{zpeeee11}
In Bhabha (M{\o}ller) scattering, electrons and positrons (only electrons)
appear in the final state.
The neutral gauge bosons are exchanged in the $s$ and $t$ ($t$ and $u$)
channels.
The resulting angular distributions are very singular for small
scattering angles. 
M{\o}ller scattering has a symmetrical angular distribution.
The angular distribution of Bhabha scattering is peaked in the forward
direction. 

Considerations similar to fermion pair production in
section~\ref{zpeeff1} show that these reactions can constrain
only the model independent parameters $a_e^N$ and $v_e^N$. 
As off--resonance fermion pair production, they are rather insensitive to
$ZZ'$ mixing.
We therefore neglect the mixing angle putting $\theta_M=0$.
\subsubsection{Cross section}\label{zpeeee12}
The Born cross section of {\it Bhabha scattering} including the $Z'$
exchange is, following the notation of reference \cite{bhabhatord},
\bq
\label{bornbhab}
\frac{d\sigma}{d c} = \frac{\pi\alpha^2}{2s}\;
\left(\nobodyfrac f_0+(\lambda_+-\lambda_-)f_1+\lambda_+\lambda_-f_2\right)
\eq
with
\ba
f_0&=&(1+c^2)\cdot G_1(s,s) + 2c\cdot G_3(s,s)
-2\frac{(1+c)^2}{1-c}\cdot [G_1(s,t)+G_3(s,t)]  \nll &&
+ 2\frac{(1+c)^2+4}{(1-c)^2}\cdot G_1(t,t)
+2\frac{(1+c)^2-4}{(1-c)^2}\cdot G_3(t,t),\nll
f_1&=&(1+c)^2H(s,s) - 4\frac{(1+c)^2}{1-c}H(s,t)
+4\frac{(1+c)^2}{(1-c)^2}H(t,t),\nll
f_2&=&-f_0 + \frac{16}{(1-c)^2}[G_1(t,t)-G_3(t,t)]
\ea
and
\ba
\label{bhabhax}
G_1(s,t)&=&\Re e\sum_{m,n=0}^N\;\chi_m(s)\chi_n^*(t)
\left[\nobodyfrac v_e(m)v_e(n)^*+a_e(m)a_e(n)^*\right]^2,\nll
G_3(s,t)&=&\Re e\sum_{m,n=0}^N\;\chi_m(s)\chi_n^*(t)
\left[\nobodyfrac v_e(m)a_e(n)^*+a_e(m)v_e(n)^*\right]^2,\nll
H(s,t)&=&\Re e\sum_{m,n=0}^N\;\chi_m(s)\chi_n^*(t)
\left[\nobodyfrac v_e(m)v_e(n)^*+a_e(m)a_e(n)^*\right]
\left[\nobodyfrac v_e(m)a_e(n)^*+a_e(m)v_e(n)^*\right]\nll
&&\mbox{\ \ with\ \ } t=-\frac{s}{2}(1-c).
\ea
The summation runs over the exchanged gauge bosons.
See section \ref{zpeeff1} for further definitions.

The Born cross section of {\it M{\o}ller scattering} including the $Z'$
exchange is \cite{moeller},
\ba
\label{moellerx}
\frac{d \sigma}{d c} &=&\frac{16\pi\alpha^2}{s}\sum_{m,n=0}^N
\frac{\frac{g_n^2g_m^2}{(4\pi\alpha)^2}}{(\mu_m^2-c^2)(\mu_n^{*2}-c^2)}\nll
&&\left\{ 4\lambda_1 (R_m^2R_n^{*2}+L_m^2L_n^{*2})\mu_m\mu_n^*
+         4\lambda_2 (R_m^2R_n^{*2}-L_m^2L_n^{*2})\mu_m\mu_n^*\right.\nll
&& \left.+\lambda_3 R_mL_mR_n^*L_n^*
\left[\mu_m\mu_n^*+(1+\mu_m\mu_n^*+2\mu_m+2\mu_n^*)c^2+c^4\right]\right\}
\ea
with
\bq
\label{leftrdef}
\mu_n=1+2\frac{m_n^2}{s} \mbox{\ \ and\ \ }
\lambda_3=1+\lambda_+\lambda_-.
\eq
Again, the summation runs over the exchanged gauge bosons.
$R_m$ and $L_m$ denote the left- and right handed couplings to electrons,
$L_m=L_e(m),\ R_m=R_e(m)$.
Further definitions can be found in section \ref{zpeeff1}.

Alternatively to formulae \req{bhabhax} and \req{moellerx}, the $Z'$
contributions can be included by form factors as explained in section
\ref{guts3plus}. 
\subsubsection{Observables}\label{zpeeee13}
Consider first {\it Bhabha scattering}.
Only contributions proportional to $f_0$ can be
measured with unpolarized beams. 
With polarized electrons, one can measure the left-right asymmetry,
\bq
A_{LR}(c)=\frac{d\sigma_L-d\sigma_R}{d\sigma_L+d\sigma_R}.
\eq
It is sensitive to contributions proportional to $f_1$.
Two polarized beams allow a measurement of the asymmetry \cite{bhabhahol},
\bq
A_{2L}(c)=\frac{d\sigma_{LL}-d\sigma_{RR}}{d\sigma_{LL}+d\sigma_{RR}}.
\eq
It is sensitive to contributions proportional to $f_2$.
This is different from fermion pair production and $W$ pair
production where two polarized beams give no new information compared
to electron polarization only. 

LEP\,2 has naturally transverse beam polarization.
Then, transverse asymmetries \cite{bhabhahol} can be considered,
\bq
A_T^\phi(c)=\frac{d\sigma^\phi-d\sigma^{\phi+\pi/2}}
                 {d\sigma^\phi+d\sigma^{\phi+\pi/2}}
\mbox{\ \ \ with\ \ \ }\frac{d\sigma^\phi}{dc}
=\int_{\phi-\pi/4}^{\phi+\pi/4}\frac{d\sigma}{d\phi dc}d\phi .
\eq

In contrast to Bhabha scattering, it is sure that in {\it M{\o}ller
scattering} both electron beams can be highly polarized.
With two polarized beams, one can measure several angular  
distributions \cite{moeller}, 
\bq
\frac{1}{\sigma}\frac{d\sigma}{dc},\ \ 
\frac{1}{\sigma^{LL}}\frac{d\sigma^{LL}}{dc},\ \ 
\frac{1}{\sigma^{RR}}\frac{d\sigma^{RR}}{dc},\ \ \
\frac{1}{\sigma^{LR}}\frac{d\sigma^{LR}}{dc},
\eq
which are all linear combinations of the three contributions \req{moellerx} 
proportional to $\lambda_1,\lambda_2$ and $\lambda_3$.

In fixed target M{\o}ller scattering, the left--right asymmetry 
\ba
\label{moelfix}
A_{LR}(e^-e^-\rightarrow e^-e^-) &=&
\frac{\sigma_L-\sigma_R}{\sigma_L+\sigma_R}=
\frac{G_\mu Q^2}{\sqrt{2}\pi\alpha}\frac{1-y}{1+y^4+(1-y)^4}(1-4s^2_W)
\nll \mbox{with\ \ } Q^2&=&y(2m_e^2+2m_eE_{beam})_{fixed\ Target}\ \ 
y=-\frac{(p'-p)^2}{(p'+p)^2}
\ea
is of special interest.
It can be measured with very high precision in future
experiments \cite{slacprop}. 
The last sequence in equation \req{moelfix} is valid  \cite{marciano}
in the limit $\sqrt{s}\ll M_Z^2$. 
$p(p')$ are the energy--momenta of one initial (final) electron.
\subsection{Radiative Corrections}\label{zpeeee2}
The generalization of the SM radiative corrections to $s$--channel $Z'$
exchange is discussed in section \ref{zpeeff2}.
No essential new problems arise due to the $Z'$ exchange in the $t$ or
$u$ channel. 
We therefore limit ourselves to give the main references to the SM processes.

QED corrections to SM Bhabha scattering can be found, for example,
in \cite{bhabhaqed,bhabhaqedweak}. 
QED corrections are universal allowing a generalization of the SM result
to the whole process including additional $Z'$ contributions.
See references
\cite{bhabhatord,bhabhaqedweak,bhabhaweak,bhabhaweakqcd} for results 
of weak corrections.
Weak corrections together with $ZZ'$ mixing could be taken into
account as in the case of fermion pair production by the 
replacements \req{weakrepl} and \req{mixrepl2} of the couplings. 
In reference \cite{bhabhafield} one finds the needed formulae.
However, such a replacement  is not necessary in Bhabha and M{\o}ller
scattering because these reactions are as insensitive to $ZZ'$ mixing
as off--resonance fermion pair production. 

QED corrections to M{\o}ller scattering are calculated in reference
\cite{moellerqed}, while the electroweak corrections to $A_{LR}$ are
calculated in reference \cite{marciano}.  
QED initial state corrections can be taken into account
\cite{moellerini} by the structure function approach \cite{kuraev}.

QCD corrections to both processes enter as virtual corrections at one
loop as described in references \cite{marciano,bhabhaweakqcd}.
\subsection{$Z'$ Constraints}\label{zpeeee3}
\subsubsection{Model independent constraints on $v_e^N$ and $a_e^N$}
\paragraph*{Estimate}
In contrast to fermion pair production, the total cross section and
simple asymmetries are not very sensitive to a $Z'$.
The best sensitivity is achieved by fits to angular distributions of
polarized cross sections.
For M{\o}ller scattering, the Cramer-Rao minimum variance bound \cite{eadie}
is given in the second reference of \cite{moeller} in the limit 
$M_Z^2\ll s\ll M_{Z'}^2,\ s_W=1/4$.
We have in our notation,
\bq
\label{cramerrao2} 
\chi^2_\infty
\approx 256\pi\frac{L}{s}\left[\nobodyfrac (R_e^N)^4+(L_e^N)^4\right]
= 32\pi\frac{L}{s}\left[\nobodyfrac(v^N_e)^4+6(v^N_ea^N_e)^2+(a^N_e)^4\right].
\eq
This bound corresponds to the sensitivity of an experiment with an infinite
number of angular bins and no systematic errors.
It can be compared with the constraints \req{bornexcl} and
\req{cramerrao1} obtained for different observables in fermion pair production.

In particular, the estimate \req{cramerrao2} predicts the widths of
the bands in the $R'_e,L'_e$ plane allowed by left- and right handed
electron scattering alone, 
\bq
\label{rlepest}
|R_e^N|,|L_e^N|<\left[\nobodyfrac
\frac{\chi^2_{\infty}s}{256\pi L}\right]^{1/4}.
\eq

\begin{figure}[tbh]
\ \vspace{1.5cm}\\
\begin{minipage}[t]{7.8cm} {
\begin{center}
\hspace{-1.7cm}
\mbox{
\epsfysize=7.0cm
\epsffile[0 0 500 500]{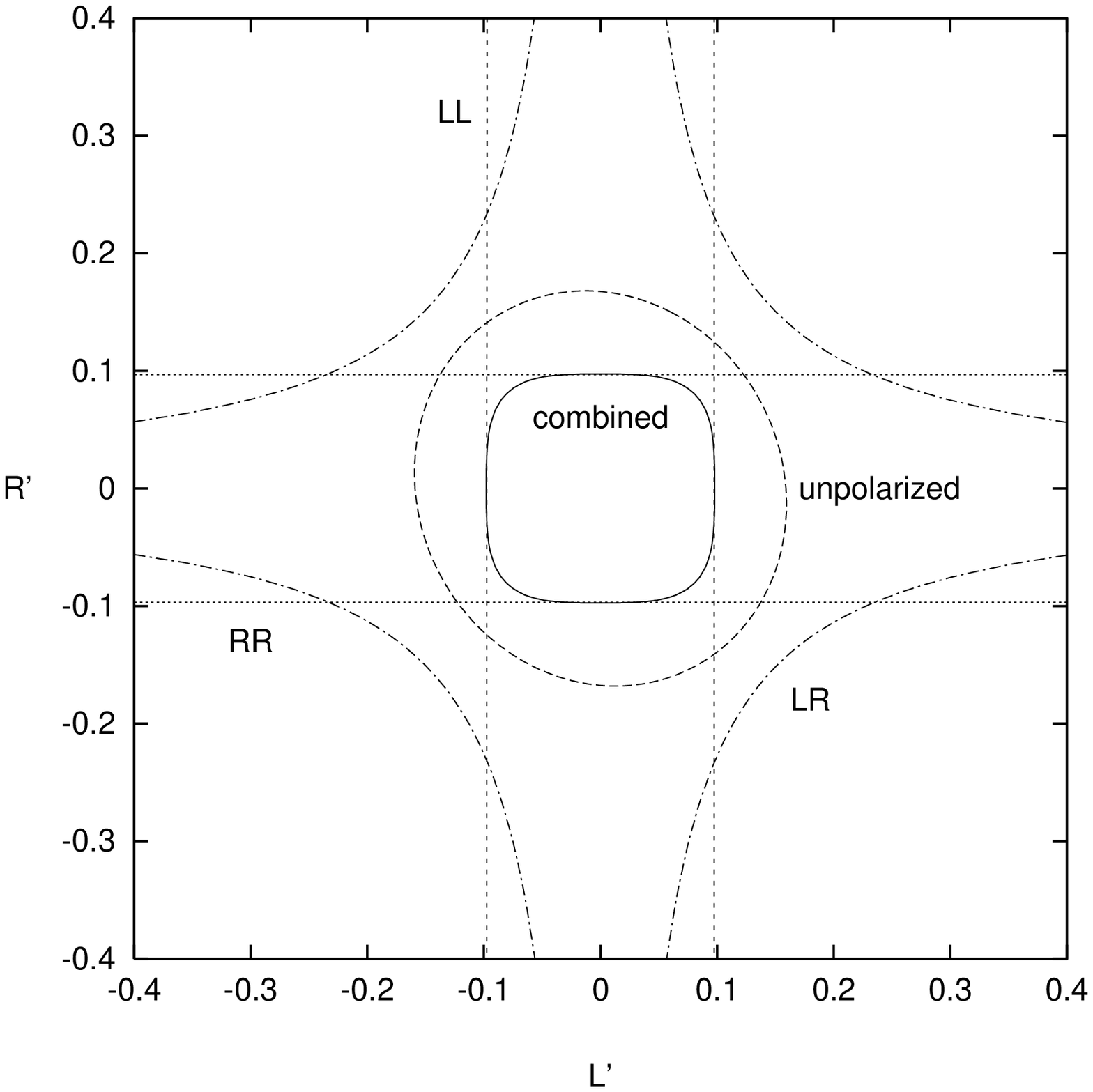}
}
\end{center}
\vspace*{-2.5cm}
\noindent
{\small\it
\begin{fig} \label{moellzp} 
Areas of the leptonic $Z'$ couplings $L'$ and $R'$, which are
excluded with 95\% confidence by M{\o}ller scattering for 
different beam polarizations.
$\sqrt{s}=500\,GeV$, $L=10\,fb^{-1}$ and $M_{Z'}=2\,TeV$ were assumed.
The results are obtained by collecting events with $|c|<0.985$ in 10
equal bins. 
This is figure~1 from the first reference of \cite{moeller}. 
\end{fig}}
}\end{minipage}
\hspace*{0.5cm}
\begin{minipage}[t]{7.8cm} {
\begin{center}
\nobody\hspace{-5.7cm}
\mbox{
\epsfysize=7.0cm
\epsffile[0 27 215 242]{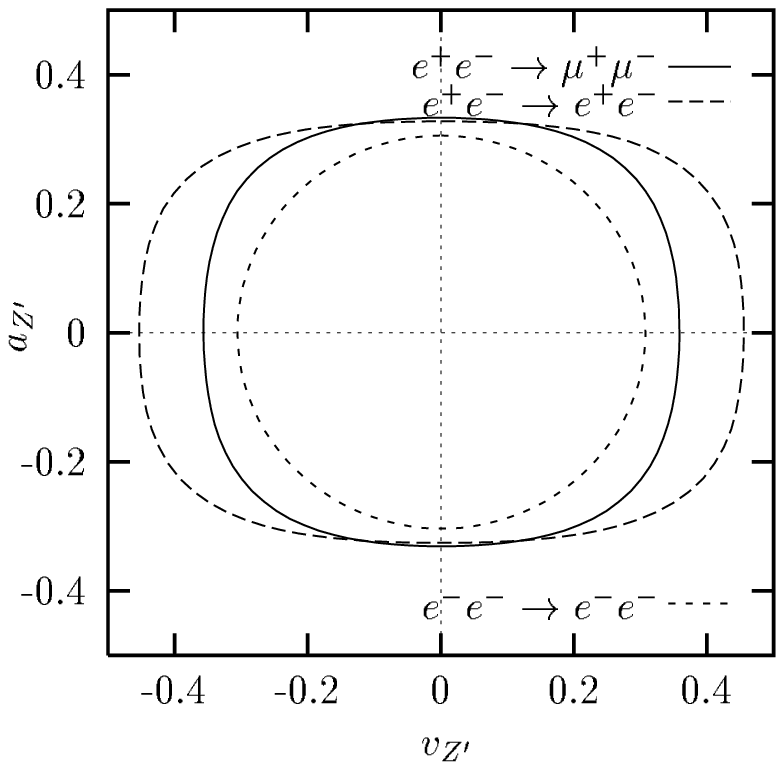}
}
\end{center}
\vspace*{-2.5cm}
\noindent
{\small\it\begin{fig} 
\label{frankbhabha} 
Exclusion limits (95\% CL) from fermion pair production, Bhabha and
M{\o}ller scattering for the couplings of the $Z'$ to leptons
including systematic errors.
The numerical input is $\sqrt{s}=500\,GeV,\
L=50\,fb^{-1}(L=25\,fb^{-1})$ for $e^+e^-(e^-e^-)$ scattering.
The electron polarization is $P_e=90\%,\ \Delta P_e/P_e=1\%,\ \Delta
L/L=0.5\%,\ |\cos\theta|<0.985,\ \Delta\theta =10mrad$.
Ten equal bins in $\cos\theta$ are chosen.
This is figure~5 from reference \cite{moellerini}.
\end{fig}}
}\end{minipage}
\end{figure}
\paragraph*{Future Constraints}
Model independent $Z'$ limits from M{\o}ller scattering are
studied in \cite{moeller} at the Born level.
Figure~\ref{moellzp} shows the regions of $Z'$ couplings to electrons, 
which could be excluded. 
$L'$ and $R'$ in figure~\ref{moellzp} are related to our conventions as
$L'=2R'_eg_2/e,\ R'=2L'_eg_2/e$.
$R'$ and $L'$ are restricted independently in experiments with both
beams right handed or both beams left handed polarized. 
The estimate \req{rlepest} is in good agreement with
figure~\ref{moellzp}. 
For one left and one right handed beam, one is sensitive to
$Z'$ models where the combination $R'L'$ exceeds a certain value. 
This property can immediately be read off from the cross section
\req{moellerx}. 
The allowed region for unpolarized beams is also shown in
figure~\ref{moellzp}. 
The distributions of two left-handed or two right-handed scattered
electrons contain almost all information on a $Z'$.
In contrast to fermion pair production, polarized beams give important
improvements already to the $Z'$ exclusion limits.

The exclusion limits from M{\o}ller scattering are compared with those
from Bhabha scattering and fermion pair production in
figure~\ref{frankbhabha}. 
We have $v_{Z'}=2v'_eg_2/e,\ a_{Z'}=2a'_eg_2/e$ in our conventions.
Under the assumptions made in \cite{moellerini}, 
M{\o}ller scattering gives the best $Z'$ constraints to the
model independent $Z'$ exclusion limits. 
After the inclusion of observables with electrons and $\tau$'s in the
final state, the exclusion limits of M{\o}ller scattering and fermion
pair production  become comparable.
The exclusion limits of Bhabha scattering would improve with polarized
positron beams.

The influence of systematic errors due to the polarization error, the
angular resolution and the luminosity error on $Z'$ exclusion limits
are studied in reference \cite{moellerini}. 

$Z'$ limits can also be obtained from $e^-\mu^-$ scattering \cite{9803450}. 
Assuming generation universality, this reaction constrains the same
parameter combination as M{\o}ller-- or Bhabha scattering. 
The exchange of neutral gauge bosons in $e^-\mu^-$ scattering is
possible only in the $t$ channel. 
However, it seems to be much more difficult to create a highly polarized muon
beam of high luminosity then an electron beam with the same properties.
\subsubsection{Model dependent constraints on $M_{Z'}$}
\paragraph*{Estimate}
An estimate of $Z'$ limits from Bhabha and M{\o}ller scattering
can be obtained by considerations similar to those which lead to the
estimate \req{offres}. 
The observable $O$ is now the relative number of 
events in a certain angular bin. 
Comparing the shift $\Delta ^{Z'}O$ due to a
$Z'$ with the SM prediction $O_{SM}$, one gets
\bq
\label{estimbhabha}
\frac{\Delta^{Z'}O}{O_{SM}}
\approx \frac{g_2^2}{g_1^2}\frac{|\Re e\{\chi_2\chi_0\}|}{|\chi_0|^2}
= \frac{g_2^2}{g_1^2}\frac{t}{t-M_{Z'}^2}.
\eq
It follows that 
\bq
\label{bhabhaest}
M_{Z'} < M_{Z'}^{lim} = 
\sqrt{s\cdot\frac{1-c}{2}}
\left(1 + \frac{g_1^2}{g_2^2}\frac{O}{\Delta O}\right)^{1/2}
= \sqrt{s\cdot\frac{1-c}{2}}\left(1 + \frac{1}{\Delta o}\right)^{1/2}
\eq
produces a signal in the observable $O$.
A similar estimate of $M_{Z'}^{lim}$ can be derived from the
constraint \req{cramerrao2}.

Comparing the estimates \req{bhabhaest} and \req{offres}, we conclude
that the $Z'$ mass limits from Bhabha and M{\o}ller scattering could be
competitive to $e^+e^-\rightarrow f\bar f$ with leptons in the final
state only. 
For completely specified models where the annihilation into quarks contributes
to $M_{Z'}^{lim}$, the mass exclusion limit from fermion pair
production is better.
\paragraph*{Future Constraints}
Future constraints on $M_{Z'}$ can be obtained from figures \ref{moellzp} and
\ref{frankbhabha} using the scaling \req{epemlim}.
The limits from  M{\o}ller scattering are better than those from
$e^+e^-\rightarrow\mu^+\mu^-$ and $e^+e^-\rightarrow e^+e^-$.
However, $e^+e^-\rightarrow f\bar f$ gives better limits on $M_{Z'}$
if observables with quarks in the final state are included.

A measurement of $A_{LR}$ \req{moelfix} in a fixed target experiment at
SLAC is expected to have the precision $\Delta A_{LR}=1.4\cdot 10^{-8}$,
while the SM prediction is $A_{LR}=1.8\cdot 10^{-7}$ \cite{marciano}.
A $Z'$ from the $E_6$ group \req{betadef} would multiply $A_{LR}$ by the
factor \cite{marciano2} 
\bq
1+7\frac{M_Z^2}{M_{Z'}^2}
\left(\cos^2\beta + \sqrt{\frac{5}{3}}\sin\beta\cos\beta\right).
\eq
Therefore, the experiment is sensitive to $M_\chi<870\,GeV$.
\subsubsection{Errors of model measurements}
%
\begin{figure}
\begin{center}
\begin{minipage}[t]{7.8cm}{
\begin{center}
\nobody\hspace{-6.0cm}
\mbox{
\epsfysize=7.0cm
\epsffile[0 0 300 300]{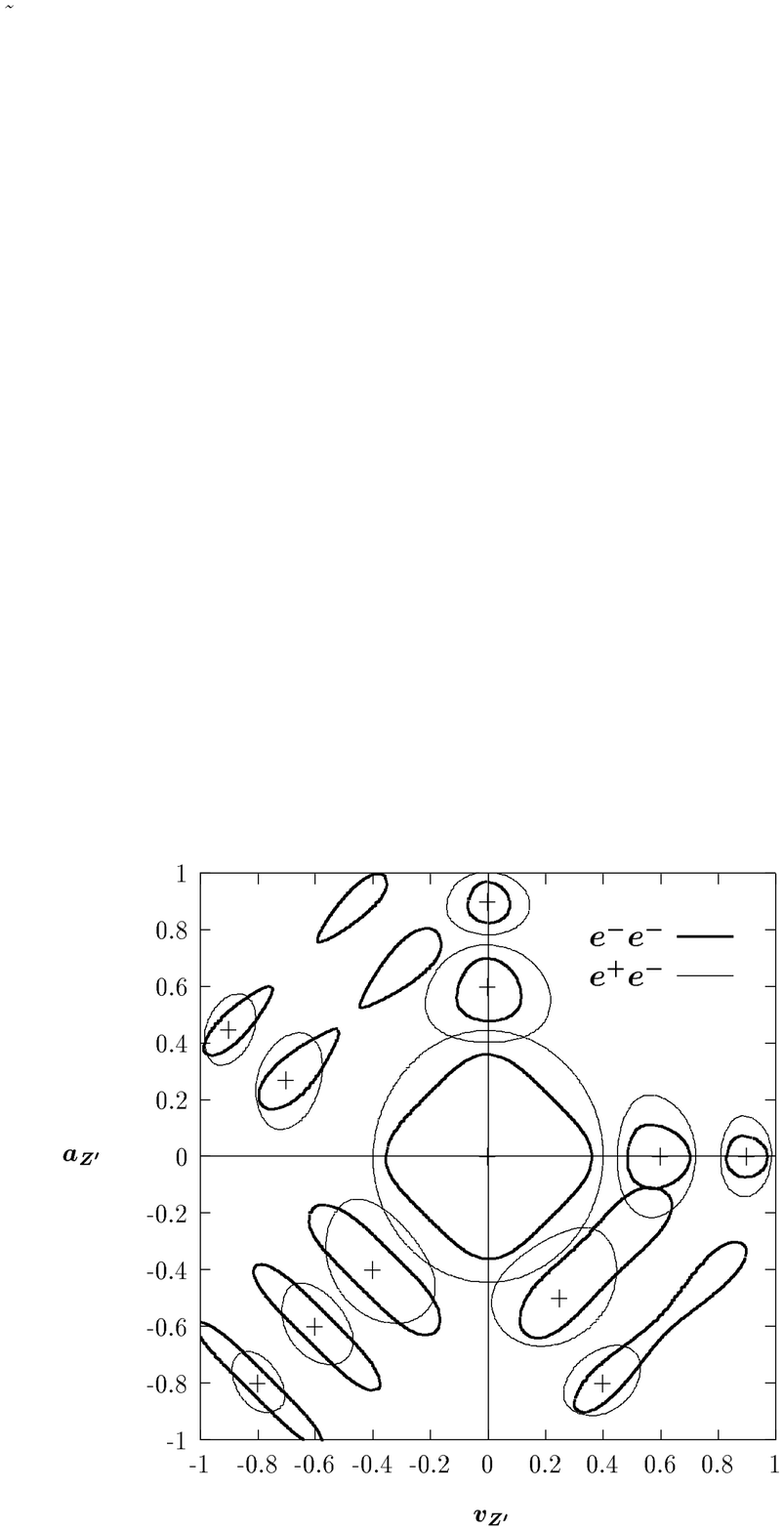}
}
\end{center}
}\end{minipage}
\end{center}
\vspace*{-2.5cm}
\noindent
{\small\it\begin{fig} 
\label{frankmod} 
  Contours of resolvability at 95\%\ confidence
  of the $Z'$ couplings
  around several possible true values 
  marked with a `+'.
  The assumptions are $\sqrt{s}=0.5\,TeV$, $L=40(20)\,fb^{-1}$ for
$e^+e^-(e^-e^-)$ collisions. 
Ten bins in the scattering angle between $10^\circ$ and $170^\circ$ are used.
The $Z'$ mass is 2\,TeV.
This is figure~5 from the third reference of \cite{moeller}.
\end{fig}}
\end{figure}

The error of a $Z'$ model measurement in Bhabha and M{\o}ller scattering
is given by the estimates \req{epemmea} and \req{epemmea2}. 
Of course, these reactions can only constrain observables involving
the $Z'$ couplings to electrons.

Figure~\ref{frankmod} shows the result of a corresponding analysis.
Fermion pair production and M{\o}ller scattering are
complementary in a model measurement.
Fermion pair production removes a sign ambiguity present in the
measurements of M{\o}ller scattering.
\section{$Z'$ search in $e^+e^-\rightarrow W^+W^-$}\label{zpeeww}
The symmetry eigenstate $Z'$ does not couple to the $W$ pair due to
the $SU(2)_L$ gauge symmetry.
The process $e^+e^-\rightarrow W^+W^-$ is sensitive to a $Z'$ only in
the case of a non-zero $ZZ'$ mixing.
The individual interferences of $W$ pair production rise proportional
to $s$ in the limit of large center--of--mass energies $\sqrt{s}$. 
In the SM, the sum of all interferences scales like $\ln s/s$ in the
limit of large $s$ due to a delicate gauge cancellation.
In the case of a non-zero $ZZ'$ mixing, the couplings of the $Z_1$ 
differ from the SM predictions for the $Z$.
Then, the gauge cancellation present in the SM is destroyed.
The result is a huge magnification of new physics effects at large energies.
Unitarity is restored at energies $s\gg M_2^2$
independently of details of the large gauge group.

Similar to the reaction $e^+e^-\rightarrow f\bar f$, it is useful to
distinguish different cases,
\ba
\label{scasesww}
\mbox{case\ 1:\ }&&s<M_2^2,\nll
\mbox{case\ 2:\ }&&s\approx M_2^2,\nll
\mbox{case\ 3:\ }&&s>M_2^2,
\ea
where $s\approx M_2^2$ means $(M_2-\Gamma_2)^2 < s < (M_2+\Gamma_2)^2$.

{\it Case 1} gives stringent {\it exclusion limits} on $ZZ'$ mixing.

{\it Case 2} allows the best exclusion limit or the most accurate {\it
measurement} of the $ZZ'$ mixing angle if a $Z'$ exists.

{\it Case 3} cannot add new information compared to the cases 1 and 2.

We assume here no mixing of the SM $W$ bosons with extra charged
gauge bosons. 
Furthermore, we neglect a possible mixing between SM fermions and
exotic fermions. 
These effects are considered, for example, in reference \cite{pankovmix}.
A measurement of the $W$ polarization would be very useful to separate $ZZ'$
and lepton mixing effects \cite{pankovpol1} and for a simultaneous
constraint of many anomalous couplings \cite{gounaris,pankovpol2}. 

An early analysis of $Z'$ effects in $W$ pair production can be found
in \cite{wwfirst}. 
\subsection{Born Approximation}\label{zpeeww1}
\subsubsection{Amplitude}
Following reference \cite{pankovnew}, we write the amplitude as ${\cal M} 
={\cal M}_t + {\cal M}_s$, where the
$s$ and $t$ channel contributions including a $Z'$ are 
\ba
\label{amplww}
{\cal M}_t^{\lambda_-} &=& \frac{\lambda_--1}{4t_\nu s_W^2}\times{\cal
T}^{\lambda_-}(s,c),\\  
{\cal M}_s^{\lambda_-} &=& \left(-\frac{g_{WW\gamma}e}{s} 
+ \frac{g_{WWZ_1}g_1\left[\nobodyfrac v_e(1)-\lambda_-a_e(1)\right]}{s-M_1^2} 
\right.\nll&&\nobody\hspace{2cm}\left.
+ \frac{g_{WWZ_2}g_2\left[\nobodyfrac
v_e(2)-\lambda_-a_e(2)\right]}{s-M_2^2}\right) 
\times{\cal G}^{\lambda_-}(s,c).\nonumber
\ea
$\lambda_-(=-\lambda_+)=\pm 1$ is the electron (positron) helicity as
defined in section \ref{zpeeff1},
$\sqrt{s}$ is the total center--of--mass energy of the $e^+e^-$ pair,
$c=\cos\theta$ where $\theta$ is the angle between the $W^+$ and the
positron and the invariant $t_\nu$ is defined in equation \req{lastww}.
The extra neutral gauge boson changes only the $s$ channel amplitude.
The functions ${\cal T}^{\lambda_-}(s,c)$ and ${\cal
G}^{\lambda_-}(s,c)$ are not important in the following discussion. 
They are the same as in the SM and can be found, for example, in
reference \cite{gounaris}. 

The amplitude of $W$ pair
production is linear in the $Z_2$ couplings to the electron. 
This makes the $W$ pair production sensitive to the absolute sign of
these couplings.
Their measurement can remove the sign ambiguity present in fermion
pair production where the $Z_2$ couplings to fermions always appear in pairs.

The coupling constants of the interactions between three gauge bosons are
\bq
g_{WW\gamma}=e,\ \ \ g_{WWZ_1}=e\frac{c_W}{s_W}c_M,\ \ \ 
g_{WWZ_2}=e\frac{c_W}{s_W}s_M.
\eq

The contribution of the extra neutral gauge boson can be absorbed 
in two $s$--dependent anomalous couplings \cite{npb429},
\bq
\label{deltagazdef}
g^*_{WW\gamma}=e(1+\delta_\gamma) \mbox{\ \ and\ \ }
g^*_{WWZ_1}=e(\cot\theta_W+\delta_Z).
\eq
The $s$ channel amplitude including a $Z'$ is then
\bq
\label{wwabsorb}
{\cal M}_s^{\lambda_-} = \left(-\frac{g^*_{WW\gamma}e}{s} 
+ \frac{g^*_{WWZ_1}g_1\left[\nobodyfrac
v_e-\lambda_-a_e\right]}{s-M_1^2} \right) 
\times{\cal G}^{\lambda_-}(s,\theta),
\eq
where $\delta_\gamma$ and $\delta_Z$ contain contributions due to the
$Z_2$ exchange and due to the $ZZ'$ mixing in the $Z_1$ exchange
\cite{pankovnew},
\ba
\label{deltagazdef2}
\delta_\gamma &=& \frac{c_W}{s_W}c_M\frac{v_eg_1}{e}
\left(\frac{a_e(1)}{a_e}-\frac{v_e(1)}{v_e}\right)\hat\chi_1 
+ \frac{c_W}{s_W}s_M\frac{v_eg_2}{e}
\left(\frac{a_e(2)}{a_e} -\frac{v_e(2)}{v_e}\right)\hat\chi_2,\nll
\delta_Z &=& 
-\frac{c_W}{s_W} + \frac{c_W}{s_W}c_M\frac{a_e(1)}{a_e}\frac{\hat\chi_1}{\hat\chi_Z} 
+ \frac{c_W}{s_W}s_M\frac{g_2a_e(2)}{g_1a_e}\frac{\hat\chi_2}{\hat\chi_Z}, 
\ea
and
\bq
\label{chiwdef}
\hat\chi_n=\frac{s}{s-m_n^2},\ \ \ n=Z,1,2. 
\eq
Because $W$ pair production is studied sufficiently far away from
the  $Z_1$ peak, we can neglect the $Z$ and $Z_1$ widths putting
$m_Z=M_Z$ and $m_1=M_1$.
We know from present measurements \cite{hollik96} that $M_Z-M_1<150\,MeV$.
This allows the approximation
\bq
\label{propapprox}
\frac{\hat\chi_1}{\hat\chi_Z} \approx 1-\frac{2M_Z(M_Z-M_1)}{s-M_Z^2}\approx
1. 
\eq
The expressions for $\delta_\gamma$ and $\delta_Z$ can then be written as
\bq
\label{wwcm}
\delta_\gamma = \frac{c_W}{s_W}\frac{v_e}{e}
\left(\frac{a_e^M}{a_e}-\frac{v_e^M}{v_e}\right)
\left(1-\frac{\hat\chi_2}{\hat\chi_Z}\right)\hat\chi_Z,\ \ \ 
\delta_Z =\frac{c_W}{s_W}\frac{a_e^M}{g_1a_e}
\left(1-\frac{\hat\chi_2}{\hat\chi_Z}\right),
\eq
where $v_e^M$ and $a_e^M$ are defined in equation \req{coupmdef}. 
The terms proportional to $\hat\chi_2$ dominate in the case $s\approx M_2^2$
but can be neglected in the case $s\ll M_2^2$.
Relation \req{wwcm} shows that measurements of $W$ pair production
below the $Z_2$ peak constrain the same combinations $a_e^M,v_e^M$ as
do measurements of fermion pair production on the $Z_1$ resonance.

Experimental constraints on the anomalous
couplings $g^*_{WW\gamma}$ and $g^*_{WWZ_1}$ bound the parameters
$\delta_\gamma$ and $\delta_Z$ in a model independent way.
Constraints on $\delta_\gamma$ and $\delta_Z$ can be interpreted as a
constraint to the combinations of $Z'$ parameters given in equation
\req{deltagazdef2}.
Far below the $Z_2$ resonance, the $Z_2$ mass, the $ZZ'$ mixing angle
and the $Z'$ couplings to fermions cannot be constrained separately. 
\subsubsection{Cross section}\label{zpeeww12}
The Born cross section with $N$ exchanged gauge bosons is, following
the notation of reference \cite{wwtord},
\ba
\label{wwsig}
\frac{d \sigma^{\lambda_-\lambda_+}}{d c} 
&=&\frac{\sqrt{\lambda}}{\pi s^2s_1s_2}\left\{
\sum_{m,n=0}^N \Re e\left\{\frac{e^4}{s^2}\tilde\chi_m(s)\tilde\chi_n^*(s)
C_T(\mbox{initial\ fermions})\right\}G_{CC3}^{33}
\right.\nll &&\left.
+\sum_{n=0}^N \Re e\left\{\frac{e^2(g_1 2c_W)^2}{8s}\tilde\chi_n^*(s)
\left[v_e^*(n)+a_e^*(n)\right]\right\}G_{CC3}^{3f}(\lambda_1+\lambda_2)
\right.\nll &&\left.
+\frac{(g_1 2c_W)^4}{32}G_{CC3}^{ff}(\lambda_1+\lambda_2)\right\}
\ea
with 
\bq
\label{lambdadefkin}
\lambda=(s-s_1-s_2)^2-4s_1s_2\mbox{\ \ and\ \ } 
\tilde\chi_n(s) = \frac{g_ng_{WWZ_n}}{4\pi\alpha}\frac{s}{s-m_n^2}.
\eq
The definition of the propagator $\tilde\chi_n(s)$ is slightly
different from \req{eq28} to absorb the coupling constants from one
triple gauge boson vertex and from one gauge boson-fermion vertex.
The invariant masses squared $s_1$ and $s_2$ of the two $W$'s are equal
to $M_W^2$ for on-shell $W$ production.
The definitions of $C_T(\mbox{initial\ fermions})$ and of the helicity
combinations $\lambda_1,\lambda_2$ are the same as introduced in
section \ref{zpeeff1}. 

The kinematic $G$--functions are \cite{wwtord,muta},
\ba
G_{CC3}^{33}(s;s_1,s_2;c)&=&
\frac{1}{32} \left[ \lambda C_1+12s_1s_2C_2\right],\nll
G_{CC3}^{3f}(s;s_1,s_2;c)&=&
\frac{1}{8} \left[ (s-s_1-s_2)C_1-
\frac{4s_1s_2[s(s_1+s_2)-C_2}{t_\nu}\right],\nll
G_{CC3}^{ff}(s;s_1,s_2;c)&=&
\frac{1}{8} \left[ C_1+\frac{4s_1s_2C_2}{t_\nu^2}\right]
\ea
with
\bq
\label{lastww}
C_1=2s(s_1+s_2)+C_2,\ \ C_2=\frac{\lambda}{4}(1-c^2),\ \ 
t_\nu=\frac{1}{2}\left(s-s_1-s_2-c\sqrt{\lambda}\right).
\eq

The notation used in equations \req{wwsig}-\req{lastww} allows a
generalization to off--shell $W$ pair production and the inclusion of
other 4-fermion background diagrams \cite{wwtord}.

In the case where the helicities $\tau_+(\tau_-)$ of the $W^+(W^-)$
can be measured, the cross sections $\frac{d
\sigma^{\lambda_-\lambda_+}_{\tau_+\tau_-}}{d c}$ should be considered.
A corresponding analysis for on-shell $W$ production can be found in references
\cite{pankovpol1,pankovpol2}. 
\subsubsection{Observables}\label{zpeeww13}
Our starting point for the construction of observables are
the differential cross sections 
$\frac{d \sigma^{\lambda_-\lambda_+}_{\tau_+\tau_-}}{d c}$.
They allow for the measurement of total cross sections and asymmetries
\cite{obswwref2,obswwref},
\ba
\label{wwobs}
\sigma_T^{\lambda_-\lambda_+}&=&\int_{-1}^{+1}d c
\frac{d \sigma^{\lambda_-\lambda_+}}{d c},\nll 
\sigma^{\lambda_-\lambda_+}_TA_{FB}^{\lambda_-\lambda_+}
&=&\int_{0}^{1}d c \frac{d \sigma^{\lambda_-\lambda_+}}{d c}
-\int_{-1}^{0}d c \frac{d \sigma^{\lambda_-\lambda_+}}{d c},\nll
\sigma_TA_{LR} &=& \sigma_T^{\lambda_-=-1}-\sigma_T^{\lambda_-=1}
\equiv\sigma_T^--\sigma_T^+\nll
\sigma_TA_{LR,FB}
&=&\int_{0}^{1}d c \frac{d \sigma^-}{d c}
-\int_{-1}^{0}d c \frac{d \sigma^-}{d c}
-\int_{0}^{1}d c \frac{d \sigma^+}{d c}
+\int_{-1}^{0}d c \frac{d \sigma^+}{d c},\nll
\sigma^{\lambda_-\lambda_+}_TA_{CE}^{\lambda_-\lambda_+}(z)
&=&\int_{-z}^{z}d c \frac{d \sigma^{\lambda_-\lambda_+}}{d c}
-\int_{z}^{1}d c \frac{d \sigma^{\lambda_-\lambda_+}}{d c}
-\int_{-1}^{-z}d c \frac{d \sigma^{\lambda_-\lambda_+}}{d c}.
\ea
We omit the indices numbering the polarizations of the $W$'s to
simplify the notation. 
The observables defined above can be understood as summed over the
$W$ polarizations or as written for fixed $W$ polarizations.
As in the reaction $e^+e^-\rightarrow f\bar f$, the helicities
$\lambda_\pm=-(+)1$ stand for a left (right) handed electron or positron.
Missing polarization indices of the electrons or positrons mean the
average over initial polarizations. 

A real detector cannot measure from $c=-1$ to $c=1$.
The correction for this effect can be trivially taken into account
in the observables \req{wwobs}.
Furthermore, an integration over only a part of the range of $c$ is
sometimes recommended to obtain maximum sensitivity to new physics as
pointed out in the second reference of \cite{pankov}.

The unpolarized cross section is dominated by the scattering of left-handed
electrons, $\sigma_T\approx \frac{1}{4}\sigma_T^{-+}$.
The cross section of right handed electrons excludes the neutrino
exchange giving a cross section, which is symmetric in the scattering
angle.
This induces a relation \cite{obswwref2} between two observables,
$A_{LR,FB}=A_{FB}$. 
Cross sections with two left handed or two right handed beams are zero.

In the LEP\,2 storage ring, the electrons and positrons have naturally
transverse polarization. 
The asimutal asymmetry $A_T$ is then an interesting alternative observable
\cite{pankovpol2},
\bq
\frac{d(\sigma_TA_T)}{dc}=2\int_0^{2\pi}d\phi_W
\frac{d^2 \sigma}{d c d\phi_W}\cos(2\phi_W).
\eq

The $W$ is an unstable particle, which can
only be identified through its decay products. 
A hadronic $W$ decay allows a measurement of the $W$'s
energy-momentum but not an identification of its charge.
A leptonic $W$ decay allows a charge identification but not a
measurement of the energy-momentum because a part
of it is carried away by the neutrino.
If one $W$ decays leptonically and one $W$ decays hadronically, the
most complete information about both $W$'s can be extracted.
Only a part of the produced $W$'s
can be reconstructed in the detector leading to an effective reduction
of the luminosity.

If the $W$ polarization can be measured, a  more detailed analysis
is possible.
For details, we refer to \cite{lep2} and \cite{maettig}.

For the definition of optimal observables \cite{prd457} in
$e^+e^-\rightarrow W^+W^-$, see reference \cite{zpc62}.
\subsection{Background and Radiative Corrections}\label{zpeeww2}
The expected accuracy of the measurement of the total $W^+W^-$ cross section
at future $e^+e^-$ colliders is about 1\%.
It has to be met by the theoretical prediction.
Therefore, radiative corrections have to be considered.
A short overview can be found in \cite{4fleike}, for details and
extensive original references see \cite{lep2}.

All present $Z'$ analyses of $W$ pair production are done in the
Born approximation. 
An analysis including all radiative corrections relevant to LEP\,2
could in principle be done with any of the codes described in
reference \cite{lep2} if the code allows a setting of the anomalous couplings
$g_{WWZ_1}^*$ and $g_{WW\gamma}^*$.
\subsubsection{Background}\label{zpeeww14}
$W$ bosons are unstable particles, which can be  detected only through their
decay products,
\bq
\label{reakww}
e^+e^-\rightarrow (W^+W^-)\rightarrow  f_1f_2f_3f_4.
\eq
We have also non-resonant (background)  processes to the same order of
perturbation theory, 
\bq
\label{reakww2}
e^+e^-\rightarrow f_1f_2f_3f_4,
\eq
which go directly to the same final state.
Their contribution has to be added coherently to the process \req{reakww}
with off--shell $W$'s to get a gauge invariant result.
Different {\tt FORTRAN} codes calculating the complete process \req{reakww2}
are compared in reference \cite{lep2}. 
\subsubsection{QED corrections}\label{zpeeww21}
QED corrections to the amplitude with $Z'$ exchange can be deduced
from the SM results.
We therefore give here only a short description of the related SM corrections.
\paragraph*{Initial state radiation}
The QED corrections to on-shell $W$ pair production are calculated in
reference \cite{wwqedon}.
One would like to separate initial state corrections from final state
corrections and the interference between them in the calculation
for off--shell $W$ pair production because it is much more involved.
Unfortunately, this cannot be done in a gauge invariant way.
The reason is a charge flow from the initial state to the final state
in $W$ pair production. 
This problem can be treated by the current splitting technique 
\cite{gentle_nunicc}, in which the
chargeless neutrino exchanged in the $t$ channel is divided into two
charge flows of opposite sign. 
Now the charge flows of the initial and the final state are separated,
and the gauge invariance of initial state QED 
corrections is ensured as it is in the case of $Z$ pair production. 

Initial state QED corrections to off--shell $W$ pair production are
calculated in \cite{gentle_nunicc}. 
They reach several \% near the $WW$ threshold.
The corrections can be split into universal
contributions, which are described by the flux function
\req{convol} or structure function \req{sfapproach} approaches with the
same functions $H_A^e(v)$ or $D(x,s)$ derived for fermion pair
production, and into non-universal contributions  depending on the
particular process. 
The non-universal contributions to off--shell $W$ and $Z$ production 
are numerically suppressed by a factor $s_1s_2/s^2$
 \cite{gentle_nunicc}. 
The initial state QED corrections to off--shell $W$ pair production
calculated in the flux function
\req{convol} or structure function \req{sfapproach} approach are
therefore a good approximation within the expected accuracy of the data.
The generalization of these SM results to cross sections including $Z_2$
exchange is straight forward. 

The radiative corrections to the background \req{reakww2} are a small
correction to a small contribution.  
QED corrections to the background are usually taken into account by
the convolutions \req{sfapproach} or \req{convol}.
\paragraph*{Coulomb singularity}
The Coulomb singularity \cite{coulomb} arises from long range electromagnetic
interactions between the produced massive charged particles.
We get the correction
\bq
\label{coul}
\sigma^{Coul} =\sigma_T\left(1+\frac{\alpha\pi}{2\lambda}\right)
\eq
for $W$ pair production, which diverges near threshold
where the velocity $\lambda$ of the $W$'s approaches zero. 
It indicates that perturbation theory is not applicable in this region. 
Fortunately, the non-zero width $\Gamma_W$ and a slight off--shell production
of the $W$'s regularize \cite{fadin} the Coulomb singularity.
Nevertheless, the numerical effect can exceed 6\% for the total cross
section near threshold  \cite{fadin}. 

Coulomb singularities also arise in QED and QCD corrections to pair
production of massive fermions \cite{jlzabl}.
There, the singularity can be avoided by a calculation in the limit of massless
fermions or by a cut on the invariant mass of the massive fermion pairs. 
Such a cut is desirable for quark pairs in any case to avoid
non-perturbative bound state regions.
\subsubsection{Weak corrections}\label{zpeeww22}
The SM one--loop correction to on-shell $W^+W^-$  production is calculated in
reference \cite{wwcorr}.
The calculation of the complete SM one--loop correction to the process
\req{reakww2} is very complex \cite{old} and not done.
If it is known in the future, the weak corrections in the presence of
$ZZ'$ mixing can be treated as described in section~\ref{zpeeff2}.
\subsubsection{QCD corrections}\label{zpeeww23}
QCD corrections give sizeable contributions to distributions and cross
sections. 
They enter the width of the $W$, where they can reach
several \%, see \cite{bakl}. 
QCD corrections  can be naively implemented multiplying the cross
section by the factor $1+\frac{\alpha_s}{\pi}$ for every gauge boson
decaying into a quark pair.
Such a procedure is only a rough guess for background diagrams and for
cross sections with kinematic cuts.

QCD corrections are calculated to $O(\alpha_s)$ for on-shell $W$ pair
production including the final $W$ polarizations and kinematic cuts in
references \cite{wwqcd}.  
The calculation for the complete process $e^+e^-\rightarrow
\mu\bar\nu_\mu u\bar d$ can be found in reference \cite{pittau}. 
The results of $O(\alpha_s)$ corrections to $e^+e^-\rightarrow q_1\bar
q_2q_3\bar q_4$, the CC10, CC11 and CC20 processes, can be found in
reference \cite{pittauall}.
See also reference \cite{9705218} for further references.

No new problems arise in QCD corrections to processes with $Z_2$ exchange. 
\subsection{$Z'$ constraints at $s<M_2^2$}\label{zpeeww3}
To order $\theta_M$, a $Z'$ would modify the cross section due to
changes in the $Z_1e^+e^-$ couplings  and due to the $Z_2$ exchange
contribution. 
Modifications due to the mass shift $\Delta
M=M_Z-M_1$ in the $Z_1$ propagator are small, see discussions above
equation \req{propapprox}, and of the order $\theta_M^2$. 
Far below the $Z_2$ resonance, the contribution of the $Z_2$ exchange
can also be neglected because it has the additional suppression factor
$s/M_2^2$. 
Therefore, the $Z'$ signal considered here arises due to the
modification of the $Z_1e^+e^-$ coupling.

All $Z'$ effects can be absorbed in the anomalous couplings
$\delta_\gamma$ and $\delta_Z$ defined in equation \req{deltagazdef2}.
A search for a $Z'$ in $W^+W^-$ production is therefore a special case
of a search for anomalous couplings \cite{gounaris,npb429,lep2ano,lcano}.
Far below the $Z_2$ peak, one has to take special care to separate $Z'$
models from other theories predicting anomalous couplings.
One hint for a $Z'$ would be a non-zero $\delta_\gamma$, which
is usually absent in other theories due to the $U(1)_{em}$ gauge invariance.
In a general data analysis, one should try to constrain many anomalous
couplings simultaneously and show that all combinations perpendicular
to $\delta_Z$ are zero. 
If at least one of these perpendicular combinations is not zero, this
will indicate that there is other new physics in addition to a $Z'$.
See reference \cite{9701359} for a corresponding analysis.

An additional check of the $Z'$ hypothesis would be a comparison with
deviations in $e^+e^-\rightarrow f\bar f$ below the $Z_2$ peak.
The final proof of the hypothesis would be a measurement at the $Z_2$ peak.

We assume in the following that all deviations from the SM are due to
a $Z'$ and ignore the possible confusion with other physics.
\subsubsection{Model independent constraints on $g^*_{\gamma WW}$, 
$g^*_{ZWW}$ or $v_e^M$, $a_e^M$}
\paragraph*{Estimate}
Consider the cross section $e^-_Re^+\rightarrow W^+W^-$.
Only the $s$--channel contributes to the scattering of right-handed electrons.
Expanding the total cross section $\sigma_T^+$ in the limit of large
$s$, we get 
\bq
\label{eewwr}
\sigma^+_T
\approx \frac{\alpha^2\pi c_W^4}{12s}\left[1
-\frac{2s}{M_Z^2}\left(\delta_\gamma -\delta_Z\frac{s_W}{c_W}\right)\right].
\eq
The first term is the leading SM contribution, the second and third
terms are the leading contributions in the parameters $\delta_\gamma$
and $\delta_Z$.
The {\it two} leading powers in $s$ have canceled in the SM term,
while only the leading power in $s$ has canceled in the contributions
proportional to  $\delta_\gamma$ and $\delta_Z$.
It follows that the observable $\sigma_T^+$ will see a signal if
\bq
\label{erwwconstr}
\frac{\Delta\sigma_T^+}{\sigma_T^+}<
\frac{\Delta^{Z'}\sigma_T^+}{\sigma_T^+}
\approx\frac{2s}{M_Z^2}\left|\delta_\gamma -\delta_Z\frac{s_W}{c_W}\right|,
\eq
where $\Delta\sigma_T^+/\sigma_T^+$ is the relative experimental error
and $\Delta^{Z'}\sigma_T^+$ is the deviation due to a $Z'$.

Similar considerations can be used for the scattering of left-handed
electrons. 
Unfortunately, the region of forward scattering $c\approx 1$ gives a
much reduced sensitivity to a $Z'$. 
In this region most of the SM events are produced (making the cross section
proportional to $\ln s/s$), while the leading contributions in  $\delta_\gamma$
and $\delta_Z$ are not logarithmically enhanced.
One can avoid this pollution effect by fits to angular distributions. 
Alternatively, the forward region can be excluded from the integration. 
The sensitivity to a $Z'$ depends on details of this procedure as
shown in figure~2 of the second reference of \cite{pankov}.
We estimate the sensitivity of the observable $\sigma_T^-$ to $Z'$
effects doing an expansion of $d\sigma_T^-(c)/dc$ around the point $c=0$,
\bq
\label{eewwl}
\frac{d\sigma^-(c=0)}{dc}
\sim\frac{1}{s}\left(\frac{1}{4}+2c_W^4\right)-\frac{s_W^2}{M_Z^2}\left[
\delta_\gamma +\delta_Z\frac{\frac{1}{2}-s_W^2}{s_Wc_W}\right].
\eq
It follows that the observable $\sigma_T^-$ will see a signal, if
\bq
\label{erwwconstrl}
\frac{\Delta\sigma_T^-}{\sigma_T^-}<
\frac{\Delta^{Z'}\sigma_T^-}{\sigma_T^-}
\approx\frac{s}{M_Z^2}\frac{s_W^2}{\left(\frac{1}{4}+2c_W^4\right)}\left|
\delta_\gamma +\delta_Z\frac{\frac{1}{2}-s_W^2}{s_Wc_W}\right|.
\eq

As expected from an inspection of the amplitude \req{wwabsorb}, the scattering
of left- and right-handed electrons constrains different combinations
of $\delta_\gamma$ and $\delta_Z$,
\bq
\frac{\Delta^{Z'}\sigma_T^\pm}{\sigma_T^\pm}\sim\left|\delta_\gamma-\delta_Z
\frac{g_1}{e}\left(\nobodyfrac v_e\mp a_e\right)\right|.
\eq
The scattering of unpolarized electrons gives constraints similar
to those from $e^-_L$ scattering.
Any single observable selected from $\sigma_T,\ \sigma_T^+$ or
$\sigma_T^-$ is  blind to an infinite band in the
$\delta_\gamma,\delta_Z$ plane.  
Combining the results from different cross sections, one is
insensitive only to a closed region in this plane.
For later use, it is instructive to give a rough estimate of the size
of this region using \req{erwwconstrl}, 
\bq
\label{deltgacon}
|\delta_\gamma|,|\delta_Z|
<\delta_\gamma^{lim},\delta_Z^{lim}\approx\frac{\frac{1}{4}+2c_W^4}{s_W^2}
\cdot\frac{\Delta\sigma_T}{\sigma_T}\frac{M_Z^2}{s}.
\approx 6.2\cdot\frac{\Delta\sigma_T}{\sigma_T}\frac{M_Z^2}{s}.
\eq

Assuming that the experimental error consisting of statistical and
systematic errors scales like the statistical error,
$\Delta\sigma/\sigma\approx 1/\sqrt{N}\approx \sqrt{s/L}$, we get a
scaling \cite{pankov} of these constraints with the center--of--mass energy
and the integrated luminosity, 
\bq
\label{wwzpscale}
\delta_\gamma^{lim},\delta_Z^{lim},|v_e^M|,|a_e^M|\sim
\sqrt{\frac{1+r^2}{sL}}.
\eq
It is the same as derived for anomalous couplings \cite{npb419}.
As before, $r$ is the ratio of the systematic and statistical errors.
At the proposed colliders, the statistical errors dominate the error of the
observable $\sigma_T^+$, while the error of $\sigma_T^-$ is usually
dominated by systematic errors depending on the cut on $c$.

\begin{figure}[tbh]
\begin{center}
\begin{minipage}[t]{7.8cm}{
\begin{center}
\hspace{-1.7cm}
\mbox{
\epsfysize=7.0cm
\epsffile[0 0 500 500]{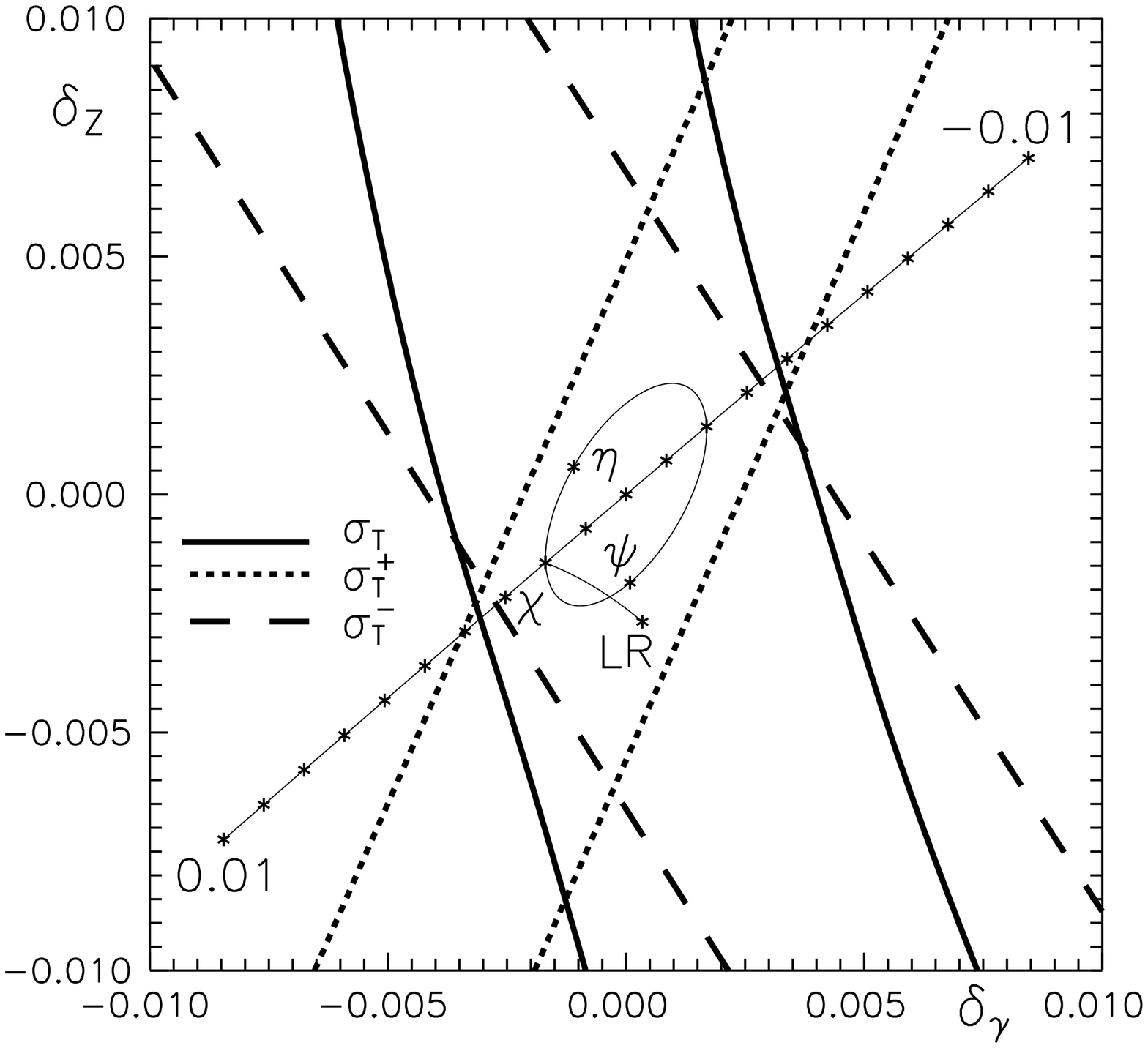}
}
\end{center}
}\end{minipage}
\end{center}
\noindent
{\small\it
\begin{fig} \label{wwfig1} 
Upper bounds (95\%\,CL) on non-standard couplings
($\delta_\gamma,\delta_Z$) from $\sigma_T^+,\sigma_T^-$ and $\sigma_T$
in $e^+e^-\rightarrow W^+W^-$.
See the text for the numerical input. 
The bands containing $(\delta_\gamma,\delta_Z)=(0,0)$ cannot be
excluded by the observables.
The thin lines (the ellipse and the line from the $\chi$ to the LR
model) are the regions of the $E_6$ and LR models for $\theta_M=0.002$.
The straight thin line is the region for $Z'=\chi$ for different values
for $\theta_M$ varied in steps of 0.001.
This an update of figure~1 of reference \cite{pankovnew}.
\end{fig}}
\end{figure}

Let us make a remark regarding the comparison with $e^+e^-\rightarrow f\bar f$.
A $Z'$ signal will arise there, if the $Z'$
interferences give deviations larger than the experimental error.
Consider the deviation due to the $\gamma Z'$ interference,
\bq
\label{erffconstr}
\Delta^{Z'}\sigma_T\approx \frac{1}{s}\cdot\chi_\gamma(s)\chi_2(s)
\approx\frac{1}{s}\cdot 1\cdot\frac{s}{s-M_2^2}
\approx\frac{1}{M_2^2}.\mbox{\ \ It follows\ \ } 
\frac{\Delta^{Z'}\sigma_T}{\sigma_T}
\approx\frac{s}{M_2^2}.
\eq
The constraint has the {\it same} dependence on $s$ as the constraints
\req{erwwconstr} and \req{erwwconstrl}.
However, the important difference is that the constraint from fermion pair
production is normalized to $M_2^2$, while the constraint from $W$
pair production is normalized to $M_Z^2$. 
The ratio $s/M_2^2$ is small far below the $Z_2$ peak, while
$s/M_Z^2$ is large at future colliders independently of $M_2^2$.
This difference is responsible for the enhanced sensitivity of $W$
pair production to the $ZZ'$ mixing angle.
\paragraph*{Present Constraints}\label{presconstrdelgaz}
Present constraints on anomalous couplings from LEP data are given in
reference \cite{ppe97125}. 
Unfortunately, the constraints given there don't allow a derivation of an
excluded region of $\delta_\gamma$ and $\delta_Z$.
\paragraph*{Future Constraints}\label{pankovanalysis}
Future constraints on $\delta_\gamma$ and $\delta_Z$
are given in reference \cite{pankovnew}.
This analysis is done at the born level and based on
$\sqrt{s}=0.5\,TeV$ and $L=50\,fb^{-1}$  assuming 90\% polarization of
the electrons, 30\% detection efficiency of the $W$ bosons and
2\% systematic errors.
The differential cross section is considered in 10 equal
bins in $c$ for $|c|<0.98$.
The resulting constraints are shown in figure~\ref{wwfig1}.
As expected from the estimates, the cross sections $\sigma_T^+$ and
$\sigma_T^-$ or $\sigma_T$ alone are insensitive to bands in the
$\delta_\gamma,\delta_Z$ plane. 
The quantitative agreement with the estimates \req{erwwconstr} and
\req{erwwconstrl} is good. 

The model independent limit on $\delta_\gamma$ and $\delta_Z$ can
be easily converted into limits on the $ZZ'$ mixing angle for
any fixed $Z'$ model.
For a fixed $\theta_M$, every model is represented by a dot in the
$\delta_\gamma,\delta_Z$ plane. 
We show the region of the $E_6$ and LR models for $\theta_M=0.002$ in
figure~\ref{wwfig1}. 
The ratio of $\delta_\gamma$ and $\delta_Z$ is determined by the 
couplings of the $Z'$ to fermions only, independent of the $ZZ'$
mixing angle and the $Z_2$ mass,
\bq
\label{wwrat}
\frac{\delta_Z}{\delta_\gamma} =\frac{e}{g_1v_e}\frac{1}{\chi_Z}
\left(1-\frac{v'_ea_e}{a'_ev_e}\right)^{-1}.
\eq
If one varies the mixing angle $\theta_M$ for a fixed model, one 
moves on a straight line in the $\delta_\gamma, \delta_Z$ plane.
The corresponding line is shown in figure~\ref{wwfig1} for $Z'=\chi$. 
Those values of $\theta_M$, for which one hits the model independent
bound, define the constraint on $\theta_M$ for that specific model.
The limits on $\theta_M$ obtained directly from a one--parameter fit
for a previously fixed model are expected to be stronger.

In the case of a deviation of $\delta_\gamma$ and $\delta_Z$ from
zero, the relation \req{wwrat} between the $Z'$ couplings to
electrons, can be tested in fermion pair production.
Such a cross check \cite{9701359} would help to verify that the
deviation is due to a $Z'$.
In the case of a disagreement, the deviation cannot be due to a $Z'$ alone.

The model independent constraints on  $\delta_\gamma$ and $\delta_Z$
can be converted into model independent constraints on $v_e^M$ and $a_e^M$
using the relations \req{wwcm}.
The mass dependence introduced by the propagator $\hat\chi_2$ can be neglected
far below the $Z_2$ resonance.
Applying this procedure to the limits presented in
figure~\ref{wwfig1}, one gets the constraints on $v_e^M,a_e^M$ shown
in figure~\ref{zpmix1}.

In contrast to measurements of fermion pair production, $W$ pair
production cannot constrain $Z'$ couplings to fermions $f\neq e$. 
Therefore, the sensitivity of $W$ pair production to
$ZZ'$ mixing is reduced for models where the $Z'$ has small couplings
to electrons. 
\subsubsection{Constraints on $\theta_M$}
%
\paragraph*{Estimate}
Combining the definition \req{wwcm} of $\delta_\gamma$ and $\delta_Z$ and
the estimate \req{deltgacon}, we derive an estimate for
$\theta_M$.
Assuming $v'_e\approx v_e$ and $a'_e\approx a_e$, we get from $\delta_Z$
\bq
\label{wwtmest}
|\theta_M|<\theta_M^{lim}
\approx 3.4\cdot\frac{\Delta\sigma_T}{\sigma_T}\cdot\frac{M_Z^2}{s}
\frac{g_1}{g_2}
\left(1-\frac{\hat\chi_2}{\hat\chi_Z}\right)^{-1}.
\eq
We have $\hat\chi_2/\hat\chi_Z\approx 0$ far below the $Z_2$ resonance.
The scaling with $s,\ L$ and $r$ is the same as for $\delta_\gamma$
and $\delta_Z$.

The estimate \req{wwtmest} can be compared with the estimate \req{thetamlim}
derived for fermion pair production at the $Z_1$ peak.
We see that the sensitivity of $W$ pair production to $\theta_M$
becomes eventually better for large energies.
\paragraph*{Future Constraints}
Limits on $\theta_M$ at future colliders are presented in reference
\cite{pankovnew}.
It is an update of the older analyses \cite{pankovpol1,pankov}.

\begin{figure}[tbh]
\ \vspace{1cm}\\
\begin{minipage}[t]{7.8cm} {
\begin{center}
\hspace{-1.7cm}
\mbox{
\epsfysize=7.0cm
\epsffile[0 0 500 500]{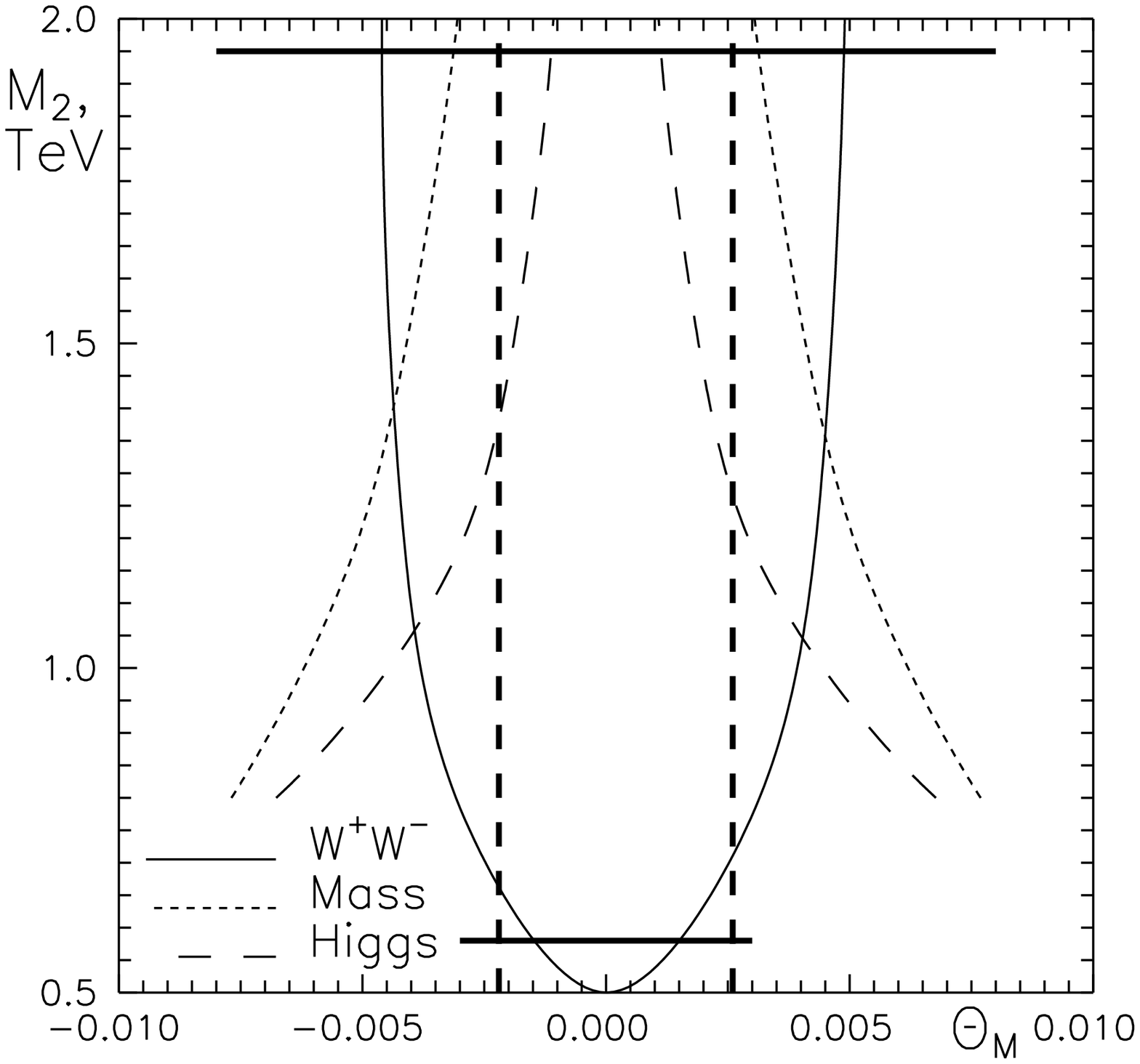}
}
\end{center}
\vspace*{-0.5cm}
\noindent
{\small\it
\begin{fig} \label{wwfig2} 
Allowed domains (95\% CL) of $\theta_M,M_2$ for the $\psi$ model.
The region below the solid curve can be excluded with
$e^+e^-\rightarrow W^+W^-$ at $\sqrt{s}=0.5\,TeV$ and $L=50\,pb^{-1}$.
See the text for further inputs  of the analysis.
The current limit on $M_2$ ($\theta_M$) and the expected exclusion
limit on $M_2$ from
$e^+e^-\rightarrow f\bar f$ are indicated by the thick solid (dashed) lines.
The thin dotted (dashed) lines correspond to the mass constraint
\req{massconstr} with $\Delta M=0.2\,GeV$  
(the Higgs constraint \req{higgsconstr}).
This is an update of figure~2 from reference \cite{pankovnew}.
\end{fig}}
}\end{minipage}
\hspace*{0.5cm}
\begin{minipage}[t]{7.8cm} {
\begin{center}
\hspace{-1.7cm}
\mbox{
\epsfysize=7.0cm
\epsffile[94 480 344 730]{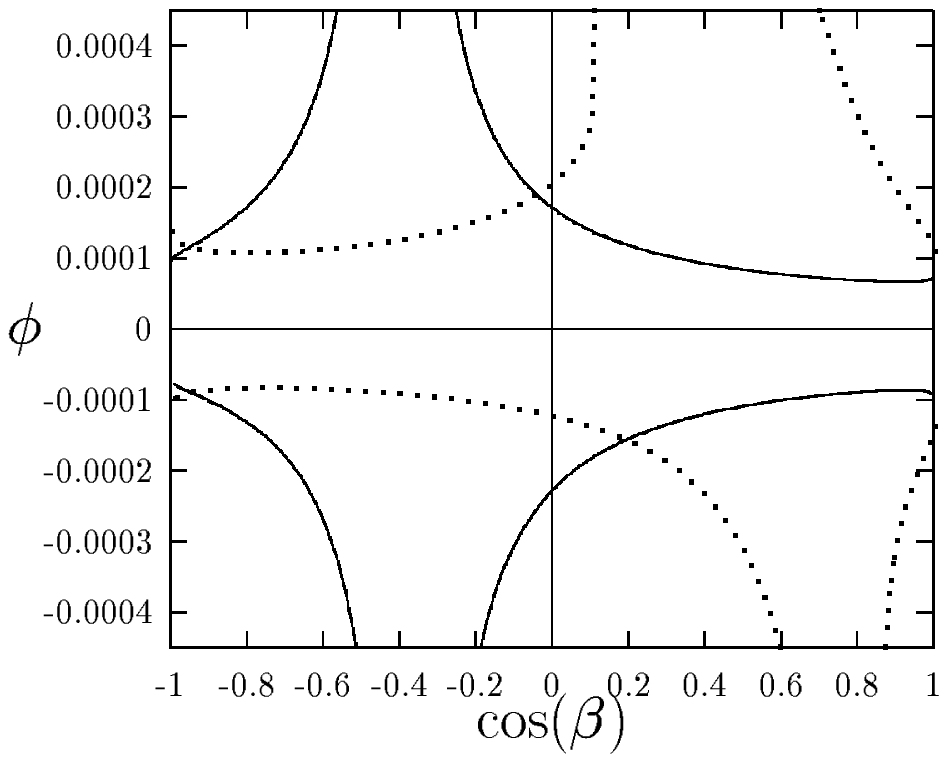}
}
\end{center}
\vspace*{-0.5cm}
\noindent
{\small\it\begin{fig} 
\label{wwfig3} 
Upper limits (95\% CL) for $\Phi=\theta_M$ as function of the $E_6$
model parameter $\cos\beta$ at $\sqrt{s}=M_2\pm \Gamma_2/2$ for 90\%
left (right) handed polarized electrons solid (dotted) solid line.
Positrons are unpolarized.
The input is $M_2=1\,TeV,\ L=50 fb^{-1}$ and a systematic error of 2\%.
I thank A.A. Pankov for providing this figure.
\end{fig}}
}\end{minipage}
\end{figure}

Figure~\ref{wwfig2} shows the future limit on $\theta_M$ as a function
of $M_2$ for $Z'=\psi$.
Remembering that the estimate \req{wwtmest} ignores details of the
$Z'$ model and is based on a crude approximation of the excluded
region, it gives a reasonable prediction of the
constraint on $\theta_M$ in the limit $M_2\rightarrow\infty$.
The present limit on $M_\psi$ and the expected improvement from
fermion pair production at the same collider are also indicated. 
The limit $M_{Z'}^{lim}$ from $e^+e^-\rightarrow f\bar f$ is obtained
from table \ref{zplimlc}. 
We took the entry of analysis \cite{lmu0296} for $20\,fb^{-1}$ with
systematic errors but scaled from $L=20\,fb^{-1}$ to $L=50\,fb^{-1}$
by relation \req{epemlim2}.
Also shown are the relations between $\theta_M$ and $M_2$ from the
mass and the (model dependent) Higgs constraint.

If $s$ approaches $M_2$, the influence of
$\hat\chi_2/\hat\chi_Z$ on $\delta_\gamma$ and $\delta_Z$, in equation
\req{wwtmest} becomes dominant leading to an additional
enhancement to be discussed in the next section. 

The limit on $\theta_M$ can be compared with the present constraint
$-0.0022<\theta_M<0.0026$ for the $\psi$ model taken from table \ref{thetmix}.
We see from figure~\ref{wwfig1} that the reaction $e^+e^-\rightarrow
W^+W^-$ at a $500\,GeV$ collider can add only little to the limits on
the $\psi$ from $e^+e^-\rightarrow f\bar f$. 
However, the estimate \req{wwtmest} predicts a considerable
improvement of the sensitivity for higher energies. 

The analysis of figure~\ref{wwfig2} can be repeated for different $Z'$ models.
The constraint on $\theta_M(M_2\rightarrow\infty)$ for different $E_6$
models is plotted as function of $\cos\beta$ in figure~4 of
reference \cite{pankovpol1}. 

\subsubsection{Constraints on $M_2$}
The interference of the $Z_2$ exchange with the SM contributions is
sensitive to the $Z_2$ mass.
Consider the change in the observable $O$ due to the  $Z_1Z_2$
interference ignoring details of the $Z'$ model
$(a'_e\approx a_e, v'_e\approx v_e)$, 
\bq
\label{mzplimww}
\frac{\Delta^{Z'}O}{O_{SM}} 
\approx\frac{g_2}{g_1}\frac{g_{WWZ_2}}{g_{WWZ_1}}\frac{s-M_1^2}{s-M_2^2}
\approx\frac{g_2}{g_1}\theta_M\frac{s}{s-M_2^2}.
\eq
We took into account that $\theta_M\ll 1$ and that $s\gg M_1^2$ at
future colliders in the last step of the approximation.
It follows that
\bq
\label{limww}
M_2<M_{Z'}^{lim} = 
\sqrt{s}\left( 1+\theta_M\frac{g_2}{g_1}\frac{O}{\Delta O}\right)^{1/2}
\eq
would give a signal in the observable $O$.
Compared to the estimate \req{offres} derived for fermion pair production,
the sensitivity to $M_2$ from $W$ pair production is suppressed by
the $ZZ'$ mixing angle. 
Remembering the discussion in section \ref{tetmicon}, $\theta_M$ is
constrained to be small by measurements at the $Z_1$ peak
independently of the model.
Therefore, the $Z'$ mass bound from $W$ pair production cannot compete with
that from fermion pair production.

The resulting indirect bounds on $M_{Z'}$, which one obtains combining
the constraint on $\theta_M$ with the Higgs constraint depend on the
extended Higgs sector. 
They are worse than the constraints from fermion pair production, compare
figure~\ref{wwfig2}. 
Figures similar to figure~\ref{wwfig2} are shown for other $E_6$ models and
$\sqrt{s}=0.5\,TeV$ and $\sqrt{s}=1\,TeV$ in reference \cite{pankovnew}. 
\subsubsection{Model measurements}
Assume that there are non--zero anomalous couplings $\delta_\gamma$
and $\delta_Z$.
Then, these couplings can be measured in future experiments.
The errors of such measurements can be estimated taking into account
that the deviations of cross sections are linear in  $\delta_\gamma$
and $\delta_Z$.
One gets
\bq
\Delta\delta_\gamma,\ \Delta\delta_Z\approx \delta_\gamma^{lim},\delta_Z^{lim}.
\eq
The estimate for the error of a $\theta_M$ measurement is 
\bq
\label{thetmlimest}
\Delta\theta_M\approx\theta_M^{lim}.
\eq
\subsection{$Z'$ measurements at $s\approx M_2^2$}\label{zpeeww4}
The production of $W$ pairs near the $Z_2$ peak is essentially different
from the production far below the resonance.
At $s\approx M_2^2$
we definitely know that there exists a $Z'$.
We also know its mass, its width and its couplings to fermions.
This information is provided best by fermion pair production on the
resonance due to the large statistics of this reaction.

Unfortunately, a description of $WW$ production has the same problem
as the description of $f\bar f$ production on the $Z_2$ resonance.
The needed radiative corrections within the GUT depend on many
unknown parameters. 
\subsubsection{Constraints on $\theta_M$}\label{zpeeww41}
The production of $W^+W^-$ pairs {\it on} the $Z_2$ resonance is suppressed
because it is proportional to $\theta_M^2$.
For $\sqrt{s}\approx M_2\pm\Gamma_2/2$, the $Z_1Z_2$ interference, being
proportional to $\theta_M$, is most sensitive to  $ZZ'$ mixing because
it is enhanced by the $Z_2$ propagator.

An estimate of the sensitivity to $\theta_M$ is given by \req{wwtmest}
as in the case $s<M_2^2$ with the important difference that the ratio
$\hat\chi_2/\hat\chi_Z$, where the width of the $Z_2$ must now be taken into
account, gives the dominant contribution \cite{pankovnew}, 
\bq
\label{wwtmest2}
|\theta_M|<\theta_M^{lim}
\approx 3.4\cdot\frac{\Delta\sigma_T}{\sigma_T}\cdot\frac{M_Z^2}{s}
\frac{g_1}{g_2}\left|\Re e\frac{\chi_Z}{\chi_2}\right|.
\eq
Compared to the off--resonance case, we have the additional enhancement
factor $|\Re e\chi_Z/\chi_2|\approx 2\Gamma_2/M_2$ with
$2\Gamma_2/M_2\approx (1/20-1/50)$ depending on the
particular $Z'$ model and on the number of the exotic 
fermion generations to which the $Z'$ can decay.
The gain in the sensitivity due to all factors in equation \req{wwtmest2}
is so large that it overcompensates
the loss in the sensitivity due to the poor statistics.
This is the reason why $W$ pair production near the $Z_2$ resonance is
much more sensitive to $\theta_M$ than fermion pair production.
See references \cite{pankov} for a further discussion of this effect.

The increase in the sensitivity for $\sqrt{s}$ approaching $M_2$ can
be seen in figure~\ref{wwfig2}.
Repeating the procedure for different $E_6$ models, one arrives at
figure~\ref{wwfig3}. 
The constraint on $\theta_M$ given there agrees with the estimate
\req{wwtmest2} derived for $E_6$ models with $2\Gamma_2/M_2=1/20$.
Figure~\ref{wwfig3} again demonstrates the essential role of beam
polarization for exclusion limits in $e^+e^-\rightarrow W^+W^-$.  
The sensitivity to $\theta_M$ becomes stronger for higher energies
according to the scaling \req{wwzpscale}.
\subsubsection{Measurements of $\theta_M$}\label{zpeeww42}
In the previous section, we assumed that there is a $Z'$ but that the
$ZZ'$ mixing angle is so small that only an upper bound can be set in
the experiment.
We now assume $\theta_M$ is large enough to give a signal.
Then, the error of a $\theta_M$--measurement is given by equation
\req{thetmlimest} where now $\theta_M^{lim}$ must be taken from the
estimate \req{wwtmest2}.
\subsection{$Z'$ Constraints at $s> M_2^2$}\label{zpeeww5}
Consider the constraint on $\theta_M$ \req{wwtmest}, which
transforms to 
\bq
\label{wwtmest3}
|\theta_M|<3.4\cdot\frac{\Delta\sigma_T}{\sigma_T}\cdot\frac{M_Z^2}{s}
\frac{g_1}{g_2}\frac{s}{M_2^2}
=3.4\cdot\frac{\Delta\sigma_T}{\sigma_T}\cdot\frac{M_Z^2}{M_2^2}
\frac{g_1}{g_2}
\eq
in the limit of high energies, $s\gg M_2^2$.
We see that there is no further enhancement of the sensitivity with
rising $s$.
{\it All} contributions to the cross sections \req{eewwr} and
\req{eewwl} are proportional to $1/s$.
Unitarity is restored independently of the details of the large gauge
group. 
This can be understood treating the $Z_1$ and $Z_2$ as massless
particles in the limit $s\gg M_2^2$. 
Then, one can consider the unmixed states $Z$ and $Z'$ instead of
$Z_1$ and $Z_2$ and remember that the $Z'$ does not couple to $W$'s. 
The resulting cross section for $e^+e^-\rightarrow W^+W^-$ 
obviously behaves like the SM for large $s$.

The estimate \req{wwtmest3} of the sensitivity to
$\theta_M$ is always much worse than case~2 where additional
enhancement factors were present.

Possible $Z'$ signals from the radiative return to the $Z_2$ resonance
cannot compete with fermion pair production due to lower statistics.
\section{$Z'$ search in other reactions}\label{zpee3}
\begin{figure}
\begin{center}
\begin{minipage}[t]{7.8cm}{
\begin{center}
\hspace{-1.7cm}
\mbox{
\epsfysize=7.0cm
\epsffile[0 0 500 500]{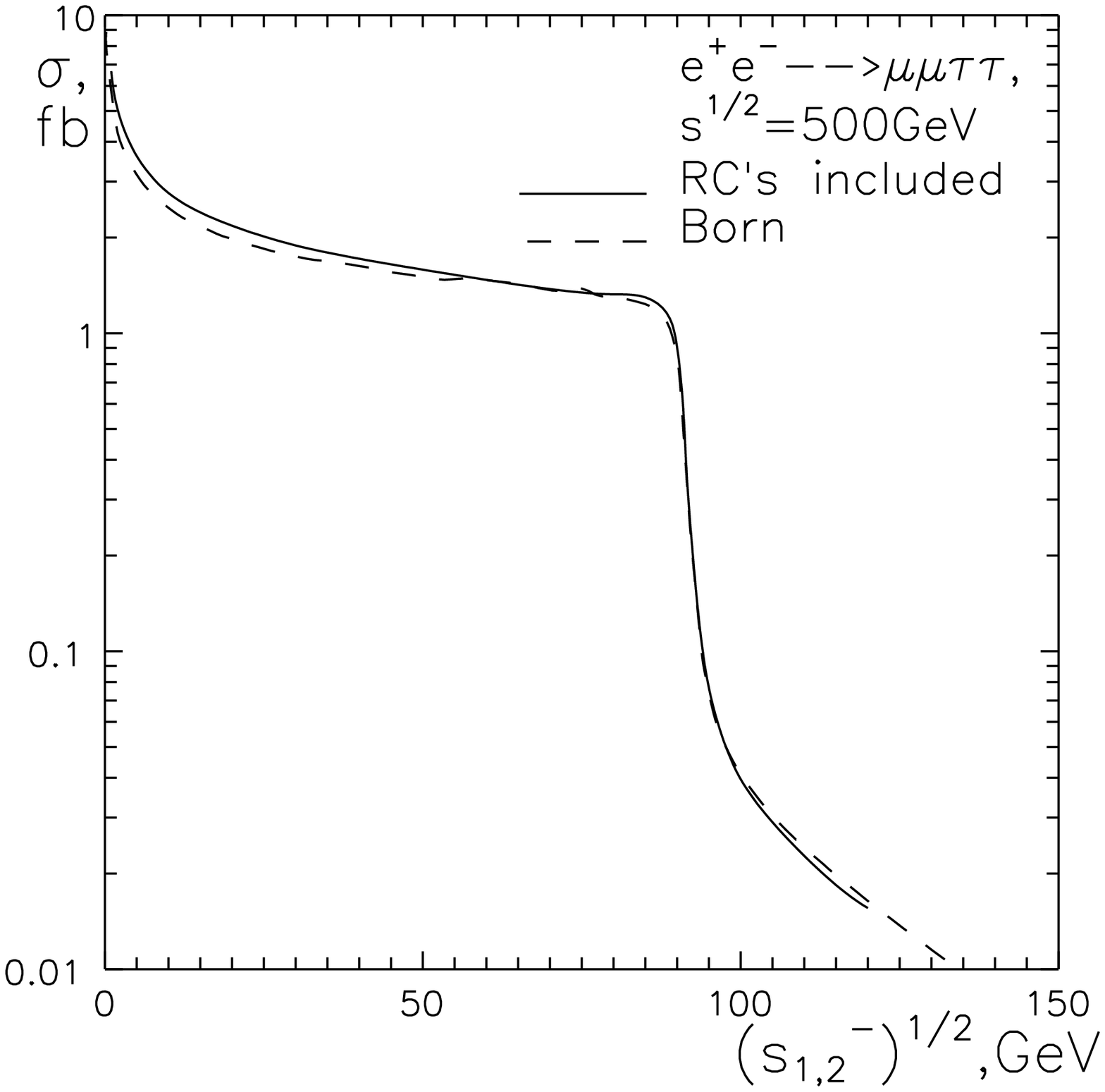}
}
\end{center}
}\end{minipage}
\end{center}
\noindent
{\small\it
\begin{fig} \label{cut4fan} 
$\sigma(e^+e^-\rightarrow \mu^+\mu^-\tau^+\tau^-)$ for
$\sqrt{s}=500\,GeV$ as function of the lower cut $s^-_{1,2}$ on the invariant
energies $s_1$ and $s_2$ of the final muon and tau pairs. The solid (dotted)
curves show total cross sections with (without) initial state QED corrections.
\end{fig}}
\end{figure}
The only SM processes in $e^+e^-$ and $e^-e^-$ collisions with two
particles in the final state not yet considered are $e^+e^-\rightarrow
ZZ$ and $e^+e^-\rightarrow ZH$. 
They are similar to $W$ pair production. 
However, they don't have the enhancement factors of new physics.
Therefore, these reactions are much less sensitive to a $Z'$ than $W$ pair
production. 

Higher order processes cannot compete in setting $Z'$ limits due to
statistics \cite{9702344}.
Some higher order processes have cross sections comparable to those of
two particle final states.
Resonating gauge bosons or collinear radiation of light particles can
be responsible for this enhancement .
However, these enhancement mechanisms do not enhance the $Z'$ contributions.
Therefore, they ``pollute'' a potential $Z'$ signal and should be removed
by appropriate kinematic cuts.
With these cuts, the resulting cross sections are too small
to compete with two particle final states.

The effect is demonstrated \cite{4fanref} in figure~\ref{cut4fan}.
A cut $s^-_{1,2}$ on the invariant masses $s_1$ and $s_2$ of the
$\mu^+\mu^-$ and $\tau^+\tau^-$ pairs suppresses first the photon
exchange ($s^-_{1,2}$ are small). 
For $s^-_{1,2}>M_Z^2$, the $Z$ exchange is suppressed too.
The resulting cross section is approximately a factor $\alpha^2$
smaller than cross sections of particle pair production.
As expected, the effect is not altered by radiative corrections. 

The reactions $e^+e^-\rightarrow Z'Z'$ and $e^+e^-\rightarrow
e^+e^-Z'$ are considered in references \cite{2zp,eezp}.
The second process is difficult to observe above the background \cite{eezp}. 
The first process needs very high energies. 
It was proposed \cite{2zp} to use this process to resolve ambiguities
in the experimental determination of the $E_6$ breaking parameter
$\theta$.
However, this ambiguity can already be resolved by measurements below
the $Z'$ peak \cite{zpsari}. 

The individual interferences of $WW$ scattering ($e^+e^-\rightarrow
e^+e^-W^+W^-$) scale like $s^2$ while the total cross section scales
in the SM like $1/s$ for very high energies.  
As in $e^+e^-\rightarrow W^+W^-$, a non-zero $ZZ'$ mixing angle would destroy
this gauge cancellation. 
Therefore, it could be interesting to investigate the potential of
$WW$ scattering for a $Z'$ search at $TeV$ colliders.
\chapter{$Z'$ search at $pp$ and $p\bar p$ colliders}\label{zppp}
%
The $Z'$ signal at hadron colliders comes from {\it direct} production.
This is a principle difference compared to $e^+e^-$ and $ep$ collisions.
Therefore, the mass of a detectable $Z'$ must be smaller
than the center--of--mass energy of the colliding protons.
In practice, the $Z'$ must be at least two times lighter because it
is produced in collisions of partons.
The $Z'$ is detected through its
decay products, which must be separated from the SM background.
Unfortunately, the background in a hadronic environment
makes it difficult or sometimes impossible to measure potential
interesting observables.

The decay of a $Z'$ to a {\it fermion pair} 
would probably give the first $Z'$ signal in hadron collisions.
The invariant mass of the final state fermion pair is centered
around the $Z'$ mass.
This allows for a good separation of the signal from the background
and for a measurement of the $Z'$ mass.
The signature was exploited in the past to measure the properties of
the SM $Z$ boson at the UA1 and UA2 experiments \cite{ua1ua2}.
Different fermions in the final state can be tagged providing
various cross sections and asymmetries as observables.
See references \cite{prd30,prd35,godfreyold,lhcproc} for some old analyses.

{\it Rare $Z'$ decays}, $Z'\rightarrow f_1\bar f_2V$ (with $V=W,Z$ and
$f_1,f_2$ are higher order processes. 
However, they are enhanced by large logarithms due to collinear and soft
radiation.
They give interesting complementary information in a $Z'$ model measurement.

{\it The $Z'$ decay to $W$ pairs} is possible only in the case of
$ZZ'$ mixing.  
Unfortunately, this decay mode suffers from the SM background of the
associated production of a $W$ and two jets \cite{plb221}.

{\it Associated production} of a $Z'$ together with another gauge
boson, $pp\rightarrow Z'V,\ V=Z,W,\gamma$ is of higher order
compared to the production of a single $Z'$.
However, for $s\gg M_Z^2$ these processes are logarithmically enhanced
similar to soft photon radiation.
They add independent information about $Z'$ models.

There are no other processes known in hadron collisions, which are
useful to add further information on a $Z'$.

\section{Born cross section of $pp(p\bar p)\rightarrow Z'\rightarrow
f\bar f$}\label{zppp11}
The considered reaction is much less sensitive to $ZZ'$ mixing than
$e^+e^-$ experiments at the $Z_1$ peak.
Therefore, any $ZZ'$ mixing effects can be neglected putting
$\theta_M=0$ and identifying the $Z_1$ with the $Z$
and the $Z_2$ with the $Z'$.

The Born cross section of the production of a  $Z'$, which decays to a
fermion pair is
\ba
\label{ppzpborn}
\sigma_A\left(pp(p\bar p)\rightarrow (\nobodyfrac\gamma,Z,Z')X\rightarrow f\bar
fX\right)\equiv \sigma_A^f \hspace{7cm}\nll
= \sum_{q} \int_0^1dx_1\int_0^1dx_2
\sigma_A(sx_1x_2;q\bar q\rightarrow f\bar f)
G_A^q(x_1,x_2,M^2_{Z'})\theta(x_1x_2s-M_\Sigma^2)\hspace*{1.2cm}\\
=\sum_{q}\frac{M}{s}\int_{M_\Sigma^2}^sdQ^2\int_{-y^{max}}^{y^{max}}dy
\sigma_A(sx_1x_2;q\bar q\rightarrow f\bar f)
G_A^q(x_1,x_2,M^2_{Z'})\theta(x_1x_2s-M_\Sigma^2),\nonumber
\ea
where $M_\Sigma$ is the sum of the
masses of the final particles,
$x_{1,2}=\sqrt{\frac{Q^2}{s}}e^{\pm y}$ and $y$ is the rapidity.
The functions $G_A^q(x_1,x_2,M^2_{Z'}),\ A=T,FB$ depend on the
structure functions of the quarks,
\ba
\label{gdists}
G_T^q(x_1,x_2,M^2_{Z'}) &=& q(x_1,M_{Z'}^2)\bar q(x_2,M_{Z'}^2)
+\bar q(x_1,M_{Z'}^2) q(x_2,M_{Z'}^2),\nll
G_{FB}^q(x_1,x_2,M^2_{Z'}) &=& q(x_1,M_{Z'}^2)\bar q(x_2,M_{Z'}^2)
-\bar q(x_1,M_{Z'}^2) q(x_2,M_{Z'}^2).
\ea
The expressions for $\sigma_A(sx_1x_2;q\bar q\rightarrow f\bar f),\ A=T,FB$
can be easily derived from equations \req{born}-\req{lambdadef}.
These formulae also contain the dependence on the
helicity of the initial and final fermions.

If the polarization of the final fermion is measurable, one can observe 
\bq
\label{pppol}
\sigma_{pol}^f = \sigma_T^{f\uparrow}-\sigma_T^{f\downarrow}.
\eq

If polarized proton beams are available \cite{rhich}, the ``left--right'' cross
section is of interest \cite{ppol}, 
\ba
\label{gdists2}
\sigma_{LR}^f &=& \sigma_L^f-\sigma_R^f,\\
G_{LR}^q(x_1,x_2,M^2_{Z'}) &=&\nll 
&&~\hspace{-3cm} 
\left[\nobodyfrac q^\uparrow(x_1,M_{Z'}^2)-q^\downarrow(x_1,M_{Z'}^2)\right]
\bar q(x_2,M_{Z'}^2)
+\left[\nobodyfrac\bar q^\uparrow(x_1,M_{Z'}^2)-
\bar q^\downarrow(x_1,M_{Z'}^2)\right] q(x_2,M_{Z'}^2).\nonumber
\ea 
$\sigma_{L,R}$ are the total production cross sections with one left
or right handed initial quark.
As in the case of $e^+e^-$ collisions, the difference of $\sigma_{FB}$
for a left and right handed quark in the initial state can be
considered \cite{ppol},
\ba
\label{gdists3}
\sigma_{LR,FB}^f &=& \sigma_{L,FB}^f-\sigma_{R,FB}^f,\\
G_{LR,FB}^q(x_1,x_2,M^2_{Z'}) &=& \nll
&&~\hspace{-3cm} 
\left[\nobodyfrac q^\uparrow(x_1,M_{Z'}^2)-q^\downarrow(x_1,M_{Z'}^2)\right]
\bar q(x_2,M_{Z'}^2)
-\left[\nobodyfrac\bar q^\uparrow(x_1,M_{Z'}^2)-
\bar q^\downarrow(x_1,M_{Z'}^2)\right]
q(x_2,M_{Z'}^2).\nonumber
\ea 
More complicated cross sections  can be measured if both proton beams
are polarized. 

\section{Higher order processes and background}\label{zppp12}
\subsection{Rare $Z'$ decays}
The decay modes $Z'\rightarrow f_1\bar f_2V$ (with $V=W,Z$ and $f_1,f_2$
ordinary fermions) \cite{plb192} are enhanced by logarithms due to
collinear and soft radiation, compare equation \req{gamrare}.
The decays $Z'\rightarrow Zl^+l^-\rightarrow l_2^+l_2^-l^+l^-$ and 
$Z'\rightarrow Wl\nu_l$ can be separated from the background
\cite{prd48425,langackerpp}. 
The first process shows only a weak dependence on the $Z'$ couplings
serving as a consistency check of the experiment, while the second
``gold-plated'' decay mode yields useful and complementary
information about the $Z'$ couplings \cite{langackerpp,prd4614,prd472}.
The rare decays where $f_1$ and $f_2$ are quarks can also be measured,
although with a larger systematic error \cite{langackerpp}.

\subsection{Associated $Z'$ production}
The production of a $Z'$ in association with another gauge boson,
$pp\rightarrow Z'V,\ V=Z,W,\gamma$ is logarithmically enhanced
for high energies similar to soft photon radiation from the initial state.
The total cross section of the partonic subprocess is \cite{1063}
\ba
\label{zpv}
\sigma_T(q\bar q\rightarrow Z'V)=\frac{g_1^2g_2^2}{4\pi s}
\left[L_q^2L'^2_q+R_q^2R'^2_q\right]
\left\{\frac{1+m_+^2}{1-m_+^2}\ln\frac{1-m_++\lambda}{1-m_+-\lambda}-2\lambda
\right\},\nll
\mbox{with\ \ } m_+=M_{Z'}^2/s+M_V^2/s \mbox{\ \ and\ \ } 
\lambda=\lambda(1,M_{Z'}^2/s,M_V^2/s),
\ea
where $\lambda$ is the kinematic function \req{lambdadefkin}.
The cross section \req{zpv} is known from electron
positron collisions \cite{baier}.

Associated $Z'$ production gives complementary information
and is free of SM backgrounds \cite{1063,prd47}. 
It can compete with other processes in the determination of the
parameters of a $Z'$ model \cite{langackerpp}.
\subsection{Radiative Corrections}
In a first approximation, the tree level $Z'$ contributions can be
added to the SM cross section.
This approach neglects radiative corrections to the new physics
treating them as a small correction to a small effect.
This approximation is probably true for a first $Z'$ discovery but has to be
checked in a $Z'$ model measurement.
In the following we briefly mention the main Standard Model corrections.
\subsubsection{QCD corrections}\label{zppp121}
QCD corrections are numerically most important.
They increase the lowest order cross section
of vector boson production at the Tevatron by $20-30\%$.
These corrections are often called $K$ factors.

Corrections to the unpolarized Drell--Yan process are known to order
$O(\alpha_s^2)$ for the invariant mass distribution \cite{ppqcd1}. 
For rapidity-- and $x$--distributions, they are calculated to order
$O(\alpha_s)$ \cite{ppqcd2} and partly to order  $O(\alpha_s^2)$ \cite{ppqcd3}.
Soft gluon contributions can be treated to all orders by
exponentiation \cite{ppqcd4}.
QCD corrections to polarized hadron scattering are known to order
$O(\alpha_s)$ \cite{ppqcd5}.
\subsubsection{QED and weak corrections}\label{zppp123}
QED corrections are model independent.
As in $e^+e^-$ collisions, initial state corrections, final state
corrections and the interference between them are separately gauge
invariant for neutral current processes.
Numerically, the dominant corrections come from final state radiation
\cite{ppqed}.
The corrections from initial state radiation and the interference
between initial and final state radiation are small after factorizing
the collinear singularities into the parton distribution functions
\cite{ppqed}. 
As in $e^+e^-$ collisions, final state corrections do not feel the
neutral gauge boson exchanged before. 

Pure weak corrections are expected to already be very small for the
Standard Model $Z$ production \cite{ppqed}.
The corrections to $Z'$ production are expected to be even smaller.
\subsection{Background}\label{zppp124}
The fermions coming from $Z'$ decay have an invariant mass, which is
peaked around $M_{Z'}$. 
Fermion pairs with the same invariant mass could also be produced by
gluon, photon or $Z$ exchange. 

The experience of existing hadron colliders shows that
$b$--quarks \cite{ppbquarks} and dijets \cite{dijet} can be detected
\cite{ppel}. 
However, the sensitivity to new gauge bosons from quark pairs is
reduced compared to 
muon and electron pairs due to the QCD background. 

The background to $\tau$ pairs is considered in \cite{taupp} and found
to be manageable for a $Z'$ originating in an $E_6$ GUT.
It is shown there that the signal can be distinguished from the
background of $W^+W^-$ production and from the background coming
from misidentified jets, which could be accidentally recognized as $\tau$
decay products. Furthermore, the background from top pairs decaying to
$\tau$ pairs can be managed. Even the background from Drell-Yan
production of $\tau$'s can be removed although this is harder
\cite{taupp} than in the case of muon pairs.
\section{Observables}\label{zppp112}
Similar to the case of $e^+e^-$ collisions, the total cross sections
$\sigma_T^f$ and different asymmetries serve
as observables, 
\bq
\label{obspp}
\sigma_T^f,\ \ \ 
A_{FB}^f=\frac{\sigma_{FB}^f}{\sigma_T^f},\ \ \ 
A_{pol}^f=\frac{\sigma_{pol}^f}{\sigma_T^f},\ \ \ 
A_{LR}^f=\frac{\sigma_{LR}^f}{\sigma_T^f},\ \ \ 
A_{LR,FB}^f=\frac{\sigma_{LR,FB}^f}{\sigma_T^f}.
\eq
Note that $A_{LR,FB}$ in our notation is $A_{FB}^{pol}$ in reference
\cite{ppol}, while we reserve $A_{pol}$ for final state polarization
asymmetries following the notation of reference \cite{taupp}.

Not all observables \req{obspp}
can be measured in a real experiment. 
In addition to the constraints mentioned in section \ref{zpeeff14},
complications arise because the signal has to be detected above
the background of the hadronic environment.

As in $e^+e^-$ collisions, $A_{pol}^f$ is independent of the couplings
to initial fermions.
In hadron collisions, all dependence on the quark structure functions
also drops out \cite{taupp}.

For lepton pairs in the final state, rapidity ratios can be defined
\cite{langackerpp}, 
\bq
\label{raprat}
r_{y1}=\frac{ \int_{-y_1}^{y_1} \frac{d\sigma_T^l}{dy}dy }
{\left(\int_{-y_{max}}^{-y_1} + \int_{y_1}^{y_{max}}\right)
\frac{d\sigma_T^l}{dy}dy},\hspace{1cm}
A_{FBy1}=\frac{\left(\int_{-y_1}^0 - \int_0^{y_1}\right)
\frac{d\sigma_{FB}^l}{dy}dy}
{\left(\int_{-y_{max}}^{-y_1} - \int_{y_1}^{y_{max}}\right)
\frac{d\sigma_{FB}^l}{dy}dy}.
\eq
The observable $r_{y1}$ is useful in distinguishing between different
$Z'$ models, while $A_{FB{y1}}$ being ``a refinement of a refinement'' is less
sensitive to different $Z'$ models \cite{langackerpp}.

Rare $Z'$ decays $Z'\rightarrow f_1f_2V$ allow for the definition of
the ratios, 
\ba
\label{rarrat}
r_{llZ}\equiv\frac{Br(Z'\rightarrow l^+l^-Z)}{Br(Z'\rightarrow l^+l^-)},\ \ \ 
r_{\nu\nu Z}\equiv\frac{Br(Z'\rightarrow \nu\bar\nu Z)}
{Br(Z'\rightarrow l^+l^-)},\ \ \ 
r_{l\nu W}\equiv\frac{Br(Z'\rightarrow l^\pm\nu W)}
{Br(Z'\rightarrow l^+l^-)}.
\ea
The index $l$ refers to a summation over $e$ and $\mu$ and the index $\nu$ in
$r_{\nu\nu Z}$ to the summation over $\nu_e,\nu_\mu$ and $\nu_\tau$. 
The ratios $r_{had Z}$ and $r_{hadW}$, where the fermions $f_1$ and
$f_2$ are quarks can be defined analogously.
$r_{l\nu W}$ depends only on the $Z'$ couplings to leptons.

The cross sections of associated $Z'$ production enter the ratios
\bq
\label{assrat}
R_{Z'V}=\frac{\sigma(pp\rightarrow Z'V)Br_2^l}
             {\sigma(pp\rightarrow Z')Br_2^l},\ \ \ 
V=Z,W,\gamma.
\eq
\section{$Z'$ constraints}\label{zppp13}
%
\subsection{Model independent constraints on $\sigma_T^f$}\label{zppp131}
%
\subsubsection{Estimate}\label{zppp1311}
The signal of extra neutral gauge bosons in hadron collisions is a
number $N_{Z'}$ of excessive fermion pairs with an invariant mass
around $M_{Z'}$. 
In the case of absence of a signal, the observable $\sigma_T^f$ is
constrained independently of the $Z'$ model.

We give here an approximation of  $\sigma_T^f$ to make the
dependence of $Z'$ exclusion limits and $Z'$ 
model measurements on the center--of--mass energy, the integrated
luminosity and model parameters transparent \cite{leikezppp}.
Such an estimate is useful to extrapolate from $Z'$ limits known for {\it one}
collider and $Z'$ model to other colliders and $Z'$ models.

Consider $\sigma_T^\mu$ neglecting the background. 
Equation \req{ppzpborn} can be approximated treating the
resonating $Z'$ propagator of $\sigma_T^\mu(Q^2;q\bar q\rightarrow f\bar f)$
in the narrow width approximation,
\bq
\frac{Q^4}{|Q^2-M^2+iM\Gamma|^2}\longrightarrow
\delta(Q^2-M^2)\frac{\pi M^4}{M\Gamma}.
\eq

We obtain 
\ba
\label{ppzpnwa}
\sigma_T^\mu = \frac{4\pi^2}{3s}\frac{\Gamma_{Z'}}{M_{Z'}}Br_2^\mu
\sum_{q} Br_2^q f^q\left(\frac{\sqrt{s}}{M_{Z'}},M_{Z'}^2 \right)
\ea
\bq
\label{fdef}
\mbox{with}\hspace{1cm}  f^q\left(r_z,M_{Z'}^2\right)
=\int_{1/r_z^2}^1\frac{dx}{x}G_T^q\left(x,\frac{1}{xr_z^2},M^2_{Z'}\right)
\mbox{\ \ and\ \ } r_z=\frac{\sqrt{s}}{M_{Z'}}.
\eq

The function $f^q\left(r_z,M_{Z'}^2\right)$ has only a very weak
dependence on $M_{Z'}^2$ in the region we are interested in. 
We therefore can drop the second argument approximating
$f^q\left(r_z,M_{Z'}^2\right)\approx f^q\left(r_z\right)$.   

The inspection of $Z'$ limits at the proposed colliders shows
\cite{godfrey} that the functions $f^q(r_z)$ are needed only in a
narrow interval of $r_z$, i.e. $3<r_z<5$ for $pp$ collisions and
$2<r_z<3.5$ for $p\bar p$ collisions.  
Under these conditions, the functions for different quarks $q=u,d$
differ mainly by a constant factor. 
Hence, we can make the following replacement in equation
\req{ppzpnwa}, 
\bq
\label{fzdef}
\sum_{q} Br_2^q f^q\left(\frac{\sqrt{s}}{M_{Z'}},M_{Z'}^2 \right)
\approx f^u\left(\frac{\sqrt{s}}{M_{Z'}}\right)
\left[ Br_2^u + \frac{1}{C_{ud}} Br_2^d \right]
\eq
with $C_{ud}=f^u(r_z)/f^d(r_z)\approx 2\ (\approx 25)$ at $pp\ (p\bar p)$
colliders.
We see that $\sigma_T^\mu$ has a reduced sensitivity to $Z'd\bar
d$ couplings. 

\begin{figure}
\begin{center}
\begin{minipage}[t]{7.8cm}{
\begin{center}
\hspace{-1.7cm}
\mbox{
\epsfysize=7.0cm
\epsffile[0 0 500 500]{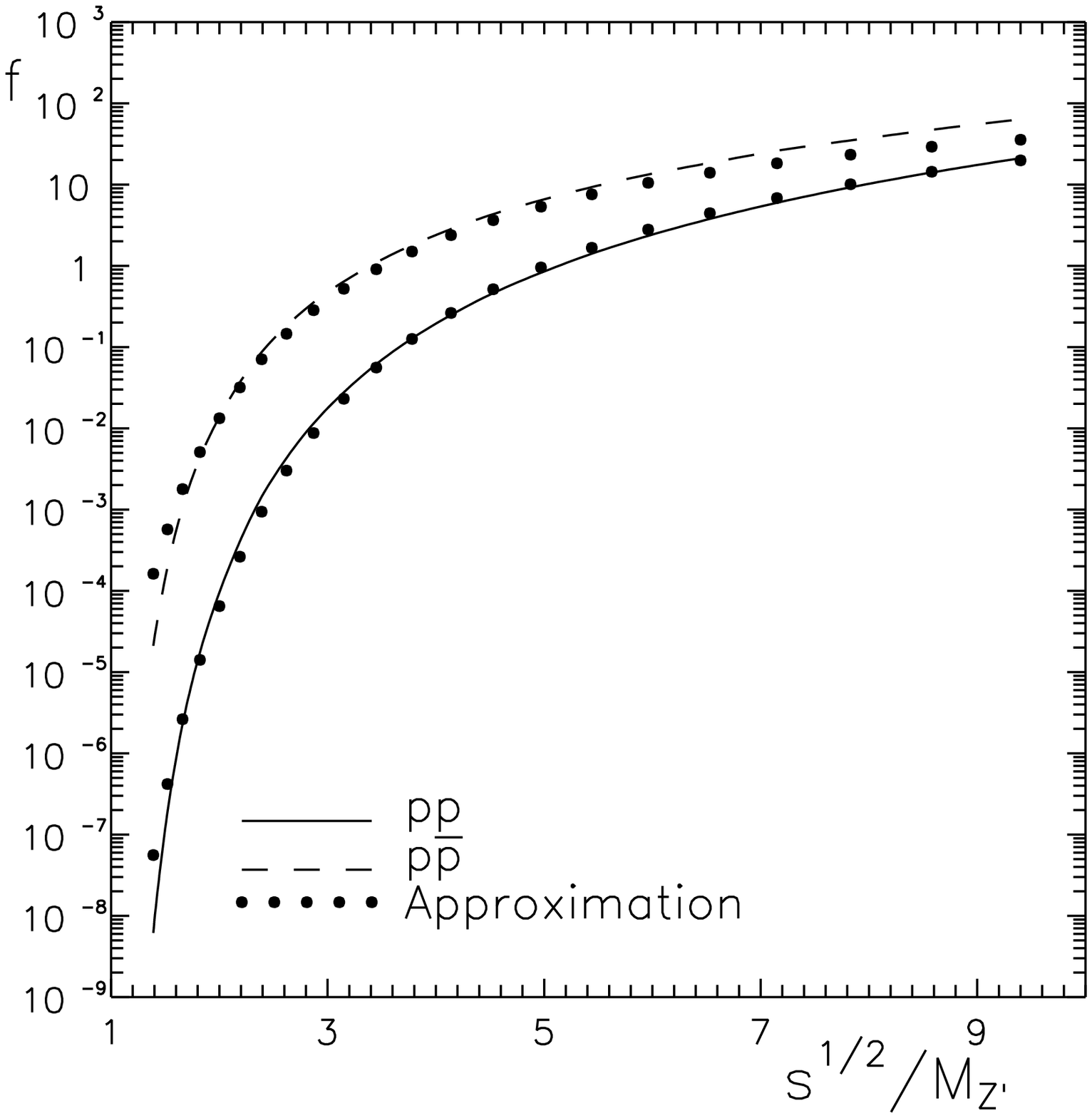}
}
\end{center}
}\end{minipage}
\end{center}
\noindent
{\small\it
\begin{fig} \label{mrsgall} 
The function $f^u(\sqrt{s}/M_{Z'},25\,TeV^2)$ and the approximation
\req{fapprox2}.  
The curves of $f^u(\sqrt{s}/M_{Z'},Q^2)$ for $Q^2=1\,TeV^2$ could not be
distinguished from $Q^2=25\,TeV^2$. 
This is figure~1 from reference \cite{leikezppp}.
\end{fig}}
\end{figure}

The integral defining the function $f^u(r_z)$ could be approximated by
the function $r_z^a(r_z-1)^b$, which takes into account the
parametrization of the structure functions.
However, we prefer an approximation by an
exponential function because it can be inverted analytically.
We get
\bq
\label{fapprox2}
f^u(r_z)\approx Ce^{-A/r_z},\ \ \ 
C=600\ (300),\ A=32\ (20)\mbox{\ \ for\ \ }pp\ (p\bar p)\mbox{\ collisions}.
\eq
The approximation \req{fapprox2} and the exact calculation \req{fdef} of
$f^u(r_z)$ are shown in figure~\ref{mrsgall}.
We use the structure functions \cite{mrs}.
The dependence of our results on this choice is negligible.
Note that the fit works satisfactorally up to $r_z=10$.
It cannot describe SM $Z$ production at the Tevatron where we have
$r_Z\approx 20$.

Collecting all approximations, $\sigma_T^\mu$  can be written as 
\ba
\label{fdef2}
\sigma_T^\mu\equiv\frac{N_{Z'}}{L}&\approx&
\frac{1}{s}c_{Z'}C\exp\left\{-A\frac{M_{Z'}}{\sqrt{s}}\right\},\\
\mbox{with}&& c_{Z'}=\frac{4\pi^2}{3}\frac{\Gamma_{Z'}}{M_{Z'}} 
Br_2^\mu\left[ Br_2^u + \frac{1}{C_{ud}} Br_2^d\right]. 
\nonumber
\ea
All details of the $Z'$ model are collected in the constant $c_{Z'}$.
For convenience, we list the value of $c_{Z'}$ for some $Z'$ models:
\bq\begin{array}{rccccc}
\mbox{Model:} & \chi  & \psi & \eta & LR    & SSM\nll
1000\cdot c_{Z'}(pp):& 1.17 & 0.572 & 0.712 & 1.35 & 2.27\\
1000\cdot c_{Z'}(p\bar p):& 0.40& 0.437 & 0.556 & 0.77 & 1.41 
\end{array}\eq 

The approximate exponential dependence of $\sigma_T^\mu$ (and of
$N_{Z'}$) on $M_{Z'}$ can be 
recognized, for instance, in figures~1 to 5 of reference \cite{prd30}.
It even holds in associated $Z'$ production, $pp\rightarrow Z'W,
pp\rightarrow Z'Z$, as can be seen from figure~3 of reference \cite{1063}. 
The reason is that associated production happens for very constrained
values of $Q^2\approx (M_{Z'}+M_V)^2$ only; for smaller $Q^2$, the
process is forbidden by kinematics, for larger $Q^2$ we have an
exponential suppression due to structure functions.

Equation \req{fdef2} is the starting point for several useful
estimates regarding $Z'$ constraints in hadron collisions.
\subsubsection{Present constraints}\label{zppp1312}
The best present constraints on $\sigma_T^f$ come from the Tevatron
experiments. 
They give a constraint on 
$\sigma_T^f\equiv\sigma_T^f(pp\rightarrow Z')\cdot Br_2^f$
as a function of the invariant mass of the final fermion pair $f\bar f$ 
in the final state. 

\begin{figure}[tbh]
\begin{center}
\ \vspace{1cm}\\
\begin{minipage}[t]{7.8cm} {
\begin{center}
\hspace{-1.7cm}
\vfill
\mbox{
\epsfysize=7.0cm
\epsffile[0 0 500 500]{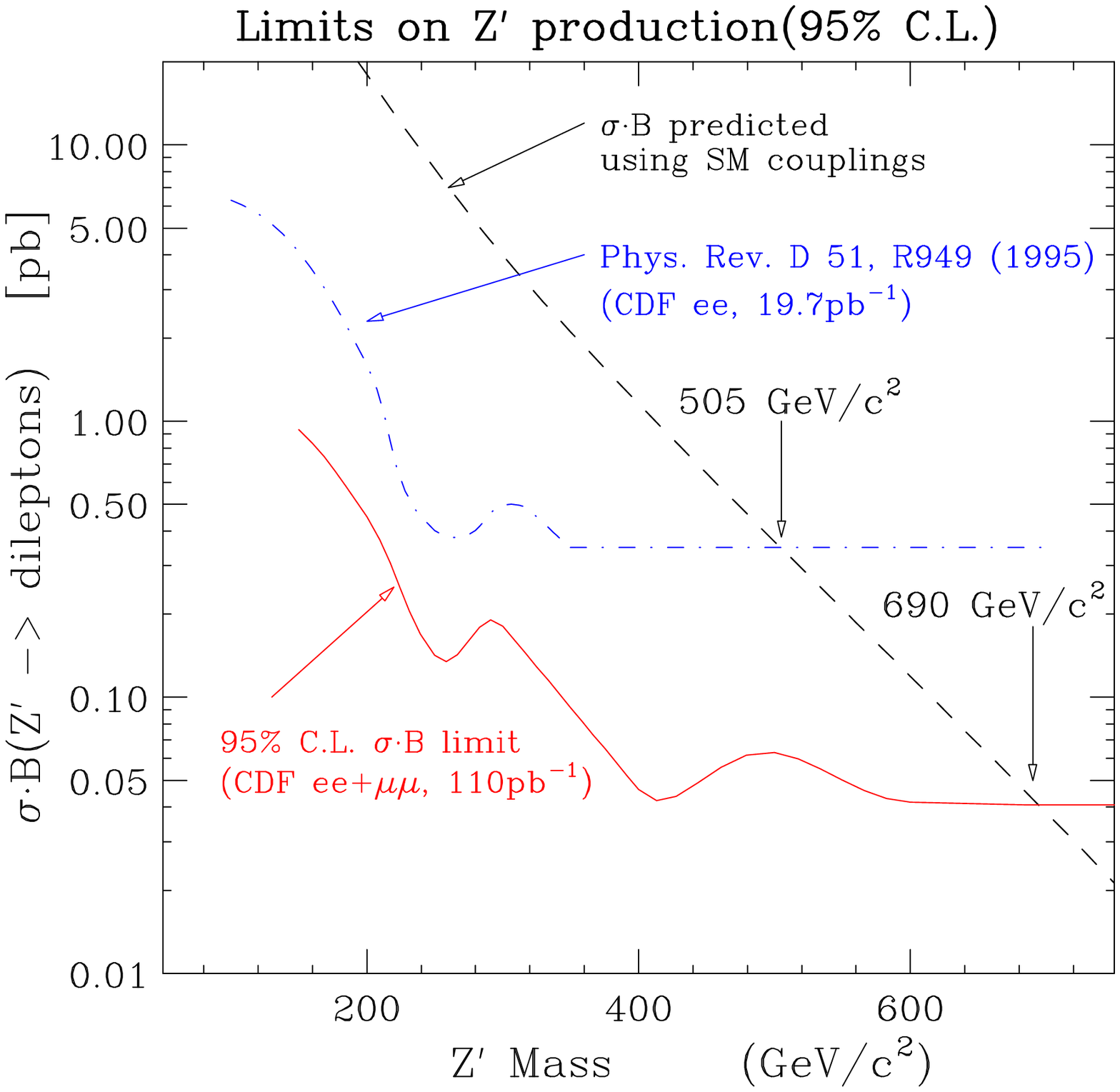}
}
\end{center}
}\end{minipage}
\end{center}
\vspace*{-2.0cm}
\noindent
{\small\it\begin{fig} 
\label{kaorism} 
The limit on $\sigma_T^\l$ as a function of the dilepton mass, as well
as the expectation for $Z'=SSM$.
This is figure~3 from reference \cite{prl79}.
\end{fig}}
\end{figure}

The constraint from CDF \cite{prl79} for electrons and muons in the
final state ($L=110\,pb^{-1}$ and $\sqrt{s}=1.8\,TeV$) is shown in figure
\ref{kaorism}.
The corresponding constraints from D0 can be found in \cite{plb385}.
The CDF data constrain $\sigma_T^l<0.04\,pb$ for large invariant masses.
The deviation of the experimental curves from this number for smaller
$Z'$ masses are due to the SM background.
The total detection efficiencies \cite{prl79} for electron and muon pairs are
$\epsilon_e\approx 47\%$ and $\epsilon_\mu\approx 20\%$.
The $K$ factor from QCD corrections is $K\approx 1.3$.

The experimental results can be confronted with the estimate \req{fdef2}.
The exponential dependence  of $\sigma_T^\l$ on $M_{Z'}$ predicted by
the estimate can clearly be seen in the figure.
Taking into account the detection efficiencies and the $K$
factor, the estimate \req{fdef2} predicts $(\sigma_T^f)^{lim}\approx
0.031\,pb$.  
This is in reasonable agreement with the exact result.

The SM background for {\it dijets} is much larger due to QCD effects.
In reference \cite{cdf1196}, events with dijets didn't allow
constraints on extra neutral gauge bosons from the $E_6$ GUT.
\subsection{Model dependent constraint on $M_{Z'}$}\label{zppp1313}
%
\subsubsection{Estimate}\label{zppp13131}
Inverting the approximation \req{fdef2}, we get a constraint on $M_{Z'}$,
\bq
\label{fdef4}
M_{Z'}>M_{Z'}^{lim}\approx\frac{\sqrt{s}}{A}
\ln\left(\frac{L}{s}\frac{c_{Z'}C}{N_{Z'}}\right)
\approx\sqrt{s}\left\{
\begin{array}{lr}
0.386 +\frac{1}{32}\ln\left(\frac{L\cdot fb}{N_{Z'}\cdot s/TeV^2}\cdot
1000c_{Z'}\right) 
& pp\\
0.583+\frac{1}{20}\ln\left(\frac{L\cdot fb}{N_{Z'}\cdot s/TeV^2}\cdot
1000c_{Z'}\right) 
& p\bar p
\end{array}\right. .
\eq
For $M_{Z'}<M_{Z'}^{lim}$ more than $N_{Z'}$ additional events are expected.

A formula similar to \req{fdef4} was quoted in reference
\cite{delaguilalog} some time ago.
However, there the dependence of $M_{Z'}^{lim}$ on $Z'$ couplings and
collider parameters is given only numerically.

Relation \req{fdef4} describes the scaling of $M_{Z'}^{lim}$ with the
center--of--mass energy and the integrated luminosity.
$M_{Z'}^{lim}$ depends on $L$ only logarithmically. 
Therefore, the dependence of $M_{Z'}^{lim}$ on detector efficiencies
or event losses due to background suppression is only marginal. 

The logarithmic dependence of $M_{Z'}^{lim}$ on $L$ can be recognized
in figure~3 of reference \cite{9609248} or in figure~2.33 in
reference \cite{prl79}. 
The reduction of $M_{Z'}^{lim}$ due to a decrease of the event rate by
a factor of two is predicted by relation \req{fdef4} to be $9\%\ (7\%)$ for
the proposed $pp\ (p\bar p)$ colliders. 
These numbers, which do not discriminate between $Z'$ models, $s$, and $L$,
are in agreement with the last line of table~1 in reference \cite{9609248}.

The model dependent constant $c_{Z'}$ enters \req{fdef4} only under the
logarithm leading to the weak model dependence of $Z'$ exclusion
limits in $pp$ and $p\bar p$ collisions.   
The physical origin of this effect is hidden in the properties of
the structure functions defining \req{fdef} the function $f(r_z)$.
Therefore, relation \req{fdef4} obtained for 
$\sigma_T^\mu$ is qualitatively true for other observables too.

Radiative corrections lead to deviations of $N_{Z'}$ from the Born
prediction.
The effect on $M_{Z'}^{lim}$ is moderate because $N_{Z'}$
enters this limit only under the logarithm. 

The scaling \req{fdef4} is the complement of relation \req{epemlim2} derived
for $e^+e^-\rightarrow f\bar f$.
The strong model dependence of $Z'$ exclusion limits from $e^+e^-$
collisions is due to their direct dependence on the square of the
coupling constants of the $Z'$ to fermions, see \req{bornexcl}.

In the last step, equation \req{fdef4} is written in a form, which makes the
logarithm nearly zero for GUT's at the proposed colliders.
We see that the numerical influence of the logarithm is suppressed by
a small prefactor.  
The constant terms 0.386 and 0.583 give an estimate for the average
sensitivity of $pp(p\bar p)$ collisions to a $Z'$ in units of $\sqrt{s}$.

For practical purposes, it is useful to rewrite equation \req{fdef4} as
\bq
\label{ppzpscale}
\frac{M_{Z'}^{lim}(s,L)}{M_{Z'}^{lim}(s_0,L_0)}\approx
\frac{\sqrt{s}}{\sqrt{s_0}}\left(1+\xi\ln\frac{s_0L}{sL_0}\right)
\mbox{\ \ with\ \ }
\hspace{1cm} \xi=\left[\ln\frac{L_0}{s_0}\frac{c_{Z'}C}{N_{Z'}}\right]^{-1},
\eq
where now all model dependence is hidden in the constant $\xi$.
Normalizing at one collider, equation \req{ppzpscale} predicts the
limits at another collider.
All $Z'$ exclusion limits published in figure~1 of reference \cite{godfrey}
can be reproduced by the estimate \req{ppzpscale} 
with an accuracy of 10\% for 
$\xi=0.13\ (0.10)$ for $pp\ (p\bar p)$ collisions. 
\subsubsection{Present constraints}\label{zppp13132}
No $Z'$ signal is found in present experiments.
This negative search result can be interpreted as exclusion limits  
$M_{Z'}^{lim}$ in different models. 

The limits quoted in reference \cite{prl79} are in a
good agreement with the prediction \req{fdef4}.
%
\begin{table}[tbh]
\begin{center}
\begin{tabular}{|lrrrrr|}\hline\rule[-2ex]{0ex}{5ex} 
Model: & $\chi$  & $\psi$ & $\eta$ & $LR$  & $SSM$\\ \hline
$M_{Z'}^{lim}/GeV$ from \cite{prl79}:& 595 & 590 & 620 & 630 &
690\rule[0ex]{0ex}{3ex}\\ 
$M_{Z'}^{lim}/GeV$ from \req{fdef4}: & 552 & 560 & 582 & 611 &
\rule[-1ex]{0ex}{3ex} 665 \\
\hline
\end{tabular}\medskip
\end{center}
{\small\it  \begin{tab}\label{cdfcomp} The lower bounds (95\% CL) on
the $Z'$ mass from the analysis \cite{prl79} compared to the estimate
\req{fdef4}. 
The finite detection efficiencies $\epsilon_e,\epsilon_\mu$ and the
$K$-factor as given in section \ref{zppp1312} are taken into account in the
estimates. 
\end{tab}} \end{table}
\subsubsection{Future constraints}\label{zppp13133}

The minimal input of the different analyses are the integrated
luminosity $L$ and the center--of--mass energy $\sqrt{s}$ of the colliding
particles ($pp$ or $p\bar p$), a list of observables entering the
fit and the number of $Z'$ events $N_{Z'}$ demanded for a signal.
If applied, kinematic cuts and radiative corrections must be
specified. 
It follows a list of different analyses.

\begin{description}
%
%
%
\item[Reference \cite{godfrey}] GODFREY95:\ \ \ 
This is a theoretical analysis for different future colliders based on
$\sigma_T^e$ and $\sigma_T^\mu$ with $N_{Z'}=10$.
No $Z'$ decays to exotic fermions are assumed.
1-loop QCD corrections are included in the $Z'$ production.
The $Z'$ decay is calculated including 2--loop QCD, 1--loop QED
corrections and top--quark decays.
We selected two scenarios to present them in table \ref{zppptab}.
The numbers are taken from table~2 of reference \cite{cvetrev} and
from figure~1 of \cite{godfrey}. 
\item[Reference \cite{9609248}] RIZZO96:\ \ \ 
This is a theoretical analysis for different future colliders based on
$\sigma_T^e$ and $\sigma_T^\mu$ with $N_{Z'}=10$.
No $Z'$ decays to exotic fermions are assumed.
No radiative corrections are included.
We selected two scenarios for our table \ref{zppptab}.
The numbers are taken from table~1 of \cite{9609248}.
\end{description}
The results of the different analyses are compared with the estimate
\req{fdef4}.
We see that the prediction \req{fdef4} agrees with the exact results
within 10\% in a wide range of $L$ and $s$.
%
\begin{table}[tbh]
\begin{center}
\begin{tabular}{|lrr|rrrrrr|}\hline\rule[-2ex]{0ex}{5ex} 
Analysis &$\frac{\sqrt{s}}{TeV}$ &$L\cdot fb$
&$\chi$&$\psi$ &$\eta$ &$LR$ &$SSM$ &estimate \req{fdef4}\\ 
\hline
\cite{godfrey} & 2($\ p\bar p$)&   10 & 1.04 & 1.05 & 1.07 & 1.10 & 1.15 & 
\rule[0ex]{0ex}{3ex} 1.06\\
\cite{godfrey} & 14($\ pp$)  &  100 & 4.38 & 4.19 & 4.29 & 4.53 & 4.80 & 4.47\\
\cite{9609248} & 60($\ pp$)&  100 & 13.3 & 12.0 & 12.3 & 13.5 & 14.4 & 13.7\\
\cite{9609248} & 200($\ pp$)& 1000 & 43.6 & 39.2 & 40.1 & 43.2 & 44.9 & 
\rule[-1ex]{0ex}{3ex} 49.3\\
\hline
\end{tabular}\medskip
\end{center}
{\small\it  \begin{tab}\label{zppptab} The lower bound on the $Z'$ mass
$M_{Z'}^{lim}$ in TeV excluded by the different analyses described in the text.
The estimate \req{fdef4} is added for $Z'=SSM$.
This is table~1 from reference \cite{leikezppp}.
\end{tab}} \end{table}
\subsection{Model independent constraint on $g_2/g_1$}
Figure~\ref{kaorism} gives constraints on $g_2/g_1$.
According to the estimate \req{fdef2}, $\sigma_T^\mu$ scales as
$\Gamma_{Z'}/M_{Z'}\sim g_2^2$. 
The missing factor depends on the $Z'$ model.
It is given in \req{g2approx1} for $Z'=SSM$ if no decays to exotic
fermions are allowed.
The resulting constraint $g_2<g_2^{lim}$ can be obtained graphically
from  figure~\ref{kaorism} by an appropriate shift of the signal cross section.
We get $g_2^{lim}\approx g_1/4.5$ for $Z'=SSM$ and $M_{Z'}<400\,GeV$. 
\subsection{Errors of Model measurements}\label{zppprecon}
%
\subsubsection{Estimate}\label{zppp1321}
\begin{figure}
\begin{center}
\begin{minipage}[t]{7.8cm}{
\begin{center}
\hspace{-1.7cm}
\mbox{
\epsfysize=7.0cm
\epsffile[0 0 500 500]{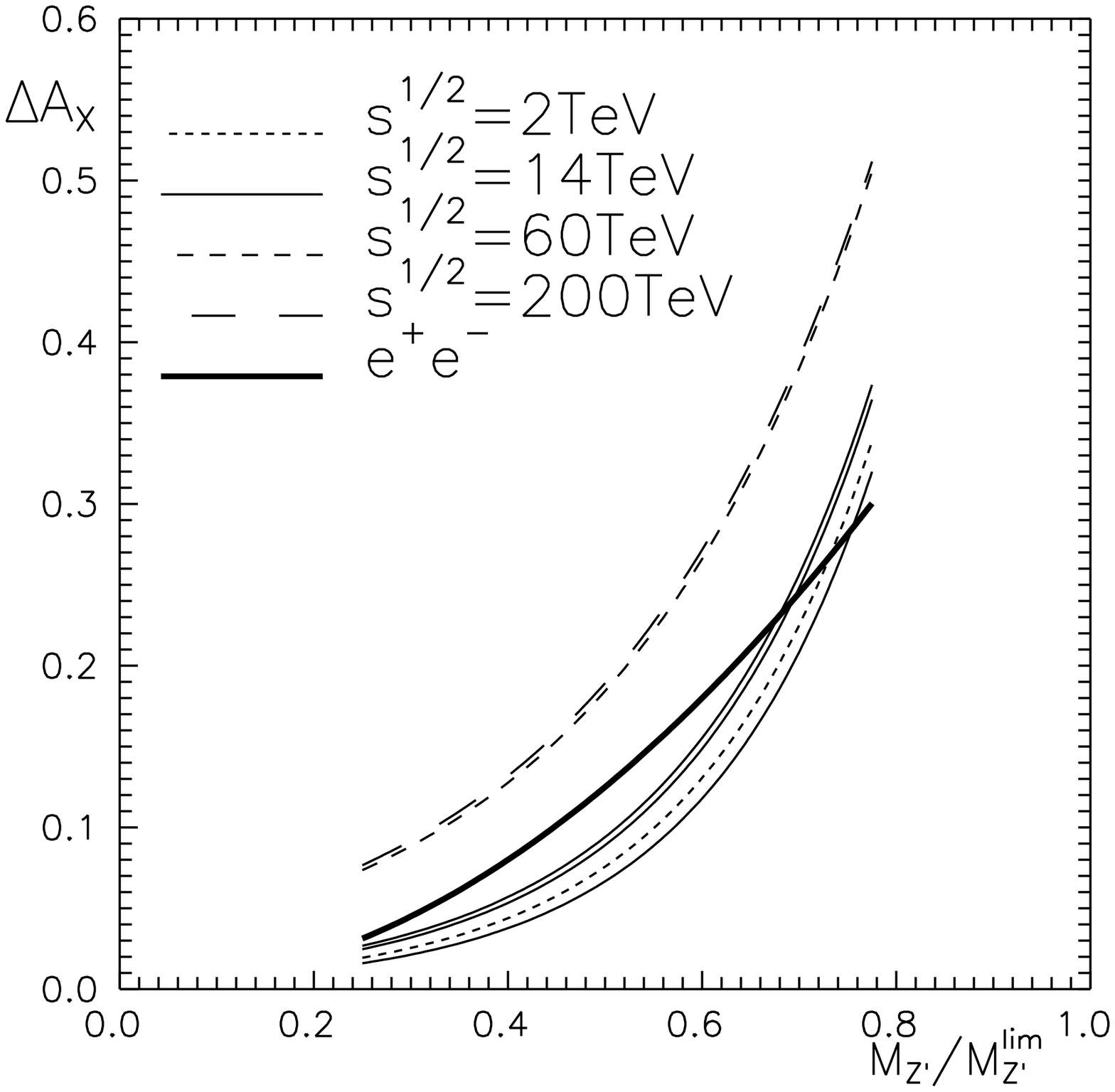}
}
\end{center}
}\end{minipage}
\end{center}
\noindent
{\small\it
\begin{fig} \label{zppf2} 
The estimate of $\Delta O$ (95\% CL) as a function of
$M_{Z'}/M_{Z'}^{lim}$ as given in equation \req{fdef7}.
$M_{Z'}^{lim}$ is the $Z'$ exclusion limit obtained from the same
observable $O$ alone.
Shown are the predictions for the scenarios listed in
table~\ref{zppptab} for $Z'=\eta$. 
For $\sqrt{s}=14\,TeV$, the dependence is shown for $Z'=\psi,\eta,SSM$
(from top to down). 
The thick solid line is obtained from relation \req{epemmea2}.
\end{fig}}
\end{figure}

Model parameters can be measured if some of the observables $O$
introduced in section \ref{zppp112} give a signal.
A reasonable model measurement requires enough events to assume that
they have a normal distribution. 
The one--$\sigma$ statistical error can then be estimated using
equation \req{fdef2}, 
\bq
\label{fdef5}
\Delta A_{FB}^l\approx\frac{1}{\sqrt{N_{Z'}}}
\approx \sqrt{\frac{s}{L}\frac{1}{c_{Z'}C}}
\exp\left\{\frac{AM_{Z'}}{2\sqrt{s}}\right\}.
\eq
Relation \req{fdef5} relies on the approximation \req{fapprox2},
which becomes inaccurate for too large $\sqrt{s}/M_{Z'}$.
The error of other observables also scales as \req{fdef5} with $s$ and $L$,
however, the prefactors differ.

Compared to the exclusion limit $M_{Z'}^{lim}$, the error of a model
measurement is much more model dependent because the influence of the
constant $c_{Z'}$ is no longer logarithmically suppressed.
The dependence on the integrated luminosity is the same as in equation
\req{epemmeas}. 
Therefore, the dependence of model measurements on systematic errors
in hadron collisions is as pronounced as in $e^+e^-$ collisions. 

Combining equations \req{fdef4} and \req{fdef5}, we can predict
$\Delta O$ for a given $M_{Z'}<M_{Z'}^{lim}$ if we know 
$M_{Z'}^{lim}$ from the observable $O$ alone for the same collider, 
\bq
\label{fdef6}
\Delta O\approx\left(N_{Z'}\right)^{-f_m/2} \cdot
\left(\frac{s}{L}\frac{1}{c_{Z'}C}\right)^{\frac{1}{2}(1-f_m)}
,\ \ \ 
f_m=M_{Z'}/M_{Z'}^{lim}.
\eq
It is the complement to the estimates \req{epemmea} and \req{epemmea2} derived
for $e^+e^-$ collisions, which relate exclusion limits and
measurements of the same confidence level. 
Relation \req{fdef6} relates exclusion limits from $N_{Z'}$ expected
$Z'$ events to one--$\sigma$ errors of model measurements.
To relate exclusion limits of 95\% confidence ($N_{Z'}=3$) to
measurements of 95\% confidence, equation \req{fdef6} modifies to
\bq
\label{fdef7}
\Delta O\approx 2\cdot3^{-f_m/2} \cdot
\left(\frac{s}{L}\frac{1}{c_{Z'}C}\right)^{\frac{1}{2}(1-f_m)}.
\eq
Both estimates \req{fdef7} and \req{epemmea2} are shown in figure~\req{zppf2}.
The estimate \req{fdef7} depends on collider parameters, while the estimate
\req{epemmea2} is universal for $e^+e^-$ collisions.
Note that both estimates do not work for too small $f_m$.
\subsubsection{Future measurements}\label{zppp1322}
The estimate \req{fdef5} can be confronted with results of the
theoretical analysis \cite{1063}, which assumes
$\sqrt{s}=14\,TeV L=100\,fb^{-1}$ and $M_{Z'}=1\,TeV$.
The measurement of $O=A_{FB}^e$ is investigated there.
Having in mind the crude approximations, which lead to the estimate
\req{fdef5}, the agreement is reasonable.

%
\begin{table}[tbh]
\begin{center}
\begin{tabular}{|lrrrrr|}\hline\rule[-2ex]{0ex}{5ex} 
Analysis &$\chi$&$\psi$ &$\eta$ &$LR$&$SSM$\\ 
\hline
$\Delta A_{FB}^e$ from \cite{1063}:& 0.007 & 0.016& 0.014& 0.006 & -\\
$\Delta O$ from \req{fdef5}:& 0.008 & 0.012& 0.011& 0.008 & 0.006\\
\hline
\end{tabular}\medskip
\end{center}
{\small\it  \begin{tab}\label{ppmeastab} 
Expected errors of measurements of $\Delta A_{FB}^e$ from reference
\cite{1063} and from estimate \req{fdef5}.
\end{tab}} \end{table}

As mentioned in section \ref{zppp112}, 
the polarization asymmetry of $\tau$'s in the final state depends on
the couplings of the $Z'$ to the $\tau$ only,
\bq
A_{pol}^\tau = v'_\tau a'_\tau/(v'^2_\tau +a'^2_\tau).
\eq
Therefore, it allows a model independent measurement of this
combination of coupling constants.
The structure functions and branching ratios of the $Z'$ influence
only the event rate and therefore the error of  $A_{pol}^\tau$.
For $E_6$ GUT's, it is estimated \cite{taupp} as 
$\Delta A_{pol}^\tau \approx 1.5/\sqrt{N_{Z'}}$. 

A measurement of the observables 
\bq
\label{ppobse}
r_{y1}, A_{FB}, A_{FBy1}, r_{l\nu W}, R_{Z'Z}, R_{Z'W}\mbox{\  and\ }
R_{Z'\gamma}
\eq
would give model independent information on the $Z'$ parameters
$\gamma_L^l,\ \gamma_L^q,\ \tilde{U}$ and $\tilde{D}$ defined in
equation \req{ppparm}.
For $M_{Z'}=1\,TeV$, the expected accuracy of such an
measurement at LHC is 5\% for $\gamma_L^l$ and between 20\% and
30\% for $\gamma_L^q, \tilde{U}$ and $\tilde{D}$ \cite{langackerpp}.
The estimate \req{fdef5} predicts errors of about 1\%.
The estimate is considerably smaller because it relies on
$A_{FB}^\mu$. 
The measurements of $\gamma_L^l,\ \gamma_L^q,\ \tilde{U}$ and
$\tilde{D}$ are based on all observables \req{ppobse}.
Unfortunately, the measurements of the observables involving
associated $Z'$ production or rare $Z'$ decays suffer from smaller
statistics.
See table~II of reference \cite{langackerpp} for details.
This explains the difference between the results in table \ref{cvetrevtab} and
the estimate.

%
\begin{table}[tbh]
\begin{center}
\begin{tabular}{|l|rrrr|}\hline\rule[-2ex]{0ex}{5ex} 
  &$\chi$&$\psi$ &$\eta$ &$LR$\\ 
\hline
$\gamma_L^l$& $0.9\pm 0.016$ & $0.5\pm 0.02$ & $0.2\pm 0.012$ & 
\rule[0ex]{0ex}{3ex} $0.36\pm 0.007$\\
$\gamma_L^q$& 0.1 & 0.5 & 0.8 & 0.04\\
$\tilde{U}$ & $1\pm 0.16$ & $1\pm 0.14$ & $1\pm 0.08$ & $37\pm 6.6$\\
$\tilde{D}$ & $9\pm 0.057$ & $1\pm 0.22$ & $0.25\pm 0.16$ & $65\pm 11$\\
\hline
\end{tabular}\medskip
\end{center}
{\small\it  \begin{tab}\label{cvetrevtab} 
Values of the parameters \req{ppparm} and its statistical error--bars
for typical models determined from probes at the LHC 
($\sqrt{s}=14\,TeV,\ L=100\,fb^{-1}$). $M_{Z'}=1\,TeV$.
This is table~3 of reference \cite{cvetrev}. 
\end{tab}} \end{table}
%
The measurements of $\gamma_L^l, \gamma_L^q, \tilde{U}$ and
$\tilde{D}$ can be used to get information on the symmetry breaking
sector \cite{prd52}.
Similar to a model measurement in $e^+e^-$ collisions described in
section \ref{zpeeff414}, a verification of the relations \req{e6check}
allows to check, whether the $Z'$ comes from the breaking of the $E_6$ or
$SO(10)$ groups.
In any case, the breaking parameters can be determined.
Under the assumptions of  reference \cite{prd52}, the statistical
errors of such a measurement are around 10\%.
The breaking parameters $\gamma_L^l, \gamma_L^q, \tilde{U}$ define the
$Z'$ couplings to SM fermions with a 16--fold sign ambiguity.
This ambiguity can be removed by measurements at $e^+e^-$ colliders.
Hadron colliders alone can reduce the sign ambiguity by collisions of
polarized beams or by a measurement of observables, which are
sensitive to the polarization in the final state as $A_{pol}^\tau$.

The measurement of $g_R/g_L$ in left--right symmetric models is
considered in reference \cite{langmix2}.
For $M_{Z'}=1\,TeV$, this ratio could be measured at the LHC with a
statistical error of about $1\%$.

If a $Z'$ signal is found in hadron collisions, $M_{Z'}$ and $\Gamma_{Z'}$
can be defined by a fit to the invariant
mass distribution of the final fermion pairs from the $Z'$ decay.
$M_{Z'}$ can be measured with an accuracy of
$\Gamma_{Z'}$ detecting only a few $Z'$ events.  
This is proven by the early measurements of the SM $W$-- and $Z$--masses
by the UA1 and UA2 experiments at CERN \cite{ua1ua2}.
For larger event numbers, the systematic errors
become important. 
See reference \cite{prd45} for details.
\chapter{$Z'$ search in other experiments}\label{otherexp}
%
There are other experiments not yet mentioned, which can give
bounds on extra neutral gauge bosons.
For completeness, we briefly comment on some of them in the next sections.
\section{$ep$ collisions}\label{zpep}
%
Neutral current electron--proton scattering occurs through photon, $Z$ or $Z'$
exchange in the $t$ channel.
The $Z'$ is detected by indirect effects similar to $e^+e^-$ collisions.
The additional contributions due to $Z'$ exchange lead to deviations
of observables from their SM predictions.
Compared to $e^+e^-$ collisions, $ep$ collisions suffer from the
hadronic background, in which these deviations must be detected.
$ep$ collisions are as insensitive to $ZZ'$ mixing as
off--resonance fermion pair production.
Therefore, we put the $ZZ'$ mixing angle $\theta_M$ to zero in this
section and identify $Z_1$ and $Z_2$ with $Z$ and $Z'$.

Some early $Z'$ analyses can be found in the references
\cite{godfreyold,plb176,plb184,prd352}. 
\subsection{Born cross section}\label{zpep11}
The {\bf amplitude} of $ep$ scattering depends only on the ratio of
the $Z'$ couplings and the $Z'$ mass.
The six couplings $a'_f,v'_f$ with $f=e,u,d$ are always involved
simultaneously because the $Z'$ must couple to the initial state.

%
The {\bf cross section} of the reaction
$e^-_{L,R}p\rightarrow e^-_{L,R}X$ including extra neutral gauge bosons is
\begin{equation}
\label{bornep}
\frac{d\sigma(e^-_{L,R})}{dxdy} 
= 2\pi \alpha^{2} \frac{s}{Q^{4}}\sum_{m,n}\sigma(m,n)
\end{equation}
with
\ba
\label{bornep2}
\sigma(m,n) &=& \chi_m(Q^{2}) \chi_n^*(Q^{2})
    \left[ [1+(1-y)]^2 F_2^{L,R}(x,Q^2) +  [1-(1-y)]^2 xF_3^{L,R}(x,Q^2) 
\right],\nll
F_2^{L,R}(x,Q^2) &=& x\left[ C_{V}(e) \pm C_{A}(e) \right] 
\sum_q\left[ C_V(q) (q(x,Q^2) + \bar q(x,Q^2)) \right],\nll
xF_3^{L,R}(x,Q^2) &=&\pm x\left[ C_{A}(e) \pm C_{V}(e) \right]
\sum_q\left[ C_A(q) (q(x,Q^2) - \bar q(x,Q^2)) \right],\nll
C_V(f) &=& v_f(m) v_f^*(n) + a_f(m) a_f^*(n),\ \ \ 
C_A(f) = v_f(m) a_f^*(n) + a_f(m) v_f^*(n).
\ea
The propagator $\chi_n(Q^2)$ is given in equation \req{eq28} and the
couplings $v_f(n)$ and $a_f(n)$ are defined in equation \req{gencoup}
and \req{vf1}. 
$q(x,Q^2)$ and $\bar q(x,Q^2)$ are structure functions of the proton.
The cross section for $e^+_{L,R}$ scattering is given by equation
\req{bornep} with the replacements $F_2^{L,R}\rightarrow F_2^{R,L},\ 
F_3^{L,R}\rightarrow -F_3^{R,L}$ in equation \req{bornep2}. 
The kinematic variables Bjorken--$x$ and $y$ are defined as
\bq
\label{herakin}
x\equiv\frac{Q^2}{2P\cdot q},\ \ \ 
y \equiv\frac{P\cdot q}{P\cdot p_e}=\frac{Q^2}{xs}
\mbox{\ \ with\ \ }Q^2=-q^2=-(p_e-p_l)^2,\mbox{\ \ and\ \ }s\equiv (p_e+P)^2.
\eq
$p_e(p_l,P)$ are the energy--momenta of the incoming electron
(scattered electron, proton).
$s$ is the center--of--mass energy squared and $Q^2$ is the momentum
transfer squared.
We treat all initial and final particles as massless.

%
Among the different {\bf observables}, the 
total cross sections are rather insensitive to a $Z'$
because the contributions from photon exchange are very large.
Charge ($A_{LL}^{-+},\ A_{RR}^{-+}$), 
polarization ($A_{LR}^{++},\ A_{LR}^{--}$) 
and mixed ($A_{LR}^{-+},\ A_{RL}^{-+}$)
asymmetries can be defined \cite{plb184,prd352}, 
\begin{equation}
\label{epobs}
A_{XY}^{mn} = \frac { d\sigma(e_X^m) - d\sigma(e_Y^n) }
                     { d\sigma(e_X^m)  + d\sigma(e_Y^n) },\ \ \
X,Y=L,R;\ m,n=+,-.
\end{equation}
Most of the systematic errors drop out in these asymmetries.
\subsection{Radiative Corrections}\label{zpep12}
Presently there are no hints for a $Z'$ at HERA.
We expect that possible new $Z'$ contributions to the cross section
are very small. 
It is therefore sufficient to take into account only the {\bf QED
corrections} in the leading log approximation
(LLA) \cite{bluleiri} to these contributions.
For the other contributions, one has to take into account the SM corrections. 
See reference \cite{fortsch,hector} for an overview of SM
corrections to $ep$ collisions and for further references.
%
QED corrections can be taken into account in a model independent
way in the LLA. 
They consist of initial and final state radiation \cite{qedinifin} and
the Compton peak \cite{compton}.

The full $O(\alpha)$ QED and {\bf weak corrections} can be found in references
\cite{npb276,heraqed}. 
See also section \ref{zpeeff22} for further references to weak corrections.
%
They can be taken into account
\cite{bluleiri} by form factors \cite{formbard} as described in
section \ref{zpeeff22}. 
As discussed in reference \cite{bluleiri}, the electroweak corrections are
of the same size as the $Z'$ effects.
Therefore, they must be taken into account in a $Z'$ analysis at
$ep$ colliders.
The $M_t$-- and $M_H$--dependence of weak corrections in presence of $Z'$
production is discussed in reference \cite{mad592}.

%
The {\bf QCD corrections} to $Z'$ production are the same as in the SM.
See references \cite{heraqcd} for a short review and for further references.
\subsection{$Z'$ constraints}\label{zpep13}
%
A {\bf model independent} $Z'$ analysis in $ep$ collisions always
involves the six $Z'$ couplings $a'_f,v'_f,\ f=e,u,d$.
If the condition \req{chargedef} is assumed, the model independent analysis can
constrain the five combinations \req{coup5ee}.
$ep$ collisions are sensitive to the relative sign of the $Z'$
couplings.
A model independent analysis for HERA is carried out in reference
\cite{herami}. 
Unfortunately, HERA cannot compete with hadron colliders \cite{herazp,9704380}
in setting $Z'$ bounds for usual GUT's.

%
%
{\bf Model dependent} $Z'$ limits at electron proton colliders can be
obtained by considerations similar to those explained for $e^+e^-$ collisions.
We get the following scaling of $M_{Z'}^{lim}$ with $L$ and $s$,
\bq
\label{zphera}
M_{Z'}^{lim}
\approx\sqrt{Q^2}\left(1+\frac{1}{\Delta o}\right)^{1/2}
\sim\left[\frac{sL}{1+r^2}\right]^{1/4}.
\eq
The scaling is the same as in $e^+e^-$ collisions. 
Therefore, the dependence on systematic errors is the same.
The difference is that the error $\Delta o$ depends on the kinematic
variable $Q^2$.
Reasonable statistics are obtained for $Q^2$ well below $s$.
As a result, $ep$ collisions are sensitive to extra neural gauge
bosons with masses comparable to the center--of--mass energy $\sqrt{s}$.
The model dependence of these bounds is large as in $e^+e^-$
collisions.

%
See reference \cite{herami2} for a first analysis of {\bf future exclusion
limits} including radiative corrections. 
A recent analyses can be found in reference \cite{heralim}.
A comparison of the numbers in table~\ref{zplimhera} with those of
table~\ref{zppptab} demonstrates that HERA cannot compete in setting mass
limits to a $Z'$ predicted in typical extended gauge theories.
The scaling \req{zphera} describes nicely the changes with the luminosity.

%
\begin{table}[tbh]
\begin{center}
\begin{tabular}{|lrrrr|}\hline
$M_{Z'}/GeV$ &$\chi$ &$\psi$ &$\eta$ &$LR$ \rule[-2ex]{0ex}{5ex}\\ 
\hline
$L=0.5\,fb^{-1}$ & 390 & 210 & 240 & 420 \\
$L=1.0\,fb^{-1}$ & 470 & 260 & 290 & 500 \\
ratio            &1.21 &1.24 &1.21 &1.19 \\
\hline
\end{tabular}\medskip
\end{center}
{\small\it  \begin{tab}\label{zplimhera} The 95\% CL predictions for
$M_{Z'}^{lim}$ from HERA with $\sqrt{s}=314\,GeV$ and the integrated
luminosities quoted in the table.
The first two rows are taken from table~3 of reference \cite{heralim}.
\end{tab}} \end{table}
%

%
Recently, the anomalous high $Q^2$ events observed at HERA
\cite{heraq2} received much attention.
As pointed out in reference \cite{9704380}, these deviations from the
SM cannot be explained by a $Z'$ coming from typical extended gauge theories,
which is compatible with the present LEP and Tevatron data.
Two models, the excited weak boson model \cite{weakboson} and the BESS model
\cite{bess} are not ruled out by the present data.

\section{Atomic parity violation}\label{atomic}
The measurement of parity--nonconserving transitions in atoms has
reached a precision \cite{science,prl74b}, which allows constraints to
extra neutral gauge bosons competitive
\cite{luo,altmix2,prd43,plb256} to those from collider
experiments.  

Parity violating transitions occur due to the exchange of vector
bosons with axial couplings.
The experimental results are usually given in terms of the weak charge,
\bq
\label{qwdef}
Q_W=-2C_{1u}(2Z+N) -2C_{1d} (Z+2N),\ \ \ C_{1q}=2a_e(1)v_q(1),\ q=u,d.
\eq
It arises due to the coherent interaction of the electron with all
$Z(N)$ protons (neutrons) in the nucleus of the considered atom.

To calculate the SM prediction of $Q_W$, radiative corrections
\cite{rcatom} must be applied,
\bq
C_{1u}=-\rho'_{PV}\left(\frac{1}{2}-\frac{4}{3}s_W^2(0)_{eff}\right),\ \ \ 
C_{1d}=-\rho'_{PV}\left(-\frac{1}{2}+\frac{2}{3}s_W^2(0)_{eff}\right).
\eq
For $\nobody^{133}_{\ 55}Cs$, we have $Q_W=-376C_{1u}-422C_{1d}$.
The resulting SM prediction \cite{9705236} for $M_t=175\,GeV$ and
$m_H=100\,GeV$ is $Q_W^{SM}=-73.04$.

This can be compared with the present experimental value \cite{science},
\bq
Q_W=-72.11(27)_{exp}(89)_{theory}. 
\eq
The agreement with the SM prediction is used to constrain possible new physics.

The exchange of extra neutral gauge bosons would give additional
contributions to parity violating transitions in atoms.
For $\nobody^{133}_{\ 55}Cs$, the predicted change in the weak charge
is \cite{luo} 
\bq
\label{atomzp}
\Delta Q_W=\left(\frac{M_Z^2}{M_1^2}-1\right)\left(Q_W^{SM}+73.8\right)
-\theta_M\frac{g_2}{g_1}\left(Q_W^{(2)}+2a_e(2)Q_W^{SM}\right)
+2\frac{M_1^2}{M_2^2}\frac{g_2^2}{g_1^2}a_e(2)Q_W^{(2)}
\eq
with $Q_W^{(2)}=-376v_u(2)-422v_d(2)$.
The first contribution is numerically negligible \cite{luo} due to an
accidental cancellation between the two contributions and due to the
present experimental constraint on $M_Z^2/M_1^2-1=\rho_{mix}-1<0.003$
already mentioned in section \ref{amvmconstr}.
The constraint \req{atomzp} can therefore be parametrized as \cite{luo}
\bq
\Delta Q_W\approx \gamma_1\theta_M+\gamma_2\frac{M_1^2}{M_2^2}.
\eq
The coefficients $\gamma_1$ and $\gamma_2$ are given for some
models in table~\ref{atomcoeff}.

The present constraints on the $ZZ'$ mixing angle $\theta_M$ from
measurements at the $Z_1$ peak are stronger then those, which would
result from atomic parity violation.
As a result, measurements of atomic parity violation constrain mainly
$M_1/M_2$. 

\begin{table}[tbh]
\begin{center}
\begin{tabular}{|lrrrr|}\hline
 &$\chi$ &$\psi$ &$\eta$ &$LR$ \rule[-2ex]{0ex}{5ex}\\ 
\hline
$\gamma_1$ & -138.0& 37.2 &  -114.0&-46.9 \\
$\gamma_2$ &   65.6&  0.0 &   -16.4& 74.7 \\
\hline
\end{tabular}\medskip
\end{center}
{\small\it  \begin{tab}\label{atomcoeff} 
Values of the parameters $\gamma_1$ and $\gamma_2$  computed for
$s_W^2=0.2334$ for different $Z'$ models.
The numbers are taken from table IX of reference \cite{luo}.
\end{tab}} \end{table}

Note however, that the weak charge $Q_W$ can receive compensating
contributions from more than one new physics source, which would relax
the $Z'$ limits. 
Such cancellation effects are explicitly demonstrated in reference
\cite{9710353}. 
\section{Neutrino scattering}
{\bf High energy neutrino scattering} experiments provided interesting
limits on $Z'$ parameters in the past.
Today, they cannot compete with the limits from collider experiments.
See references \cite{e6,e6add} for a review and references to older
experiments. 
Recent results of neutrino scattering experiments can be found in
references \cite{nuscat,plb332}, a recent review is given in reference
\cite{9707015}. 

Neutrino--electron and neutrino--nucleon scattering experiments measure the
couplings of the neutrino to the $Z$ boson \cite{9707015},
\bq
v_\nu=-0.035\pm0.012\pm 0.012,\ \ \ a_\nu=-0.503\pm 0.006\pm 0.016.
\eq
These measurements are complementary to experiments at the $Z_1$  peak
because they measure the couplings at much lower center--of--mass energies.
The weak-- and QED corrections to neutrino scattering are given in
the first reference of \cite{qedinifin} and in references \cite{prd22b}.

As $e^+e^-$ collisions at the $Z_1$ peak, the agreement of these
measurements with the SM prediction constrains physics beyond the SM.
The constraints on extra neutral gauge bosons obtained by the CHARM II
Collaboration \cite{plb332} are given in table~\ref{nutab}.
Comparisons of the excluded ranges in the $\theta_M-M_{Z'}$ plane with
L3 measurements can be found in figure~46 of reference \cite{9707015}.

The CHARM, CCFR and CDHS collaborations quote model independent results on
neu\-tri\-no--nucleon scattering \cite{9707015}.
These results can be converted into constraints on different extra neutral
gauge bosons using the formalism of reference \cite{e6add}. 
The resulting numbers for an unconstrained Higgs sector are given in
table~\ref{nutab}. 

\begin{table}[tbh]
\begin{center}
\begin{tabular}{|lrrrr|}\hline
 &$\chi$ &$\psi$ &$\eta$ &$LR$ \rule[-2ex]{0ex}{5ex}\\ 
\hline
reference \cite{plb332}              & 262 & 135 & 100 & 253 \\
table~XI of reference \cite{9707015} & 215 &  54 &  87 &  \\
figures~2 of reference \cite{9712215}& 500 & 155 & 190 & 220 \\
\hline
\end{tabular}\medskip
\end{center}
{\small\it  \begin{tab}\label{nutab} 
Present 95\% CL limits on $M_{Z'}$ in GeV in different $Z'$ models
from neutrino scattering experiments.
\end{tab}} \end{table}

Recently, {\bf low energy neutrino scattering} experiments
($\nu_ee\rightarrow \nu_ee$) are proposed \cite{9712215}.
The proposal foresees to place a strong neutrino source in the center
of a neutrino detector.
Such a neutrino source with an activity of $1.67\pm 0.03\,MCi$ based
on $\nobody^{51}$Cr was already used to calibrate the GALLEX neutrino
experiment. 
A $\bar\nu_e$ source based on $\nobody^{147}$Pm is proposed to have an
activity of $5-15\,MCi$ \cite{9712215}. 
All these sources emit neutrinos with energies well below $1\,MeV$.
New detectors are proposed to measure the small recoil energy of the
scattered electrons with high precision \cite{9712215}.

The experiment measures the neutrino couplings.
This information can be used to set limits on extra neutral gauge bosons. 
The possible constraints on the $ZZ'$ mixing angle cannot compete with
the present LEP measurements, while the possible bounds on $M_{Z'}$
are interesting.
They are shown in figures~1 and 2 of reference \cite{9712215}.
We produce mass limits from figure~2 of that reference neglecting
systematic errors and present them in table~\ref{nutab}.
\section{Cosmology}
The number of light neutrinos interacting with the SM $Z$ boson is
known from experiments at the $Z_1$--resonance to be
$N_\nu=2.989\pm0.012$ \cite{warschau}. 

GUT's containing extra neutral gauge bosons also predict the existence
of additional (right handed) neutrinos.
The number of these neutrinos, which do not interact with the SM $Z$ boson,
is not constrained by the experiments at LEP and SLAC.

The big bang nucleosynthesis of neutrons and the related {\bf abundance of
$\nobody^4$He} in the universe is sensitive to any particles with a mass
lighter or about $1\,MeV$ \cite{schramm}, the mass difference between
the proton and the neutron.
Assuming a primeval $\nobody^4$He abundance $Y_P=0.242\pm0.003$, one
gets \cite{schramm} a 95\% CL interval $N_\nu=3.0 - 3.7$ assuming the
D+$\nobody^3$He lower bound to the baryon density, and  $N_\nu=3.0 -
3.2$ assuming  (D/H)$_P=(2.5\pm0.75)\cdot 10^{-5}$ \cite{prd55b}.

This measurement of $N_\nu$ can be interpreted as a constraint on
theories predicting light neutrinos and light $Z's$ \cite{plb240}.
The resulting bounds on $M_{Z'}$ are stronger than those from collider
experiments but they contain more assumptions on the model.

Constraints on extra neutral gauge bosons from the {\bf supernova SN 1987A}
are considered in references \cite{sn1987a,sn1987b,sn1987c,sn1987d}.
Models of stellar collapse predict an energy release of $4\cdot 10^{46}W$. 
The measured neutrino events suggest an energy release exceeding
$2\cdot 10^{46}W$ within 10 seconds.  
Therefore, at most $2\cdot 10^{46}W$ could be emitted by other particles.
In particular, additional neutrinos lighter than about $50\,MeV$, the
core temperature, would carry away a part of the energy. 
They can be produced through nucleon-nucleon bremsstrahlung
($NN\rightarrow NN\nu_R\bar\nu_R$) in presence of a light $Z'$.

The agreement between the models of stellar collapse and the neutrino
observation puts constraints on extra neutral gauge bosons if there are
additional light neutrinos present in the model.
These constraints are considerably stronger than present collider
limits except for models where the $Z'$ coupling to right--handed
neutrinos vanishes \cite{sn1987a}.
Of course, these limits are based on more model assumptions than the
collider constraints.
A non-zero $ZZ'$ mixing would only strengthen the limits \cite{sn1987b}.
The influence of radiative corrections on the limits is small \cite{sn1987b}.
\chapter{Summary and conclusions} \label{concl}
%
In this review, we have investigated the phenomenology of extra
neutral gauge bosons. 
We have considered in detail the $Z'$ constraints, which can be
obtained at $e^+e^-,\ e^-e^-$ and $\mu^+\mu^-$ colliders.
At these machines, fermion pair production, Bhabha scattering,
M{\o}ller scattering and $W$ pair production can contribute to a
$Z'$ analysis.
The constraints from lepton colliders are compared with those from
$pp$ and $p\bar p$ colliders.

In the case of the absence of a $Z'$ signal, lepton and hadron colliders
give complementary $Z'$ constraints. 
Lepton colliders give the best constraints on the $ZZ'$ mixing angle
and on weakly interacting $Z'$s.
The exclusion limits from lepton colliders are almost insensitive to
$\Gamma_2/M_2$ but they are rather model dependent. 
Hadron colliders give the best present constraints on the $Z'$ mass
for $Z'$s predicted in popular GUT's.
These constraints are rather insensitive to the $Z'$ model.
They become worse for an enlarged $Z_2$ width, which can arise if
decays to exotic fermions are kinematically allowed.
The complementary role of lepton and hadron colliders is demonstrated
in table~\ref{tabconcl}.

\begin{center}
\begin{tabular}{|lrrrrr|}\hline
 &$\chi$ &$\psi$ &$\eta$ &$LR$ &$SSM$ \rule[-2ex]{0ex}{5ex}\\ 
\hline
Tevatron 1997 \cite{prl79} & 595 & 590 & 620 & 630 & 690\\
LEP 1997 \cite{sr1997} & 300 & 220 & 230 & 310 & 520\\
end of LEP ($190\,GeV,\ 0.5\,fb^{-1}$) \cite{lmu0296}&990&560&620&1100& 1500\\
Tevatron after run II ($2\,TeV,\ 2\,fb^{-1}$) &940&930&970&970&1040\\
\hline
\end{tabular}\medskip
\end{center}
{\small\it  \begin{tab}\label{tabconcl} 
Present and future limits on $M_{Z'}$ (95\% CL) for different $E_6$
models and the SSM in GeV. 
The last line is obtained from the present Tevatron bounds using relation
\req{ppzpscale}.
\end{tab}} 

In the case of a $Z'$ signal, lepton and hadron colliders are
complementary in a $Z'$ model measurement. 
The proposed hadron colliders with unpolarized beams measure the
couplings of the $Z'$ to SM fermions with smaller errors than the
proposed lepton colliders but with a 16--fold sign ambiguity.
Lepton colliders with polarized electron beams measure the same
couplings with a 2--fold sign ambiguity only.

The $Z'$ limits from electron--proton colliders cannot compete with
those from lepton and hadron colliders.

We emphasized the importance of model independent and model dependent $Z'$
analyses.
Both analyses are complementary.
Model independent constraints are useful to restrict any present and
future $Z'$ model.
For this universality, one has to pay the price that not always all
observables are useful for a model independent analysis and that model
independent constraints from different reactions are not always comparable.
On the other hand, additional model assumptions bias the limits.
However, as far as they are consistent with the data, they help to tighten
the exclusion limit or to reduce the error of the measurement of the
remaining model parameters. 

We reviewed the status of the radiative corrections,
discussed the importance of the different radiative corrections in
detail, and described how they can be included in theories including a $Z'$.
QED corrections can be calculated in a model independent way.
QCD corrections to $Z'$ processes are the same as in the SM.
Weak corrections to the new $Z'$ contributions cannot be calculated
independently of the model.
We assume that the $Z'$ effects arise first at the tree level and not
in loops.  
Then, the higher--order corrections including new GUT particles are a
small correction to a small effect and can be neglected.
However, they cannot be neglected in precision measurements at the $Z_2$ peak. 

Computer programs with these corrections are required for $Z'$ analyses.
We listed officially released {\tt FORTRAN} programs relevant for a $Z'$
search and indicated where these programs have already been used in an
analysis of experimental data. 

In contrast to $Z'$ model measurements, $Z'$ exclusion limits are
rather robust against details of systematic errors.

For different reactions, we discussed kinematic cuts, which enhance
the sensitivity to extra neutral gauge bosons.

We now comment on the $Z'$ limits from different processes in more detail.
They are collected in table~\ref{zpconcl}, which summarizes the main
results of the different sections of this review in a telegraphic style.
 
The first column refers to the considered reaction. 
If necessary, different cases of the center--of--mass energy are
distinguished. 
In the next three columns, the status of the $Z'$ search is indicated.
The simplest analysis could be done at the Born level.
The next step would be the investigation of radiative corrections
needed to meet the accuracy of future data. 
We put a $+$ there if the radiative corrections are known with an
accuracy comparable or better than the expected experimental errors.
This column is more subjective because neither the future experimental
errors nor the magnitude of the radiative corrections are precisely
known in advance.  
The existence of officially released computer programs containing all
radiative corrections needed for a direct fit to data is indicated in
the fourth column. 

The last column of the table contains typical bounds on different
model parameters or combinations of them and a scaling of these bounds
with the integrated luminosity and the center--of--mass energy.  
Of course, the input of this column is only representative depending on the
assumptions and limitations not given in the table but described in
the corresponding sections of this review. 
The scaling with the luminosity assumes that the systematic error
decreases proportional to the statistical error.

It follows a short comment on every row:
$Z'$ effects in fermion pair production at the $Z_1$ peak arise mainly
through deviations in the couplings of the mass eigenstate $Z_1$ to
fermions compared to the SM prediction for the $Z$ boson.
On--resonance $Z_1$ production gives the best present limits on the
$ZZ'$ mixing angle $\theta_M$.  
The number in the table is a typical experimental bound for GUT's. 
It is almost independent of the $Z'$ mass.
Without assumptions on the $Z'$ model, only $v_f^M$ and $a_f^M$, which are
the product of the $ZZ'$ mixing angle and the $Z'$
couplings, can be constrained.
The present constraints on $v_f^M$ and  $a_f^M$ with $f=u,d$ could only
be improved by fermion pair production at the $Z_2$ resonance.
The strongest improvements on the present limits on  $v_e^M$ and
$a_e^M$ will come from $W$ pair production at $e^+e^-$ or $\mu^+\mu^-$
colliders at $TeV$ energies. 

\begin{center}
\begin{tabular}{|l|cccll|}\hline &&&&& \\[0.5ex] 
Reaction, &{\footnotesize Born}     &{\footnotesize Main}
&{\footnotesize Program}   &
\multicolumn{2}{l|}{Typical 95\% CL constraint} \\ 
c.m. energy &{\footnotesize Analysis} &{\footnotesize RC's}
&{\footnotesize with RC's} &
\multicolumn{2}{l|}{and scaling with c.m. energy $s$ }\\ 
         &{\footnotesize exists}   &{\footnotesize known}
&{\footnotesize exists}    & 
\multicolumn{2}{l|}{and integrated luminosity $L$} \\ 
\hline\hline\rule[-1.3ex]{0ex}{4ex}
%
$e^+e^-\rightarrow f\bar f$&&&&&\\ \cline{1-1} \rule[-1ex]{0ex}{3.5ex} 
\ \hfill $s\approx M_1^2$ & + & + & + & 
$|a_e^M|<0.0005,|v_e^M|<0.001$ & $\sim L^{-1/2}$
\\ \rule[-1ex]{0ex}{3.5ex}
&&&&
$|a_q^M|,|v_q^M|<0.02,\ q=c,b$ & $\sim L^{-1/2}$
\\ \rule[-1ex]{0ex}{3.5ex}
&&&&
$|\theta_M|< 0.003$ & $\sim L^{-1/2}$
\\ \rule[-1ex]{0ex}{3.5ex}
\ \hfill $M_1^2<s< M_2^2$ & + & + & + & 
$|a_l^N|,|v_l^N|<0.01 $ & $\sim (sL)^{-1/4}$\\ \rule[-1ex]{0ex}{3.5ex}
&&&&
$M_2>M_{Z'}^{lim}= (3$ to $8)\sqrt{s}$ & $\sim (sL)^{1/4}$\\
\rule[-1ex]{0ex}{3.5ex} 
&&&& $g_2<g_1/7$ for $M_2<\sqrt{2s}$ & $\sim (sL)^{-1/4}$\\ 
\rule[-1ex]{0ex}{5ex}
&&&&
$\frac{\Delta g_2}{g_2},\frac{\Delta M_2}{M_2}
\approx \frac{1}{2}\left(M_2/M_{Z'}^{lim}\right)^2$ &
$\sim (sL)^{-1/2}$
\\ \rule[-1ex]{0ex}{3.5ex}
\ \hfill $s\approx M_2^2$ & + & $-$ & $-$ & 
$g_2<g_1/140$ & $\sim L^{-1/4}$
\\ \rule[-1ex]{0ex}{3.5ex}
&&&&
$\Delta\Gamma_2,\Delta M_2\approx \Delta E_{beam}$&\\ \rule[-1.5ex]{0ex}{4ex}
&&&&
$\Delta a_f(2)/a_f(2),\Delta v_f(2)/v_f(2)$ & $\sim L^{-1/2}$\\
\rule[-1.5ex]{0ex}{4ex} 
\ \hfill $s>M_2^2$ & $+$ & $-$ & $-$ & 
\multicolumn{2}{l|}{ $g_2<g_1/24$\ \ \mbox{for}\ $M_{Z'}\approx\sqrt{s}\hfill 
\sim 
\left(\frac{\Delta E_\gamma}{E_\gamma}\frac{s}{M_{Z'}^2}\right)^{1/4}$}\\  
%
\hline\hline\rule[-1.3ex]{0ex}{4ex}
$e^\pm e^-\rightarrow e^\pm e^-$ & + & $+$ & $-$ &
\multicolumn{2}{l|}{see $e^+e^-\rightarrow f\bar f,\ M_1^2<s<M_2^2$ }\\
%
\hline\hline\rule[-1.3ex]{0ex}{4ex}
$e^+e^-\rightarrow W^+W^-$ &&&&&\\ \cline{1-1} \rule[-1ex]{0ex}{3.5ex}
\ \hfill $s< M_2^2$ & + & + & $+$ &
$|a_e^M|,|v_e^M|<0.00025\,TeV/\sqrt{s}$ &$\sim (sL)^{-1/2}$
\\ \rule[-1.5ex]{0ex}{4ex}
&&&&
$|\theta_M|< 0.001\,TeV/\sqrt{s}$ & $\sim (sL)^{-1/2}$
\\ \rule[-1.5ex]{0ex}{4ex}
\ \hfill $s\approx M_2^2$ & + & $-$ & $-$ &
$|\theta_M|<0.0001\,TeV/\sqrt{s}$ & $\sim L^{-1/2}$
\\ 
%
\hline\hline\rule[-1.3ex]{0ex}{4ex}
$pp,p\bar p\rightarrow Z'X$ &&&&&\\ \cline{1-1} \rule[-1ex]{0ex}{3.5ex}
\ \hfill $Z'\rightarrow l\bar l,b\bar b$ & + & + & + &
$\sigma_T^\mu<3/L$ & $\sim L^{-1}$\\ \rule[-1ex]{0ex}{3.5ex} 
&&&&
\multicolumn{2}{l|}{ 
$M_2>M_{Z'}^{lim}\approx
\sqrt{s}\left(0.4+\frac{1}{32}\ln\frac{L\cdot fb\cdot
1000c_{Z'}}{3s/TeV^2}\right)\ (pp)$}\\ 
\rule[-1.5ex]{0ex}{4ex}
&&&&
\multicolumn{2}{l|}{ 
$M_2>M_{Z'}^{lim}\approx
\sqrt{s}\left(0.6+\frac{1}{20}\ln\frac{L\cdot fb\cdot
1000c_{Z'}}{3s/TeV^2}\right) \ (p\bar p)$}\\ 
%
\hline\hline\rule[-1.3ex]{0ex}{4ex}
$pe^\pm,\bar pe^\pm\rightarrow e^\pm X$ & + & + & + 
& $M_2>(0.7$ to $1.6)\sqrt{s}$
&$\sim (sL)^{-1/4}$\\ 
\hline
\end{tabular}\medskip
\end{center}
{\small\it  \begin{tab}\label{zpconcl}
Summary table of the $Z'$ limits.
\end{tab}}

The $Z'$ effects in off--resonance fermion pair production arise
mainly through interferences of the $Z'$ amplitude with the SM amplitudes. 
Off--resonance fermion pair production is sensitive to $Z'$ masses
considerably larger than the center--of--mass energy. 
The limits have a strong dependence on the $Z'$ couplings to fermions.
They are insensitive to $ZZ'$ mixing.
The exclusion limits in the table are given for typical GUT's and for
colliders with $L=80\,fb^{-1}s/TeV^2$. 
If no information on the $Z'$ model is available, one can only
constrain the parameters $v_f^N$ and $a_f^N$, which are proportional to
ratios of the $Z'$ couplings and the $Z'$ mass.
If the $Z'$ couplings to all SM fermions are very small,
the $Z'$ could escape detection even for energies not far below
its mass.
However, it is hard to obtain such $Z'$s in a GUT. 
The errors of $Z'$ {\it model measurement} and of $Z'$ {\it exclusion} limits 
scale differently with the integrated luminosity and center--of--mass energy.

Experiments on top of the $Z_2$ resonance would certainly allow the most
accurate measurements of the $Z_2$ mass, of the $Z_2$ width and of the
$Z_2$ couplings to SM fermions.
In such measurements, muon colliders are clearly favored against
electron positron colliders because they have a much smaller beam
energy spread.
The accuracy of the measurements of the $Z_2$ couplings
to fermions is expected to be comparable to the precision presently
achieved at the $Z_1$ resonance.

As mentioned before, a weakly coupled $Z'$ can be missed in
experiments below its resonance.
If its couplings are not too small, such a $Z'$ can be observed in
experiments above its resonance. 
In those experiments, the $Z'$ signal arises through the hard
photons, which come from the radiative return to the $Z_2$ resonance.
These photons appear by the same mechanism, which is responsible for
the hard photons from the radiative return to the $Z_1$ resonance at
LEP\,2 energies.
The $Z'$ limit from experiments above the $Z_2$ resonance is sensitive
to the photon energy resolution. 

Bhabha and M{\o}ller scattering set bounds on the model independent
parameters $v_e^N$ and $a_e^N$, which are comparable to off--resonance
fermion pair production.  
Of course, Bhabha and M{\o}ller scattering can only constrain the $Z'$
couplings to electrons.
Model assumptions link $Z'$ couplings to leptons and quarks.
In model dependent analyses, fermion pair production profits from its
additional observables with quarks in the final state.
Therefore, the model dependent $Z'$ limits from fermion pair
production are better than those from Bhabha and M{\o}ller scattering.

$W$ pair production is very sensitive to changes of the $Z_1$
couplings to fermions. 
Such changes destroy the gauge cancellation present in the SM.
The result are large factors, which amplify the $Z'$ effects.
$W$ pair production can give the best model independent constraints on
the parameters $v_e^M,a_e^M$.
For models where the $Z'$ couplings to electrons are not zero, the
resulting bounds on $\theta_M$ are better than those from fermion pair
production. 
The best limits can be obtained in measurements
near the $Z_2$ peak where the $Z_2\gamma$ and $Z_2Z_1$ interferences dominate.
At energies above the $Z_2$ resonance, unitarity is restored and $Z'$
effects are no longer enhanced by large factors.

Proton colliders can see a $Z'$ if it is directly produced by a
quark--anti--quark pair.
Therefore, these colliders can detect only $Z'$s with masses considerably
smaller than the center--of--mass energy of the colliding protons.
The $Z'$ mass exclusion limits are rather insensitive to details of the
model as far as the signal can be separated from the background.
This becomes harder for $Z'$s with small couplings 
or for $Z'$s with small branching ratios to SM fermions.
The $Z'$  exclusion limits scale non-symmetrically with the
center--of--mass energy and with the integrated luminosity.
To improve the limits, an increase of the center--of--mass
energy is favored against an increase of the luminosity.
For a first $Z'$ discovery, $Z'$ decays to muon pairs are the
favored process.
In the case of a $Z'$ signal, there are many other useful observables,
which can help to measure the model parameters.
The errors of $Z'$ model measurements at hadron colliders scale with
the integrated luminosity as in $e^+e^-$ collisions but have an
enhanced sensitivity to the center--of--mass energy.

Some other experiments can provide $Z'$ limits.
We could only briefly comment on $ep$ collisions,
atomic parity violation, neutrino scattering and cosmology.

Many different bounds on extra neutral gauge bosons can be obtained
from various experiments.
In the foreseeable future, we  shall learn from the new experiments
whether the $Z$ boson has one or several massive partners as predicted
by most unified theories.
Let's hope for surprises.
\hfill\vspace{1cm} 

\begin{flushleft}{\Large\bf Acknowledgments}\end{flushleft}\vspace{0.5cm}

It is a pleasure to thank Tord Riemann for many years of fruitful
collaboration. With him, I started to work on extra neutral gauge bosons.
He encouraged me to write this review.
I had innumerable discussions with him.
Many of his ideas entered this review.  
I'm happy to thank Sabine Riemann for several years of pleasant
collaboration, during which I learned a lot of details on experiments.
I'm grateful to Wolfgang Hollik, who always had time to discuss with me
questions on extra neutral gauge bosons and on radiative corrections.
Further, I would like to thank Francesco del Aguila and Claudio
Verzegnassi for many discussions and continuous encouragement.
I'm grateful to G. Altarelli, K. Ellis, S. Godfrey, P. Langacker,
P. Minkovski, T.G. Rizzo and P. Zerwas, for discussions of parts of
this paper, valuable hints and warm hospitality and 
F. Berends, M. Cveti\v{c}, W.T. Giele, N. Lockeyer, K. Maeshima,
A.A. Pankov, M. Peskin, M. Zra\l ek for interesting discussions. 
I thank S. Riemann for providing the figures~\ref{stiefel} and
\ref{leptexcl} and  A.A. Pankov for providing figure~\ref{wwfig3}. 
I benefited greatly from H. Fritzsch and R. R\"uckl due to their continuous
support, many discussions and due to the stimulating working conditions at
their institute.
I thank S. Godfrey for the careful reading of the manuscript and for
his many useful comments.  

Finally, I would like to thank my wife Ines for her patience
while this paper was written and for hints concerning the manuscript.

This work was partially supported by 
the German Federal Ministry of Research and Technology under contract 
No. 05 GMU93P, 
the Deutsche Forschungsgemeinschaft, 
and the EC contracts CHRX-CT-92-0004 and CHRX-CT940579.
\vspace{1cm}\\
%

%
%
\begin{appendix}
%

\chapter{Notation}
For the convenience of the reader, we collect the main notation in
the following tables.
The notation appearing in chapter one has mainly to do with the
definition of the $Z'$ parameters.
It follows the notation of kinematic parameters and observables
introduced in chapter two.
We conclude with the notation relevant in the remaining chapters.
%
\begin{center}
\begin{tabular}{|l|l|}\hline 
Symbol\rule[-2ex]{0ex}{5ex} &Meaning\hspace{3cm} Chapter 1\\ 
\hline\rule[-1ex]{0ex}{4.5ex}
$Z'$ & Vector particle associated with the extra $U'(1)$ group in 
\req{gutgauge}. 
\\ \rule[-1ex]{0ex}{3.5ex} 
$\chi,\psi,\eta,LR,SSM$ & Particular $Z'$ models, see sections
\ref{guts3} and \ref{veccoup}.
\\ \rule[-1ex]{0ex}{3.5ex} 
$\theta_M$ & Mixing angle between the symmetry eigenstates $Z$ and $Z'$.
\\  \rule[-1ex]{0ex}{3.5ex} 
$Z_1,Z_2$ & Mass eigenstates resulting from the mixing of the $Z$ and $Z'$.
\\  \rule[-1ex]{0ex}{3.5ex} 
$\Gamma_n^f,\Gamma_n^W,\Gamma_n^{ffV}$ & Partial decay widths of the
$Z_n$ as defined in section \ref{zndecaywidth}.
\\  \rule[-1ex]{0ex}{3.5ex} 
$M_{Z'},\Gamma_{Z'}$ & Mass and width of the $Z'$.
\\  \rule[-1ex]{0ex}{3.5ex} 
$M_n,\Gamma_n$ & Masses and total widths of the $Z_n,\ n=0,1,2,\ Z_0=\gamma$.
\\  \rule[-1ex]{0ex}{3.5ex} 
$m_n^2$ & Complex mass, $\ m_n^2=M_n^2-i\Gamma_nM_n$.
\\  \rule[-1ex]{0ex}{3.5ex} 
$Br_n^f$ & Branching ratio of the $Z_n$ decay to $f\bar f,\
Br_n^f=\Gamma_n^f/\Gamma_n$. 
\\  \rule[-1ex]{0ex}{3.5ex} 
$e,g_1,g_2$ & Coupling strengths of the $\gamma,Z$ and $Z'$
to fermions, see \\ &
equation \req{eq31}, $e\approx 0.31,\ g_1=g=e/(2c_Ws_W)\approx 0.37$.  
\\  \rule[-1ex]{0ex}{3.5ex} 
$v_f,a_f\ (v'_f,a'_f)$&Vector and axial vector couplings of the $Z(Z')$ to the
fermion $f$.
\\  \rule[-1ex]{0ex}{3.5ex} 
$L_f,R_f\ (L'_f,R'_f)$ & Left and right handed couplings of the $Z(Z')$
to the fermion $f$.
\\  \rule[-1ex]{0ex}{3.5ex} 
$v_f(n),a_f(n)$ & Vector and axial vector couplings of the $Z_n$ to the fermion
$f$.
\\  \rule[-1ex]{0ex}{3.5ex} 
$L_f(n),R_f(n)$ & Left and right handed couplings of the $Z_n$ to the fermion
$f$.
\\  \rule[-1ex]{0ex}{3.5ex} 
$\epsilon_A,P_V^e,P_L^q,P_R^u,P_R^d$ & Sign--dependent coupling
combinations, see equation \req{coup5ee}.
\\  \rule[-2ex]{0ex}{4.5ex} 
$\epsilon_A,\gamma_L^l,\gamma_L^q,\tilde{U},\tilde{D}$ &
Sign--independent coupling combinations, see equation \req{ppparm}.
\\ 
\hline\end{tabular}\medskip
\end{center}
{\small\it  \begin{tab}\label{notap}
Main conventions and notation.
\end{tab}}
%
%

\begin{center}
\begin{tabular}{|l|l|}\hline 
Symbol\rule[-2ex]{0ex}{5ex} &Meaning\hspace{3cm} Chapter 2\\ 
\hline\rule[-1ex]{0ex}{4.5ex}
$s(s')$ & Centre-of-mass energy squared of the considered (sub)process.
\\ \rule[-1ex]{0ex}{3.5ex} 
$L$ & Integrated luminosity
\\ \rule[-1ex]{0ex}{3.5ex} 
$\lambda_1, \lambda_2, \lambda_3$ & Helicity combinations
\req{lambdadef}, \req{leftrdef} of the initial particles.
\\  \rule[-1ex]{0ex}{3.5ex} 
$c=\cos\theta$ & $\theta$ is the angle between the initial electron
and the final\\ 
&  fermion $f$ (or final $W^-$).
\\ \rule[-1ex]{0ex}{3.5ex} 
$\chi_n(s)$ & Propagator as defined in equation \req{eq28}.
\\  \rule[-1ex]{0ex}{3.5ex} 
$\hat\chi_n(s)$ & Propagator in $e^+e^-\rightarrow e^+e^-$ as defined
in equation \req{chiwdef}. 
\\  \rule[-1ex]{0ex}{3.5ex} 
$v^N_f,a^N_f$ & Normalized couplings of the $Z'$, as defined in
equation \req{normcoup}.
\\  \rule[-1ex]{0ex}{3.5ex} 
$v^M_f,a^M_f$ & Mixing dependent couplings \req{coupmdef} of the $Z'$.
\\  \rule[-1ex]{0ex}{3.5ex} 
$\delta_\gamma,\delta_Z$ & Shifts \req{deltagazdef} in the couplings
$g_{WW\gamma}$ and $g_{WWZ}$ due to a $Z'$.
\\  \rule[-1ex]{0ex}{3.5ex} 
$O,\Delta O,O_{SM}$ & Some observable (total or differential cross
section or\\  
& asymmetry), its experimental error and its SM prediction.
\\  \rule[-1ex]{0ex}{3.5ex} 
$r$ & Ratio of the systematic and statistical errors, see \req{rdef}
\\  \rule[-1ex]{0ex}{3.5ex} 
$\Delta o$ & $g_2$--depending error, of an observable, see equation
\req{onres}. 
\\  \rule[-1ex]{0ex}{3.5ex} 
$\Delta^{Z'} O$ & Shift of the observable $O$ from its SM prediction
due to a $Z'$.
\\  \rule[-1ex]{0ex}{3.5ex} 
$\sigma_T^f$ & Total cross section of fermion pair production in
different\\
              & processes, see equations \req{born} and \req{ppzpborn}.
\\  \rule[-1ex]{0ex}{3.5ex} 
$A^f_{FB},A_{LR},A^f_{pol}$ & Forward--backward, left--right and
polarization asymmetries\\
       &  for fermion pair production in different processes, 
\\  \rule[-1ex]{0ex}{3.5ex} 
          & see equations \req{observables1}, \req{obspp} and \req{epobs}.
\\  \rule[-1ex]{0ex}{3.5ex} 
$A^f_{LR,FB},A^f_{pol,FB},A^f_{LR,pol}$ & Combined asymmetries,
see equations \req{observables2} and \req{obspp}.
\\  \rule[-1ex]{0ex}{3.5ex} 
$R^{had},A_{LR}^{had},R_b,R_c$ & Ratios and asymmetries as defined in
equation \req{ohad}.
\\  \rule[-1ex]{0ex}{3.5ex} 
$M_{Z'}^{lim}$ & Largest $M_2$, which can be detected.
\\  \rule[-1ex]{0ex}{3.5ex} 
$f_m$ & Ratio of $M_2$ and $M_{Z'}^{lim}$, see equation \req{epemmea}.
\\  \rule[-1ex]{0ex}{3.5ex} 
$\theta_M^{lim}$ & Smallest mixing angle, which can be detected.
\\  \rule[-1.5ex]{0ex}{3.5ex} 
$g_2^{lim}$ & Smallest coupling strength, which can be detected.
\\ \rule[-1.5ex]{0ex}{3.5ex} 
$\delta_\gamma^{lim},\delta_Z^{lim}$ & Smallest shifts $\delta_\gamma$
and $\delta_Z$, which can be detected, see \req{deltgacon}.
\\ \rule[-1.5ex]{0ex}{3.5ex} 
$H_A^e(v)$ & Flux function \req{convol} describing QED initial state
corrections. 
\\ \rule[-1ex]{0ex}{3.5ex} 
$\Delta$ & Cut on the photon energy in units of the beam energy.
\\ \rule[-1ex]{0ex}{3.5ex} 
$\Delta^+,\Delta^-$ & Boundaries of the considered range of the photon energy.
\\ \rule[-2ex]{0ex}{4.5ex} 
$\Delta E_\gamma/E_\gamma$ & Photon energy resolution of the experiment.
\\ 
\hline\end{tabular}\medskip
\end{center}
{\small\it  \begin{tab}\label{notap2}
Continuation of conventions and notation.
\end{tab}}
%

\begin{table}[tbh]
\begin{center}
\begin{tabular}{|l|l|}\hline 
Symbol\rule[-2ex]{0ex}{5ex} &Meaning\hspace{3cm} Chapter 2\\ 
\hline\rule[-1ex]{0ex}{4.5ex}
$F_A(q^2)$ &Form factor taking into account the vacuum polarization of the\\
&photon, see equation \req{weakrepl}.
\\ \rule[-1ex]{0ex}{3.5ex} 
$x_f,y_f$ & Form factors of the $ZZ'$ mixing, see equation \req{formix}. 
\\ \rule[-1ex]{0ex}{3.5ex} 
$\kappa_f,\kappa_{ef},\rho_{ef}$ & Form factors taking into account the weak
corrections,\\ \rule[-2ex]{0ex}{4.5ex} 
& see section \ref{zpeeff22}. 
\\ \rule[-1ex]{0ex}{3.5ex} 
$\kappa_f^m,\kappa_{ef}^m,\rho_{ef}^m$ & Form factors \req{formmix} taking into
account the $ZZ'$ mixing.
\\ \rule[-1ex]{0ex}{3.5ex} 
$\kappa_f^M,\kappa_{ef}^M,\rho_{ef}^M$ & Form factors \req{mixrepl2} taking
into account the weak corrections,\\ \rule[-2ex]{0ex}{4.5ex} 
& and the $ZZ'$ mixing. 
\\ 
\hline 
\rule[-2ex]{0ex}{5ex} &\hspace{3cm} Chapters 3 and 4\\ 
\hline\rule[-1ex]{0ex}{4.5ex}
$r_{y1},A_{FBy1}$ & Rapidity ratios as defined in equation \req{raprat}.
\\  \rule[-1ex]{0ex}{3.5ex} 
$r_{llZ},r_{\nu\nu Z},r_{l\nu W}$ & Ratios from rare $Z'$ decays, 
see equation \req{rarrat}.
\\  \rule[-1ex]{0ex}{3.5ex} 
$R_{Z'V},V=Z,W,\gamma$ & Ratios from associated $Z'$ production, 
see equation \req{assrat}.
\\  \rule[-1ex]{0ex}{3.5ex} 
$q_f(x,Q^2)$ & Structure function defining the distribution of partons $q_f$
in the $p(\bar p)$.
\\ \rule[-1ex]{0ex}{3.5ex} 
$c_{Z'}$ & Model dependent constant defined in equation \req{fdef2}.
\\ \rule[-1ex]{0ex}{3.5ex} 
$N_{Z'}$ & Number of expected $Z'$ events, compare equation \req{fdef2}.
\\ \rule[-1ex]{0ex}{3.5ex} 
$f^q(r_z,Q^2)$ & Integrated product of two structure functions, see
equation \req{fdef}.
\\ \rule[-1ex]{0ex}{3.5ex} 
$f^u(r_z)$ & Approximation \req{fapprox2} of $f^q(r_z,Q^2)$ for large
$Q^2$ and $q=u$. 
\\ \rule[-1ex]{0ex}{3.5ex} 
$C_{ud}$ & Is defined below equation \req{fzdef}, $C_{ud}=f^u(r_Z)/f^d(r_z)$.
\\ \rule[-1ex]{0ex}{3.5ex} 
$x,y,Q^2$ & Kinematic variables \req{herakin} in $ep$ scattering.
\\ \rule[-1ex]{0ex}{3.5ex} 
$A_{XY}^{mn}$ & Asymmetries in \req{epobs} $ep$ scattering.
\\ \rule[-2ex]{0ex}{4.5ex} 
$Q_W$ & Weak charge \req{qwdef} measured in atomic parity violation.
\\ 
\hline\end{tabular}\medskip
\end{center}
{\small\it  \begin{tab}\label{notap3}
Continuation of conventions and notation.
\end{tab}}
\end{table}
\chapter{Available {\tt FORTRAN} programs for $Z'$ fits}
%
Several computer programs are available allowing either theoretical
investigations or direct fits to data.
All officially released programs for $Z'$ analyses known to the author
are collected in table \ref{programs}.
In the first column, the name of the program is listed. The second
column contains references related to the program. 
If there exists a program description, its reference is printed in
{\bf bold}.
References to original papers describing the underlying
physics of the program are printed in roman.
Examples of references, where the program was used in theoretical or
experimental $Z'$ studies, are printed in {\it italics}.
In the last column the location of the program is given.
It follows a short description of every program listed in the table.

\begin{description}
\item[{\tt ZCAMEL}]
The program describes the pair production of massless fermions in
$e^+e^-$ collisions including the full $O(\alpha)$ QED corrections and
the exponentiation of soft photons radiated from the initial state.
It is a fast stand-alone program designed for theoretical studies. 
Use {\tt ZEFIT} for fits to data. 
\item[{\tt ZEFIT}]
The program must be used together with {\tt ZFITTER}. 
It describes the fermion pair production in $e^+e^-$ collisions. 
{\tt ZEFIT} contains all additional $Z'$ contributions needed for fits
to data.
Special attention is paid to the simultaneous treatment of
electroweak corrections and $ZZ'$ mixing needed for direct fits to
data distributed around the $Z$ peak.
It is designed for fits to data above the $Z$ peak too.

\item[{\tt ZFITTER}] describes fermion pair production in $e^+e^-$
collisions.  
It contains all known SM corrections needed for fits to data. 
It is designed for SM studies. 
It is required by {\tt ZEFIT} and distributed together with this code.
The description of {\tt ZFITTER} is needed to work with {\tt ZEFIT}.
This is the reason why we list this code in table~\ref{programs}.
\item[{\tt GENTLE/4fan}]
The stand-alone program describes the production of four fermions in
$e^+e^-$ collisions.  
It contains all known SM corrections needed for fits to data. 
It is originally designed for SM studies. 
It allows a study of $Z'$ effects in $W$ pair production using the
branch of anomalous couplings, see appendix C of the program description.
An on--line description can found at http://www.ifh.de/theory/publist.html .
\item[{\tt PYTHIA}]
is a general-purpose event generator for a multitude of
processes in $e^+e^-,\ ep$ and $pp$ physics. 
The emphasis is on the detailed modeling of hadronic final states,
i.e. QCD parton showers, string fragmentation and secondary
decays. 
The electroweak description is normally restricted to improved
Born--level formulae.
It contains physics beyond the SM as supersymmetry, extra neutral
gauge bosons or leptoquarks.
Pythia was used to obtain the present $Z'$ limits at hadron colliders.
\end{description}
%
\begin{table}[tbh]
\begin{center}
\begin{tabular}{|l|l|l|}\hline 
Program\rule[-2ex]{0ex}{5ex} &Main References & Location \\ 
\hline
{\tt ZCAMEL}\rule[-1ex]{0ex}{3.5ex} 
& \cite{zet,zefb}, {\it \cite{zpsari}} &
{\footnotesize ftp://gluon.hep.physik.uni-muenchen.de/zcamel.f }
 \rule[-1ex]{0ex}{3.5ex} \\
{\tt ZEFIT} & \cite{zefit}, {\it \cite{L31961,zmixl3,zmixdelphi,lmu0296}} &
{\footnotesize http://www.ifh.de/\symbol{126}riemanns/uu/zefit5\_0.uu}
 \rule[-1ex]{0ex}{3.5ex} \\
{\tt ZFITTER} & {\bf \cite{zfitter}}, \cite{666,formbard,zfitterplus},
{\it \cite{zfitterexp}}& 
distributed together with {\tt ZEFIT}
 \rule[-1ex]{0ex}{3.5ex} \\
{\tt GENTLE/4fan} & {\bf \cite{gentle}},
\cite{wwtord,gentle_nunicc,npb477}, {\it \cite{l3mw}} &
{\footnotesize http://www.ifh.de/\symbol{126}biebel/gentle2.f}
\rule[-1ex]{0ex}{3.5ex} \\
{\tt PYTHIA} v.6.1 & {\bf \cite{pythiaman}}, \cite{pythiaphys},
{\it\cite{pythiaexp}} &
{\footnotesize http://www.thep.lu.se/tf2/staff/torbjorn}
 \rule[-1ex]{0ex}{3.5ex} \\
\hline
\end{tabular}\medskip
\end{center}
{\small\it  \begin{tab}\label{programs}
Officially released {\tt FORTRAN} programs designed for a $Z'$ search.
\end{tab}}
\end{table}
\end{appendix}

\begin{thebibliography}{99}
\bibitem{zfitterexp} The code {\tt ZFITTER} is used by all four LEP
collaborations.
\bibitem{pythiaexp} Pythia is used by many collaborations at $e^+e^-,\
ep$ and hadron colliders.
\end{thebibliography}
\end{document}